\documentclass[pre,twocolumn,preprintnumbers,amsmath,amssymb,floatfix,longbibliography,superscriptaddress]{revtex4-2}
\usepackage{graphicx}
\usepackage{epstopdf}
\usepackage{psfrag}
\usepackage{hyperref}
\usepackage{multirow}
\usepackage[sort&compress]{natbib}
\usepackage{amssymb}
\usepackage{amsmath}
\usepackage{amsthm}
\usepackage{bbold}
\usepackage{bbm}
\usepackage{enumerate}
\usepackage{textcomp}
\usepackage{verbatim}
\usepackage{soul}
\usepackage{float}
\usepackage{leftindex}

\usepackage{orcidlink}

\newcommand{\vare}{\varepsilon}

\newcommand{\rmi}{{\rm i}}

\DeclareMathOperator{\diag}{diag}

\newcommand{\bk}{{\boldsymbol{k}}}

\newcommand{\bd}[1]{{\boldsymbol{#1}}}
\newcommand{\beq}{\begin{equation}\begin{aligned}}
\newcommand{\eeq}{\end{aligned}\end{equation}}

\renewcommand{\vec}[1]{\boldsymbol{#1}}

\usepackage{xcolor}
\usepackage{cancel}

\definecolor{AC}{rgb}{1, 0.2, 0.7}
\definecolor{SD}{rgb}{1,0,1}

\begin{document}

\hypersetup{pdftitle={title}}
\title{Exact critical-temperature bounds for two-dimensional Ising models}

\author{Davidson Noby Joseph\,\orcidlink{0009-0008-2420-0951}}
\affiliation{Department of Physics, University of Alberta, Edmonton, Alberta, Canada}
\affiliation{Theoretical Physics Institute, University of Alberta, Edmonton, Alberta, Canada}

\author{Igor Boettcher\,\orcidlink{0000-0002-1634-4022}}
 \affiliation{Department of Physics, University of Alberta, Edmonton, Alberta, Canada}
 \affiliation{Theoretical Physics Institute, University of Alberta, Edmonton, Alberta, Canada}
 \affiliation{Quantum Horizons Alberta, University of Alberta, Edmonton, Alberta, Canada}

\begin{abstract}
We derive exact critical-temperature bounds for the classical ferromagnetic Ising model on two-dimensional periodic tessellations of the plane. For any such tessellation or lattice, the critical temperature is bounded from above by a universal number that is solely determined by the largest coordination number on the lattice. Crucially, these bounds are tight in some cases such as the Honeycomb, Square, and Triangular lattices. We prove the bounds using the Feynman--Kac--Ward formalism, confirm their validity for a selection of over two hundred lattices, and construct a two-dimensional lattice with 24-coordinated sites and high critical temperature.
\end{abstract}

\maketitle
\section{Introduction}
The occurrence of ordered phases like ferromagnetism or superconductivity below a critical temperature, $T_{\rm c}$, is a many-body phenomenon pivotal for technological applications \cite{ChaikinLubenskyBook}. To fully harness the ordered states, a sufficiently high $T_{\rm c}$ is typically required to avoid the need for sophisticated cryogenic environments. The ongoing hunt for superconducting materials with a transition at room temperature exemplifies this \cite{BednorzMueller,FLORESLIVAS20201,Pickett23}. Unfortunately, predicting $T_{\rm c}$ for a given system is genuinely hard, as it is a non-universal and non-perturbative quantity---unlike critical exponents, for instance, which can be estimated perturbatively \cite{ZinnJustinBook,herbutbook}. The lack of simple criteria where to look for materials with large $T_{\rm c}$ is a major obstacle to designing future solid-state technology.

In this work, we give such a simple criterion for the classical two-dimensional (2D) ferromagnetic Ising model \cite{McCoyWu,Baxter}, which played a crucial role in the advancement of many-body physics in the last 100 years \cite{Ising25,KramersWannier41,Onsager44,KaufmanOnsager49,Yang52,FisherHartwig68,Kadanoff69,Polyakov70,Wilson71,Polyakov84,Smirnov06,Chelkak14}. The model describes classical spin or magnetic variables, $s_i$, placed on the sites $i$ of a lattice, that take values $\pm1$ and are assigned the energy
\begin{align}
 \label{intro1} H = - J \sum_{\langle i,j\rangle} s_i s_j.
\end{align}
Here the sum is over nearest-neighboring sites on the lattice and $J>0$ is the ferromagnetic exchange energy. For any 2D periodic tessellation of the plane, the system features a second-order phase transition at a $T_{\rm c}$ that can be computed exactly with the Feynman--Kac--Ward--formalism. A recent computation  of $T_{\rm c}$ for all 1248 $k$-uniform 2D lattices with $k\leq 6$ (to be defined below) showed that none of these lattices has a $T_{\rm c}$ above the one of the Triangular lattice \cite{Portillo}. On the other hand, some Laves lattices are known to have a higher $T_{\rm c}$ \cite{Codello2010,IsingArch}. This raises two questions: What types of lattices have high $T_{\rm c}$, and are arbitrarily high $T_{\rm c}$ possible in principle?

We formulate our criterion in terms of an exact upper bound on $T_{\rm c}$ that is true for all 2D periodic tessellations of the plane. Written in terms of the temperature variable $t=\tanh(\beta J)$, with $\beta=1/T$ (we set Boltzmann's constant $k_{\rm B}=1$), it reads
\begin{align}
 \label{intro2} t_{\rm c} \geq t_{\rm c,min} = \tan\Bigl(\frac{\pi}{2q_{\rm max}}\Bigr).
\end{align}
Herein, $q_{\rm max}$ is the highest coordination number of any site on the lattice, defined as the number of its nearest neighbors. In terms of $T_{\rm c}$, the bound reads
\begin{align}
 \label{intro3} \frac{T_{\rm c}}{J} \leq \frac{T_{\rm c,max}}{J} = \frac{1}{\text{artanh}(t_{\rm c,min})} = \frac{2}{\log \Bigl(\frac{1+t_{\rm c,min}}{1-t_{\rm c,min}}\Bigr)},
\end{align}
with $\log(\dots)$ the natural logarithm. Crucially, these bounds are {\it tight}, i.e. they can be saturated for certain $q_{\rm max}$. The Honeycomb, Square, and Triangular lattices with $T_{\rm c}/J=1.519$, $2.269$, and $3.641$ saturate their bounds with $q_{\rm max}=3$, $4$, and $6$, respectively \cite{Onsager44,RevModPhys.17.50,HusimiSyozi50,Houtappel50,Wannier50,NewellMontroll53}. In contrast, the mean-field-theory bound $T_{\rm c}/J\leq q_{\max}$ determined by the largest eigenvalue of the lattice adjacency matrix \cite{LynnLee} is significantly higher. (Note that our bound asymptotically yields $T_{\rm c,max}/J\simeq (2/\pi)q_{\rm max}$ for large $q_{\rm max}$.) 

The outline of the paper is as follows: In Sec. \ref{Sec:KW}, we introduce the Feynman--Kac--Ward formalism together with the Kac-Ward matrix $W(\bk)$, which gives an exact expression for the thermodynamic free energy per site on any planar lattice. Next, in Sec. \ref{Sec:bounds}, we show how $W(\bd{0})$ is related to the critical temperature and derive an exact bound in terms of the largest root of polynomials $P_{q}(u)$ that depend on the coordination numbers $q$. These polynomials are defined through reference matrices $\Phi_0^{(q)}$, which we introduce in Sec. \ref{Sec:Ref}. Finally, in Secs. \ref{Sec:Nums} and \ref{Sec:Eng}, we numerically confirm our bound with a selection of lattices and discuss new avenues to construct both low- and high-$T_{\rm c}$ planar periodic lattices. Appendix \ref{AppPlots} contains plots of the Archimedean and their dual lattices, whereas the proofs of the exact bound are given in Apps. \ref{Pfkac3} and \ref{AppBound}. Details of the reference matrices $\Phi_0(q)$ and the polynomials $P_q(u)$ are collected in Apps. \ref{refPhi0} and \ref{polyq}. Explicit calculations of some Kac-Ward matrices are presented in Apps. \ref{W(k)}, with a more comprehensive collection in the Supplemental Material \cite{SI}.

\section{Kac--Ward matrices}
\label{Sec:KW}
The Kac--Ward matrices are an efficient way to compute the free energy of the Ising model on periodic tessellations of the plane \cite{PhysRev.88.1332,Feynman,Kardar,Codello2010,IsingArch}. For instance, Onsager's original solution of the square-lattice Ising model \cite{Onsager44} is widely regarded as a tour de force of mathematical physics \cite{Kardar}, whereas the Kac--Ward matrix \cite{PhysRev.88.1332} can be computed in a few steps as shown in App. \ref{AppSquare}. Periodic tessellations consist of the placement of a unit cell with $N_{\rm u}$ sites along a 2D Bravais lattice. Define the coordination number $q_i$ of a site $i$ as the number of edges that leave $i$ to connect it to neighboring sites, and denote the average coordination number of the tessellation by $\bar{q}$. 
The thermodynamic free energy per site, $f$, is given by
\begin{equation}\label{kac1}
 \begin{aligned}
 \beta f(t) ={}& -\log 2 +\frac{\bar{q}}{4} \log (1-t^2) \\
 &-\frac{1}{2N_{\rm u}}\int_{\vec{k}}\log \det (\mathbb{1}-tW(\vec{k})),
 \end{aligned}
\end{equation}
with $W(\vec{k})$ the Kac--Ward matrix, $\vec{k}=(k_1,k_2)\in \text{BZ}=[0,2\pi)^2$ the crystal momentum, and $\int_{\vec{k}}=\int_{\text{BZ}}\frac{d^2k}{(2\pi)^2}$ the integral over the Brillouin zone \cite{Kardar}.

To obtain the matrix $W(\vec{k})$, 
we label sites by Latin indices $i,j,\cdots$ and directed edges by Greek indices $\mu,\nu,\cdots$. Here, a directed edge $\mu$ from site $i$ to $j$ is denoted by $\mu=(i,j)$, with $\mu^{-1}=(j,i)$ the edge in reversed direction. We define the origin and terminus of $\mu$ as $\text{ori}(\mu)=i$ and $\text{ter}(\mu)=j$, respectively. Each directed edge $\mu$ encloses an orientation angle $\vartheta_\mu$ with a reference axis. We denote by $\vartheta_{\mu\nu}=[\vartheta_\nu-\vartheta_\mu]$ the change in orientation between two directed edges, with the brackets $[\dots]$ indicating that $\vartheta_{\mu\nu}$ is projected onto the interval $(-\pi,\pi]$.

We label the $N_{\rm u}$ sites of the unit cell by Latin indices $a,b,\dots \in\{1,\dots, N_{\rm u}\}$. Importantly, each site $a$ of the unit cell is a site $i$ of the full lattice, and there are $q_a$ directed edges $\mu=(a,j)$ originating from $a$, where $j$ runs over the neighbors of $a$. In total, there are $Q=\sum_{a=1}^{N_{\rm u}}q_a=\bar{q}N_{\rm u}$ directed edges originating from unit cell sites. (Here we used that $\bar{q} = \frac{1}{N_{\rm u}}\sum_{a=1}^{N_{\rm u}} q_a$.) For each directed edge $\mu=(a,j)$ originating from a unit cell site, we associate $\mu\leftrightarrow (a,b_j)$, where $b_j$ is the unit cell site related to $j$ by a translation on the Bravais lattice. The elements of the $Q \times Q$-dimensional Kac--Ward matrix $W(\vec{k})$ are then given by
\begin{align}
 \label{kac2} W_{\mu\nu}(\vec{k}) = \begin{cases} e^{\rmi \vartheta_{\mu\nu}/2}e^{-\rmi \vec{k}\cdot\hat{\vec{n}}_\mu} & \text{ter}(\mu)=\text{ori}(\nu),\ \mu\neq \nu^{-1} \\ 0 & \text{else} \end{cases},
\end{align}
where $e^{-\rmi \vec{k}\cdot\hat{\vec{n}}_\mu}$ is the Bloch phase picked up along the directed edge $\mu$. The condition $\mu\neq \nu^{-1}$ removes backtracking and the phase factor $e^{\rmi \vartheta_{\mu\nu}/2}$ implements Whitney's theorem, both included to avoid overcounting of loops in the Ising partition function \cite{Kardar}.

In this work, we show that the $Q\times Q$-dimensional Kac--Ward matrix $W(\vec{k})$ consists of blocks $W_{ab}(\vec{k})$ of size $q_a\times q_b$ given by a product of three matrices according to
\begin{align}
  \label{kac3} W_{ab}(\vec{k}) =E_{a;b}B_a(\vec{k})\Phi_{ab}.
\end{align}
Herein, the Bloch matrices $B_a(\vec{k})$ and edge-connectivity matrices $E_{a;b}$ are diagonal with entries
\begin{align}
 \label{kac4} (B_a(\vec{k}))_{\mu\mu} &= e^{-\rmi \vec{k}\cdot\vec{\hat{n}}_\mu},\\
 \label{kac5} (E_{a;b})_{\mu\mu} &= \begin{cases} 1 & \mu=(a,b) \\ 0 & \text{else} \end{cases}.
\end{align}
The matrices $E_{a;b}$ list the directed edges going from $a$ to $b$. To compute the matrix $\Phi_{ab}$, denote by $\vec{\theta}_a=(e^{\rmi \vartheta_1},\dots,e^{\rmi \vartheta_{q_a}})$ the $q_a$-component orientation vector whose entries are the phases $e^{\rmi \vartheta_\mu}$ of edges $\mu$ leaving site $a$. Then construct the rectangular $q_a\times q_b$ matrix $\phi_{ab} = \vec{\theta}_a^* \vec{\theta}_b^T$ and define
\begin{align}
 \label{kac7} (\Phi_{ab})_{\mu\nu} = \begin{cases} \sqrt{(\phi_{ab})_{\mu\nu}} & \text{if } (\phi_{ab})_{\mu\nu}\neq -1 \\ 0 & \text{if } (\phi_{ab})_{\mu\nu}=-1\end{cases}.
\end{align}
We choose the branch cut of the square root along the negative real axis. The condition $(\phi_{ab})_{\mu\nu}\neq -1$ eliminates backtracking and the square root yields the Whitney factor. We prove Eq. (\ref{kac3}) in App. \ref{Pfkac3} and explicitly compute $W(\vec{k})$ for four selected lattices in App. \ref{W(k)}. In the Supplemental Material (SM) \cite{SI}, we derive $W(\bk)$ for over twenty additional lattices.

\renewcommand{\arraystretch}{1.5}
\begin{table*}[t!]
\begin{tabular}{|c|c|c|c|}
\hline
 \ $q$ \ & \ $P_q(u)$ \ & \ $t_{\rm c,min}^{(q)}$ \ & \ $T_{\rm c,max}^{(q)}/J$ \ \\
\hline\hline
2 & $(u-1)^2$  & $1$ & $0$ \\
\hline
3 & $u(u-3)^2$  & $\frac{1}{\sqrt{3}}=0.57735$ & $\frac{2}{\log(2+\sqrt{3})} = 1.51865$ \\
\hline
4 & $(u^2-6u+1)^2$ & $\sqrt{2}-1=0.414214$ & $\frac{2}{\log(1+\sqrt{2})} = 2.26919$ \\
\hline
5 & $u(u^2-10u+5)^2$ & $\frac{1}{\sqrt{5+2\sqrt{5}}}=0.32492$ & $\frac{2}{\log(\sqrt{5}-1+\sqrt{5-2\sqrt{5}})} = 2.96615$ \\
\hline
6 & $(u^3-15u^2+15u-1)^2$ & $2-\sqrt{3} = 0.267949$ & $\frac{2}{\log \sqrt{3}} = 3.64096$ \\
\hline
7 & $u(u^3-21u^2+35u-7)^2$ & 0.228243   & 4.30412 \\
\hline
8 & $(u^4-28u^3+70u^2-28u+1)^2$ & $\frac{1}{1+\sqrt{2}+\sqrt{2(2+\sqrt{2})}} = 0.198912$ & $\frac{2}{\log(\sqrt{2}-1+\sqrt{4-2\sqrt{2}})} = 4.96032$ \\
\hline
9 & $u(u^4-36u^3+126u^2-84u+9)^2$  & 0.176327 & 5.61201 \\
\hline
10 & $(u^5-45u^4+210u^3-210u^2+45u-1)^2$ & \ $1+\sqrt{5}-\sqrt{5+2\sqrt{5}} =  0.158384$ \ & $\frac{2}{\log\sqrt{1+2/\sqrt{5}}}= 6.26060$ \\
\hline
11 & $u(u^5-55u^4+330u^3-462u^2+165u-11)^2$  & 0.143778 & 6.90696 \\
\hline
12 & \ $(u^6-66u^5+495u^4-924u^3+495u^2-66u+1)^2$ \ & $\sqrt{6}+\sqrt{2}-\sqrt{3}-2 = 0.131652$ & \ $\frac{2}{\log[2+\sqrt{6}-\sqrt{5+2\sqrt{6}}]} = 7.55167$ \ \\
\hline
\end{tabular}
    \caption{The polynomials $P_q(u)$ appear as factors of the characteristic polynomial of $W_0^\dagger W_0$ in Eq. (\ref{ex5}), where $q$ runs over the coordination numbers of the unit cell sites. The largest roots of $P_q(u)$ imply critical temperature bounds $t_{\rm c,min}^{(q)}$ and $T_{\rm c,max}^{(q)}$ for each $q$, listed here. For a given lattice, the upper bound $T_{\rm c,max}$ in Eq. (\ref{intro3}) is determined by the entry $q=q_{\rm max}$, with $q_{\rm max}$ the largest coordination number on the lattice. Explicit formulas for $P_q(u)$ are given in Eqs. (\ref{em31}).}
\label{TabPq}
\end{table*}
\renewcommand{\arraystretch}{1}

\section{Exact critical-temperature bounds}
\label{Sec:bounds}
For a given lattice, the critical value of $t$ is determined by the point where the argument of the logarithm in Eq. (\ref{kac1}) vanishes. This requires $\boldsymbol{k}=0$ and
\begin{align}
 \label{ex1} \mbox{det}(\mathbbm{1}-t_{\rm c}W_0)=0,
\end{align}
where we defined $W_0=W(\textbf{0})$. The matrix $W_0$ is not Hermitian, but has a unique eigenvalue $w_{\rm max}\in[1,\infty)$, and $t_{\rm c}=1/w_{\rm max}$ \cite{CDC}. Denote by $u_{\rm max}$ the largest eigenvalue of the Hermitian matrix $W_0^\dagger W_0$. Since $w_{\rm max}\leq \sqrt{u_{\rm max}}$, we conclude that
\begin{align}
 \label{ex2} t_{\rm c} \geq t_{\rm c,min} = \frac{1}{\sqrt{u_{\rm max}}}.
\end{align}
We show in the following that $u_{\rm max}$ is entirely determined by the value of $q_{\rm max}$. We emphasize the unusual nature of this finding: While $T_{\rm c}$ related to the largest eigenvalue of $W_0$ is a genuinely non-universal quantity that depends on the details of the underlying lattice, the bound through $u_{\rm max}$ as the largest eigenvalue of $W_0^\dagger W_0$ is universal and only depends on $q_{\rm max}$.

The complete proof of the universal bound and the computation of $u_{\rm max}$ is presented in App. \ref{AppBound}. Here we summarize the key steps. We use the explicit form in Eq.\ (\ref{kac3}) to write the blocks of $W_0$ as $(W_0)_{ab} = E_{a;b}\Phi_{ab}$. Using the definition of the matrices $E_{a;b}$, we find that $W_0^\dagger W_0$ has the block-diagonal form
\begin{align}
 W_0^\dagger W_0 = \text{diag}(\mathcal{B}_1, \mathcal{B}_2, \dots, \mathcal{B}_{N_{\rm u}})
\end{align}
with $q_a\times q_a$-dimensional blocks $\mathcal{B}_a$. The blocks of $W_0$ satisfy $\mathcal{B}_a = V_a^\dagger [\Phi_0^{(q_a)}]^2 V_a$, where $\Phi_0^{(q)}$ is an explicit reference matrix defined below that only depends on $q$, and $V_a$ is a unitary matrix that depends on the site $a$. Consequently, the characteristic polynomial of $W_0^\dagger W_0$ factorizes as
\begin{align}
  \label{ex5} \mbox{det}(u\mathbb{1} - W_0^\dagger W_0) = \prod_{a=1}^{N_{\rm u}} \mbox{det}(u \mathbb{1}-\mathcal{B}_a)= \prod_{a=1}^{N_{\rm u}}P_{q_a}(u),
\end{align}
where the reference determines
\begin{align}
 \label{ex5b} P_q(u) = \mbox{det}(u\mathbb{1} - [\Phi_0^{(q)}]^2).
\end{align}
We list some polynomials $P_q(u)$ and their roots in Tab.\ \ref{TabPq}. The largest eigenvalue $u_{\rm max}$ of $W_0^\dagger W_0$ is the largest root of Eq.\ (\ref{ex5}), which is fully determined by the largest value of $q_a$ (i.e. $q_{\rm max}$). The surprising, exact formula (\ref{ex5}) is the main result of this work. It is unexpected, because the right-hand side is independent of any details of the lattice except the coordination numbers.

\section{Reference matrices for each $q$}
\label{Sec:Ref}
We now construct the reference matrix $\Phi_0=\Phi_0^{(q)}$ for a site with coordination number $q$ and compute $u_{\rm max}$. Denote by $\omega= e^{2\pi \rmi /q}$ the $q$-th root of unity. 
For even or odd $q$, $\Phi_0$ is the $q\times q$ matrix with entries
\begin{align}
 \label{ex9} (\Phi_0)_{\mu\nu} &\stackrel{q\text{ even}}{=} \begin{cases} \sqrt{\omega^{(\nu-\mu)\ \text{mod}\ q}} & \text{if }\omega^{(\nu-\mu)\ \text{mod}\ q}\neq -1 \\ 0 & \text{else} \end{cases},\\
 \label{ex10} (\Phi_0)_{\mu\nu} &\stackrel{q\text{ odd}}{=} \begin{cases} \sqrt{-\omega^{(\nu-\mu)\ \text{mod}\ q}} & \text{if }\omega^{(\nu-\mu)\ \text{mod}\ q}\neq 1 \\ 0 & \text{else} \end{cases}.
\end{align}

It corresponds to a $q$-coordinated site with equal angles between all edges. The matrix $\Phi_0$ is Hermitian  and circulant. 
\begin{figure*}[t!]
\includegraphics[width=0.85\linewidth]{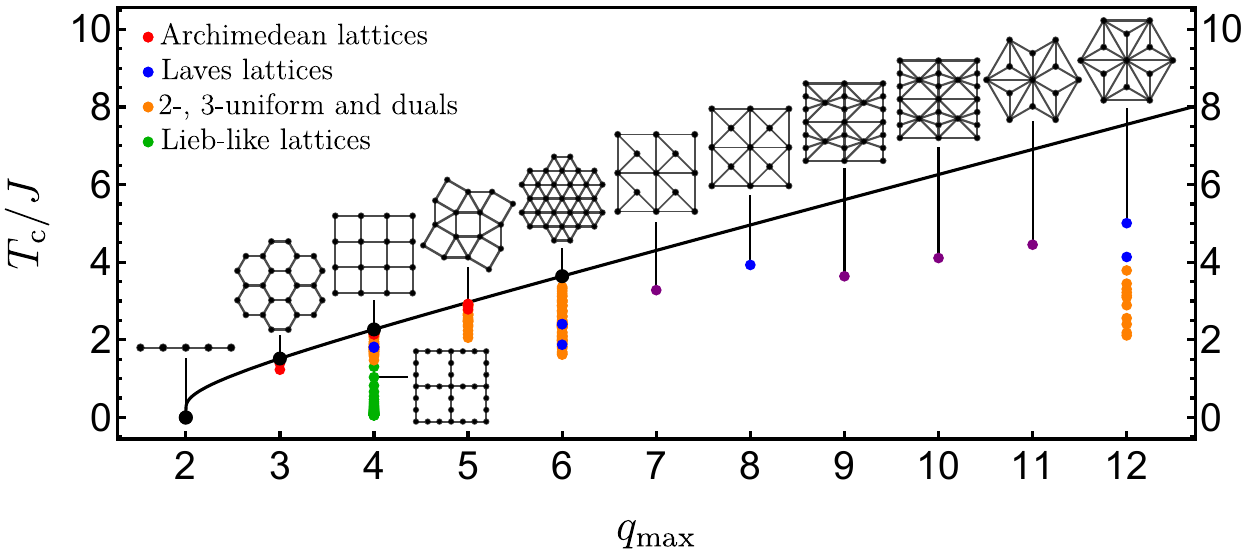}
\caption{Critical temperature of the ferromagnetic Ising model, $T_{\rm c}$, on various 2D periodic tessellations versus their maximal coordination number, $q_{\rm max}$. The solid line is the exact upper bound from Eq. (\ref{intro3}). For each $q_{\rm max}$, we show a subunit of the tessellation with the largest value of $T_{\rm c}/J$ among the lattices considered here. The Honeycomb, Square, and Triangular lattices saturate their bounds, while all other lattices considered here lie strictly below. The plot encompasses all $k$-uniform lattices with $k\leq 3$ and their dual lattices, with $k=1$ corresponding to the Archimedean and Laves lattices. Purple dots constitute a selection of lattices with $q_{\rm max}=7,9,10,11$. Lieb-like lattices are the sequence of lattices where each edge of a Square lattice is replaced by a chain of $2^n$ two-coordinated sites, with $0\leq n\leq 50$, yielding arbitrarily low $T_{\rm c}/J \sim 1/n$ as $n$ increases. The one-dimensional chain with $q_{\rm max}=2$ and $T_{\rm c}=0$ is included for completeness. The plot comprises data from 236 lattices.}
\label{FigTc}
\end{figure*}

Since $\Phi_0$ is Hermitian, we have $\Phi_0^\dagger \Phi_0 = \Phi_0^2$. In order to find the largest eigenvalue of $\Phi_0^\dagger \Phi_0$, it is thus sufficient to find the largest eigenvalue $\lambda_{\rm max}(q)$ of $\Phi_0$. Since $\Phi_0$ is circulant, its largest eigenvalue corresponds to the all-ones eigenvector $\vec{v}=\frac{1}{\sqrt{q}}(1,\dots,1)^T$ and we have $\lambda_{\rm max}(q)= \vec{v}^T \Phi_0 \vec{v} = \cot(\frac{\pi}{2q})$, as shown in App. \ref{Appmax}. Thus the largest root of the polynomial $P_q(u)$ is $\lambda_{\rm max}(q)^2$. Since $\lambda_{\rm max}(q)$ increases with $q$, the largest root of the product $\prod_a P_{q_a}(u)$ appearing in Eq.\ (\ref{ex5}) is determined by the largest coordination number $q_{\rm max}$. Hence, the largest eigenvalue of $W_0^\dagger W_0$ is $u_{\rm max}=\lambda_{\rm max}(q_{\rm max})^2$ and we arrive at
\begin{align}
 \label{ex12} t_{\rm c,min}= \frac{1}{\lambda_{\rm max}(q_{\rm max})} = \tan\Bigl(\frac{\pi}{2q_{\rm max}}\Bigr),
\end{align}
which together with Eq.\ (\ref{ex2}) gives the exact bound stated in Eq. (\ref{intro2}). The remaining eigenvalues of $\Phi_0$ are computed in App. \ref{polyq}, together with the explicit expressions
\begin{align}
 \nonumber P_q(u) &= (-1)^q\Bigl[\frac{1}{2}\Bigl((1+\sqrt{-u})^q+(-1+\sqrt{-u})^q\Bigr)\Bigr]^2\\
 \label{em31} &= \begin{cases} \Bigl(\sum_{k=0}^{q/2} \binom{q}{2k} (-u)^k\Bigr)^2 & q\ \text{even} \\ u\Bigl(\sum_{k=0}^{(q-1)/2} \binom{q}{2k+1} (-u)^k\Bigr)^2 & q\ \text{odd} \end{cases},
\end{align}
derived from Chebyshev polynomials.

\section{Numerical confirmation}
\label{Sec:Nums}
We confirm the validity of the exact bound (\ref{intro3}) for all $k$-uniform lattices with $k\leq 3$ and their dual (``d'') lattices. A $k$-uniform lattice is a periodic 2D tessellation that uses regular polygons such that there are $k$ distinct types of vertices \cite{GalebachWebpage,Sanchez1,Sanchez2,SanchezWebpage,GomJau-Hogg}. The eleven Archimedean lattices shown in Fig. (\ref{FigAL}) correspond to $k=1$ \cite{Grunbaum2016-rs,1fvj-91v6}, whereas there are 20 and 61 lattices for $k=2$ and $k=3$, respectively. All these lattices have $q_{\rm max}\leq 6$. 
To construct their dual lattices, a vertex is placed into each face of the original lattice, with edges connecting neighboring faces. The resulting dual lattices use $k$ types of tiles and have $q_{\rm max}^{(\rm d)}\leq 12$. The duals of the Archimedean lattices shown in Fig. (\ref{FigLL}) are called Laves lattices \cite{Grunbaum2016-rs}. 
The critical value $t_{\rm c}^{(\rm d)}$ of the dual lattice is obtained from the original $t_{\rm c}$ via \cite{Codello2010,IsingArch}
\begin{align}
 \label{num1} t_{\rm c}^{(\rm d)} = \frac{1-t_{\rm c}}{1+t_{\rm c}}.
\end{align}
We determine $t_{\rm c}$ from Eq. (\ref{kac3}) for $k=1$ in App. \ref{W(k)} and the SM \cite{SI}, and from the data of Ref. \cite{Portillo} for $k=2,3$.  We plot the resulting values of $T_{\rm c}/J$ vs. $q_{\rm max}$ in Fig. \ref{FigTc}. The full data set is given in App. \ref{Tcdata}.

We observe from the data presented in Fig. \ref{FigTc} that only the Honeycomb, Square, and Triangular lattices saturate their bounds. Within the set of lattices considered, the highest $T_{\rm c}$ for $q_{\rm max}=5$ is achieved for the Archimedean SrCuBO (or Shastry--Sutherland \cite{SRIRAMSHASTRY19811069}) lattice  with $T_{\rm c}/T_{\rm c,max}=98.7\%$. The largest values for $q_{\rm max}=8$ and $q_{\rm max}=12$ are obtained for the Laves-CaVO and Laves-Star (or Asanoha) lattices, with $T_{\rm c}/J=3.93100$ and $T_{\rm c}/J=5.00795$, respectively, corresponding to $79.2\%$ and $66.3\%$ of $T_{\rm c,max}$. Within a family of lattices with the same value of $q_{\rm max}$, we observe that those with a smaller value of the average coordination number $\bar{q}$ tend to have lower values of $T_{\rm c}$, but for two lattices with the same $q_{\rm max}$, $\bar{q}_1 < \bar{q}_2 \nRightarrow T_{\rm c,1}<T_{\rm c,2}$, see Tab. V in App. \ref{Tcdata}. We also observe that for the lattices with $q_{\rm max}>6$ considered here, the discrepancy between $T_{\rm c,max}$ and $T_{\rm c}$ increases with $q_{\rm max}$, suggesting that it is more difficult to come close to the bound for large $q_{\rm max}$.

Restricting to $k$-uniform lattices and their duals does, of course, not exhaust the possibilities of periodic tessellations of the plane, and not all values of $q_{\rm max}$ are attained. To have a representative lattice for each $q_{\rm max}\leq 12$, we constructed lattices with $q_{\rm max}=7,9,10,11$ that are included in Fig. \ref{FigTc} and compute their $T_{\rm c}$ the SM \cite{SI}. While the values of $T_{\rm c}/J$ of this selection are fairly high (comparable to or higher than the Triangular lattice), we emphasize that these representative lattices were not constructed to be special or optimal.

\section{$T_{\rm c}$ engineering}
\label{Sec:Eng}
In view of our analytical and numerical findings, it is natural to ask whether it is possible to come arbitrarily close to $T_{\rm c,max}$ for each $q_{\rm max}$, or whether we can make $T_{\rm c}$ arbitrarily small. The answer to the simpler, second question is in the affirmative. For every $q_{\rm max}$, we can create lattices with arbitrarily small $T_{\rm c}$. For this, take any lattice with a given value of $q_{\rm max}$ and replace \textit{every} edge by a one-dimensional (1D) chain of $m$ 2-coordinated sites, which does not change the value of $q_{\rm max}$. This replaces $t\to t^{m+1}$ in the free energy per site $f(t)$, as shown in App. \ref{Lieb}. Accordingly, $t_{\rm c}\to t_{\rm c}^{1/(m+1)}\to 1$ as $m\to \infty$ and $T_{\rm c}\to 0$ logarithmically in $m$. The reason for $T_{\rm c}$ decreasing in this process is that the lattice becomes effectively more 1D. 
As an example, consider the Square lattice and add $m$ sites along each edge. For $m=1$, this gives the Lieb lattice with $t_{\rm c} = (t_{\rm c,\square})^{1/2}=(\sqrt{2}-1)^{1/2}$ and $T_{\rm c}/J=1.30841$, with $t_{\rm c,\square}$ the Square lattice result. As $m\to \infty$, we get a sequence of Lieb-like lattices with $q_{\rm max}=4$ and $T_{\rm c}/J \sim 1/\log(m)$, discussed in Ref. \cite{PhysRevB.96.174407} as planar lacunary lattices, shown in Fig. \ref{FigTc}.

\begin{figure}[t!]
\includegraphics[width=0.9\linewidth]{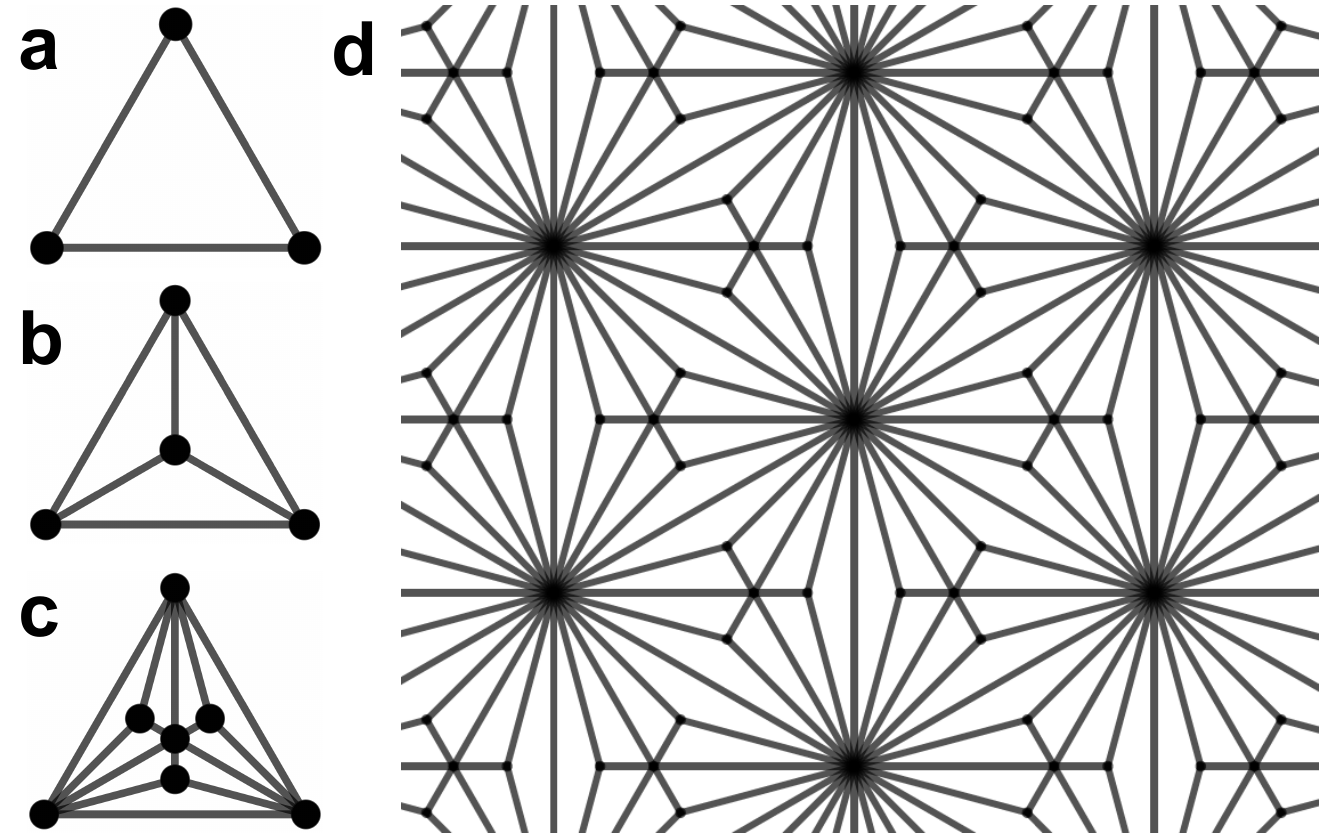}
\caption{Starting from the Triangular lattice (one tile shown in \textbf{a}, $q_{\rm max}=6$), the Laves-Star lattice (\textbf{b}, $q_{\rm max}=12$) is obtained by adding a 3-coordinated site into each triangle. Iterating this step once again (\textbf{c}), we obtain the Compass-Rose lattice (\textbf{d}) with $q_{\rm max}=24$, the name inspired by the 24 classical wind directions. Its critical temperature $T_{\rm c}/J=6.49190$ is computed in the SM \cite{SI}. All three lattices have a high $T_{\rm c}$ for their value of $q_{\rm max}$. They also have average coordination number $\bar{q}=6$, which is the maximal possible value for 2D periodic tessellations, as we derive in App. \ref{Tcdata}.}
\label{Fig2}
\end{figure}

To construct lattices with large $T_{\rm c}$, we require at least one vertex with a large coordination number $q_{\rm max}$ in the unit cell. However, to embed this vertex into a proper periodic tessellation of the plane, we need to introduce additional vertices of lesser coordination into the unit cell. We empirically find that having vertices of small coordination reduces $T_{\rm c}$. Consequently, making a lattice with a large $T_{\rm c}$ appears to be a trade-off between having a large $q_{\rm max}$ while, at the same time, introducing as little low-coordinated sites as possible. To exemplify designing lattices with large $q_{\rm max}$ as a route to large $T_{\rm c}$, we construct in Fig. \ref{Fig2} a periodic 2D lattice with $q_{\rm max}=24$ and comparably high $T_{\rm c}/J=6.49190$. The high value of $T_{\rm c}/J$ in this case can be explained by using the star-triangle relation \cite{Baxter} to construct lattices with increasingly large critical temperature from the triangular lattice, as will be discussed in a broader setting elsewhere \cite{InPrep}. Note also that the subunit of a partitioned triangle is called an Apollonian network \cite{apollo1,apollo2,apollo3}.

Another way to find lattices with large $T_{\rm c}$ is to start from a lattice with low $T_{\rm c}$ (and $t_{\rm c}$ close to unity) and construct the dual lattice. Because of Eq. (\ref{num1}), the dual lattice will have a small $t_{\rm c}$ and high  $T_{\rm c}$. For example, the Star lattice has low $T_{\rm c}$, but the dual Laves-Star has a large $T_{\rm c}$. In this construction, however, the original lattice cannot contain any 2-coordinated sites, since the corresponding dual lattice would then have vertices that are connected by multiple edges. Such dual lattices are not admissible lattices for Ising models, where each edge $(i,j)$ corresponds to a spin-spin-exchange $s_is_j$. Hence our construction of Lieb-like lattices with arbitrarily small  $T_{\rm c}$ does not produce admissible dual lattices with $T_{\rm c}$ arbitrarily close to the bound $T_{\rm c,max}$. 

\section{Conclusion and Outlook}
In this work, we used Kac--Ward matrices to derive an exact bound on the critical temperature $T_{\rm c}$ of the 2D Ising model on any periodic tessellation of the plane. The bound only depends on the maximal coordination number $q_{\rm max}$ of the lattice. We confirmed the bound numerically, constructed examples with large $q_{\rm max}$, and showed that $T_{\rm c}$ is not bounded from below. The list of lattices we considered is by no means exhaustive. Instead, our findings motivate the graph theoretic challenge of finding the lattice with the largest $T_{\rm c}$ for any $q_{\rm max}$. It is striking that the Honeycomb, Square, and Triangular lattices, which use only one type of vertex and one type of tile (i.e. they are $\{p,q\}$ lattices \cite{PhysRevB.105.125118}), are the only lattices found to saturate their bounds. Since there are no further $\{p,q\}$ lattices in the Euclidean plane \cite{PhysRevB.105.125118}, this could suggest that lattices with other $q_{\rm max}$ are strictly below the bound. In future work, these observations should be tested by exhausting even larger sets of lattices or by using exact theorems from graph theory.

The study of non-Euclidean tessellations of the hyperbolic plane recently garnered great interest because of their experimental realization \cite{kollar2019hyperbolic,zhang2022observation,Lenggenhager2022,huang2024hyperbolic,fleury2024,Xu25}. The ferromagnetic Ising model on hyperbolic $\{p,q\}$ lattices features an order-disorder-transition at a critical temperature $T_{\rm c}$ \textit{above} the bound for $q_{\rm max}=q$ \cite{Rietman92,Krcmar08,Breuckmann20,Nussinov25}. This is not in contradiction to our exact bounds, which were derived for periodic tessellations of the Euclidean plane. Furthermore, the critical temperature on the Bethe or $\{\infty,q\}$ lattice in the zero-field limit is $T_{\rm c}/J= 2/\log(\frac{q}{q-2})$ \cite{Baxter}, which is also above the bound for $q_{\rm max}=q$, and exceeds the hyperbolic-lattice values. Extending our exact bound to non-Euclidean tessellations will shed light on the interplay between graph geometry and phase transitions in many-body systems in the future.

Some of the lattices with large coordination number considered in this work, such as the ones displayed in Fig. \ref{Fig2}, may appear artificial and fine-tuned in the sense that they have equal exchange coupling $J$ on all bonds, while at the same time having varying bond distances in the plane. However, in recent years an increasing number of experimental platforms has emerged where the vertex location (and thus bond distance) is immaterial and the couplings in planar networks are tunable. Such platforms include topoelectrical circuits \cite{zhang2022observation,Lenggenhager2022}, topological photonics \cite{huang2024hyperbolic}, or circuit quantum electrodynamics \cite{kollar2019hyperbolic}, as exemplified in the realization of aforementioned hyperbolic lattices, which also appear fine-tuned at first sight. In addition, spatial photonic Ising machines have shown the potential to encode Ising Hamiltonians with arbitrary couplings and connectivities \cite{PhysRevLett.134.203801}. Furthermore, the rapid progress in metal-organic frameworks promises another avenue for realizing designer lattices \cite{mof1,mof2,mof3}. At last, with the growing interest in similar classical statistical systems such as Hopfield models \cite{Hopfield} in the context of machine learning theory, a purely graph-theoretic perspective on questions of optimal connections in networks is emerging as well. Summarizing, we believe that the study of extremality conditions in planar lattices and networks offers many promising future research directions.

\section{Acknowledgments}
The authors thank David Feder, Joseph Maciejko, Frank Marsiglio, and Connor M. Walsh for insightful comments. They acknowledge funding from the Natural Sciences and Engineering Research Council of Canada (NSERC) Discovery Grants RGPIN-2021-02534 and DGECR2021-00043.

\begin{appendix}

\section{Plots of Archimedean and Laves lattices}\label{AppPlots}
\begin{figure*}[h!]
\includegraphics[width=0.9\linewidth]{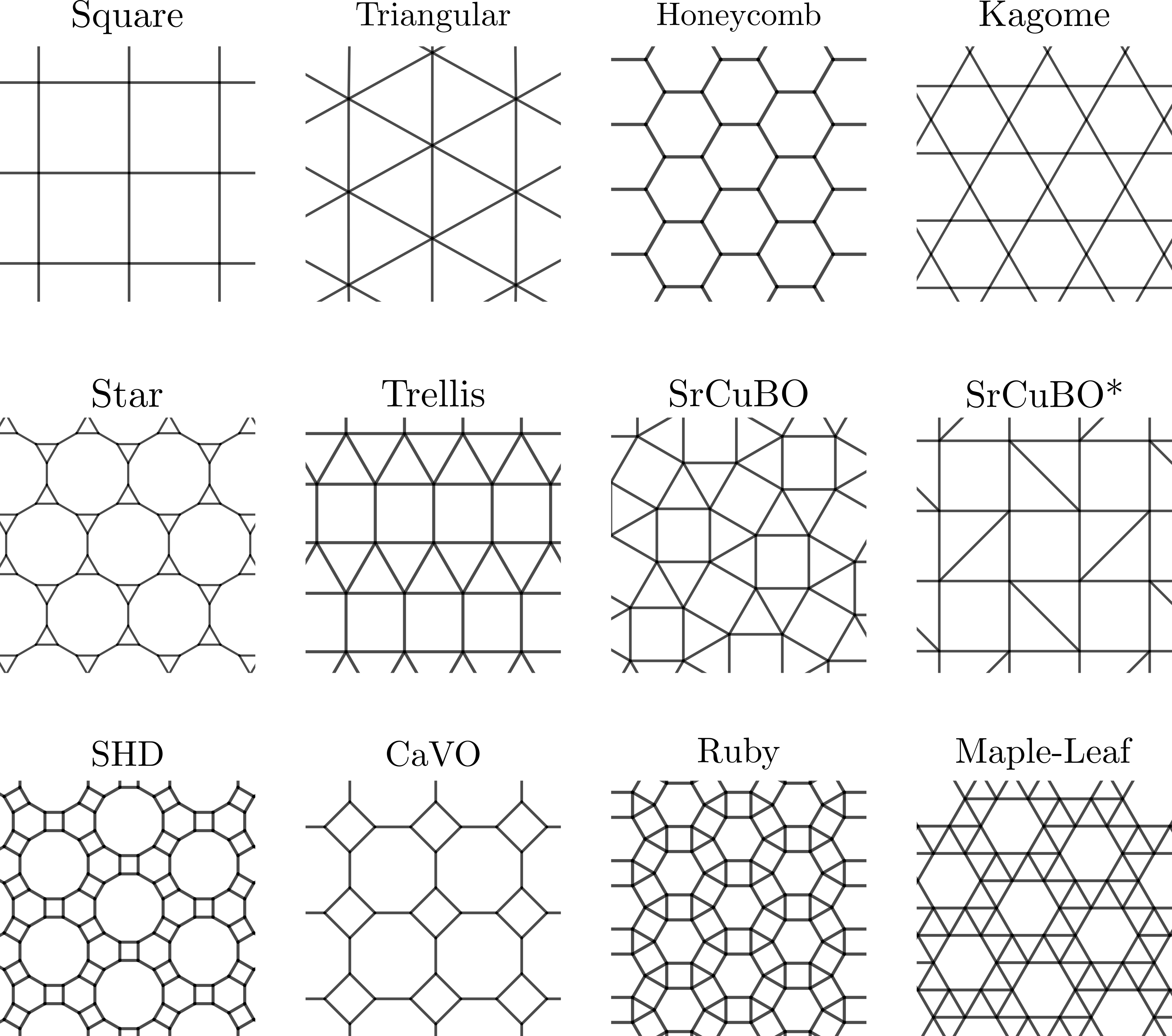}
\caption{The eleven Archimedean lattices are the 1-uniform periodic tessellations of the plane by regular polygons. Names follow common
conventions in the literature. Note that the Ruby lattice is also referred to as Bounce lattice. The Honeycomb lattice is topologically equivalent to the Brickwork lattice discussed below. Similarly, the SrCuBO lattice is topologically equivalent to the lattice SrCuBO$^*$ shown here, which appears in the quantum spin model by Shastry and Sutherland \cite{SRIRAMSHASTRY19811069}.}
\label{FigAL}
\end{figure*}

\begin{figure*}[h!]
\includegraphics[width=0.9\linewidth]{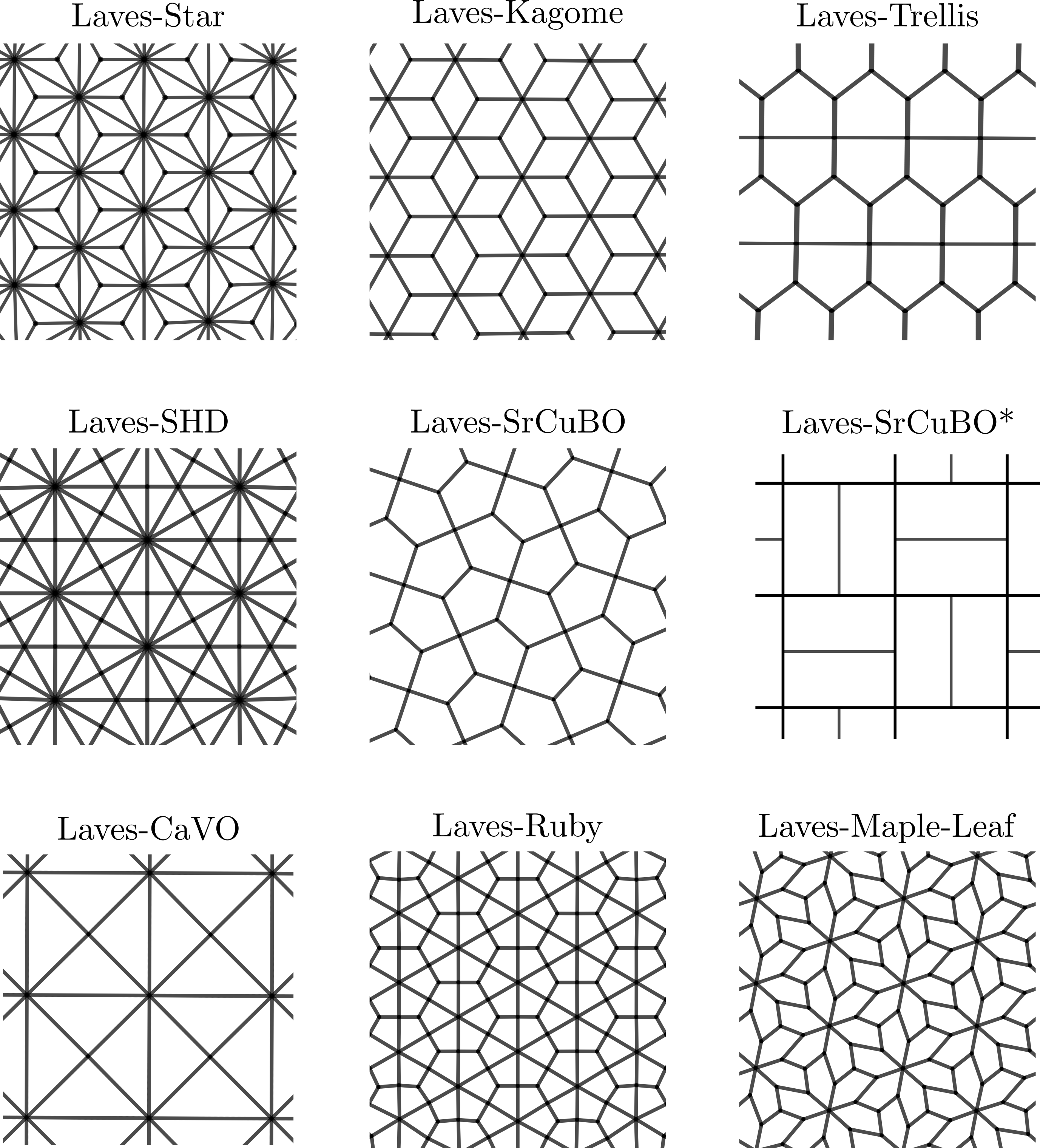}
\caption{The Laves lattices are the dual lattices to the Archimedean lattices. They use only one type of tile, since the Archimedean lattices only use one type of vertex. We exclude the dual lattices of the Honeycomb, Square, and Triangular lattices, since they are again Archimedean lattices. Since the Star (and SHD) and CaVO lattices employ regular 12- and 8-gons, it follows that the Laves-Star (and Laves-SHD) and Laves-CaVO lattices have vertices with coordination numbers 12 and 8, respectively. The Laves-SrCuBO lattice is topologically equivalent to the Laves-SrCuBO$^*$ lattices shown here. The Laves-Star lattice is also known as Asanoha or hemp-leaf lattice.}
\label{FigLL}
\end{figure*}

The eleven Archimedean lattices and their dual Laves lattices are shown in Figs. \ref{FigAL} and \ref{FigLL}.

\section{Proof of $W(\vec{k})$ formula}\label{Pfkac3}
To prove that Eq. (\ref{kac3}) is equivalent to Eq. (\ref{kac2}), we first choose two directed edges $\mu_a=(a,c)$ and $\nu_b=(b,d)$ whose origins lie within the unit cell and show that 
\begin{align}
 \label{em1}  W_{\mu_a\nu_b}(\bk) &=  (E_{a;b}B_{a}(\bk)\Phi_{ab})_{\mu_a\nu_b}\\
 \label{em2} &= (E_{a;b}B_{a}(\bk))_{\mu_a\mu_a}(\Phi_{ab})_{\mu_a\nu_b}.
\end{align} 
Indeed, since $E_{a;b}$ and $B_{a}(\bk)$ are diagonal when expressed in terms of the directed edges that leave site $a$, we have
\begin{equation}
    \label{em3}
    (E_{a;b}B_{a}(\bk))_{\mu_a\mu_a}=\begin{cases}
         e^{-\rmi \bk \cdot \hat{\vec{n}}_{\mu_a}} & {\rm ter}(\mu_a)=b\\
         0 & \text{else} 
    \end{cases}.
\end{equation}
Importantly, this expression vanishes when there is no directed edge from $a$ to $b$. Further using Eq. (\ref{kac7}) we arrive at
\begin{align}
\label{em4} &W_{\mu_a\nu_b}(\bk) =\begin{cases} e^{-\rmi \bk \cdot \hat{\vec{n}}_{\mu_a}} (\Phi_{ab})_{\mu_a\nu_b} & {\rm ter}(\mu_a)=b \\ 0 & \text{else} \end{cases}\\
  \label{em5}  &=\begin{cases}
        e^{-\rmi \bk \cdot \hat{\vec{n}}_{\mu_a}}e^{\rmi \vartheta_{\mu_a\nu_b}/2} & {\rm ter}(\mu_a)=b, e^{\rmi\vartheta_{\mu_a\nu_b}}\neq -1\\ 
        0 & \text{else}
    \end{cases}.
\end{align}
Finally note that $\text{ter}(\mu_a)=b$ is equivalently to $\rm{ter}(\mu_a)={\rm ori}(\nu_b)$, and $e^{\rmi\vartheta_{\mu_a\nu_b}}\neq -1$ is equivalent to $\mu_a \neq \nu_b^{-1}$. The matrix element from Eq. (\ref{em1}) is thus given by
\begin{align}
 \nonumber  &W(\bk)_{\mu_a\nu_b}(\bk)\\
  \label{em6}   &=\begin{cases}
        e^{-\rmi \bk \cdot \hat{\vec{n}}_{\mu_a}}e^{\rmi \vartheta_{\mu_a\nu_b}/2} &  \text{ter}(\mu_a)=\text{ori}(\nu_b),\ \mu_a\neq \nu_b^{-1}\\
        0 & \text{else}
    \end{cases}.
\end{align}
Since the choice of $\mu_a$ and $\nu_b$ was arbitrary, we recover Eq. (\ref{kac2}) for $W_{\mu\nu}(\vec{k})$.\\

\section{Proof of $T_{\rm c}$ bound}\label{AppBound} 
In the following, we prove formula (\ref{ex5}) for the factorization of the characteristic polynomial of $W_0^\dagger W_0$. This implies the $T_{\rm c}$ bound (\ref{ex12}) due to the properties of the reference matrices $\Phi_0^{(q_a)}$, as outlined in the main text. While the proof is rather technical, we invite the reader to follow the explicit steps for the exemplary Kagome or CaVO lattices covered in App. (\ref{W(k)}), and several more in the SM \cite{SI}.

Define $M_k^{(a)} = \Phi_{ka}^\dagger E_{k;a}$ so that
\begin{align}
 \label{em7} W_0^\dagger &= \begin{pmatrix} M_1^{(1)} & M_2^{(1)} & \dots & M_{N_{\rm u}}^{(1)} \\ M_1^{(2)} & M_2^{(2)} & \dots & M_{N_{\rm u}}^{(2)} \\ \vdots  & \vdots & \ddots & \vdots \\ M_1^{(N_{\rm u})} & M_2^{(N_{\rm u})} & \dots & M_{N_{\rm u}}^{(N_{\rm u})}\end{pmatrix}.
\end{align}
The matrices $E_{a;b}$ have the properties
\begin{align}
 \label{kac6} \sum_{a=1}^{N_{\rm u}} E_{a;b} &= \mathbb{1},\ E_{a;b}E_{a;c} =\begin{cases} E_{a;b} & b=c \\ 0 & b\neq c \end{cases},
\end{align}
which follow from their definition. Using the latter property, we find that $W_0^\dagger W_0$ has the block-diagonal form
\begin{align}
 \label{em8} W_0^\dagger W_0 = \text{diag}(\mathcal{B}_1, \mathcal{B}_2, \dots, \mathcal{B}_{N_{\rm u}}),
\end{align}
 with $q_a\times q_a$-dimensional blocks $\mathcal{B}_a= \sum_{k=1}^{N_{\rm u}}M_k^{(a)} [M_k^{(a)}]^\dagger$. We now show that each block $\mathcal{B}_a$ is unitarily equivalent to the reference $(\Phi_0^{(q_a)})^2$, where $q_a$ is the coordination number of unit cell site $a$. That is, we show that
\begin{align}
 \label{em9} \mathcal{B}_a = V_a^\dagger [\Phi_0^{(q_a)}]^2 V_a,
\end{align} 
where $V_a$ are unitary matrices, and thus
\begin{align}
 \label{em10} \mbox{det}(u\mathbb{1} - \mathcal{B}_a) = \mbox{det}[u\mathbb{1} - (\Phi_0^{(q_a)})^2] = P_{q_a}(u).
\end{align}
Together with the block-diagonal form of $W_0^\dagger W_0$ this implies Eq. (\ref{ex5}) and hence the $T_{\rm c}$ bound.

To show Eq. (\ref{em9}), consider a site $a$ with coordination number $q_a$. Since there are only $q_a$ edges arriving at $a$, only $q_a$ diagonal entries of the set of matrices $\{E_{k;a}\}_{k=1,\dots N_{\rm u}}$ are nonzero. This implies that, for fixed $a$, the set of matrices $\{M_k^{(a)}\}_{k=1,\dots,N_{\rm u}}$ has only $q_a$ nonvanishing columns, which we label $\vec{Y}_\mu^{(a)}$ with $\mu=1,\dots,q_a$. We fix the $\mu$-ordering of the $\vec{Y}_\mu^{(a)}$ such that the $\mu$th entry of column $\vec{Y}_\mu^{(a)}$ is zero. We then have
\begin{align}
 \label{em12} \mathcal{B}_a= \sum_{k=1}^{N_{\rm u}}M_k^{(a)} [M_k^{(a)}]^\dagger = \sum_{\mu=1}^{q_a} \vec{Y}_\mu^{(a)}[\vec{Y}_\mu^{(a)}]^\dagger.
\end{align}
Here we used the fact that if a matrix $A$ has columns $\vec{A}_i$ according to $A=(\vec{A}_1\cdots\vec{A}_n)$, then $AA^\dagger =\sum_{i=1}^n \vec{A}_i \vec{A}_i^\dagger$.

Denote the columns of the reference matrix $\Phi_0^{(q_a)}$ by $\vec{X}_\mu=\vec{X}_\mu^{(q_a)}$ according to $\Phi_0^{(q_a)} = (\vec{X}_1\dots\vec{X}_{q_a})$. They satisfy
\begin{align}
 X_\mu(\alpha) = \begin{cases} \sqrt{-\theta_0(\mu)\theta_0^*(\alpha)} & \alpha\neq \mu \\ 0 & \alpha = \mu\end{cases},
\end{align}
where $\alpha=1,\dots,q$ labels the vector components. Examples for low $q_a$ are given in the App. \ref{refPhi0}. We now show explicitly that for each $a$ we have
\begin{align}
 \label{em14} \vec{Y}_\mu^{(a)} = (V_a)_{\mu\mu} V_a^\dagger \vec{X}_\mu,
\end{align}
where $V_a$ are unitary matrices and (as throughout the work) no sum over $\mu$ is implied. To construct $V_a$, write
\begin{align}
 \label{em15} \vec{\theta}_a = U_a \vec{\theta}_0^{(q_a)}
\end{align}
for each site $a$, with $U_a$ a diagonal unitary matrix that rotates $\vec{\theta}_a$ into the reference vector for $q_a$ given by
\begin{align}
 \label{em16} \vec{\theta}_0=\vec{\theta}_0^{(q_a)} = (1,\omega,\dots,\omega^{q_a-1})^T
\end{align}
with $\omega = e^{2\pi \rmi/q_a}$. Then define $V_a=\sqrt{U_a}$, which is again unitary and diagonal. From the explicit expressions for $\Phi_{ka}$ and $E_{k;a}$ in $M_k^{(a)} = \Phi_{ka}^\dagger E_{k;a}$, it follows that the nonvanishing columns of the set of matrices $\{M_k^{(a)}\}_k$ for fixed $a$ are of the form
\begin{align}
 \label{em17} Y_\mu^{(a)}(\alpha) &= \begin{cases} \sqrt{\theta_{k'}(\mu')\theta_a^*(\alpha)} & \text{if}\ \theta_{k'}(\mu') \neq -\theta_a(\alpha) \\ 0 & \text{if}\ \theta_{k'}(\mu') = -\theta_a(\alpha)\end{cases},
\end{align}
where the complex phase $\theta_{k'}(\mu')$ is fixed by $\mu$ and $a$ and identical for the vector $\vec{Y}_\mu^{(a)}$. (The index $\alpha=1,\dots,q_a$ labels the components of $\vec{Y}_\mu^{(a)}$.) The condition for the appearance of 0 implies that $\theta_{k'}(\mu')=-\theta_a(\alpha_0)$ for some $\alpha_0$. Hence we have
\begin{align}
 \label{em18} Y_\mu^{(a)}(\alpha) &= \begin{cases} \sqrt{-\theta_a(\alpha_0)\theta_a^*(\alpha)} & \text{if}\ \theta_a(\alpha_0) \neq \theta_a(\alpha) \\ 0 & \text{if}\ \theta_a(\alpha_0) = \theta_a(\alpha)\end{cases}\\
 \label{em19} &= \begin{cases} \sqrt{-\theta_a(\alpha_0)\theta_a^*(\alpha)} & \text{if}\  \alpha \neq \alpha_0 \\ 0 & \text{if}\ \alpha=\alpha_0 \end{cases}\\
 \label{em20} &= \begin{cases} \sqrt{-\theta_a(\mu)\theta_a^*(\alpha)} & \text{if}\  \alpha \neq \mu \\ 0 & \text{if}\ \alpha= \mu \end{cases}.
\end{align} 
In the last line, we used that the zero-entry occurs in the $\mu$th component of $\vec{Y}_\mu^{(a)}$. Now use $\vec{\theta}_a=U_a\vec{\theta}_0$ to write
\begin{align}
 \nonumber Y_\mu^{(a)}(\alpha) &=\begin{cases} \sqrt{-(U_a)_{\mu\mu}(U_a^*)_{\alpha\alpha}\theta_0(\mu)\theta_0^*(\alpha)} & \text{if}\  \alpha \neq \mu \\ 0 & \text{if}\ \alpha = \mu \end{cases}\\
 \nonumber  &=(V_a)_{\mu\mu} (V_a^*)_{\alpha\alpha}\begin{cases} \sqrt{-\theta_0(\mu)\theta_0^*(\alpha)} & \text{if}\  \alpha \neq \mu \\ 0 & \text{if}\ \alpha = \mu \end{cases}\\
 \label{em23}   &=(V_a)_{\mu\mu} (V_a^*)_{\alpha\alpha}X_\mu(\alpha).
\end{align}
This is Eq (\ref{em14}). Since $(V_a)_{\mu\mu}$ is just a complex phase, we arrive at
\begin{align}
 \label{em25} \mathcal B_a = V_a^\dagger \Bigl(\underbrace{\sum_{\mu=1}^q \vec{X}_\mu \vec{X}_\mu^\dagger}_{\Phi_0^2}\Bigr) V_a = V_a^\dagger \Phi_0^2 V_a
\end{align} 
with $\Phi_0=\Phi_0^{(q_a)}$, which is Eq. (\ref{em9}).\\

\section{Lieb-like lattices}\label{Lieb} 
Consider for concreteness a Square lattice (``$\square$''). The partition function is proportional to
\begin{align}
 \label{ll1} Z_{\square}(t) \propto \sum_{\{s_i\}} \prod_{\langle i,j\rangle\in\square} (1+t s_i s_j).
\end{align}
Nonzero contributions to the sum result from closed loops on the lattice, with each vertex and its spin variable appearing an even number of times. Now replace \textit{every} edge by a chain of $m$ two-coordinated sites. For a given edge $(i,j)$, we have $t s_i s_j \to t^{m+1} s_i s_{k_1}\cdots s_{k_m}s_j$ in Eq. (\ref{ll1}). Nonzero contributions now result from the same closed loops on the (Bravais) Square lattice and each newly inserted site appearing twice, with $s_{k}^2=1$. As a result, the counting of loops is unchanged, but the effective value of $t$ is replaced according to
\begin{align}
 \label{ll2} Z_{\text{Lieb-like}}(t)  \propto \sum_{\{s_i\}} \prod_{\langle i,j\rangle\in \square} (1+t^{m+1} s_i s_j).
\end{align}
Thus the critical value $t_{\rm c}$ of the Lieb-like lattice is obtained from $\mbox{det}(\mathbb{1}-t_{\rm c}^{m+1}W_{0,\square})=0$, with $W_{0,\square}$ the Square-lattice matrix, yielding $t_{\rm c}=(t_{\rm c,\square})^{1/(m+1)}$. Clearly, the same reasoning applies to any starting lattice as long as \textit{every} edge is replaced by a chain of $m$ two-coordinated sites.

\section{Explicit expressions for reference matrices $\Phi_0$}
\label{refPhi0}

In this section, we collect the expressions for the reference matrices $\Phi_0=\Phi_0^{(q)}$ for a selection of small, relevant values of $q$ and determine the eigenvalues of $\Phi_0^2$ and the polynomials
\begin{align}
 P_q(u) = \mbox{det}(u\mathbb{1}_u - [\Phi_0^{(q)}]^2).
 \end{align}
Throughout this section, we define $\omega=e^{2\pi\rmi /q}$.

The equation for the $q\times q$ matrix $\Phi_0$ is given by
\begin{align}
 (\Phi_0^{(q)})_{\mu\nu} &\stackrel{q\text{ even}}{=} \begin{cases} \sqrt{\omega^{(\nu-\mu)\ \text{mod}\ q}} & \text{if }\omega^{(\nu-\mu)\ \text{mod}\ q}\neq -1 \\ 0 & \text{else} \end{cases},\\
 (\Phi_0^{(q)})_{\mu\nu} &\stackrel{q\text{ odd}}{=} \begin{cases} \sqrt{-\omega^{(\nu-\mu)\ \text{mod}\ q}} & \text{if }\omega^{(\nu-\mu)\ \text{mod}\ q}\neq 1 \\ 0 & \text{else} \end{cases}.
\end{align}
For the definition of the complex square root with the branch cut along the negative real axis, note that if $z= r e^{\rmi \phi}$ is a complex number with $r>0$ and arbitrary $\phi\in \mathbb{R}$, define by $[\phi]$ the projection of $\phi$ onto the interval $(-\pi,\pi]$. Then $\sqrt{z}$ is defined as
\begin{align}
 \sqrt{z} := \sqrt{r}\ e^{\rmi [\phi]/2}.
\end{align}
For an alternative definition of $\Phi_0$, we construct the reference vector $\vec{\theta}_0=(1,\omega,\dots,\omega^{q-1})^T$ and define
\begin{align}
\phi_0^{(q)}=(-1)^q\vec{\theta}_0^*\vec{\theta}_0^T.
\end{align}
From $\phi_0^{(q)}$, we obtain the matrix $\Phi_0^{(q)}$ by replacing any entries $\phi^{(q)}_{0,\mu\nu}=-1$ with $0$, and the other entries $\phi^{(q)}_{0,\mu\nu}\neq -1$ with $\sqrt{\phi^{(q)}_{0,\mu\nu}}$. Both ways of computing $\Phi_0^{(q)}$ yield the same result. \\

\noindent \underline{$q=3$.} For $q=3$, we have $\omega = e^{2\pi\rmi/3}$ and the corresponding reference vector is
\begin{align}
\vec{\theta}_0^{(q=3)}=(1,\omega,\omega^2)^T=(1,e^{2\pi\rmi/3},e^{-2\pi\rmi/3})^T.
\end{align}
We have
\begin{align}
 \phi_0^{(q=3)} &= -\vec{\theta}_0^*\vec{\theta}_0^T = \begin{pmatrix} -1 & -\omega & -\omega^2 \\ -\omega^2 & -1 & -\omega \\ -\omega & -\omega^2 & -1\end{pmatrix}\\
  &= \begin{pmatrix} -1 & e^{-\rmi\pi/3} & e^{\rmi\pi/3} \\ e^{\rmi\pi/3} & -1 & e^{-\rmi\pi/3} \\ e^{-\rmi\pi/3} & -e^{\rmi\pi/3} & -1\end{pmatrix}
\end{align}
and
\begin{align}
 \nonumber  \Phi_0^{(q=3)} &= \begin{pmatrix} 0 & \sqrt{-\omega} & \sqrt{-\omega^2} \\ \sqrt{-\omega^2} & 0 & \sqrt{-\omega} \\ \sqrt{-\omega} & \sqrt{-\omega^2} & 0 \end{pmatrix} \\ \label{ref3b}& = \begin{pmatrix} 0 & e^{-\rmi \pi/6} & e^{\rmi \pi/6} \\ e^{\rmi \pi/6} & 0 & e^{-\rmi \pi/6} \\ e^{-\rmi\pi/6} & e^{\rmi \pi/6} & 0 \end{pmatrix}.
\end{align}
The matrix $\Phi_0^\dagger=\Phi_0$ has $q=3$ nonvanishing columns according to
\begin{align}
\Phi_0^{(q=3)}=(\vec{X}_1\ \vec{X}_2\ \vec{X}_3),
\end{align}
which are given by
 \begin{equation}
\begin{aligned}
 \vec{X}_1^{(q=3)} &= \begin{pmatrix} 0 \\ e^{\rmi \pi/6} \\ e^{-\rmi \pi/6}\end{pmatrix},\ \vec{X}_2^{(q=3)} = \begin{pmatrix} e^{-\rmi \pi/6} \\ 0 \\ e^{\rmi \pi/6} \end{pmatrix},\\
 \vec{X}_3^{(q=3)} &= \begin{pmatrix} e^{\rmi \pi/6} \\ e^{-\rmi \pi/6} \\ 0 \end{pmatrix}.
\end{aligned}
\end{equation}
They satisfy
\begin{align}
 X_\mu(\alpha) = \begin{cases} \sqrt{-\theta_0(\mu)\theta_0^*(\alpha)} & \alpha\neq \mu \\ 0 & \alpha = \mu\end{cases}
\end{align}
and
\begin{align}
 \Phi_0^2 = \Phi_0\Phi_0^\dagger =\sum_{\mu=1}^3 \vec{X}_\mu\vec{X}_\mu^\dagger.
\end{align}
We have
\begin{align}
 P_3(u) &=\mbox{det}(u \mathbb{1}_3 - \Phi_0^2)= u(u-3)^2.
\end{align}\\

\noindent \underline{$q=4$.} For $q=4$, we have $\omega = \rmi$ and the reference vector is given by
\begin{align}
 \label{ref4c} \vec{\theta}_0^{(q=4)}=(1,\omega,\omega^2,\omega^3)^T=(1,\rmi,-1,-\rmi)^T.
\end{align}
Thus we have
\begin{equation}\label{ref4e}
\begin{aligned}
  \phi_0^{(q=4)} &= \vec{\theta}_0^*\vec{\theta}_0^T = \begin{pmatrix} 1 & \omega & \omega^2 & \omega^3 \\ \omega^3 & 1 & \omega &\omega^2 \\ \omega^2 & \omega^3 & 1 & \omega \\ \omega & \omega^2 & \omega^3 & 1 \end{pmatrix} \\
  &= \begin{pmatrix} 1 & \rmi & -1 & -\rmi \\ -\rmi & 1 & \rmi & -1 \\ -1 & -\rmi & 1 & \rmi \\ \rmi & -1 & -\rmi & 1 \end{pmatrix}
\end{aligned}
\end{equation}
and
\begin{equation}\label{ref4b}
\begin{aligned}
  \Phi_0^{(q=4)} &= \begin{pmatrix} 1 & \sqrt{\omega} & 0 & \sqrt{\omega^3} \\ \sqrt{\omega^3} & 1 & \sqrt{\omega} & 0 \\ 0 & \sqrt{\omega^3} & 1 & \sqrt{\omega} \\ \sqrt{\omega} & 0 & \sqrt{\omega^3} & 1 \end{pmatrix} \\
 &=\begin{pmatrix} 1 & e^{\rmi \pi/4} & 0 & e^{-\rmi \pi/4} \\ e^{-\rmi \pi/4} & 1 & e^{\rmi \pi/4} & 0 \\ 0 & e^{-\rmi \pi/4} & 1 & e^{\rmi \pi/4} \\ e^{\rmi \pi/4} & 0 & e^{-\rmi \pi/4} & 1 \end{pmatrix}.
\end{aligned}
\end{equation}
The matrix $\Phi_0^\dagger=\Phi_0$ contains $q=4$ nonzero columns vectors according to
\begin{align}
 \Phi_0^{(q=4)} = (\vec{X}_3\ \vec{X}_4\ \vec{X}_1\ \vec{X}_2),
\end{align}
which are given by
\begin{equation}\label{ref4g}
\begin{aligned}
   \vec{X}_1^{(q=4)} &= \begin{pmatrix}  0  \\  e^{\rmi \pi/4}  \\  1\\  e^{-\rmi \pi/4} \end{pmatrix},\ \vec{X}_2^{(q=4)} = \begin{pmatrix}  e^{-\rmi \pi/4} \\  0 \\  e^{\rmi \pi/4} \\  1 \end{pmatrix},\\
 \vec{X}_3^{(q=4)} &= \begin{pmatrix} 1  \\ e^{-\rmi \pi/4}  \\ 0  \\ e^{\rmi \pi/4} \end{pmatrix},\ \vec{X}_4^{(q=4)} = \begin{pmatrix} e^{\rmi \pi/4} \\  1  \\  e^{-\rmi \pi/4} \\  0  \end{pmatrix}.
\end{aligned}
\end{equation}
The vectors are ordered such that the $\mu$-component of $\vec{X}_\mu$ is 0. We have
\begin{align}
 X_\mu(\alpha) = \begin{cases} \sqrt{-\theta_0(\mu)\theta_0^*(\alpha)} & \alpha\neq \mu \\ 0 & \alpha = \mu\end{cases}
\end{align}
and
\begin{align}
 \label{ref4h} \Phi_0^2 = \Phi_0\Phi_0^\dagger = \sum_{\mu=1}^4 \vec{X}_\mu \vec{X}_\mu^\dagger.
\end{align}
We find that
\begin{align}
 \label{ref4f} P_4(u) &=\mbox{det}(u \mathbb{1}_4 - \Phi_0^2)= (u^2-6u+1)^2.
\end{align} \\

\noindent \underline{$q=5$.} 
For $q=5$, we have $\omega = e^{2\pi\rmi/5}$ and the reference vector is
\begin{align}
\nonumber \vec{\theta}_0^{(q=5)} &=(1,\omega,\omega^2,\omega^3,\omega^4)^T \\&=(1,e^{2\pi\rmi/5},e^{4\pi\rmi/5},e^{-4\pi\rmi/5},e^{-2\pi\rmi/5})^T.
\end{align}
From this we find
\begin{align}
 \nonumber &\phi_0^{(q=5)} = -\vec{\theta}_0^*\vec{\theta}_0^T = \begin{pmatrix} -1 & -\omega & -\omega^2 & -\omega^3 & -\omega^4  \\ -\omega^4 &-1 & -\omega & -\omega^2 & -\omega^3 \\ -\omega^3 & -\omega^4 &-1 & -\omega & -\omega^2 \\ -\omega^2 & -\omega^3 & -\omega^4 &-1 & -\omega \\ -\omega & -\omega^2 & -\omega^3 & -\omega^4 &-1   \end{pmatrix} \\
 &=\begin{pmatrix} -1 & e^{-3\pi\rmi/5} & e^{-\rmi\pi/5} & e^{\rmi\pi/5} & e^{3\pi\rmi/5}  \\ e^{3\pi\rmi/5} &-1 & e^{-3\pi\rmi/5} & e^{-\rmi\pi/5} & e^{\rmi\pi/5} \\ e^{\rmi\pi/5} & e^{3\pi\rmi/5} &-1 & e^{-3\pi\rmi/5} & e^{-\rmi\pi/5} \\ e^{-\rmi\pi/5} & e^{\rmi\pi/5} & e^{3\pi\rmi/5} &-1 & e^{-3\pi\rmi/5} \\ e^{-3\pi\rmi/5} & e^{-\rmi\pi/5} & e^{\rmi\pi/5} & e^{3\pi\rmi/5} &-1   \end{pmatrix}
\end{align}
and
\begin{align}
 \nonumber &\Phi_0^{(q=5)} = \begin{pmatrix} 0 & \sqrt{-\omega} & \sqrt{-\omega^2} & \sqrt{-\omega^3} & \sqrt{-\omega^4} \\ \sqrt{-\omega^4} & 0 & \sqrt{-\omega} & \sqrt{-\omega^2} & \sqrt{-\omega^3} \\  \sqrt{-\omega^3} & \sqrt{-\omega^4} & 0 & \sqrt{-\omega} & \sqrt{-\omega^2} \\ \sqrt{-\omega^2} & \sqrt{-\omega^3} & \sqrt{-\omega^4} & 0 & \sqrt{-\omega} &  \\ \sqrt{-\omega} & \sqrt{-\omega^2} & \sqrt{-\omega^3} & \sqrt{-\omega^4} & 0 \end{pmatrix} \\
 &= \begin{pmatrix} 0 & e^{-3\pi\rmi/10} & e^{-\pi\rmi/10} & e^{\rmi \pi/10} & e^{3\pi\rmi/10} \\ e^{3\pi\rmi/10} & 0 & e^{-3\pi\rmi/10} & e^{-\pi\rmi/10} & e^{\rmi \pi/10} \\  e^{\rmi \pi/10} & e^{3\pi\rmi/10} & 0 & e^{-3\pi\rmi/10} & e^{-\pi\rmi/10} \\ e^{-\pi\rmi/10} & e^{\rmi \pi/10} & e^{3\pi\rmi/10} & 0 & e^{-3\pi\rmi/10} &  \\ e^{-3\pi\rmi/10} & e^{-\pi\rmi/10} & e^{\rmi \pi/10} & e^{3\pi\rmi/10} & 0 \end{pmatrix}.
\end{align}
The matrix $\Phi_0^\dagger=\Phi_0$ has $q=5$ nonvanishing columns according to
\begin{align}
 \Phi_0^{(q=5)} = (\vec{X}_1\ \vec{X}_2\ \vec{X}_3\ \vec{X}_4\ \vec{X}_5),
\end{align}
which are given by
\begin{align}
 \vec{X}_1^{(q=5)} &= (0,e^{3\pi\rmi/10},e^{\pi\rmi/10},e^{-\pi\rmi/10},e^{-3\pi\rmi/10})^T,\\
 \vec{X}_2^{(q=5)} &= (e^{-3\pi\rmi/10},0,e^{3\pi\rmi/10},e^{\pi\rmi/10},e^{-\pi\rmi/10})^T,\\
 \vec{X}_3^{(q=5)} &= (e^{-\pi\rmi/10},e^{-3\pi\rmi/10},0,e^{3\pi\rmi/10},e^{\pi\rmi/10})^T,\\
 \vec{X}_4^{(q=5)} &= (e^{\pi\rmi/10},e^{-\pi\rmi/10},e^{-3\pi\rmi/10},0,e^{3\pi\rmi/10})^T,\\
 \vec{X}_5^{(q=5)} &= (e^{3\pi\rmi/10},e^{\pi\rmi/10},e^{-\pi\rmi/10},e^{-3\pi\rmi/10},0)^T.
\end{align}
They satisfy
\begin{align}
 X_\mu(\alpha) = \begin{cases} \sqrt{-\theta_0(\mu)\theta_0^*(\alpha)} & \alpha\neq \mu \\ 0 & \alpha = \mu\end{cases}
\end{align}
and
\begin{align}
 \Phi_0^2 = \Phi_0\Phi_0^\dagger =\sum_{\mu=1}^5 \vec{X}_\mu\vec{X}_\mu^\dagger.
\end{align}
We have
\begin{align}
 P_5(u) &=\mbox{det}(u \mathbb{1}_5 - \Phi_0^2)= u(u^2-10u+5)^2.
\end{align} \\

\noindent \underline{$q=6$.} 
For $q=6$, we have $\omega = e^{\rmi \pi/3}$ and the reference vector is
\begin{equation}
\begin{aligned}
\vec{\theta}_0^{(q=6)} & =(1,\omega,\dots,\omega^5)^T\\&  =(1,e^{\rmi \pi/3},e^{2\pi\rmi/3},-1,e^{4\pi\rmi/3},e^{5\pi \rmi/3})^T .
 \end{aligned}
 \end{equation}
We have
\begin{align} 
\nonumber &\Phi_0^{(q=6)} = \begin{pmatrix} 1 & \sqrt{\omega} & \omega & 0 & \sqrt{\omega^4} & \sqrt{\omega^5} \\ \sqrt{\omega^5} & 1 & \sqrt{\omega} & \omega & 0 & \sqrt{\omega^4} \\ \sqrt{\omega^4} & \sqrt{\omega^5} & 1 & \sqrt{\omega} & \omega & 0 \\ 0 & \sqrt{\omega^4} & \sqrt{\omega^5} & 1 & \sqrt{\omega} & \omega \\ \omega & 0 & \sqrt{\omega^4} & \sqrt{\omega^5} & 1 & \sqrt{\omega} \\ \sqrt{\omega} & \omega & 0 & \sqrt{\omega^4} & \sqrt{\omega^5} & 1 \end{pmatrix} \\
 &= \begin{pmatrix} 1 & e^{\rmi \pi/6} & e^{\rmi \pi/3} & 0 & e^{-\rmi \pi/3} & e^{-\rmi \pi/6} \\ e^{-\rmi \pi/6} & 1 & e^{\rmi \pi/6} & e^{\rmi \pi/3} & 0 & e^{-\rmi \pi/3} \\ e^{-\rmi \pi/3} & e^{-\rmi \pi/6} & 1 & e^{\rmi \pi/6} & e^{\rmi \pi/3} & 0 \\ 0 & e^{-\rmi \pi/3} & e^{-\rmi \pi/6} & 1 & e^{\rmi \pi/6} & e^{\rmi \pi/3} \\ e^{\rmi \pi/3} & 0 & e^{-\rmi \pi/3} & e^{-\rmi \pi/6} & 1 & e^{\rmi \pi/6} \\ e^{\rmi \pi/6} & e^{\rmi \pi/3} & 0 & e^{-\rmi \pi/3} & e^{-\rmi \pi/6} & 1 \end{pmatrix}. 
\end{align}
The matrix $\Phi_0^\dagger=\Phi_0$ has $q=6$ nonvanishing columns according to
\begin{align}
 \Phi_0^{(q=6)} = (\vec{X}_4\ \vec{X}_5\ \vec{X}_6\ \vec{X}_1\ \vec{X}_2\ \vec{X}_3),
\end{align}
which are given by
\begin{align}
 \vec{X}_1^{(q=6)} &= (0, e^{\rmi \pi/3}, e^{\rmi \pi/6}, 1, e^{-\rmi \pi/6}, e^{-\rmi \pi/3})^T,\\
 \vec{X}_2^{(q=6)} &= (e^{-\rmi \pi/3}, 0, e^{\rmi \pi/3}, e^{\rmi \pi/6}, 1, e^{-\rmi \pi/6})^T,\\
 \vec{X}_3^{(q=6)} &= (e^{-\rmi \pi/6},e^{-\rmi \pi/3}, 0, e^{\rmi \pi/3}, e^{\rmi \pi/6}, 1)^T,\\
 \vec{X}_4^{(q=6)} &= (1, e^{-\rmi \pi/6}, e^{-\rmi \pi/3}, 0, e^{\rmi \pi/3}, e^{\rmi \pi/6})^T,\\
 \vec{X}_5^{(q=6)} &= (e^{\rmi \pi/6}, 1, e^{-\rmi \pi/6}, e^{-\rmi \pi/3}, 0, e^{\rmi \pi/3})^T,\\
 \vec{X}_6^{(q=6)} &= (e^{\rmi \pi/3}, e^{\rmi \pi/6}, 1, e^{-\rmi \pi/6}, e^{-\rmi \pi/3}, 0)^T.
\end{align}
The vectors are ordered such that the $\mu$-component of $\vec{X}_\mu$ is 0. They satisfy
\begin{align}
 X_\mu(\alpha) = \begin{cases} \sqrt{-\theta_0(\mu)\theta_0^*(\alpha)} & \alpha\neq \mu \\ 0 & \alpha = \mu\end{cases}
\end{align}
and
\begin{align}
 \Phi_0^2 = \Phi_0\Phi_0^\dagger =\sum_{\mu=1}^6 \vec{X}_\mu\vec{X}_\mu^\dagger.
\end{align}
We have
\begin{align}
 P_6(u) &=\mbox{det}(u \mathbb{1}_6 - \Phi_0^2)= (u^3-15u^2+15u-1)^2.
\end{align} \\

\noindent \underline{$q=8$.}
For $q=8$ we have $\omega = e^{\rmi \pi/4}$. The corresponding reference vector is
\begin{align}
 \nonumber \vec{\theta}_0^{(q=8)} &= (1,\omega,\dots,\omega^7)^T \\&=(1,e^{\rmi\pi/4},\rmi,e^{3\pi\rmi/4},-1,e^{-3\pi\rmi/4},-\rmi,e^{-\rmi\pi/4})^T.
\end{align}
We have
\begin{align}
 &\Phi_0= \begin{pmatrix} 1 & \sqrt{\omega} & \sqrt{\omega^2} & \sqrt{\omega^3} & 0 & \sqrt{\omega^5} & \sqrt{\omega^6} & \sqrt{\omega^7} \\ \sqrt{\omega^7}  & 1 & \sqrt{\omega} & \sqrt{\omega^2} & \sqrt{\omega^3} & 0 & \sqrt{\omega^5} & \sqrt{\omega^6} \\ \sqrt{\omega^6} &  \sqrt{\omega^7}  & 1 & \sqrt{\omega} & \sqrt{\omega^2} & \sqrt{\omega^3} & 0 & \sqrt{\omega^5} \\  \sqrt{\omega^5} & \sqrt{\omega^6} &  \sqrt{\omega^7}  & 1 & \sqrt{\omega} & \sqrt{\omega^2} & \sqrt{\omega^3} & 0 \\ 0 & \sqrt{\omega^5} & \sqrt{\omega^6} &  \sqrt{\omega^7}  & 1 & \sqrt{\omega} & \sqrt{\omega^2} & \sqrt{\omega^3} \\ \sqrt{\omega^3} & 0 & \sqrt{\omega^5} & \sqrt{\omega^6} &  \sqrt{\omega^7}  & 1 & \sqrt{\omega} & \sqrt{\omega^2} \\ \sqrt{\omega^2} & \sqrt{\omega^3} & 0 & \sqrt{\omega^5} & \sqrt{\omega^6} &  \sqrt{\omega^7}  & 1 & \sqrt{\omega}  \\ \sqrt{\omega} & \sqrt{\omega^2} & \sqrt{\omega^3} & 0 & \sqrt{\omega^5} & \sqrt{\omega^6} &  \sqrt{\omega^7}  & 1   \end{pmatrix}.
\end{align}
The matrix $\Phi_0^\dagger=\Phi_0$ contains $q=8$ nonzero column vectors according to
\begin{align}
 \Phi_0^{(q=8)} = (\vec{X}_5\ \vec{X}_6\ \vec{X}_7\ \vec{X}_8\ \vec{X}_1\ \vec{X}_2\ \ \vec{X}_3\ \vec{X}_4),
\end{align}
which are given by 
\begin{align}
 \vec{X}_1^{(q=8)} &= (0,e^{\frac{3 \pi\rmi}{8}},e^{\frac{\rmi \pi}{4}},e^{\frac{\rmi \pi}{8}},1,e^{\frac{-\rmi \pi}{8}}, e^{\frac{-\rmi \pi}{4}}, e^{\frac{-3\pi\rmi }{8}})^T,\\
 \vec{X}_2^{(q=8)} &= (e^{\frac{-3\pi\rmi }{8}},0,e^{\frac{3 \pi\rmi}{8}},e^{\frac{\rmi \pi}{4}},e^{\frac{\rmi \pi}{8}},1,e^{\frac{-\rmi \pi}{8}}, e^{\frac{-\rmi \pi}{4}})^T,\\
 \vec{X}_3^{(q=8)} &= (e^{\frac{-\rmi \pi}{4}},e^{\frac{-3\pi\rmi }{8}},0,e^{\frac{3 \pi\rmi}{8}},e^{\frac{\rmi \pi}{4}},e^{\frac{\rmi \pi}{8}},1,e^{\frac{-\rmi \pi}{8}})^T,\\
 \vec{X}_4^{(q=8)} &= (e^{\frac{-\rmi \pi}{8}},e^{\frac{-\rmi \pi}{4}},e^{\frac{-3\pi\rmi }{8}},0,e^{\frac{3 \pi\rmi}{8}},e^{\frac{\rmi \pi}{4}},e^{\frac{\rmi \pi}{8}},1)^T,\\
 \vec{X}_5^{(q=8)} &= (1,e^{\frac{-\rmi \pi}{8}}, e^{\frac{-\rmi \pi}{4}}, e^{\frac{-3\pi\rmi }{8}},0,e^{\frac{3 \pi\rmi}{8}},e^{\frac{\rmi \pi}{4}},e^{\frac{\rmi \pi}{8}})^T,\\
 \vec{X}_6^{(q=8)} &= (e^{\frac{\rmi \pi}{8}},1,e^{\frac{-\rmi \pi}{8}}, e^{\frac{-\rmi \pi}{4}}, e^{\frac{-3\pi\rmi }{8}},0,e^{\frac{3 \pi\rmi}{8}},e^{\frac{\rmi \pi}{4}})^T,\\
 \vec{X}_7^{(q=8)} &= (e^{\frac{\rmi \pi}{4}},e^{\frac{\rmi \pi}{8}},1,e^{\frac{-\rmi \pi}{8}}, e^{\frac{-\rmi \pi}{4}}, e^{\frac{-3\pi\rmi }{8}},0,e^{\frac{3 \pi\rmi}{8}})^T,\\
 \vec{X}_8^{(q=8)} &= (e^{\frac{3 \pi\rmi}{8}},e^{\frac{\rmi \pi}{4}},e^{\frac{\rmi \pi}{8}},1,e^{\frac{-\rmi \pi}{8}}, e^{\frac{-\rmi \pi}{4}}, e^{\frac{-3\pi\rmi }{8}},0)^T.
\end{align}
The vectors are ordered such that the $\mu$-component of $\vec{X}_\mu$ is 0. We have
\begin{align}
 X_\mu(\alpha) = \begin{cases} \sqrt{-\theta_0(\mu)\theta_0^*(\alpha)} & \alpha\neq \mu \\ 0 & \alpha = \mu\end{cases}
\end{align}
and
\begin{align}
 \Phi_0^2 = \Phi_0 \Phi_0^\dagger = \sum_{\mu=1}^8 \vec{X}_\mu \vec{X}_\mu^\dagger.
\end{align}
We obtain the polynomial
\begin{equation}
\begin{aligned}
 P_8(u) &=\mbox{det}(u \mathbb{1}_8 - \Phi_0^2) \\
 &= (u^4-28u^3+70u^2-28u+1)^2.
\end{aligned}
\end{equation}

\section{Eigenvalues of $\Phi_0$ and polynomial $P_q(u)$}\label{polyq}
\noindent In the following, we compute the eigenvalues of $\Phi_0^{(q)}$ and show that
\begin{align}
\det(u\mathbbm{1}- (\Phi_0^{(q)})^2)= (1+u)^q T_q\left(\sqrt{\frac{u}{1+u}}\right)^2.
\end{align}
We first compute the exact eigenvalues of $\Phi_0$ by exploiting its circulant nature, and then we express the eigenvalues in terms of the Chebyshev polynomials of the first kind to derive the expression for $P_q(u)$.

\subsection{Largest eigenvalue $\lambda_{\rm max}$} \label{Appmax}
Since $\Phi_0$ is circulant, its largest eigenvalue $\lambda_{\rm max}(q)$ corresponds to the all-ones eigenvector $\vec{v}=\frac{1}{\sqrt{q}}(1,\dots,1)^T$. For even $q$ we have
 \begin{align}
 \nonumber &\lambda_{\rm max}(q)= \vec{v}^T \Phi_0 \vec{v} = \frac{1}{q}\sum_{\mu,\nu=1}^q (\Phi_0)_{\mu\nu} = \sum_{\mu=1}^q (\Phi_0)_{1\mu}\\
\label{em26} &=  \Bigl(\sum_{\mu=1}^q \sqrt{\omega^{\mu-1}}\Bigr)-\rmi = 1+ 2\sum_{\mu=2}^{q/2} \cos\Bigl(\frac{(\mu-1)\pi}{q}\Bigr),
 \end{align}
while for odd $q$ we similarly have
\begin{align}
 \lambda_{\rm max}(q) &=\sum_{\mu=1}^q (\Phi_0)_{1\mu} = 2\sum_{\mu=1}^{\frac{q-1}{2}}\sin\Bigl(\frac{ \mu \pi}{q}\Bigr).
\end{align}
Using Lagrange's identities
\begin{align}
 \label{em28} \sum_{k=1}^n \cos k \theta &= \frac{-\sin(\frac{1}{2}\theta)+\sin[(n+\frac{1}{2})\theta]}{2\sin(\frac{1}{2}\theta)},\\
 \sum_{k=1}^n \sin k \theta &= \frac{\cos(\frac{1}{2}\theta)-\cos[(n+\frac{1}{2})\theta]}{2\sin(\frac{1}{2}\theta)},
\end{align}
we verify that both expressions equal
\begin{align}
 \label{em29} \lambda_{\rm max}(q) = \cot\Bigl(\frac{\pi}{2q}\Bigr),
\end{align}
which is the result given in the main text.

\subsection{All eigenvalues}
 Since $\Phi_0^{(q)}$ is a circulant Hermitian matrix of size $q\times q$, its eigenvalues $\lambda_j^{(q)}$ for $j=0,1,\dots,q-1$ are given by
\begin{equation}
    \lambda_j^{(q)}= \sum_{k=0}^{q-1} c_k (\omega^j)^k,
\end{equation}
where $(c_0, c_1,\dots, c_{q-1})$ is the first row of $\Phi_0^{(q)}$ and $\omega=e^{2\pi \rmi /q}$. In order to choose a branch that makes $\Phi_0^{(q)}$ explicitly Hermitian, we rewrite the first row of $\Phi_0^{(q)}$  as
\begin{equation}
    (1, \sqrt{\omega},\dots,\sqrt{\omega^{\frac{q}{2}-1}},0, \sqrt{\omega^{-\left(\frac{q}{2}-1\right)}},\sqrt{\omega^{-\left(\frac{q}{2}-2\right)}},\dots, \sqrt{\omega^{-1}})
\end{equation}
for even $q$ and
\begin{equation}
    (0, \rmi \sqrt{\omega}, \dots, \rmi \sqrt{ \omega^{\frac{q-1}{2}}},-\rmi \sqrt{ \omega^{-\left(\frac{q-1}{2}\right)}},\dots, -\rmi \sqrt{\omega^{-1}})
\end{equation}
for odd $q$. When $q$ is even, we find that the eigenvalues are given by the sum
\begin{equation}
    \lambda_j^{(q {\text{ even})}}= 1 +\sum_{k=1}^{\frac{q}{2}-1} \left((\omega^{\frac{1}{2}+j})^k + (\omega^{-(\frac{1}{2}+j)})^k\right).
\end{equation}
Let $\omega^{j+\frac{1}{2}}= e^{\rmi \theta}$ where $\theta= \frac{2\pi}{q}(j+\frac{1}{2})$. Then we get 
\begin{equation}
    \lambda_j^{(q {\text{ even})}}= 1 +2\sum_{k=1}^{\frac{q}{2}-1} \cos(k \theta)
\end{equation}
with $j=0,\dots,q-1$. This can be simplified using Lagrange's identity to yield
\begin{align}
    \nonumber \lambda_j^{(q {\text{ even})}}&=1+ \frac{-\sin\left(\frac{\pi}{q}(j+\frac{1}{2})\right)}{\sin\left(\frac{\pi}{q}(j+\frac{1}{2})\right)}\nonumber \\&+\frac{\sin\left((\frac{q}{2}\frac{1}{2})\frac{2\pi}{q}(j+\frac{1}{2})\right)}{\sin\left(\frac{\pi}{q}(j+\frac{1}{2})\right)},\\
   \nonumber \implies \lambda_j^{(q {\text{ even})}}&= (-1)^j \frac{\cos\left(\frac{\pi}{q} \left(j+\frac{1}{2}\right)\right)}{\sin\left(\frac{\pi}{q}(j+\frac{1}{2})\right)} \\&= (-1)^{j}\cot\left(\frac{\pi}{q}\left(j+\frac{1}{2}\right)\right).
\end{align}
On the other hand, applying Lagrange's identity to 
\begin{equation}
    \lambda_j^{(q {\text{ odd})}}= i \sum_{k=1}^{\frac{q-1}{2}} (e^{\rmi \theta}- e^{-\rmi \theta})=2    \sum_{k=1}^{\frac{q-1}{2}} \sin( k \theta)
\end{equation}
gives 
\begin{equation}
    \lambda_j^{(q {\text{ odd})}}=- \cot\left(\frac{\pi}{q}\left(j+\frac{1}{2}\right)\right)
\end{equation}
for $j=0,\dots,q-1$. Both eigenvalues are of the same magnitude and their squares are identical. Thus for all $q$ we have the decomposition
\begin{equation}\label{exactdet}
\begin{aligned}
    P_q(u) &= \det(u\mathbbm{1}- (\Phi_0^{(q)})^{2})\\
     &=\prod_{j=0}^{q-1}\left(u- \cot\left(\frac{\pi}{q}\left(j+\frac{1}{2}\right)\right)^2\right).
\end{aligned}
\end{equation}

\subsection{Chebyshev factorization} Recall that the roots of the Chebyshev polynomial of the first kind, $T_q(x)$, are given by
\begin{equation}
    x_{j}= \cos\left( \frac{\pi}{q}\left(j+\frac{1}{2}\right)\right),\ j=0,1,\dots,q-1,
\end{equation}
where the argument $\frac{\pi}{q}\left(j+\frac{1}{2}\right)$ is identical to that of the cotangent in Eq. (\ref{exactdet}). We use this observation to rewrite Eq. (\ref{exactdet}) in terms of the Chebyshev roots $x_j$. Recall that if $x_j$ are known for $j=0,1,\dots, q-1$, then the Chebyshev polynomials $T_q(x)$ for $q>0$ with largest weight $2^{q-1}$ can be constructed explicitly as 
\begin{equation}\label{chbfact}
    T_q(x)=2^{q-1} \prod_{j=0}^{q-1} (x-x_j).
\end{equation}
We further define $T_0(x)=1$. The Chebyshev polynomials also satisfy $\forall x\in \mathbb{R}$
\begin{equation}\label{chbfact2}
   T_q(-x)= (-1)^qT_q(x),\ T_q(1)=1.
\end{equation}
 Rewriting Eq. (\ref{exactdet}) in terms of $x_j$ gives
\begin{equation}\label{exactdet2}
    P_q(u)=\prod_{j=0}^{q-1}\left(u- \frac{x_j^2}{1-x_j^2}\right).
\end{equation}
To write Eq. (\ref{exactdet2}) in terms of Eq. (\ref{chbfact}), let $u= \frac{x^2}{1-x^2}$. Then, Eq. (\ref{exactdet2}) becomes
\begin{equation}
\begin{aligned}
    P_q(u)&=\prod_{j=0}^{q-1}\frac{x^2-x_j^2}{(1-x^2)(1-x_j^2)}\\
     &=\prod_{j=0}^{q-1}\Bigl(\frac{1}{1-x^2} \ \frac{(x+x_j)(x-x_j)}{(1+x_j)(1-x_j)}\Bigr)\\
     &=\prod_{j=0}^{q-1}\Bigl(\frac{1}{1-x^2} \ \frac{(-x-x_j)(x-x_j)}{(-1-x_j)(1-x_j)}\Bigr),
\end{aligned}
\end{equation}
where $x^2=\frac{u}{1+u}$ is an implicit function of $u$. Using
\begin{align}
    \prod_{j=0}^{q-1}\frac{1}{(1-x^2)}&= (1-x^2)^{-q}=(1+u)^q
\end{align}
and relations (\ref{chbfact}) and (\ref{chbfact2}), we obtain
\begin{align}
 P_q(u) &= (1+u)^q \frac{T_q(-x)T_q(x)}{T_q(-1)T_q(1)} = (1+u)^q T_q(x)^2.
\end{align}
Thus we arrive at
\begin{equation}
    P_q(u)=(1+u)^{q} T_q\left(\sqrt{\frac{u}{1+u}}\right)^2,
\end{equation}
which is the claimed result.

\subsection{Power series representation}
Using the identity of the Chebyshev polynomials
\begin{align}
 T_q(x) &= \frac{1}{2}\Bigl[ (x-\sqrt{x^2-1})^q+(x+\sqrt{x^2-1})^q\Bigr],
\end{align}
we obtain ($u>0$)
\begin{align}
 \nonumber &P_q(u) =(1+u)^q T_q(x)\\
 \nonumber &=(1+u)^q \Bigl(\frac{1}{2}\Bigr)^2 \Bigl[ (x+\sqrt{x^2-1})^q+(x-\sqrt{x^2-1})^q\Bigr]^2\\
 \nonumber &=(1+u)^q \Bigl(\frac{1}{2}\Bigr)^2 \Biggl[ \Biggl(\sqrt{\frac{u}{1+u}}+\sqrt{\frac{-1}{1+u}}\Biggr)^q\\
 \nonumber &+\Biggl(\sqrt{\frac{u}{1+u}}-\sqrt{\frac{-1}{1+u}}\Biggr)^q\Biggr]^2\\
 \nonumber &=(1+u)^q \Bigl(\frac{1}{2}\Bigr)^2 \Biggl[ \Biggl(\frac{\sqrt{u}+\rmi}{\sqrt{1+u}}\Biggr)^q+\Biggl(\frac{\sqrt{u}-\rmi}{\sqrt{1+u}}\Biggr)^q\Biggr]^2\\
 \nonumber &=(1+u)^q \Bigl(\frac{1}{2}\Bigr)^2 \Biggl[ \Biggl(\frac{\rmi \sqrt{-u}+\rmi}{\sqrt{1+u}}\Biggr)^q+\Biggl(\frac{\rmi \sqrt{-u}-\rmi}{\sqrt{1+u}}\Biggr)^q\Biggr]^2\\
 &=(-1)^q\Bigl[\frac{1}{2}\Bigl((1+\sqrt{-u})^q+(-1+\sqrt{-u})^q\Bigr)\Bigr]^2.
\end{align}
This can be expanded using the binomial formula to  arrive at
\begin{align}
 P_q(u) &= (-1)^q\Biggl[ \frac{1}{2} \sum_{k=0}^q \binom{q}{k} (-u)^{k/2} [1+(-1)^q(-1)^k]\Biggr]^2.
\end{align}
For even $q$ this yields
\begin{align}
 P_q(u) &=\Biggl(\sum_{k=0}^{q/2} \binom{q}{2k} (-u)^k\Biggr)^2,
\end{align}
and for odd $q$ this yields
\begin{align}
 P_q(u) &=u\Biggl(\sum_{k=0}^{(q-1)/2} \binom{q}{2k+1} (-u)^k\Biggr)^2.
\end{align}

\section{Mean-field bound on $T_{\rm c}$}
\label{MF}
In this section, we derive the mean-field (MF) bound for the critical temperature given by 
\begin{equation}
    \frac{T_{\rm c}}{J} < \frac{T_{\rm c}^{\rm (MF)}}{J}= q_{\rm max}, \label{MF1}
\end{equation}
where $q_{\max}$ is the maximum coordination number of the corresponding two-dimensional periodic tiling of the plane. We remark that the bound in Eq. (\ref{MF1}) is a special case of a more general result stated in terms of the spectral radius of the interaction matrix $\mathcal{J}$ of the non-uniform Ising model \cite{LynnLee}. In our case, $\mathcal{J}= J A$, where $A$ is the adjacency matrix of the underlying lattice. The adjacency matrix is a symmetric matrix that tracks the connectivity between sites $i$ and $j$ of the lattice: $A_{ij}=A_{ji}=1$ if and only if $i$ and $j$ are connected, else it is zero. The adjacency matrix allows us to rewrite the Ising Hamiltonian in Eq. (\ref{intro1}) as
\begin{equation}
    H= -\frac{J}{2}\sum_{i,j} A_{ij} s_i s_j, \label{MF2}
\end{equation}
where the factor of $\frac{1}{2}$ is included to avoid double counting. The mean-field bound is derived in three parts: First, we decouple the spin at site $i$ as the sum of its thermal average, $\langle s_i \rangle$, and the fluctuation about its mean, $\delta s_i$. Then, for the mean-field approximation, we set the cross fluctuations $\delta s_i \delta s_j\to 0$ in the Hamiltonian, producing the mean-field Hamiltonian $H^{\rm (MF)}$.  Finally, the resultant linear mean-field Hamiltonian can be used to self-consistently compute the mean-field critical temperature. To proceed, denote the thermal average or local magnetization at site $i$ by $m_i=\langle s_i\rangle$. Then we have
\begin{align}
    s_i&=m_i +\delta s_i,\\ 
    \delta s_i&= s_i-m_i.
\end{align}
The product $s_i s_j$ in the Ising Hamiltonian can be expressed in terms of $m_i$ and $s_i$ as
\begin{align}
    \nonumber s_is_j&= m_i m_j + m_i \delta s_j  + m_j \delta s_i+\delta s_i \delta s_j\\
    \nonumber &=m_i m_j + m_i (s_j-m_j) + m_j (s_i-m_i)+\delta s_i \delta s_j\\
    &\approx m_i s_j + m_j s_i - m_i m_j, \label{MF3}
\end{align}
in the limit  $\delta s_i \delta s_j\to0$. Replacing the product $s_is_j$ in Eq. (\ref{MF2}) with Eq. (\ref{MF3}) gives the mean-field Hamiltonian 
\begin{equation}
    H^{(\rm MF)}=H_0 - J \sum_{j} h_j s_j,
\end{equation}
where $H_0=\frac{J}{2}\sum_{i,j}m_i m_j$ is a constant energy offset with respect to the spin variables $\{s_1,s_2,\dots\}:=\{s_k\}_{k=1}^{N}$ and 
\begin{equation}
    h_j:=\sum_{i}A_{i j} m_i \label{MF4}
\end{equation} is the effective external magnetic field at site $j$. Computing the local magnetization at site $i$ using $H^{\rm(MF)}$, we arrive at
\begin{align}
    \nonumber m_i&= \frac{\sum_{\{s_k\}}s_i e^{-\beta H^{\rm(MF)}}}{\sum_{\{s_k\}}e^{-\beta H^{\rm(MF)}}}\\ 
    \nonumber &= \frac{2 \sinh(\beta J h_i)\sum_{\{s_k\}_{k\neq i}}\prod_{j\neq i} e^{\beta J h_j s_j}}{2\cosh(\beta J h_i)\sum_{\{s_k\}_{k\neq i}}\prod_{j\neq i} e^{\beta J h_j s_j}}\\
    &= \tanh(\beta J h_i),
\end{align}
which gives the self-consistency condition
\begin{equation}
    m_i= \tanh\Bigl( \beta J \sum_j A_{ij} m_j\Bigr). \label{MF5}
\end{equation}
Linearizing Eq. (\ref{MF5}) about the paramagnetic phase $\bd{m}=(m_1,m_2,m_3,\dots)^T=\bd{0}$ gives the eigenvalue problem
\begin{equation}
    \bd{m}= \beta J A\bd{m}\implies \det(\mathbbm{1}- \beta J A)=0. \label{MF6}
\end{equation}
The smallest value of $\beta J>0$ that satisfies this equation, denoted $\beta_{\rm c}^{(\rm MF)} J$, is the inverse of the largest eigenvalue of $A$. The largest eigenvalue of $A$ is the maximum coordination number on the lattice $q_{\max}$, i.e
\begin{equation}
    \beta_{\rm c}^{(\rm MF)}J=\frac{1}{q_{\max}}\iff \frac{T_{\rm c}^{\rm (MF)}}{J}= q_{\rm max}.
\end{equation}
Since the mean-field critical temperature is an upper bound to the critical temperature $T_{\rm c}$, the resultant inequality Eq. (\ref{MF1}) follows. 

For periodic lattices, we may exploit translation invariance and reduce Eq. (\ref{MF6}) in terms of the Bloch adjacency matrices $A(\vec{k})$ defined in Ref. \cite{1fvj-91v6}. If the unit cell contains $N_{\rm u}$ sites, then
\begin{equation}
    \bd{m}_{\rm u}(\bk)=\beta J A(\bk) \bd{m}_{\rm u}(\bk),
\end{equation}
where $\bk\in {\rm BZ}$ and $\bd{m}_{\rm u}=(m_1,m_2,\dots,m_{N_{\rm u}})$ is the local magnetization of sites within the unit cell. As an example, consider the regular Honeycomb lattice with constant coordination number $q=3$ and Bloch adjacency matrix
\begin{equation}
    A^{(\rm H)}(\boldsymbol{k})  =\begin{pmatrix} 0 & 1+e^{-\rmi k_1}+e^{-\rmi k_2} \\ 1+e^{\rmi k_1}+e^{\rmi k_2} & 0 \end{pmatrix},
\end{equation}
with eigenvalues 
\begin{equation}
    \pm |1+e^{\rmi k_1}+e^{\rmi k_2}|
\end{equation}
that have a global maximum over the Brillouin zone at $\bk=0$ of $\vare_{\rm H}(\bd{0})=3=q$. Therefore, the mean-field bound on the Honeycomb lattice gives $T_c^{\rm H}/J< 3$. Note that for regular lattices, $q_{\rm max}=\bar{q}$, and so $T_{\rm c}/J < \bar{q}$ is true for regular lattices. It is surprising that this inequality holds even for the $k$-uniform lattices considered in this work. However, for non-regular lattices, this inequality is not guaranteed by the mean-field approximation. As a counter example, consider the Compass-Rose (CR) lattice with $\bar{q}=6$. Since $T_{\rm c}^{\rm (\rm CR)}/J=6.4>\bar{q}$, the inequality $T_{\rm c}/{J}<\bar{q}$ fails. However, the mean-field inequality still remains valid since ${T_{\rm c}^{\rm (\rm CR)}}/{J}< q_{\rm max}=24.$

\section{Kac--Ward matrices for selected lattices}\label{W(k)}

In this section, we present explicit expressions for the Kac-Ward matrix $W(\bk)$ for the  Square, Honeycomb, Kagome, and CaVO lattices. For these lattices, we also make the connection to the reference matrix $\Phi_0^2$ explicit, which supplements the general proof presented in the App. (\ref{refPhi0}). For each lattice considered, we verify the decomposition 
\begin{align}
\det(u\mathbbm{1}_Q-W_{0}^\dagger W_0)=\prod_{a=1}^{N_{\rm u}}P_{q_{a}}(u).
\end{align}
The dispersion relations on the Square and Triangular lattice are defined as
\begin{align}
 \vare_\square(\vec{k}) &= -2(\cos k_1+\cos k_2),\\
 \vare_\Delta(\vec{k}) &=-2 [\cos k_1+\cos k_2+\cos (k_1-k_2)],
\end{align}
respectively.

\subsection{Square lattice ($q=4$)}\label{AppSquare}

\subsubsection{Matrix $W(\vec{k})$} 

\noindent The Square lattice has coordination number $q=4$ and $N_{\rm u}=1$ site in the unit cell. In this section we consider the Square lattice with the usual right angles according to the following scheme with $\chi=\frac{\pi}{2}$:\\
\begin{figure}[h!]
    \includegraphics[width=7cm]{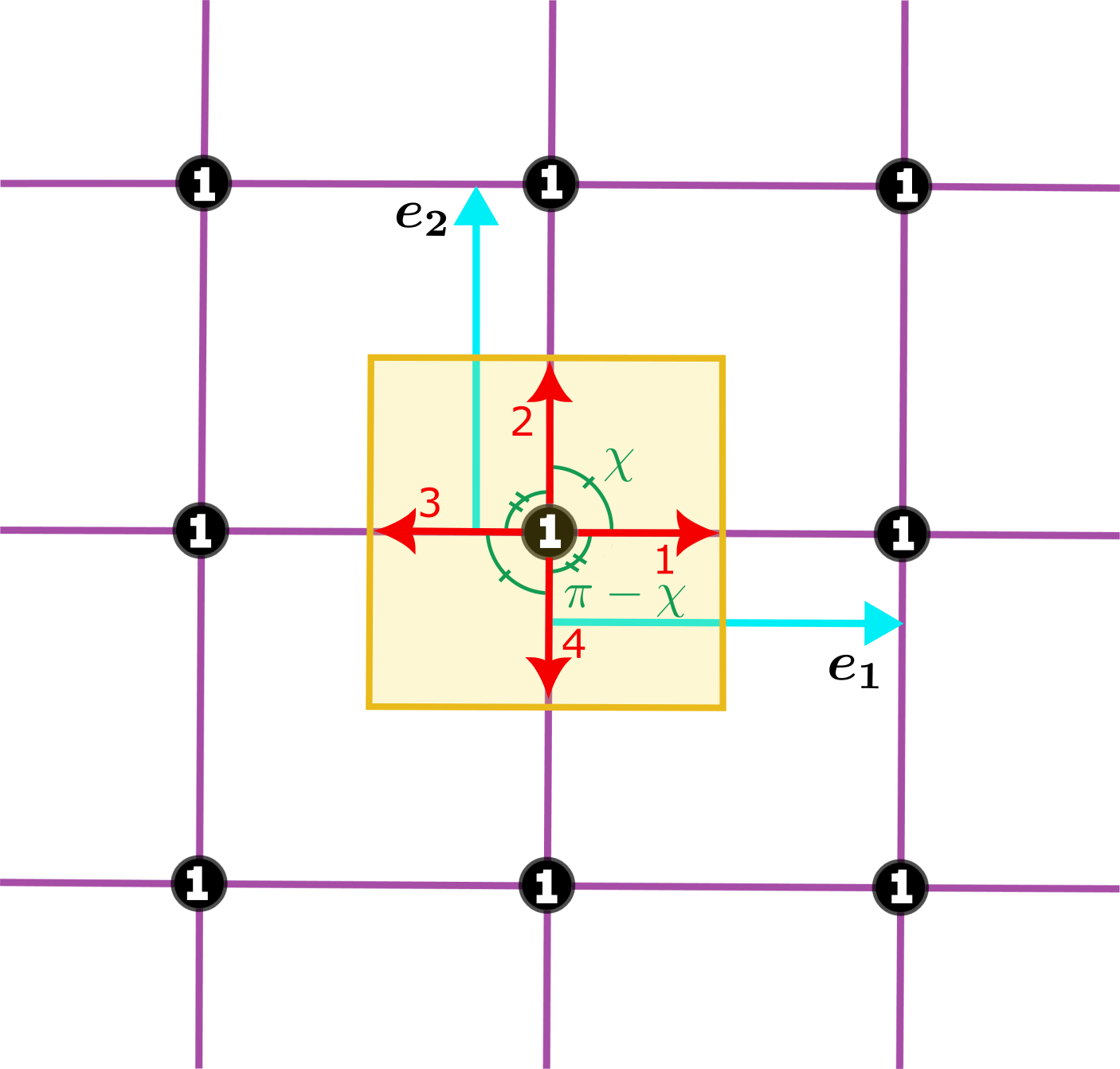}
    \label{FigSquareLabels}
\end{figure}\\
We have 
\begin{align}
 B_1(\vec{k}) = \begin{pmatrix} e^{-\rmi k_1} & & & \\ & e^{-\rmi k_2} & & \\ & & e^{\rmi k_1} & \\ & & & e^{\rmi k_2}\end{pmatrix}
\end{align} 
and $E_{1;1}=\mathbb{1}_4$. We have
\begin{align}
 \vec{\theta}_1 = \vec{\theta}_0^{(q=4)} = (1,\rmi,-1,-\rmi)
\end{align}
and
\begin{align}
 \phi_{11} &= \vec{\theta}_1^*\vec{\theta}^T = \phi_0^{(q=4)} = \begin{pmatrix} 1 & \rmi & -1 & -\rmi \\ -\rmi & 1 & \rmi & -1 \\ -1 & -\rmi & 1 & \rmi \\ \rmi & -1 & -\rmi & 1 \end{pmatrix},
 \end{align}
 \begin{align}
 \Phi_{11} &= \Phi_0^{(q=4)} = \begin{pmatrix} 1 & e^{\rmi \pi/4} & 0 & e^{-\rmi \pi/4} \\ e^{-\rmi \pi/4} & 1 & e^{\rmi \pi/4} & 0 \\ 0 & e^{-\rmi \pi/4} & 1 & e^{\rmi \pi/4} \\ e^{\rmi \pi/4} & 0 & e^{-\rmi \pi/4} & 1 \end{pmatrix}.
\end{align}
Note that $\Phi_{11}=\Phi_{11}^\dagger$ is Hermitian. By construction, $\phi_0$ and $\Phi_0$ coincide with the $q=4$ references $\phi_0^{(q=4)}$ and $\Phi_0^{(q=4)}$ from Eqs. (\ref{ref4e}) and (\ref{ref4b}). We arrive at
\begin{align}
 \nonumber &W(\vec{k})=B_1(\vec{k})\Phi_{11}\\
 &=\begin{pmatrix} e^{-\rmi k_1} & e^{-\rmi k_1+\rmi \pi/4} & 0 & e^{-\rmi k_1-\rmi \pi/4} \\ e^{-\rmi k_2-\rmi \pi/4} & e^{-\rmi k_2} & e^{-\rmi k_2+\rmi \pi/4} & 0 \\ 0 & e^{\rmi k_1-\rmi \pi/4} & e^{\rmi k_1} & e^{\rmi k_1+\rmi \pi/4} \\ e^{\rmi k_2+\rmi \pi/4} & 0 & e^{\rmi k_2-\rmi \pi/4} & e^{\rmi k_2} \end{pmatrix}.
\end{align} 
We have
\begin{align}
 \mbox{det}(\mathbb{1} - t W(\vec{k})) = (1+t^2)^2+t(1-t^2)\vare_\square(\vec{k}).
\end{align}

\subsubsection{Matrices $W_0$ and $\mathcal{B}$} 

\noindent For $\vec{k}=0$ we have
\begin{align}
 W_0 = \Phi_0 = \begin{pmatrix} 1 & e^{\rmi \pi/4} & 0 & e^{-\rmi \pi/4} \\ e^{-\rmi \pi/4} & 1 & e^{\rmi \pi/4} & 0 \\ 0 & e^{-\rmi \pi/4} & 1 & e^{\rmi \pi/4} \\ e^{\rmi \pi/4} & 0 & e^{-\rmi \pi/4} & 1 \end{pmatrix},
\end{align}
and so
\begin{align}
 \mathcal{B} = W_0^\dagger W_0 = \Phi_0^2.
\end{align}
The corresponding characteristic polynomial is 
\begin{align}
 \det(u \mathbb{1}_{q} - W_0^\dagger W_0) = \det(u\mathbb{1}_q - \mathcal{B}) = (u^2-6u+1)^2.
\end{align}

\subsection{Honeycomb lattice ($q=3$)}

\subsubsection{Matrix $W(\vec{k})$} 

\noindent The Honeycomb lattice has coordination number $q=3$ and $N_{\rm u}=2$ sites in the unit cell. When embedded as a regular tiling (i.e. with hexagons with equal-length sides), we label the unit cell sites $a\in\{1,2\}$ and edges $\mu\in\{1,2,3\}$ according to the schematic below.\\
\begin{figure}[h!]
    \includegraphics[width=8cm]{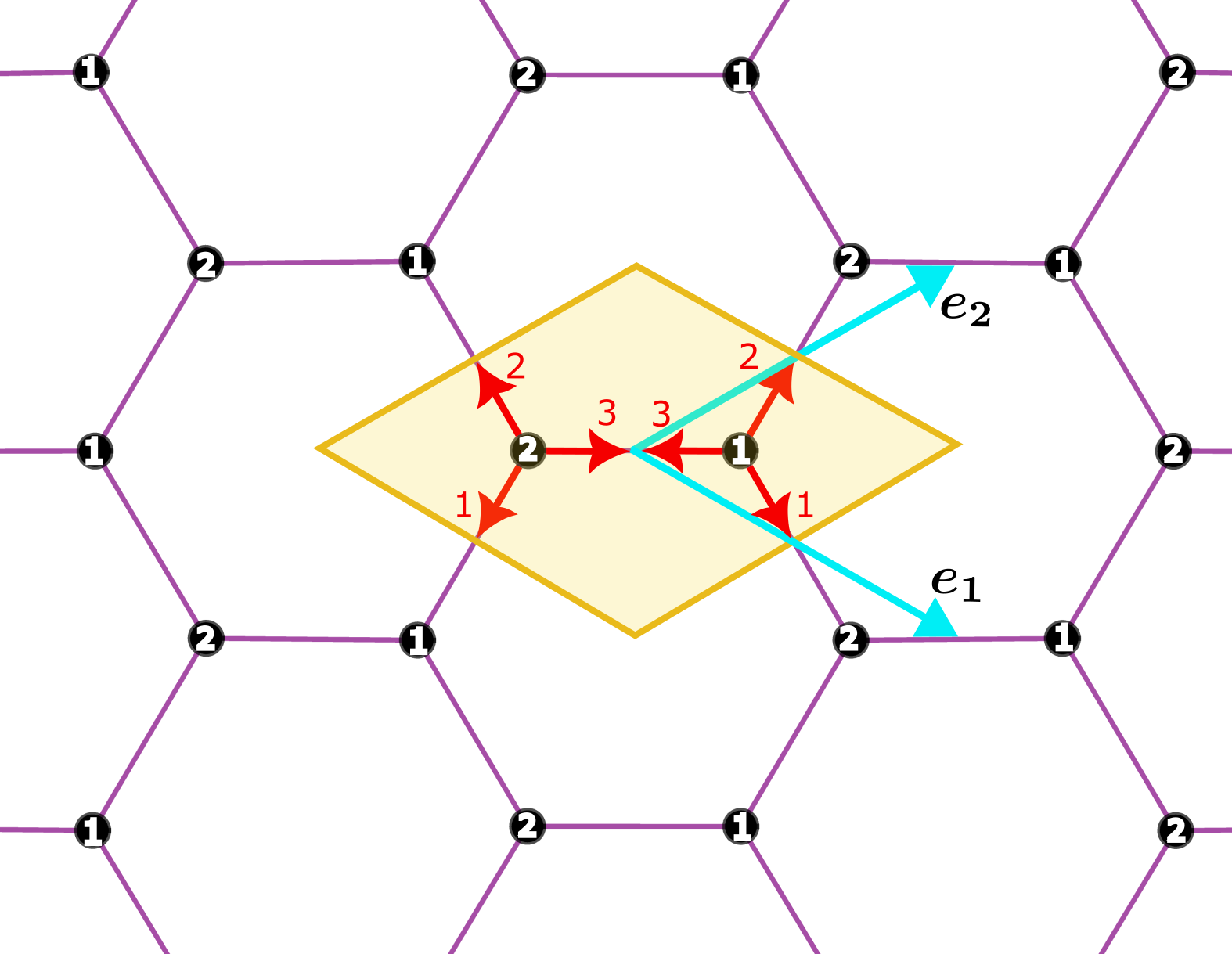}
     \label{FigHoneycombLabels}
\end{figure}\\
We have
\begin{align}
 W(\vec{k}) = \begin{pmatrix} 0 & E_{1;2}B_1(\vec{k})\Phi_{12} \\ E_{2;1} B_2(\vec{k}) \Phi_{21} & 0 \end{pmatrix}.
\end{align}
The Bloch matrices for the unit-cell sites read
\begin{align}
 B_1(\vec{k}) &= \begin{pmatrix} e^{-\rmi k_1} & & \\ & e^{-\rmi k_2} & \\ & & 1 \end{pmatrix},\ B_2(\vec{k}) = \begin{pmatrix} e^{\rmi k_1} & & \\ & e^{\rmi k_2} & \\ & & 1 \end{pmatrix}.
\end{align}
The edge-connectivity matrices are
\begin{align}
 E_{1;2}= E_{2;1} = \mathbb{1}_3.
\end{align}
We have
\begin{equation}
\begin{aligned}
 \vec{\theta}_1 &= \vec{\theta}_0 = (e^{-\pi\rmi/3},e^{\pi\rmi/3},-1)^T,\\
 \vec{\theta}_2 &= - \vec{\theta}_0 = (e^{2\pi\rmi/3},e^{-2\pi\rmi/3},1)^T
\end{aligned}
\end{equation}
with $\vec{\theta}_0=\vec{\theta}_0^{(q=3)}$ the $q=3$ reference, so that
\begin{align}
 \Phi_{12} &= \Phi_{21}=\Phi
\end{align}
with
\begin{align}
 \Phi = \Phi_0^{(q=3)} = \begin{pmatrix} 0 & e^{-\pi\rmi/6}  & e^{\pi \rmi/6} \\  e^{\pi\rmi/6} & 0 & e^{-\pi\rmi/6} \\ e^{-\pi\rmi/6} & e^{\pi\rmi/6} & 0 \end{pmatrix}.
\end{align}
Note that $\Phi^\dagger=\Phi$ is Hermitian and agrees with the $q=3$ reference $\Phi_0^{(q=3)}$ from Eq. (\ref{ref3b}). We arrive at
\begin{align}
 W(\vec{k}) = \begin{pmatrix} 0 & B_1(\vec{k})\Phi_0 \\ B_2(\vec{k})\Phi_0 & 0 \end{pmatrix}
\end{align}
with
\begin{align}
 \mbox{det}(\mathbb{1}-tW(\vec{k})) = 1+3t^4+(1-t^2)t^2\vare_\Delta(\vec{k}).
\end{align}

\subsubsection{Matrices $W_0$ and $\mathcal{B}_a$}

\noindent For $\vec{k}=0$ we find
\begin{align}
 W_0 = \begin{pmatrix} 0 & \Phi_0 \\ \Phi_0 & 0 \end{pmatrix}
\end{align}
and thus
\begin{align}
 W_0^\dagger W_0 = \begin{pmatrix} \mathcal{B} & 0 \\ 0 & \mathcal{B} \end{pmatrix}
\end{align}
with $q\times q$ block
\begin{align}
 \mathcal{B} = \Phi_0^2 = \begin{pmatrix} 2 & e^{\pi\rmi/3} & e^{-\pi\rmi/3} \\ e^{-\pi\rmi/3} & 2 & e^{\pi\rmi/3} \\ e^{\pi\rmi/3} & e^{-\pi\rmi/3} & 2 \end{pmatrix}.
\end{align}
The characteristic polynomial of $W_0^\dagger W_0$ is
\begin{equation}
\begin{aligned}
 \det(u\mathbb{1}-W_0^\dagger W_0) &= \Bigl[\det(u\mathbb{1}-\mathcal{B})\Bigr]^{N_{\rm u}} \\
 &= [u(u-3)^2]^2=u^2(u-3)^4.
\end{aligned}
\end{equation}
The matrices $\mathcal{B}$ are identical to $[\Phi_0^{(q=3)}]^2$.

\subsection{Kagome lattice ($q=4$)}

\subsubsection{Matrix $W(\vec{k})$} 

\noindent The Kagome lattice has coordination number $q=4$ and $N_{\rm u}=3$ sites in the unit cell. We label the sites in the unit cell by $a\in\{1,2,3\}$ and the directed edges at each site by $\mu\in\{1,\dots,4\}$ according to the following schematic:
\begin{figure}[h!]
    \includegraphics[width=6cm]{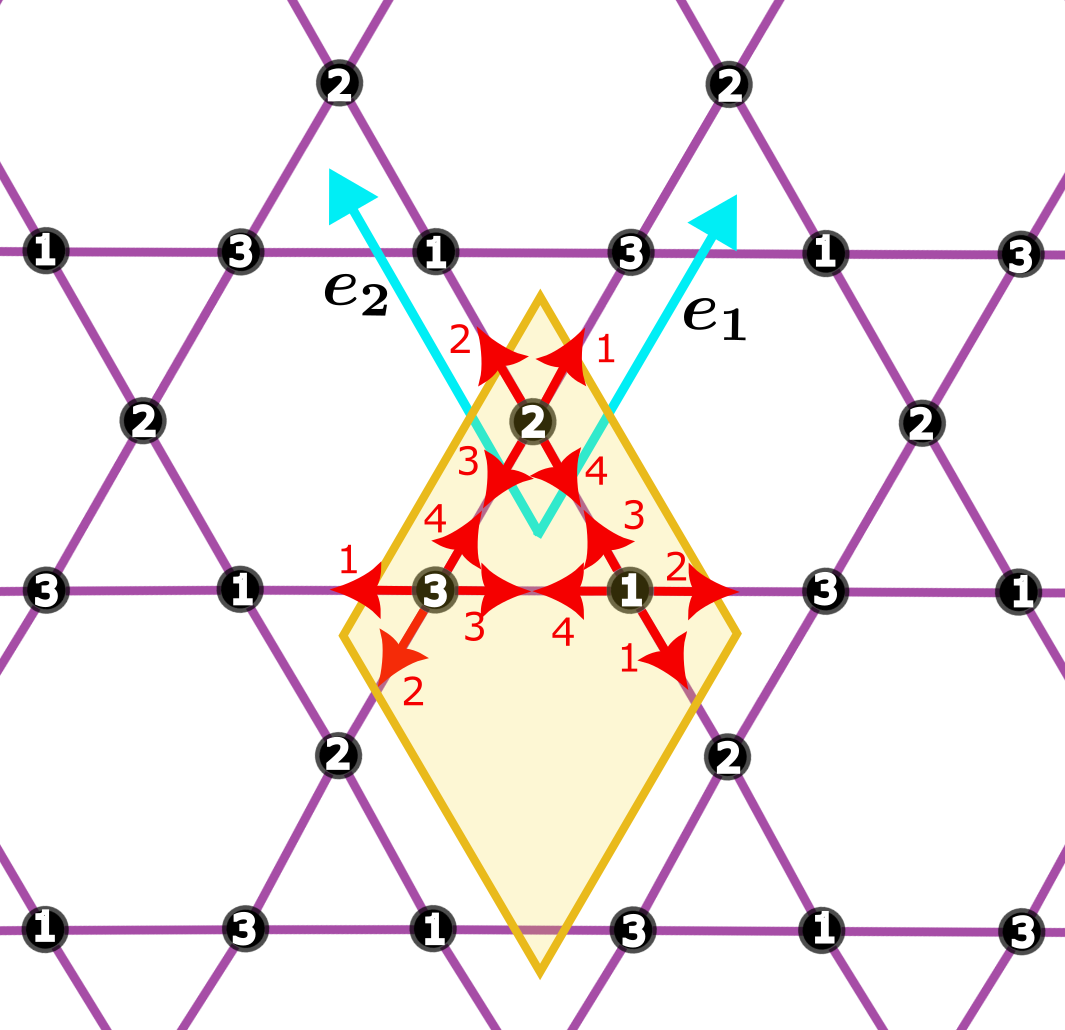}
     \label{FigKagomeLabels}
\end{figure}\\
We have
\begin{align}
 \label{kag2} W(\vec{k}) = \begin{pmatrix} 0 & E_{1;2}B_1(\textbf{k}) \Phi_{12} & E_{1;3}B_1(\textbf{k}) \Phi_{13} \\ E_{2;1} B_2(\textbf{k})\Phi_{21} & 0 & E_{2;3} B_2(\textbf{k})\Phi_{23} \\ E_{3;1} B_3(\textbf{k})\Phi_{31} & E_{3;2} B_3(\textbf{k})\Phi_{32} & 0 \end{pmatrix}.
\end{align}
For the Bloch matrices at the unit cell sites we find
\begin{equation}
\begin{aligned}
 B_1(\vec{k}) &=\text{diag}(e^{\rmi k_2},e^{-\rmi(k_1-k_2)},1,1),\\
 B_2(\vec{k}) &= \text{diag}(e^{-\rmi k_1},e^{-\rmi k_2},1,1),\\
 B_3(\vec{k}) &=\text{diag}(e^{\rmi (k_1-k_2)},e^{\rmi k_1},1,1).
\end{aligned}
\end{equation}
The edge-connectivity matrices are
\begin{equation}
\begin{aligned}
  E_{1;2} &= E_{2;3} = E_{3;1} =\vare_1,\\
  E_{1;3} &= E_{2;1} = E_{3;2} = \vare_2
\end{aligned}
\end{equation}
with
\begin{align}
 \vare_1 &= \text{diag}(1,0,1,0),\ \vare_2=\text{diag}(0,1,0,1).
\end{align} 
Define $g=e^{2\pi\rmi/3}$ so that
\begin{equation}
\begin{aligned}
  \vec{\theta}_1 &= (e^{-\pi \rmi /3},1,e^{2 \pi\rmi /3},-1)^T,\\
   \vec{\theta}_2 &= g \vec{\theta}_1,\ \vec{\theta}_3 = g^2 \vec{\theta}_1.
\end{aligned}
\end{equation}
This yields
\begin{equation}
\begin{aligned}
  \Phi_{12} &= \Phi_{23}=\Phi_{31} =: \Phi,\\
  \Phi_{21} &= \Phi_{32} = \Phi_{13} = \Phi^\dagger
\end{aligned}
\end{equation}
with
\begin{align}
 \label{kag10} \Phi = \begin{pmatrix} e^{\pi\rmi/3} & 0 & e^{-\pi\rmi/6} & 1 \\ e^{\pi\rmi/6} & e^{\pi\rmi/3} & e^{-\pi\rmi/3} & e^{-\pi\rmi/6} \\ e^{-\pi\rmi/6} & 1 & e^{\pi\rmi/3} & 0 \\ e^{-\pi\rmi/3} & e^{-\pi\rmi/6} & e^{\pi\rmi/6} & e^{\pi\rmi/3} \end{pmatrix}.
\end{align}
We then arrive at
\begin{align}
 \label{kag11} W(\vec{k}) = \begin{pmatrix} 0 & \vare_1 B_1(\textbf{k}) \Phi & \vare_2 B_1(\textbf{k}) \Phi^\dagger \\ \vare_2B_2(\textbf{k})\Phi^\dagger & 0 & \vare_1 B_2(\textbf{k})\Phi \\ \vare_1 B_3(\textbf{k})\Phi & \vare_2 B_3(\textbf{k})\Phi^\dagger & 0 \end{pmatrix}.
\end{align}
We have
\begin{align}
 \nonumber &\mbox{det}(\mathbb{1}-tW(\vec{k})) = (1+t)^4\Bigl[1-4t+10t^2-16t^3+22t^4 \\
 \label{kag12} &-16t^5+10t^6-4t^7+t^8+(1-t)^2(1+t^2)t^2\vare_\Delta(\vec{k})\Bigr].
\end{align}

\subsubsection{Matrices $W_0$ and $\mathcal{B}_a$} 

\noindent For $\vec{k}=0$, we find 
\begin{equation}\label{kag13}
\begin{aligned}
  W_0 =\begin{pmatrix} 0 & \vare_1 \Phi & \vare_2 \Phi^\dagger \\ \vare_2\Phi^\dagger & 0 & \vare_1 \Phi \\ \vare_1 \Phi & \vare_2 \Phi^\dagger & 0 \end{pmatrix}.
\end{aligned}
\end{equation}
Using $\vare_1^2=\vare_1$, $\vare_2^2=\vare_2$, and $\vare_1\vare_2=\vare_2\vare_1=0$ we arrive at
\begin{align}
 \label{kag15} W_0^\dagger W_0 = \text{diag}(\mathcal{B},\mathcal{B},\mathcal{B})
\end{align}
with $q\times q$ block
\begin{align}
 \label{kag16} \mathcal{B} = \Phi^\dagger \vare_1\Phi +\Phi\vare_2\Phi^\dagger.
\end{align} 
The characteristic polynomial of $W_0^\dagger W_0$ reads
\begin{equation}\label{kag17}
\begin{aligned}
  \det(u\mathbb{1}-W_0^\dagger W_0) &= \Bigl[ \det(u \mathbb{1}_q-\mathcal{B})\Bigr]^{N_{\rm u}}\\
  &= (u^2-6 u+1)^6.
\end{aligned}
\end{equation}

\subsubsection{Relation to reference $\Phi_0^2$}

To make the connection to the reference matrix $\Phi_0^{(q=4)}$, we write $\vec{\theta}_a=U_a\vec{\theta}_0$, with $\vec{\theta}_0=(1,\rmi,-1,-\rmi)$ from Eq. (\ref{ref4c}) and diagonal unitary matrices
\begin{equation}
\begin{aligned}
 U_1 &= \text{diag}(e^{-\rmi \pi/3},-\rmi,e^{-\rmi \pi/3},-\rmi),\\
 U_2 &= gU_1 =\text{diag}(e^{\rmi \pi/3},e^{\rmi \pi/6},e^{\rmi \pi/3},e^{\rmi \pi/6}),\\
 U_3 &= g^2U_1 =\text{diag}(-1,e^{5\pi\rmi/6},-1,e^{5\pi\rmi/6}).
 \end{aligned}
\end{equation}
Further define $V_a$ so that $V_a^2=U_a$ according to
\begin{equation}
\begin{aligned}
 V_1 &= \text{diag}(e^{-\rmi \pi/6},e^{-\rmi \pi/4},e^{-\rmi \pi/6},e^{-\rmi \pi/4}),\\
 V_2 &= \text{diag}(e^{\rmi\pi/6},e^{\rmi \pi/12},e^{\rmi\pi/6},e^{\rmi \pi/12})= e^{\rmi \pi/3}V_1,\\
 V_3 &= \text{diag}(\rmi,e^{5\pi\rmi/12},\rmi,e^{5\pi\rmi/12})= e^{2\pi\rmi/3} V_1.
\end{aligned}
\end{equation}
Consider the matrices $M_k^{(a)}=\Phi_{ka}^\dagger E_{k;a}$ such that $\mathcal{B}_a=\sum_{k=1}^{N_{\rm u}} M_k^{(a)}[M_k^{(a)}]^\dagger$. For the unit cell sites $a=1,2,3$ we have
\begin{equation}
\begin{aligned}
 M_1^{(1)} &= 0,\ M_2^{(1)} = (\vec{0}\ \vec{Y}_1\ \vec{0}\ \vec{Y}_3),\ M_3^{(1)} = (\vec{Y}_2\ \vec{0}\ \vec{Y}_4\ \vec{0}),\\
 M_1^{(2)} &= (\vec{Y}_2\ \vec{0}\ \vec{Y}_4\ \vec{0}),\ M_2^{(2)} = 0,\ M_3^{(2)} = (\vec{0}\ \vec{Y}_1\ \vec{0}\ \vec{Y}_3),\\
 M_1^{(3)} &= (\vec{0}\ \vec{Y}_1\ \vec{0}\ \vec{Y}_3),\ M_2^{(3)} = (\vec{Y}_2\ \vec{0}\ \vec{Y}_4\ \vec{0}),\ M_3^{(3)} = 0.
\end{aligned}
\end{equation}
For each $a$, the set of matrices $\{M^{(a)}_k\}_{k=1,2,3}$ contain exactly $q=4$ nonzero column vectors $\vec{Y}_\mu^{(a)}= \vec{Y}_\mu$ given by
\begin{equation}
\begin{aligned}
 \vec{Y}_1 &= \begin{pmatrix} 0 \\ e^{\rmi \pi/3} \\ 1 \\ e^{-\rmi \pi/6} \end{pmatrix},\  \vec{Y}_2 =\begin{pmatrix} e^{-\rmi \pi/3} \\ 0 \\ e^{\rmi \pi/6} \\ 1 \end{pmatrix},\\
 \vec{Y}_3&=\begin{pmatrix} 1 \\ e^{-\rmi \pi/6} \\ 0 \\ e^{\rmi\pi/3}\end{pmatrix},\ \vec{Y}_4 = \begin{pmatrix} e^{\rmi\pi/6} \\ 1 \\ e^{-\rmi \pi/3} \\ 0 \end{pmatrix}.
\end{aligned}
\end{equation}
We have
\begin{align}
a=1,2,3:\ \mathcal{B}_a = \sum_{k=1}^{N_{\rm u}} M_k^{(a)}[M_k^{(a)}]^\dagger=\sum_{\mu=1}^4 \vec{Y}_\mu \vec{Y}_\mu^\dagger.
\end{align}
The vectors $\vec{Y}_\mu$ satisfy
\begin{align}
 \nonumber a=1,2,3:\ &Y_\mu^{(a)}(\alpha) = \begin{cases} \sqrt{-\theta_a(\mu)\theta_a^*(\alpha)} & \alpha\neq \mu \\ 0 & \alpha = \mu\end{cases} \\
  &= \begin{cases} (V_a)_{\mu\mu}(V_a^*)_{\alpha\alpha} \sqrt{-\theta_0(\mu)\theta_0^*(\alpha)} & \alpha\neq \mu \\ 0 & \alpha = \mu\end{cases},
\end{align}
which is equivalent to
\begin{align}
 a=1,2,3:\ \vec{Y}_\mu = (V_a)_{\mu\mu} V_a^\dagger \vec{X}_\mu
\end{align}
with the $q=4$ reference vectors $\vec{X}_\mu=\vec{X}_\mu^{(q=4)}$. Consequently,
\begin{align}
 a=1,2,3:\ \mathcal{B}_a = V_a^\dagger \Bigl(\sum_{\mu=1}^q \vec{X}_\mu \vec{X}_\mu^\dagger\Bigr)V_a = V_a^\dagger \Phi_0^2 V_a.
\end{align}
Hence  $\mathcal{B}_a$ is unitarily equivalent to $\Phi_0^2$ and 
\begin{equation}
\begin{aligned}
\det(u\mathbb{1}-W_0^\dagger W_0) &= \Bigl[ \det(u \mathbb{1}_q-\Phi_0^2)\Bigr]^{N_{\rm u}}\\
&= [(u^2-6u+1)^2]^3\\&= (u^2-6u+1)^6.
\end{aligned}
\end{equation}

\subsection{CaVO lattice ($q=3$)}

\subsubsection{Matrix $W(\vec{k})$}

The CaVO lattice has coordination number $q=3$ and $N_{\rm u}=4$ sites in the unit cell. We label the sites in the unit cell by $a\in\{1,2,3,4\}$ and the local edges at each site by $\mu\in\{1,2,3\}$ according to the following schematic:
\begin{figure}[h!]
    \includegraphics[width=8cm]{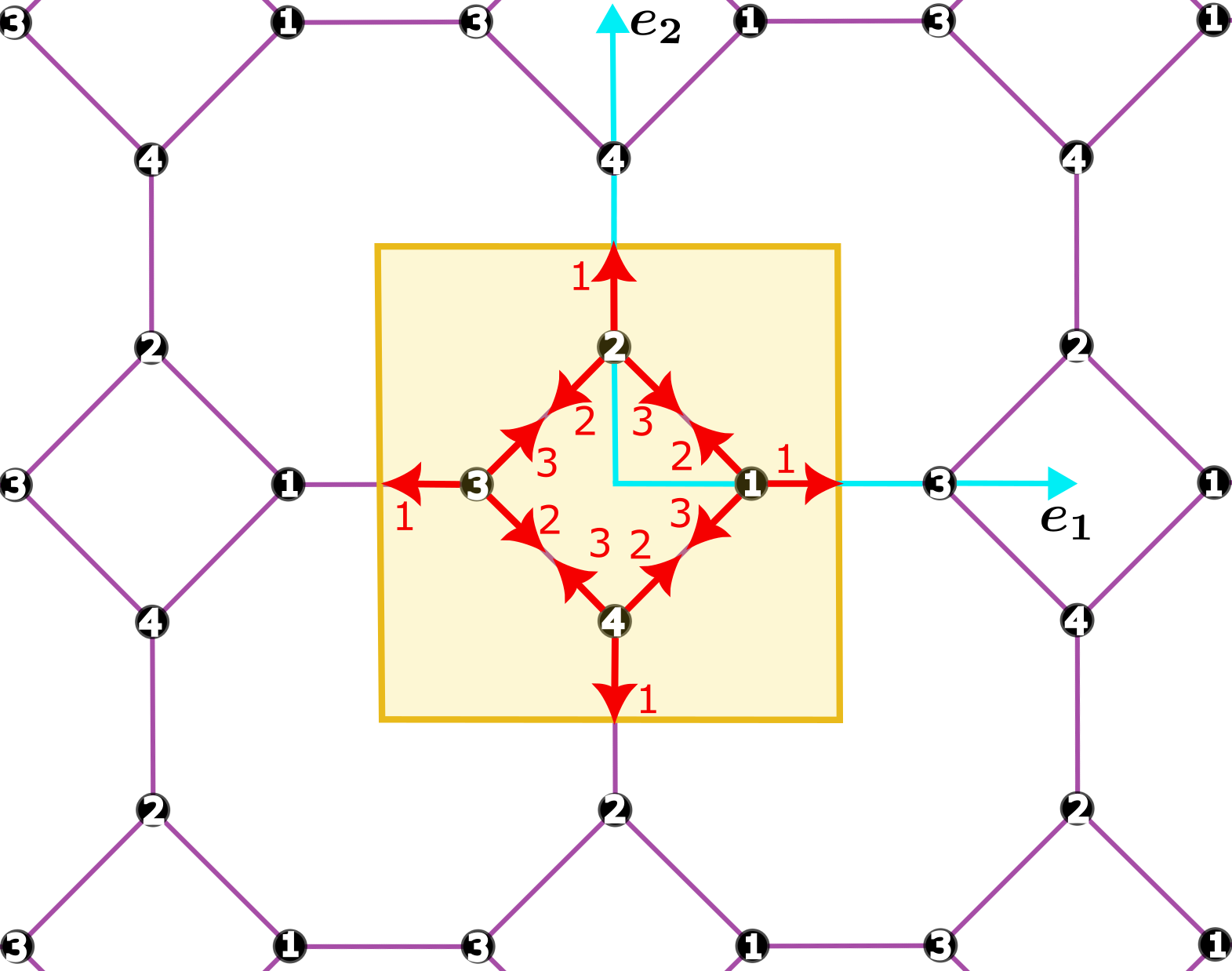}
     \label{FigCaVOLabels}
\end{figure}\\
The matrix $W(\vec{k})$ has the nonzero components
\begin{align}
  W(\vec{k}) &=\begin{pmatrix} 0 & * & * & * \\ * & 0 & * & *  \\ * & * & 0  & * \\ * & * & * & 0 \end{pmatrix}.
\end{align}
For the Bloch matrices at the unit cell sites we have
\begin{align}
  \nonumber B_1(\vec{k}) &= \begin{pmatrix} e^{-\rmi k_1} & & \\ & 1 &  \\ & & 1 \end{pmatrix},\ B_2(\vec{k}) = \begin{pmatrix} e^{-\rmi k_2} & & \\ & 1 &  \\ & & 1 \end{pmatrix},\\
  B_3(\vec{k}) &= \begin{pmatrix} e^{\rmi k_1} & & \\ & 1 &  \\ & & 1 \end{pmatrix},\ B_4(\vec{k}) = \begin{pmatrix} e^{\rmi k_2} & & \\ & 1 &  \\ & & 1 \end{pmatrix},
\end{align}
The edge-connectivity matrices are
\begin{equation}
\begin{aligned}
  E_{1;3} &= E_{2;4} = E_{3;1} = E_{4;2} = \vare_1,\\
  E_{1;2} &= E_{2;3} = E_{3;4}=E_{4;1} =  \vare_2,\\
  E_{1;4} &= E_{2;1} = E_{3;2} = E_{4;3} =   \vare_3.
\end{aligned}
\end{equation}
with
\begin{align}
 \vare_1 = \begin{pmatrix} 1 & & \\ & 0 & \\ & & 0 \end{pmatrix},\ \vare_2 = \begin{pmatrix} 0 & & \\ & 1 & \\ & & 0 \end{pmatrix},\ \vare_3 = \begin{pmatrix} 0 & & \\ & 0 & \\ & & 1 \end{pmatrix}.
\end{align}
Define $g=\rmi$ so that
\begin{align}
  \vec{\theta}_a &= g^{a-1}(1,e^{3\pi\rmi /4},e^{-3\pi\rmi /4})^T.
\end{align}
This yields
\begin{equation}
\begin{aligned}
 \Phi_{13} &=\Phi_{24} = \Phi_{31} = \Phi_{42} = \Phi,\\
  \Phi_{12} &= \Phi_{23}=\Phi_{34}=\Phi_{41} =\Psi,\\
  \Phi_{14} &= \Phi_{21} = \Phi_{32} = \Phi_{43} =\Psi^\dagger
\end{aligned}
\end{equation}
with
\begin{align}
   \Phi &= \Phi^\dagger = \begin{pmatrix} 0 & e^{-\pi\rmi/8} & e^{\pi\rmi/8} \\ e^{\pi \rmi/8} & 0 & e^{-\pi\rmi/4} \\ e^{-\pi\rmi /8} & e^{\pi\rmi/4} & 0 \end{pmatrix},\\
   \Psi &= \begin{pmatrix} e^{\pi\rmi/4} & e^{-3\pi\rmi/8} & e^{-\pi\rmi/8} \\ e^{-\pi\rmi/8} & e^{\pi\rmi/4} & 0 \\ e^{-3\pi\rmi/8} & 1 & e^{\pi\rmi/4} \end{pmatrix}.
\end{align}
We have
\begin{align}
W(\vec{k}) &=  \begin{pmatrix} 0 & \vare_2 B_1\Psi & \vare_1 B_1\Phi & \vare_3 B_1\Psi^\dagger \\ \vare_3 B_2\Psi^\dagger & 0 & \vare_2 B_2\Psi & \vare_1 B_2\Phi \\ \vare_1 B_3\Phi & \vare_3 B_3\Psi^\dagger & 0 &  \vare_2 B_3\Psi \\ \vare_2 B_4\Psi & \vare_1 B_4\Phi & \vare_3 B_4\Psi^\dagger  & 0 \end{pmatrix}.
\end{align}
The characteristic polynomial is
\begin{equation}
\begin{aligned}
 &\mbox{det}(\mathbb{1}- t W(\vec{k})) = 1+2t^4+8t^6+5t^8 \\
  &+2t^3(1-t^4)\vare_\square(\vec{k})-4t^4(1-t^2)^2\cos k_1 \cos k_2.
\end{aligned}
\end{equation}

\subsubsection{Matrices $W_0$ and $\mathcal{B}$}

For $\vec{k}=0$ we find that
\begin{align}
W_0^\dagger W_0 =\begin{pmatrix} \mathcal{B} & 0 & 0 & 0 \\ 0 & \mathcal{B} & 0 & 0 \\ 0 & 0 & \mathcal{B} & 0 \\ 0 & 0 & 0 & \mathcal{B} \end{pmatrix}
\end{align}
with
\begin{align}
 \mathcal{B} = \Phi \vare_1 \Phi +\Psi^\dagger \vare_2\Psi +\Psi \vare_3 \Psi^\dagger.
\end{align}
We have
\begin{equation}
\begin{aligned}
 \det(u\mathbb{1}-W_0^\dagger W_0) &=\Bigl[\det(u\mathbb{1}-\mathcal{B})\Bigr]^{N_{\rm u}} \\
  &= \Bigl[u^2(u-3)^4\Bigr]^2 = u^4(u-3)^8.
\end{aligned}
\end{equation}

\subsubsection{Relation to reference $\Phi_0^2$}

The vectors $\vec{\theta}_a$ are related to the $q=3$ reference $\vec{\theta}_0=(1,e^{2\pi\rmi/3},e^{-2\pi\rmi/3})^T$ according to
\begin{align}
 \vec{\theta}_a = U_a \vec{\theta}_0
\end{align}
with
\begin{equation}
\begin{aligned}
 U_1 &= \text{diag}(1,e^{\rmi\pi/12},e^{-\rmi\pi/12}),\\
 U_2 &= \rmi U_1,\ U_3 = -U_1,\ U_4 = -\rmi U_1.
\end{aligned}
\end{equation}
We define $V_a=\sqrt{U_a}$ so that
\begin{equation}
\begin{aligned}
 V_1 &= \text{diag}(1,e^{\pi\rmi/24},e^{-\rmi\pi/24}),\\
 V_2 &= \text{diag}(e^{\pi\rmi/4},e^{7\pi\rmi/24},e^{5\pi\rmi/24})= e^{\rmi\pi/4}V_1,\\
 V_3 &= \text{diag}(\rmi,-e^{-11\pi\rmi/24},e^{11\pi\rmi/24})= \rmi V_1,\\
 V_4 &= \text{diag}(e^{-\pi\rmi/4},e^{-5\pi\rmi/24},e^{-7\rmi\pi/24})= e^{-\rmi\pi/4} V_1.
\end{aligned}
\end{equation}
Consider the matrices $M_k^{(a)}=\Phi_{ka}^\dagger E_{k;a}$ such that $\mathcal{B}_a= \sum_{k=1}^{N_{\rm u}} M_k^{(a)} [M_k^{(a)}]^\dagger$. The nonvanishing matrices are
\begin{widetext}
\begin{align}
 a=1:\ M_2^{(a)} &= (\vec{0}\ \vec{0}\ \vec{Y}_2),\ M_3^{(a)} = (\vec{Y}_1\ \vec{0}\ \vec{0}),\ M_4^{(a)} = (\vec{0}\ \vec{Y}_3\ \vec{0}),\\
 a=2:\ M_1^{(a)} &= (\vec{0}\ \vec{Y}_3\ \vec{0}),\ M_3^{(a)} = (\vec{0}\ \vec{0}\ \vec{Y}_2),\ M_4^{(a)} = (\vec{Y}_1\ \vec{0}\ \vec{0}),\\
 a=3:\ M_1^{(a)} &= (\vec{Y}_1\ \vec{0}\ \vec{0}),\ M_2^{(a)} = (\vec{0}\ \vec{Y}_3\ \vec{0}),\ M_4^{(a)} = (\vec{0}\ \vec{0}\ \vec{Y}_2),\\
 a=4:\ M_1^{(a)} &= (\vec{0}\ \vec{0}\ \vec{Y}_2),\ M_2^{(a)} = (\vec{Y}_1\ \vec{0}\ \vec{0}),\ M_3^{(a)} = (\vec{0}\ \vec{Y}_3\ \vec{0}).
\end{align}
\end{widetext}
For each $a$, up to re-ordering, these matrices have the same three nonvanishing columns $\vec{Y}_\mu^{(a)} = \vec{Y}_\mu$ given by
\begin{align}
 \vec{Y}_1 = \begin{pmatrix} 0 \\ e^{\rmi \pi/8} \\ e^{-\rmi \pi/8} \end{pmatrix},\ \vec{Y}_2 = \begin{pmatrix} e^{-\rmi \pi/8} \\ 0 \\ e^{\rmi\pi/4} \end{pmatrix},\ \vec{Y}_3 = \begin{pmatrix} e^{\rmi\pi/8} \\ e^{-\rmi \pi/4} \\ 0 \end{pmatrix}.
\end{align}
We have
\begin{align}
 a=1,\dots,4:\ \mathcal{B} = \sum_{k=1}^{N_{\rm u}} M_k^{(a)}[M_k^{(a)}]^\dagger = \sum_{\mu=1}^3 \vec{Y}_\mu \vec{Y}_\mu^\dagger.
\end{align}
The vectors $\vec{Y}_\mu$ satisfy
\begin{align}
 \nonumber a=1,2,3,4:\ &Y_\mu^{(a)}(\alpha) = \begin{cases} \sqrt{-\theta_a(\mu)\theta_a^*(\alpha)} & \alpha\neq \mu \\ 0 & \alpha = \mu\end{cases} \\
  &= \begin{cases} (V_a)_{\mu\mu}(V_a^*)_{\alpha\alpha} \sqrt{-\theta_0(\mu)\theta_0^*(\alpha)} & \alpha\neq \mu \\ 0 & \alpha = \mu\end{cases},
\end{align}
which is equivalent to
\begin{align}
 a=1,\dots,4:\ \vec{Y}_\mu = (V_a)_{\mu\mu} V_a^\dagger \vec{X}_\mu.
\end{align}
Thus we have
\begin{align}
 a=1,\dots,4:\ \mathcal{B}_a = V_a^\dagger \Bigl(\sum_{\mu=1}^q \vec{X}_\mu \vec{X}_\mu^\dagger\Bigr)V_a = V_a^\dagger \Phi_0^2 V_a.
\end{align}
This shows that $\mathcal{B}$ is unitary equivalent to $\Phi_0^2$. We conclude that
\begin{equation}
\begin{aligned}
 \det(u \mathbb{1} - W_0^\dagger W_0) &= \Bigl[\det(u \mathbb{1}_3-\mathcal{B})\Bigr]^{N_{\rm u}} \\
 &= \Bigl[\det(u \mathbb{1}_3-\Phi_0^2)\Bigr]^{N_{\rm u}} \\
 &= P_3(u)^4 = u^4(u-3)^8.
\end{aligned}
\end{equation}

\section{$T_{\rm c}$ data}
\label{Tcdata}
In this section, we collect the data for $T_{\rm c}$ for the $k$-uniform lattices with $k\leq 3$, their dual lattices, and a selection of other lattices (Lieb-like lattices and representatives with fixed $q_{\rm max}$). This data constitutes the basis for Fig. 1 in the main text. 

The 1-uniform lattices are the eleven Archimedean lattices, and their dual lattices are the Laves lattices. The Archimedean and Laves lattices represent particularly important lattices with some of the largest values of $T_{\rm c}$, see Tab. \ref{tcSome}. Their Kac--Ward matrices, free energy densities, and critical temperatures are computed in the next section. For the 20 and 61 lattices with $k=2$ and $k=3$, respectively, we follow the labeling of Galebach \cite{GalebachWebpage} and Sanchez et al. \cite{Sanchez1,Sanchez2,SanchezWebpage} given by t$k$.$xxx$, where $xxx\in\{1,\dots,\mathcal{F}_k\}$ with $\mathcal{F}_1=11,$ $\mathcal{F}_2=20$, $\mathcal{F}_3=61$, adding ``dual" for the dual lattices. Alternative labelings for $k$-uniform lattices, like the Cundy--Rollet or GomJau--Hogg notation, can be found in Ref. \cite{GomJau-Hogg}.

\renewcommand{\arraystretch}{1.5}
\begin{table}[t!]
\begin{tabular}{|c|c|c|c|c|}
\hline
 \multicolumn{5}{|c|}{Archimedean Lattices}\\
 \hline\hline
 \ Lattice \ & \ $q_{\rm max}$ \ & \ $\bar{q}$ \ & \ $t_{\rm c}$ \ & \ $T_{\rm c}/J$ \  \\
\hline\hline
 \ Triangular \ & 6  & 6  &  0.26795  &  3.64095 	\\
\hline
 \ SrCuBO \ & 5  & 5  & \ 0.32902 \  &  2.9263 \\
\hline
 \ Trellis \ & 5  & 5  & $1/3$   &  2.88542 \\
\hline 
 \ Maple-Leaf \ & 5  & 5  & 0.34430   &  2.7858 \\
\hline
\ Square \ & 4  & 4  & 0.41421   &  2.26921 	\\
\hline
\ Kagome \ & 4  & 4  & 0.43542   &   2.14332 \\
\hline
 \ Ruby \ & 4  & 4  & 0.43542   &   2.14332 \\
\hline
 \ Honeycomb \ & 3  & 3  & \ 0.57735 \  &  \ 1.51865 \  	\\
\hline
 \ CaVO \ & 3  & 3  & 0.60123   & 1.4387 \\
\hline
 \ SHD \ & 3  & 3  & 0.61661   &  1.38982 \\
\hline
 \ Star \ & 3  & 3  & 0.67070   &  1.23151 \\
\hline\hline
 \multicolumn{5}{|c|}{Laves Lattices}\\
 \hline\hline
 \ Lattice \ & \ $q_{\rm max}$ \ & \ $\bar{q}$ \ & \ $t_{\rm c}$ \ & \ $T_{\rm c}/J$ \  \\
\hline\hline
 Laves-Star \ & 12  & 6  & 0.19711   &  5.00704 	\\
\hline
 Laves-SHD \ & 12  & 6  & 0.23716   &  4.13629 	\\
\hline
 Laves-CaVO \ & 8  & 6  & 0.24904   &  3.93102 	\\
\hline
Laves-Kagome \ & 6  & 4  & 0.39332   &  2.40546 	\\
\hline
 Laves-Ruby \ & 6  & 4  & 0.39332  &  2.40546 	\\
\hline
 \ Laves-Maple-Leaf \ & 6  & 3.33  & 0.48777   &  1.87572 	\\
\hline
 Laves-Trellis \ & 4  & \ 3.33 \  & $1/2$   &  1.82048 	\\
\hline
 Laves-SrCuBO \ & 4  & 3.33  & \ 0.50486  \ &  1.79917	\\
\hline
\end{tabular}
    \caption{Critical temperatures for Archimedean and their dual Laves lattices. The data is ordered according to decreasing values of $T_{\rm c}/J$, which coincides here with decreasing values of $q_{\rm max}$.}
\label{tcSome}
\end{table}
\renewcommand{\arraystretch}{1}

The data of critical temperatures for the $k$-uniform lattices together with their dual lattices for $k\geq 3$ is presented in Tabs. \ref{BigTab1} and \ref{BigTab2}, sorted according to $T_{\rm c}/J$ for each value of $q_{\rm max}$. The data for the individual 1-, 2-, 3-uniform lattices and their duals is contained in the SM \cite{SI}. The critical values $t_{\rm c}$ and $T_{\rm c}$ for $k\in\{2,3\}$ are taken from Ref. \cite{Portillo}. To determine the critical values of the dual lattice, $t_{\rm c}^{(\rm d)}$ and $T_{\rm c}^{(\rm d)}$, we use the exact formula
\begin{align}
 \label{data1} t_{\rm c}^{(\rm d)} = \frac{1-t_{\rm c}}{1+t_{\rm c}}.
\end{align}
We obtain the value of $q_{\rm max}$ for each $k$-uniform lattice from the plot of the tessellation given by \cite{SanchezWebpage}. Furthermore, the value of $q_{\rm max}^{(\rm d)}$ of the dual lattice, which corresponds to the largest number of sides of any polygon in the original lattice, can equally be inferred from these plots. In order to examine how the critical temperature $T_{\rm c}$ for a fixed value of $q_{\rm max}$ is correlated to the value of the average coordination number $\bar{q}$, we arrange all lattices considered in Tabs. \ref{BigTab1} and \ref{BigTab2} according to groups of $q_{\rm max}$ and with decreasing $T_{\rm c}$. In this table, we also include the $T_{\rm c}$-bounds, some Lieb-like lattices and representatives with $q_{\rm max}=7,9,10,11$. We observe that a small value of $\bar{q}$ typically implies that $T_{\rm c}$ is lower. However, no strict rule can be formulated. For instance, the 2-uniform lattices t2.018 and t2.019 both have $q_{\rm max}=6$, with
\begin{equation}\label{data1b}
\begin{aligned}
 \text{t2.018}:\ &\bar{q}=5.14,\ T_{\rm c}/J=3.02267,\\
 \text{t2.019}:\ &\bar{q}=5.25,\ T_{\rm c}/J=2.98574.
\end{aligned}
\end{equation}
Clearly, t2.018 has a larger critical temperature despite having a smaller value of $\bar{q}$.

The values of $q_{\rm max}$ and $q_{\rm max}^{(\rm d)}$ are sufficient to test the validity of the exact bound on $T_{\rm c}$. To examine the influence of the average coordination, $\bar{q}$, we note that the values of $\bar{q}$ for all $k$-uniform tessellations with $k\leq 6$ have been determined in Ref. \cite{Portillo}. The average coordination number of the dual lattice, $\bar{q}^{(\rm d)}$, follows from the exact formula
\begin{align}
 \label{data2} \bar{q}^{(\rm d)} = \frac{2\bar{q}}{\bar{q}-2},
\end{align}
where $\bar{q}$ is the average coordination number of the original lattice. Note that this is equivalent to
\begin{align}
 \label{data3} (\bar{q}-2)(\bar{q}^{(\rm d)}-2) =4\ \Leftrightarrow\ \frac{1}{\bar{q}}+\frac{1}{\bar{q}^{(\rm d)}}=\frac{1}{2}.
\end{align}
To derive this formula, note that the unit cell of the original lattice can be drawn on the surface of a torus, with genus $g=1$. The corresponding Euler characteristic is given by
\begin{equation}
   \label{data4}  V-E+F=2(1-g)=0,
\end{equation}
where $V=N_{\rm u}$ is the number of unit cell sites, $E=Q/2$ is the number of edges, and $F=F_{\rm u}$ is the number of faces of this pattern. By the construction of the dual lattice, the number of edges of the unit cell of the dual lattices $E^{(\rm d)}=E$, while the number of sites in the unit cell of the dual lattice, $N_{\rm u}^{(\rm d)}$, satisfies $N_{\rm u}^{(\rm d)}=F_{\rm u}$. We arrive at
\begin{align}
 \label{NuNu} N_{\rm u}+ N_{\rm u}^{(\rm d)} = \frac{Q}{2}.
\end{align}
The average coordination numbers of the original and dual lattice are given by
\begin{align}
 \label{data5} \bar{q} &= \frac{2E}{N_{\rm u}} = \frac{Q}{N_{\rm u}},\\
 \label{data6} \bar{q}^{(\rm d)} &= \frac{2E^{(\rm d)}}{N_{\rm u}^{(d)}}= \frac{Q}{N_{\rm u}^{(d)}},
\end{align}
respectively. Dividing Eq. (\ref{NuNu}) by $Q$ thus yields the claimed result.

Note that $\bar{q}\leq 6$ for any periodic tessellation of the plane. Indeed, when embedding a finite number of unit cells onto a torus, we have for the number of its vertices, $N$, edges, $E$, and faces, $F$, that
\begin{align}
 \bar{q} N = 2E = \bar{p}F,\ N-E+F=0,
\end{align}
with $\bar{q}$ and $\bar{p}$ the average coordination number and sides of each face. We have $\bar{p}\geq 3$, because each polygon needs to have at least three sides. Thus we have
\begin{align}
 \nonumber 2E &= \bar{p} F \geq 3 F = 3(E-N)\\
\nonumber \implies 3N &\geq E\\
\implies \bar{q} &= \frac{2E}{N} \leq 6.
\end{align}

The dual lattice of a $\{p,q\}$ lattice is the $\{q,p\}$ lattice. In the Euclidean plane, the only $\{p,q\}$ lattices are the Triangular ($\{3,6\})$, Square ($\{4,4\}$), and Honeycomb lattices ($\{6,3\}$). They satisfy
\begin{align}
 \label{data7}  (p-2)(q-2)=4\ \Leftrightarrow\ \frac{1}{p}+\frac{1}{q} = \frac{1}{2}.
\end{align} 
It is easy to show that if a $\{p,q\}$ lattice satisfies the bound on $t_{\rm c}$, then so does the dual lattice. Indeed, since $q_{\rm max}=q$ for a $\{p,q\}$ lattice, the bound is saturated when
\begin{align}
 \label{data8} t_{\rm c} = \tan\Bigl(\frac{\pi}{2q}\Bigr).
\end{align}
For the dual lattice we then have
\begin{align}
 \label{data9} t_{\rm c}^{(\rm d)} = \frac{1-t_{\rm c}}{1+t_{\rm c}}.
\end{align} 
But since $q^{(\rm d)}_{\rm max}=p$, we have
\begin{align}
  \nonumber \tan\Bigl(\frac{\pi}{2q_{\rm max}^{(\rm d)}}\Bigr) &= \tan\Bigl(\frac{\pi}{2p}\Bigr)= \tan\Bigl(\frac{\pi}{4}-\frac{\pi}{2q}\Bigr) \\\nonumber &= \frac{1-\tan({\frac{\pi}{2q})}}{1+\tan(\frac{\pi}{2q})}= \frac{1-t_{\rm c}}{1+t_{\rm c}} \\ \label{data10}  &= t_{\rm c}^{(\rm d)},
\end{align}
which shows that the dual lattice also saturates the bound. Here we used
\begin{align}
 \label{data11} \tan\Bigl(\frac{\pi}{4}-x\Bigr) = \frac{1-\tan x}{1+\tan x}.
\end{align}

\renewcommand{\arraystretch}{1.5}
\begin{table*}[t!]
\begin{tabular}{|c|c|c|c||c|c|c|c||c|c|c|c|}
\hline
 \multicolumn{12}{|c|}{$T_{\rm c}$ vs. $q_{\rm max}$ and $\bar{q}$}\\
 \hline\hline
 Lattice & \ $q_{\rm max}$ \ & $\bar{q}$ & $T_{\rm c}/J$ & Lattice & \ $q_{\rm max}$ \ & $\bar{q}$ & $T_{\rm c}/J$ & Lattice & \ $q_{\rm max}$ \ & $\bar{q}$ & $T_{\rm c}/J$ \\
\hline
\multicolumn{4}{|c||}{$q_{\rm max}=24$} & t3.049 & 6 & 5.60 & 3.31774 & t3.029dual & 6 & 4 & 2.40546 \\
\hline
\textbf{Bound} & \textbf{24} &  & \ \textbf{15.2352} \ & t3.059 & 6 & 5.60 & \ 3.28050 \ & t3.032dual & 6 & 4 & \ 2.38534 \ \\
\hline
 \ Compass-Rose \ & 24 & 6 & 6.4919 & t2.015 & 6 & 5.50 & 3.24161 & t3.026dual & 6 & 4 & 2.38160 \\
\hline
\multicolumn{4}{|c||}{$q_{\rm max}=12$} & t2.020 & 6 & 5.50 & 3.19352 & t3.030dual & 6 & 4 & 2.38091 \\
\hline
\textbf{Bound} & \textbf{12} &  & \textbf{7.55167} & t3.058 & 6 & 5.45 & 3.15438 & t2.005dual & 6 & 4 & 2.37704 \\
\hline
Laves-Star & 12 & 6 & 5.00705 & t2.014 & 6 & 5.33 & 3.11852 & t2.007dual & 6 & 4 & 2.37279 \\
\hline
Laves-SHD & 12 & 6 & 4.13629 & t3.052 & 6 & 5.33 & 3.11852 & t3.028dual & 6 & 4 & 2.36929 \\
\hline
t2.002dual & 12 & 4.67 & 3.78874 & t3.021 & 6 & 5.20 & 3.02327 & t2.006dual & 6 & 4 & 2.36585 \\
\hline
t2.001dual & 12 & 5 & 3.44942 & t3.051 & 6 & 5.20 & 3.02327 & t3.048 & 6 & 4.57 & 2.35965 \\
\hline
t3.006dual & 12 & 4.36 & 3.30674 & t2.018 & 6 & 5.14 & 3.02267 & t3.022dual & 6 & 4 & 2.35515 \\
\hline
t3.007dual & 12 & 4.29 & 3.22177 & t2.019 & 6 & 5.25 & 2.98574 & t3.018dual & 6 & 4 & 2.34040 \\
\hline
t3.008dual & 12 & 4.29 & 3.14440 & t3.057 & 6 & 5.08 & 2.97755 & t3.017dual & 6 & 4 & 2.33958 \\
\hline
t3.009dual & 12 & 4.22 & 3.09617 & t3.054 & 6 & 5.08 & 2.96503 & t2.010dual & 6 & 3.75 & 2.33920 \\
\hline
t3.005dual & 12 & 4.50 & 2.89276 & t3.061 & 6 & 5.14 & 2.89594 & t3.012dual & 6 & 4 & 2.32402 \\
\hline
t2.013dual & 12 & 3.75 & 2.55978 & t3.015 & 6 & 5 & 2.88539 & t3.011dual & 6 & 4 & 2.32391 \\
\hline
t3.035dual & 12 & 3.82 & 2.40309 & t3.020 & 6 & 5 & 2.88539 & t3.027dual & 6 & 3.82 & 2.29967 \\
\hline
t3.048dual & 12 & 3.56 & 2.18343 & t3.044 & 6 & 5.14 & 2.88539 & t2.010 & 6 & 4.29 & 2.20203 \\
\hline
t3.047dual & 12 & 3.45 & 2.11463 & t3.002dual & 6 & 4.50 & 2.85309 & t3.025dual & 6 & 3.78 & 2.19710 \\
\hline
\multicolumn{4}{|c||}{$q_{\rm max}=11$} & t3.001dual & 6 & 4.40 & 2.75758 & t2.012dual & 6 & 3.60 & 2.16262 \\
\hline
\textbf{Bound} & \textbf{11} &  & \textbf{6.90696} & t3.014 & 6 & 4.80 & 2.75304 & t3.034dual & 6 & 3.65 & 2.11811 \\
\hline
Rep-11 & 11 & 5.67 & 4.45218 & t3.037 & 6 & 4.80 & 2.74330 & t3.024dual & 6 & 3.65 & 2.10808 \\
\hline
\multicolumn{4}{|c||}{$q_{\rm max}=10$} & t3.041 & 6 & 5 & 2.73428 & t3.033dual & 6 & 3.60 & 2.08791 \\
\hline
\textbf{Bound} & \textbf{10} &  & \textbf{6.26060} & t3.036 & 6 & 4.80 & 2.69714 & t2.008dual & 6 & 3.60 & 2.08020 \\
\hline
Rep-10 & 10 & 6 & 4.10901 & t3.043 & 6 & 4.80 & 2.62159 & t3.042dual & 6 & 3.53 & 2.07789 \\
\hline
\multicolumn{4}{|c||}{$q_{\rm max}=9$} & t3.004dual & 6 & 4 & 2.61316 & t2.009dual & 6 & 3.60 & 2.04869 \\
\hline
\textbf{Bound} & \textbf{9} &  & \textbf{5.61201} & t3.038 & 6 & 4.80 & 2.59669 & t2.013 & 6 & 4.29 & 2.02213 \\
\hline
Rep-9 & 9 & 5.60 & 3.63901 & t3.040 & 6 & 4.80 & 2.58840 & t3.040dual & 6 & 3.43 & 2.00175 \\
\hline
\multicolumn{4}{|c||}{$q_{\rm max}=8$} & t3.003dual & 6 & 4 & 2.51475 & t3.038dual & 6 & 3.43 & 1.99595 \\
\hline
\textbf{Bound} & \textbf{8} &  & \textbf{4.96032} & t3.042 & 6 & 4.62 & 2.48560 & t3.004 & 6 & 4 & 1.98457 \\
\hline
Laves-CaVO & 8 & 6 & 3.93100 & t2.011dual & 6 & 4 & 2.47816 & t3.043dual & 6 & 3.43 & 1.97883 \\
\hline
\multicolumn{4}{|c||}{$q_{\rm max}=7$}   & t3.024 & 6 & 4.42 & 2.44774 & t3.045dual & 6 & 3.43 & 1.97017 \\
\hline
\textbf{Bound} & \textbf{7} &  & \textbf{4.30412} & t3.047 & 6 & 4.75 & 2.43973 & t3.046dual & 6 & 3.43 & 1.97017 \\
\hline
Rep-7 & 7 & 5 & 3.28204 & t3.039dual & 6 & 4 & 2.43961 & t3.036dual & 6 & 3.43 & 1.92939 \\
\hline
\multicolumn{4}{|c||}{$q_{\rm max}=6$} & t3.034 & 6 & \ 4.42 \ & 2.43550 & t3.041dual & 6 & 3.33 & 1.90637 \\
\hline
 \ \textbf{Triangular} \ & \textbf{6} & \textbf{6} & \textbf{3.64096} & t3.031dual & 6 & 4 & 2.41037 & t3.037dual & 6 & 3.43 & 1.90091 \\
\hline
t3.050 & 6 & 5.67 & 3.36949 & \ Laves-Kagome \ & 6 & 4 & 2.40546 & \ Laves-Maple-Leaf \ & 6 & 3.33 & 1.87572 \\
\hline
t3.060 & 6 & \ 5.67 \ & 3.33919 & Laves-Ruby & 6 & 4 & 2.40546 & t3.056dual & 6 & \ 3.33 \ & 1.83009 \\
\hline
\end{tabular}
    \caption{Critical temperatures  of all lattices considered, grouped according to $q_{\rm max}$. We observe that there is a trend that low $\bar{q}$ implies a low $T_{\rm c}$ within each group, but $\bar{q}_1<\bar{q}_2 \nRightarrow T_{\rm c,1}<T_{\rm c,2}$ (see the counterexample in Eq. (\ref{data1b})). Rep-$q_{\rm max}$ are the representative lattices with $q_{\rm max}=7,9,10,11$ constructed in Sec. S4 below.}
\label{BigTab1}
\end{table*}
\renewcommand{\arraystretch}{1}

\renewcommand{\arraystretch}{1.5}
\begin{table*}[t!]
\begin{tabular}{|c|c|c|c||c|c|c|c||c|c|c|c|}
\hline
 \multicolumn{12}{|c|}{$T_{\rm c}$ vs. $q_{\rm max}$ and $\bar{q}$}\\
 \hline\hline
 Lattice & \ $q_{\rm max}$ \ & $\bar{q}$ & $T_{\rm c}/J$ & Lattice & \ $q_{\rm max}$ \ & $\bar{q}$ & $T_{\rm c}/J$ & Lattice & \ $q_{\rm max}$ \ & $\bar{q}$ & $T_{\rm c}/J$ \\
\hline\hline
\multicolumn{4}{|c||}{$q_{\rm max}=6$} & \multicolumn{4}{|c||}{$q_{\rm max}=4$} & t3.055dual & 4 & 3.33 & 1.80477 \\
\hline
t3.044dual & 6 & \ 3.27 \ & 1.82048 & \textbf{Square} & \textbf{4} & \textbf{4} & \ \textbf{2.26919} \ & t2.016dual & 4 & \ 3.33 \ & 1.80439 \\
\hline
t3.061dual & 6 & 3.27 & 1.81490 & t3.011 & 4 & 4 & 2.21622 & \ Laves-SrCuBO \ & 4 & 3.33 & 1.79917 \\
\hline
t2.019dual & 6 & 3.23 & 1.76945 & t3.012 & 4 & 4 & 2.21611 & t3.054dual & 4 & 3.30 & 1.77963 \\
\hline
t3.058dual & 6 & 3.16 & 1.69272 & t3.017 & 4 & 4 & 2.20167 & t3.057dual & 4 & 3.30 & 1.77346 \\
\hline
t2.020dual & 6 & 3.14 & 1.67633 & t3.018 & 4 & 4 & 2.20092 & t2.018dual & 4 & 3.27 & 1.75174 \\
\hline
t3.059dual & 6 & 3.11 & 1.64161 & t3.022 & 4 & 4 & 2.18749 & t3.021dual & 4 & 3.25 & 1.75146 \\
\hline
t3.006 & 6 & 3.69 & 1.63157 & t2.006 & 4 & 4 & 2.17788 & t3.051dual & 4 & 3.25 & 1.75146 \\
\hline
t3.060dual & 6 & 3.09 & 1.61942 & t3.028 & 4 & 4 & 2.17481 & t3.009 & 4 & 3.80 & 1.71804 \\
\hline
\multicolumn{4}{|c||}{$q_{\rm max}=5$} & \ House (t2.007) \ & 4 & 4 & 2.17171 & t2.014dual & 4 & 3.20 & 1.70818 \\
\hline
\textbf{Bound} & \textbf{5} &  & \ \textbf{2.96615} \ & t2.005 & 4 & 4 & 2.16795 & t3.052dual & 4 & 3.20 & 1.70818 \\
\hline
SrCuBO & 5 & 5 & 2.92626 & t3.030 & 4 & 4 & 2.16455 & t3.008 & 4 & 3.75 & 1.69698 \\
\hline
t2.016 & 5 & 5 & 2.91611 & t3.026 & 4 & 4 & 2.16395 & t3.007 & 4 & 3.75 & 1.66481 \\
\hline
t3.055 & 5 & 5 & 2.91537 & t3.032 & 4 & 4 & 2.16067 & t2.015dual & 4 & 3.14 & 1.65685 \\
\hline
t2.017 & 5 & 5 & 2.91019 & Kagome & 4 & 4 & 2.14332 & t3.049dual & 4 & 3.11 & 1.62742 \\
\hline
t3.053 & 5 & 5 & 2.90236 & Ruby & 4 & 4 & 2.14332 & t3.050dual & 4 & 3.09 & 1.60832 \\
\hline
Trellis & 5 & 5 & 2.88539 & t3.029 & 4 & 4 & 2.14332 & t2.001 & 4 & 3.33 & 1.58017 \\
\hline
t3.056 & 5 & 5 & 2.86742 & t3.031 & 4 & 4 & 2.13914 & t2.002 & 4 & 3.50 & 1.47614 \\
\hline
t3.023 & 5 & 4.89 & 2.83467 & t3.039 & 4 & 4 & 2.11473 & Lieb & 4 & 2.67 & 1.30841 \\
\hline
 \ Maple-Leaf \ & 5 & 5 & 2.78584 & t3.010dual & 4 & 3.71 & 2.09578 & Lieb-2 & 4 & 2.40 & 1.03886 \\
\hline
t3.019 & 5 & 4.80 & 2.75304 & t2.011 & 4 & 4 & 2.08372 & \multicolumn{4}{|c|}{$q_{\rm max}=3$} \\
\hline
t2.004 & 5 & 4.67 & 2.66764 & t3.016dual & 4 & 3.67 & 2.06443 & \textbf{Honeycomb} & \textbf{3} & \textbf{3} & \ \textbf{1.51865} \ \\
\hline
t3.013 & 5 & 4.67 & 2.66764 & t2.003dual & 4 & 3.60 & 2.01922 & CaVO & 3 & 3 & 1.43870 \\
\hline
t3.045 & 5 & 4.80 & 2.63441 & t2.004dual & 4 & 3.50 & 1.94826 & SHD & 3 & 3 & 1.38983 \\
\hline
t3.046 & 5 & 4.80 & 2.63441 & t3.013dual & 4 & 3.50 & 1.94826 & Star & 3 & 3 & 1.23151 \\
\hline
t2.003 & 5 & 4.50 & 2.56381 & t3.014dual & 4 & 3.43 & 1.89505 &  &  &  &  \\
\hline
t2.009 & 5 & 4.50 & 2.52372 & t3.019dual & 4 & 3.43 & 1.89505 &  &  &  &  \\
\hline
t3.016 & 5 & 4.40 & 2.50297 & t3.001 & 4 & \ 3.67 \ & 1.89234 &  &  &  &  \\
\hline
t2.008 & 5 & 4.50 & 2.48264 & t3.023dual & 4 & 3.38 & 1.84801 &  &  &  &  \\
\hline
t3.033 & 5 & 4.50 & 2.47285 & t3.002 & 4 & 3.60 & 1.83787 &  &  &  &  \\
\hline
t3.010 & 5 & 4.33 & 2.46298 & \ Laves-Trellis \ & 4 & 3.33 & 1.82048 &  &  &  &  \\
\hline
t2.012 & 5 & 4.50 & 2.38312 & t3.015dual & 4 & 3.33 & 1.82048 &  &  &  &  \\
\hline
t3.025 & 5 & 4.25 & 2.34456 & t3.020dual & 4 & 3.33 & 1.82048 &  &  &  &  \\
\hline
t3.027 & 5 & 4.20 & 2.23925 & t3.005 & 4 & 3.60 & 1.81658 &  &  &  &  \\
\hline
t3.035 & 5 & 4.20 & 2.14534 & t3.053dual & 4 & 3.33 & 1.81154 &  &  &  &  \\
\hline
t3.003 & 5 & 4 & 2.05546 & t2.017dual & 4 & 3.33 & 1.80746 &  &  &  &  \\
\hline
\end{tabular}
    \caption{Continuation of Tab. \ref{BigTab1}. For the Lieb-like lattices Lieb-$m$, we only show the critical temperatures for $m=1,2$. The remaining values for $m\geq 2$ shown in Fig. 1 of the main text follow from $t_{{\rm c},m}=(\sqrt{2}-1)^{1/(m+1)}$. We further refer to Lieb-1 simply as the Lieb lattice.}
\label{BigTab2}
\end{table*}
\renewcommand{\arraystretch}{1}

\clearpage

\end{appendix}

\bibliography{refs_ising}

\clearpage
\begin{widetext}

\section*{Supplemental Material for ``Exact critical-temperature bounds for two-dimensional Ising models''}
\label{SI}

\noindent This Supplemental Material contains explicit computations of $W(\vec{k})$ and $T_{\rm c}$ for 25 lattices. For the Brickwall, tilted Square, Ruby, House, Trellis, Triangular, and Laves-CaVO lattices we also make the connection to the reference matrix $\Phi_0^2$ explicit. We further include the data for $T_{\rm c}/J$ for all $k$-uniform lattices and their duals for $k\leq 3$, grouped according to $k$.

\subsection{Brickwork lattice ($q=3$)}

\subsubsection{Matrix $W(\vec{k})$} The Brickwork lattice is topologically equivalent to the Honeycomb lattice. It has $N_{\rm u}=2$ sites in the unit cell and coordination number $q=2$. We label the unit cell sites $a\in\{1,2\}$ and edges $\mu\in\{1,2,3\}$ according to the following schematic:
\begin{figure}[h!]
    \includegraphics[width=8cm]{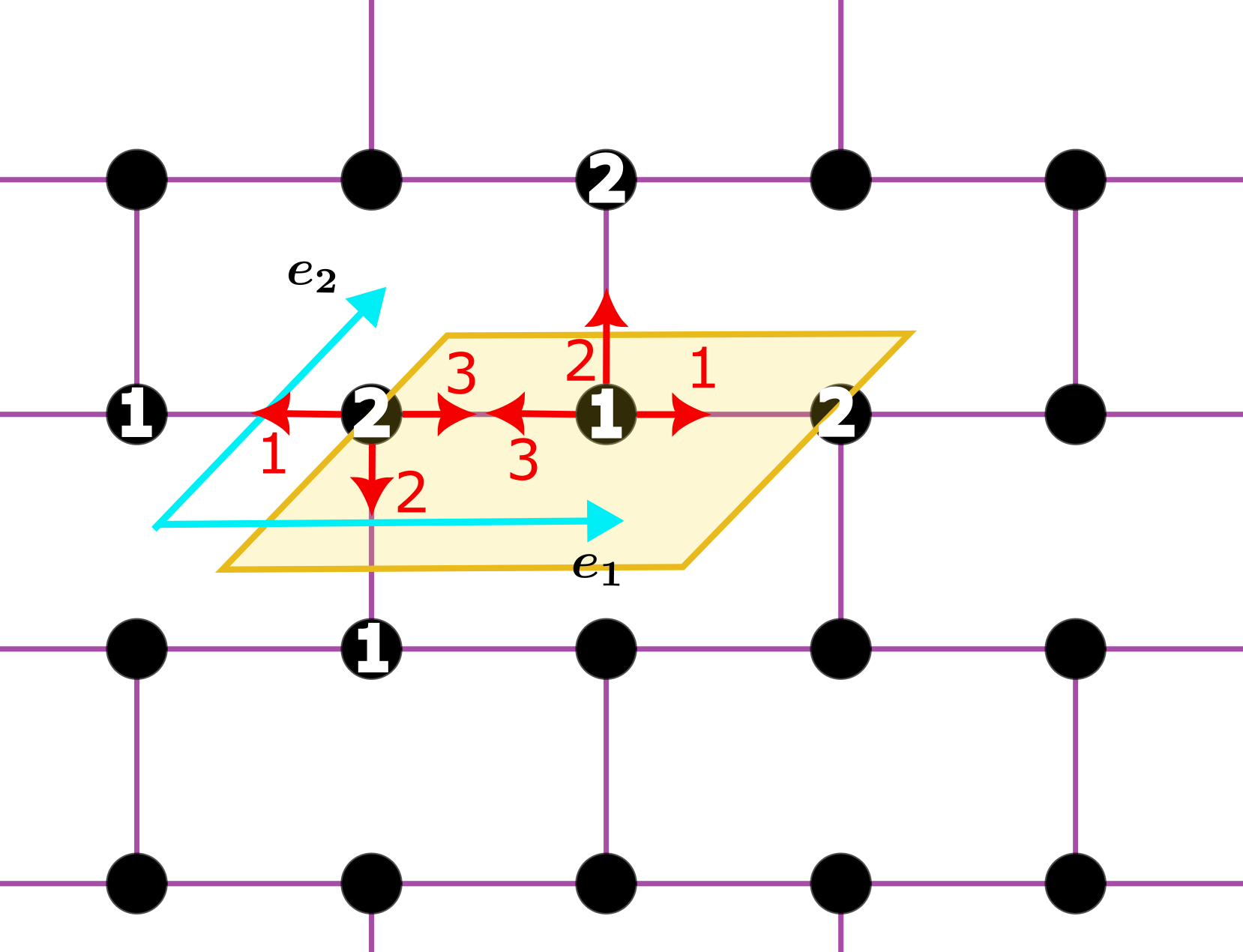}
     \label{FigBrickworkLabels}
\end{figure}\\
Our labeling leads to identical expressions for $B_{1,2}(\vec{k})$ and $E_{1;2}=E_{2;1}=\mathbb{1}_3$ as for the Honeycomb lattice. However, for the angles we now have
\begin{align}
 \vec{\theta}_1 &= (1,\rmi ,-1)^T,\\
  \vec{\theta}_2&= -\vec{\theta}_1 = (-1,-\rmi ,1)^T,
\end{align}
so that
\begin{align}
 \Phi_{12} = \Phi_{21}  = \begin{pmatrix} 0 & e^{-\pi\rmi/4} & 1 \\ e^{\pi\rmi/4} & 0 & e^{-\pi\rmi/4} \\ 1 & e^{\pi\rmi/4} & 0 \end{pmatrix} =: \Phi.
\end{align}
Note that $\Phi=\Phi{}^\dagger$ is Hermitian. We have
\begin{align}
 W(\vec{k})=\begin{pmatrix} 0 & B_1(\vec{k})\Phi \\ B_2(\vec{k})\Phi & 0 \end{pmatrix}
\end{align}
and
\begin{align}
 \mbox{det}(\mathbb{1}-tW(\vec{k})) = 1+3t^4+(1-t^2)t^2\vare_\Delta(\vec{k}).
\end{align}
This expression is identical to the Honeycomb lattice one.

\subsubsection{Matrices $W_0$ and $\mathcal{B}$} We have
\begin{align}
 W_0^\dagger W_0 = \begin{pmatrix} \mathcal{B} & 0 \\ 0 & \mathcal{B} \end{pmatrix}
\end{align}
with
\begin{align}
 \mathcal{B}= \Phi^2 = \begin{pmatrix} 2 & e^{\pi\rmi/4} & -\rmi \\ e^{-\pi\rmi/4} & 2 & e^{\pi\rmi/4} \\ \rmi & e^{-\pi\rmi/4} & 2\end{pmatrix}.
\end{align}
The resulting characteristic polynomial of $W_0^\dagger W_0$ is
\begin{align}
 \mbox{det}(u\mathbb{1}-W_0^\dagger W_0) &= \Bigl[\mbox{det}(u\mathbb{1}-\mathcal{B})\Bigr]^{N_{\rm u}}=u^2(u-3)^4.
\end{align}

\subsubsection{Relation to reference $\Phi_0^2$} The relation between $\vec{\theta}_{1,2}$ for the Brickwork and Honeycomb lattices is
\begin{align}
 \vec{\theta}_1^{(\rm BW)} &= U\vec{\theta}_1^{(\rm HC)},\\
 \vec{\theta}_2^{(\rm BW)} &= U\vec{\theta}_2^{(\rm HC)}
\end{align}
with
\begin{align}
 U  &= \begin{pmatrix} e^{\rmi \pi/3} & & \\ & e^{\rmi \pi/6} & \\ & & 1 \end{pmatrix}.
\end{align}
We define $V=\sqrt{U}$ so that
\begin{align}
 V = \begin{pmatrix} e^{\rmi \pi/6} & & \\ & e^{\rmi \pi/12} & \\ & & 1 \end{pmatrix}.
\end{align}
Consider the matrices $M_k^{(a)}=\Phi_{ka}^\dagger E_{k;a}$ for the Brickwork lattice. They are given by
\begin{align}
 a=1:\ M_1^{(1)} &= \begin{pmatrix} 0 & 0 & 0 \\ 0 & 0 & 0 \\ 0 & 0 & 0 \end{pmatrix},\ M_2^{(a)} =  \begin{pmatrix} 0 & e^{-\rmi \pi/4} & 1 \\ e^{\rmi \pi/4} & 0 & e^{-\rmi \pi/4} \\ 1 & e^{\rmi\pi/4} & 0 \end{pmatrix},\\
 a=2:\ M_1^{(1)} &= \begin{pmatrix} 0 & e^{-\rmi \pi/4} & 1 \\ e^{\rmi \pi/4} & 0 & e^{-\rmi \pi/4} \\ 1 & e^{\rmi\pi/4} & 0 \end{pmatrix},\ M_2^{(a)} =  \begin{pmatrix} 0 & 0 & 0 \\ 0 & 0 & 0 \\ 0 & 0 & 0 \end{pmatrix}.
\end{align}
For each $a$, the matrices contain exactly $q=3$ nonvanishing columns $\vec{Y}_\mu$ given by
\begin{align}
 \vec{Y}_1 = \begin{pmatrix} 0 \\ e^{\rmi\pi/4} \\ 1\end{pmatrix},\ \vec{Y}_2 = \begin{pmatrix} e^{-\rmi \pi/4} \\ 0 \\ e^{\rmi \pi/4} \end{pmatrix},\ \vec{Y}_3 = \begin{pmatrix} 1 \\ e^{-\rmi \pi/4} \\ 0 \end{pmatrix}.
\end{align}
They are related to the reference vectors $\vec{X}_3^{(q=3)}$ through
\begin{align}
 \vec{Y}_\mu = S_\mu \vec{X}_\mu
\end{align} 
with
\begin{align}
 S_1 &=e^{\rmi \pi/6} V^\dagger,\ S_2 = e^{-\rmi \pi/12}S_1,\ S_3 = e^{-\rmi \pi/6} S_1.
\end{align}
Thus we have
\begin{align}
\mathcal{B} = \sum_{k=1}^{N_{\rm u}} M_k^{(a)} [M_k^{(a)}]^\dagger = \sum_{\mu=1}^2 \vec{Y}_\mu \vec{Y}_\mu^\dagger = S_1\Bigl(\sum_{\mu=1}^2 \vec{X}_\mu \vec{X}_\mu^\dagger\Bigr) S_1^\dagger = S_1 \Phi_0^2 S_1^\dagger = V^\dagger \Phi_0^2V.
\end{align}
This shows that $\mathcal{B}$ is unitarily related to the $q=3$ reference $\Phi_0^2$. We conclude that
\begin{align}
 \mbox{det}(u\mathbb{1}-W_0^\dagger W_0) = \Bigl[ \mbox{det}(u\mathbb{1}_3-\mathcal{B})\Bigr]^{N_{\rm u}} = \Bigl[ \mbox{det}(u\mathbb{1}_3-\Phi_0^2)\Bigr]^{N_{\rm u}} = [P_3(u)]^2 = u^2 (u-3)^4.
\end{align}

\clearpage
\subsection{Star lattice ($q=3$)}

\subsubsection{Matrix $W(\vec{k})$} The Star lattice consists of a regular triangle and two regular dodecagons meeting at a vertex. Thus, it has coordination number $q=3$ with $N_{\rm u}=6$ sites in the unit cell. We label the sites in the unit cell by $a\in\{1,\dots,6\}$ and the local edges at each site by $\mu\in\{1,2,3\}$ according to the following schematic:
\begin{figure}[h!]
    \centering
    \includegraphics[width=0.65\linewidth]{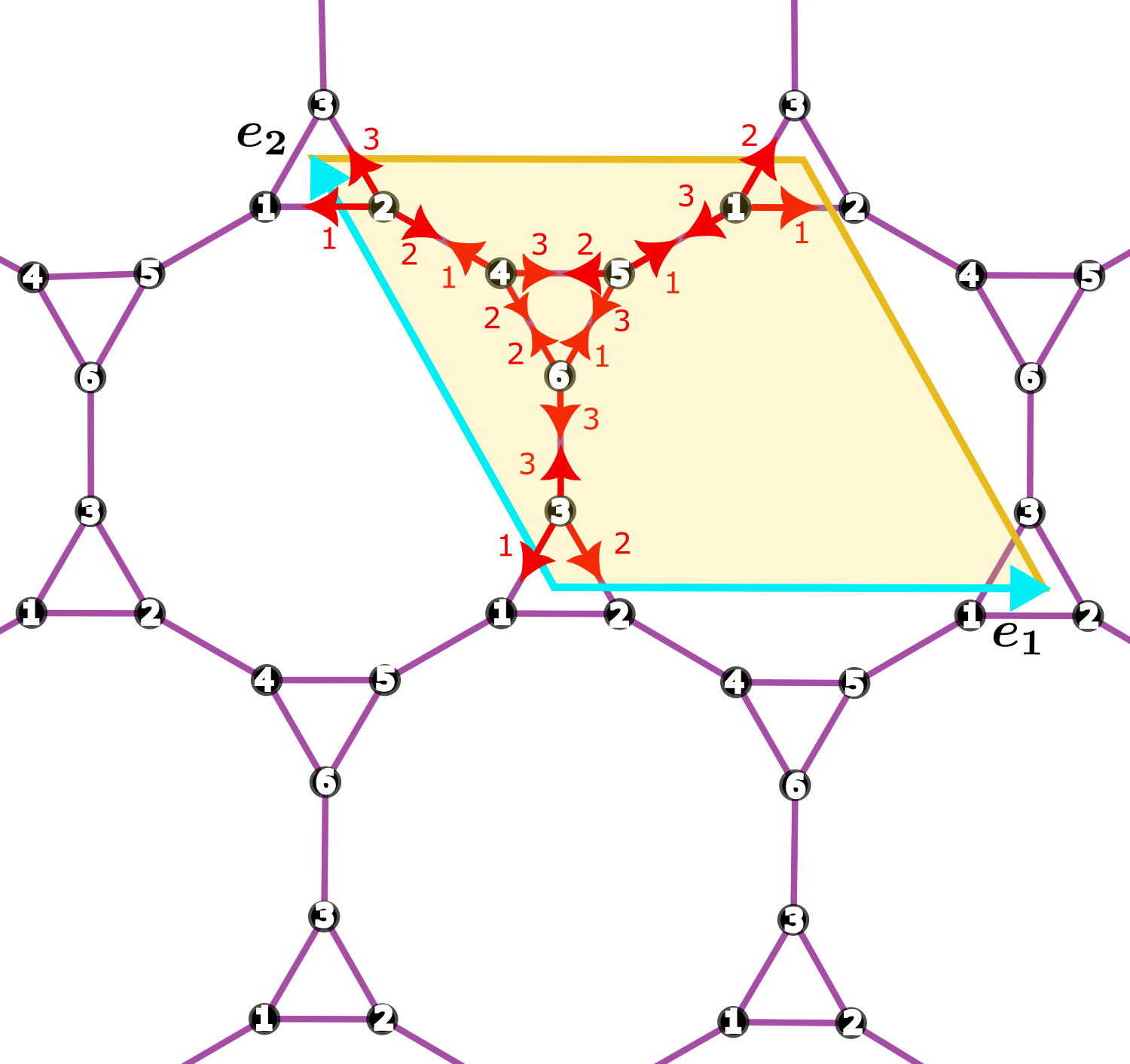}
\end{figure}\\
The nonzero entries of $W(\vec{k})$ are
\begin{equation}W(\vec{k})=\begin{pmatrix}
 0 & * & * & * & 0 & 0 \\
 * & 0 & * & 0 & * & 0 \\
 * & * & 0 & 0 & 0 & * \\
 * & 0 & 0 & 0 & * & * \\
 0 & * & 0 & * & 0 & * \\
 0 & 0 & * & * & * & 0 \end{pmatrix}.
\end{equation}
The Bloch matrices for the six sites are given by 
\begin{align} 
B_1(\boldsymbol{k})&=\begin{pmatrix}
 e^{-\rmi k_1} & 0 & 0 \\
 0 & e^{-\rmi \left(k_1+k_2\right)} & 0 \\
 0 & 0 & 1 \end{pmatrix},\ B_2(\boldsymbol{k})= \begin{pmatrix}
 e^{-\rmi k_2} & 0 & 0 \\
 0 & e^{\rmi k_1} & 0 \\
 0 & 0 & 1 \end{pmatrix},\ B_3(\boldsymbol{k})= \begin{pmatrix}
 e^{-\rmi \left(-k_1-k_2\right)} & 0 & 0 \\
 0 & e^{\rmi k_2} & 0 \\
 0 & 0 & 1 \end{pmatrix} \\
B_4(\boldsymbol{k}) &=\begin{pmatrix}
 1 & 0 & 0 \\
 0 & 1 & 0 \\
 0 & 0 & 1 \end{pmatrix},\ B_5(\boldsymbol{k})=\begin{pmatrix}
 1 & 0 & 0 \\
 0 & 1 & 0 \\
 0 & 0 & 1 \end{pmatrix},\ B_6(\boldsymbol{k})=\begin{pmatrix}
 1 & 0 & 0 \\
 0 & 1 & 0 \\
 0 & 0 & 1 \end{pmatrix}.\end{align}
The edge-connectivity matrices are given by
\begin{equation}E_{1;2}=E_{2;3}=E_{3;1}=E_{4;5}=E_{5;6}=E_{6;4}=\left(
\begin{array}{ccc}
 1 & 0 & 0 \\
 0 & 0 & 0 \\
 0 & 0 & 0 \\
\end{array}
\right)=\varepsilon_1,\end{equation}
\begin{equation}E_{2;1}=E_{3;2}=E_{1;3}=E_{5;4}=E_{6;5}=E_{4;6}=\left(
\begin{array}{ccc}
 0 & 0 & 0 \\
 0 & 1 & 0 \\
 0 & 0 & 0 \\
\end{array}
\right)=\varepsilon_2,\end{equation}
\begin{equation}E_{4;1}=E_{5;2}=E_{6;3}=E_{1;4}=E_{2;5}=E_{3;6}=\left(
\begin{array}{ccc}
 0 & 0 & 0 \\
 0 & 0 & 0 \\
 0 & 0 & 1 \\
\end{array}
\right)=\varepsilon_3.\end{equation}
Since the tiling consists of regular dodecagons and a triangle, the angles between edges are uniquely determined. Let $\alpha=e^{\rmi\frac{\pi}{3}},\omega=e^{\rmi\frac{5\pi}{6}}$ and $g=e^{\rmi\frac{2\pi}{3}}$, then 
\begin{equation}
\boldsymbol{\theta}_a=g^{a-1}(1,\alpha,\alpha\omega)^T 
\end{equation}
for $a=1,2,3$ and
\begin{equation}
\boldsymbol{\theta}_b=g^{b-4}(-1,-\alpha,-\alpha\omega)^T
\end{equation} for $b=4,5,6$.
Computing $\det(\mathbbm{1}-t W(\bk))$ yields
\begin{align}
   \nonumber \det(\mathbbm{1}-tW(\boldsymbol{k}))&=(1+t)^4 \left[1-4 t+10 t^2-16 t^3+19 t^4-16 t^5+10 t^6-4 t^7+4 t^8+ t^4 (1-t) \left(1-t+2 t^2\right)\varepsilon_\Delta(\bk)\right].
\end{align}
The critical temperature is a root of the equation
\begin{align}
   \det(\mathbbm{1}-tW_0)=(1+t)^4 \left(-1+2 t-3 t^2+2 t^3+2 t^4\right)^2,
\end{align}
which gives
\begin{equation}
    t_c = 0.670698.
\end{equation}
The characteristic polynomial of $W_0^\dagger W_0$ is given by
\begin{equation}
    \det(u \mathbbm{1}- W_0^\dagger W_0)=(u -3)^{12} u ^6.
\end{equation}

\clearpage
\subsection{SHD lattice ($q=3$)}

\subsubsection{Matrix $W(\vec{k})$} The SHD lattice consists of a regular dodecagon, a regular square, and a regular hexagon meeting at ech vertex. Thus, it has coordination number $q=3$ with $N_{\rm u}=12$ sites in the unit cell. We label the sites in the unit cell by $a\in\{1,\dots,12\}$ and the local edges at each site by $\mu\in\{1,2,3\}$ according to the following schematic:
\begin{figure}[h!]
    \centering
    \includegraphics[width=0.6\linewidth]{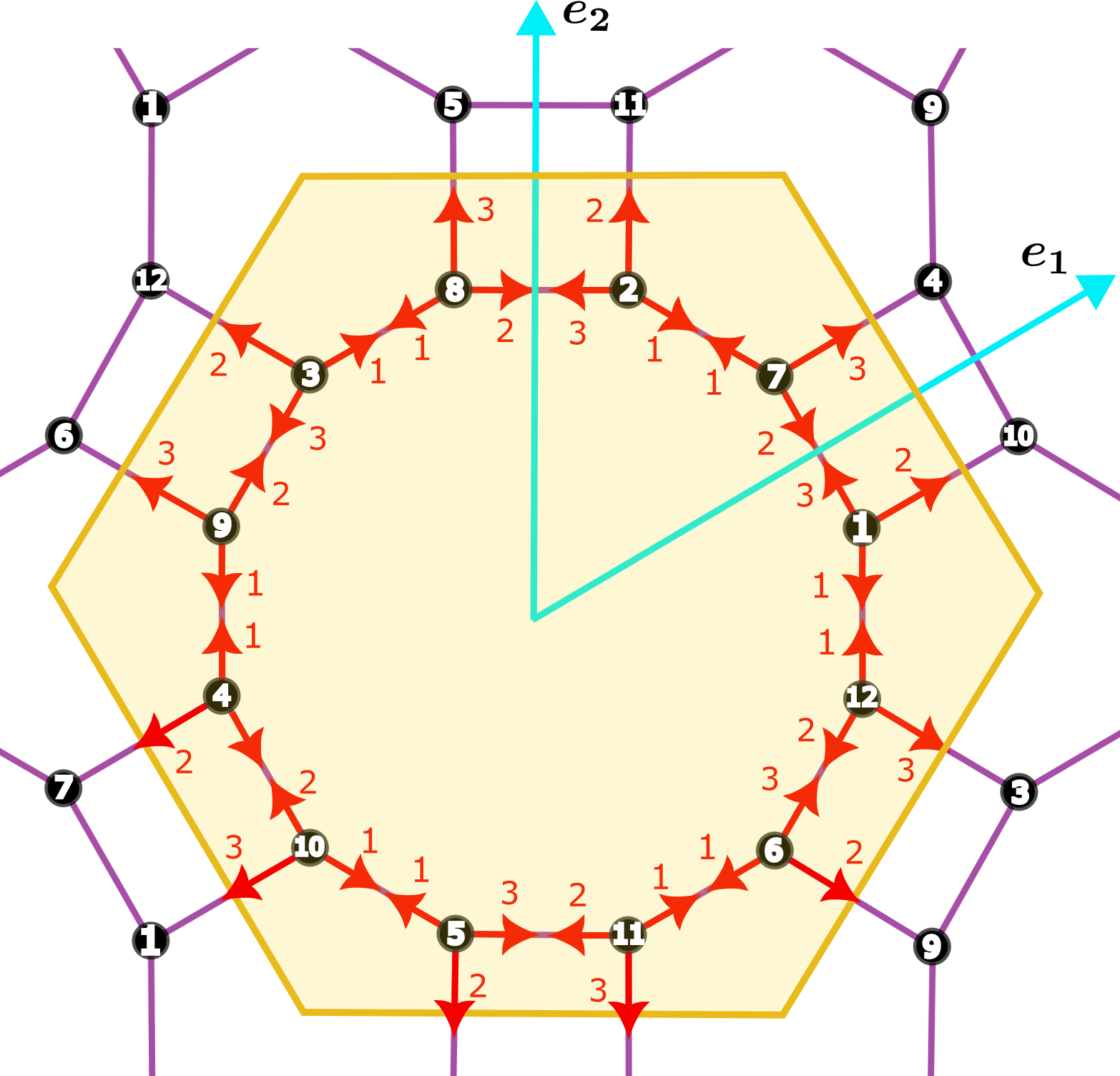}
\end{figure}\\
The nonzero entries of $W(\vec{k})$ are given by
\begin{equation} W(\vec{k})=\left(\begin{array}{cccccccccccc}
0 & 0 & 0 & 0 & 0 & 0 & * & 0 & 0 & * & 0 & * \\
 0 & 0 & 0 & 0 & 0 & 0 & * & * & 0 & 0 & * & 0 \\
 0 & 0 & 0 & 0 & 0 & 0 & 0 & * & * & 0 & 0 & * \\
 0 & 0 & 0 & 0 & 0 & 0 & * & 0 & * & * & 0 & 0 \\
 0 & 0 & 0 & 0 & 0 & 0 & 0 & * & 0 & * & * & 0 \\
 0 & 0 & 0 & 0 & 0 & 0 & 0 & 0 & * & 0 & * & * \\
 * & * & 0 & * & 0 & 0 & 0 & 0 & 0 & 0 & 0 & 0 \\
 0 & * & * & 0 & * & 0 & 0 & 0 & 0 & 0 & 0 & 0 \\
 0 & 0 & * & * & 0 & * & 0 & 0 & 0 & 0 & 0 & 0 \\
 * & 0 & 0 & * & * & 0 & 0 & 0 & 0 & 0 & 0 & 0 \\
 0 & * & 0 & 0 & * & * & 0 & 0 & 0 & 0 & 0 & 0 \\
 * & 0 & * & 0 & 0 & * & 0 & 0 & 0 & 0 & 0 & 0 \\
\end{array}\right),\end{equation}
and the Bloch matrices are given by

\begin{align}B_1(\boldsymbol{k})&=\left(
\begin{array}{ccc}
 1 & 0 & 0 \\
 0 & e^{-\rmi k_1} & 0 \\
 0 & 0 & 1 \\
\end{array}
\right),\ B_2(\boldsymbol{k})=\left(
\begin{array}{ccc}
 1 & 0 & 0 \\
 0 & e^{-\rmi k_2} & 0 \\
 0 & 0 & 1 \\
\end{array}
\right),\ B_3(\boldsymbol{k})=\left(
\begin{array}{ccc}
 1 & 0 & 0 \\
 0 & e^{-\rmi \left(k_2-k_1\right)} & 0 \\
 0 & 0 & 1 \\
\end{array}
\right),\ B_4(\boldsymbol{k})=\left(
\begin{array}{ccc}
 1 & 0 & 0 \\
 0 & e^{\rmi k_1} & 0 \\
 0 & 0 & 1 \\
\end{array}
\right),\\
B_5(\boldsymbol{k})&=\left(
\begin{array}{ccc}
 1 & 0 & 0 \\
 0 & e^{\rmi k_2} & 0 \\
 0 & 0 & 1 \\
\end{array}
\right),\ B_6(\boldsymbol{k})=\left(
\begin{array}{ccc}
 1 & 0 & 0 \\
 0 & e^{-\rmi \left(k_1-k_2\right)} & 0 \\
 0 & 0 & 1 \\
\end{array}
\right),\ B_7(\boldsymbol{k})=\left(
\begin{array}{ccc}
 1 & 0 & 0 \\
 0 & 1 & 0 \\
 0 & 0 & e^{-\rmi k_1} \\
\end{array}
\right),\ B_8(\boldsymbol{k})=\left(
\begin{array}{ccc}
 1 & 0 & 0 \\
 0 & 1 & 0 \\
 0 & 0 & e^{-\rmi k_2} \\
\end{array}
\right),\\
B_9(\boldsymbol{k})&=\left(
\begin{array}{ccc}
 1 & 0 & 0 \\
 0 & 1 & 0 \\
 0 & 0 & e^{-\rmi \left(k_2-k_1\right)} \\
\end{array}
\right),\ B_{10}(\boldsymbol{k})=\left(
\begin{array}{ccc}
 1 & 0 & 0 \\
 0 & 1 & 0 \\
 0 & 0 & e^{\rmi k_1} \\
\end{array}
\right),\ B_{11}(\boldsymbol{k})=\left(
\begin{array}{ccc}
 1 & 0 & 0 \\
 0 & 1 & 0 \\
 0 & 0 & e^{\rmi k_2} \\
\end{array}
\right),\ B_{12}(\boldsymbol{k})=\left(
\begin{array}{ccc}
 1 & 0 & 0 \\
 0 & 1 & 0 \\
 0 & 0 & e^{-\rmi \left(k_1-k_2\right)} \\
\end{array}
\right).\end{align}
The edge-connectivity matrices are given by
\begin{align}
\nonumber &E_{1;10}=E_{2;11}=E_{3;12}=E_{4;7}=E_{5;8}=E_{6;9}=E_{7;1}=E_{8;2}=E_{9;3}=E_{10;4}=E_{11;5}=E_{12;6}=\left(
\begin{array}{ccc}
 0 & 0 & 0 \\
 0 & 1 & 0 \\
 0 & 0 & 0 \\
\end{array}
\right)=\varepsilon_1,
\end{align}
\begin{align}
\nonumber &E_{10;1}=E_{11;2}=E_{12;3}=E_{7;4}=E_{8;5}=E_{9;6}=E_{1;7}=E_{2;8}=E_{3;9}=E_{4;10}=E_{5;11}=E_{6;12}=\left(
\begin{array}{ccc}
 0 & 0 & 0 \\
 0 & 0 & 0 \\
 0 & 0 & 1 \\
\end{array}
\right)=\varepsilon_2,
\end{align}
\begin{align}
\nonumber &E_{1;12}=E_{6;11}=E_{5;10}=E_{4;9}=E_{3;8}=E_{2;7}=E_{12;1}=E_{11;6}=E_{10;5}=E_{9;4}=E_{8;3}=E_{7;2}=\left(
\begin{array}{ccc}
 1 & 0 & 0 \\
 0 & 0 & 0 \\
 0 & 0 & 0 \\
\end{array}
\right)=\varepsilon_3.
\end{align}
Let $\alpha=e^{\rmi\frac{2\pi}{3}},\beta=e^{\rmi\frac{\pi}{6}},\omega=e^{\rmi\frac{\pi}{2}},\sigma=e^{\rmi\frac{5\pi}{6}}$ and $g=e^{\rmi\frac{\pi}{3}}$. The phases at each site are then given by
\begin{equation}
\boldsymbol{\theta}_a=g^{a-1}(-\omega,-\omega \alpha,-\omega^2\alpha)^T
\end{equation}
for $a=1,\dots,6$ and
\begin{equation}
\boldsymbol{\theta}_b=g^{b-8}(\alpha\beta,\alpha\beta\sigma,\alpha\beta\sigma\omega)^T
\end{equation}
for $b=7,\dots,12$.
Then, we have 
\begin{align}
   \nonumber \det(\mathbbm{1}-tW(\boldsymbol{k}))={}& 1 + t^4(6 + 4t^2 + 27t^4 + 24t^6 + 182t^8 + 324t^{10} + 837t^{12} + 992t^{14} + 1188t^{16} + 384t^{18} + 127t^{20}) \\ 
 \nonumber &-2t^{10}(1-t^2)^4(-2 - 3t^2 - 4t^4 + t^6)\varepsilon_\Delta(2k_1-k_2,k_1-2k_2) \\ 
 \nonumber &+ 2t^6(1-t^2)^2(1 + t^2 + 2t^4)(3 + 6t^2 + 19t^4 + 25t^6 + 38t^8 + 5t^{10})\varepsilon_\Delta(\bk) \\
 &- t^{12}(1-t^2)^6\varepsilon_\Delta(2\bk). 
\end{align}
The critical temperature is the root of the equation
\begin{align}
   \det(\mathbbm{1}-tW_0)&=\left(1+t^4 \left(3-16 t^2-9 t^4-48 t^6+5 t^8\right)\right)^2=0 \\ \implies t_c &=0.616606.
\end{align}
The characteristic polynomial of $W_0^\dagger W_0$ is given by
\begin{equation}
    \det(u\mathbbm{1}-W_0^\dagger W_0)=u^{12}(u -3)^{24} .
\end{equation}

\clearpage

\subsection{Square lattice with tilt ($q=4$)}

\subsubsection{Matrix $W(\vec{k})$} We now consider the Square lattice according to the schematic in the previous section, but allow for an arbitrary tilt angle
\begin{align}
 \chi\in(0,\pi).
\end{align}
Of course, this lattice is equivalent to the right-angle Square lattice and has the same Ising free energy. However, it is an instructive case to consider. We have
\begin{align}
 \vec{\theta}_1 = (1,e^{\rmi \chi},-1,-e^{\rmi \chi}),
\end{align}
so that
\begin{align}
 \phi_{11} &= \begin{pmatrix} 1 & e^{\rmi \chi} & -1 & -e^{\rmi \chi} \\ e^{-\rmi \chi} & 1 & - e^{-\rmi \chi} & -1 \\ -1 & -e^{\rmi \chi} & 1 & e^{\rmi \chi} \\ -e^{-\rmi \chi} & -1 & e^{-\rmi \chi} & 1\end{pmatrix}= \begin{pmatrix} 1 & e^{\rmi \chi} & -1 & e^{-\rmi(\pi-\chi)} \\ e^{-\rmi \chi} & 1 & e^{\rmi(\pi-\chi)} & -1 \\ -1 & e^{-\rmi(\pi-\chi)} & 1 & e^{\rmi \chi} \\ e^{\rmi(\pi-\chi)} & -1 & e^{-\rmi \chi} & 1\end{pmatrix},\\
 \Phi_{11} &= \begin{pmatrix} 1 & e^{\rmi \chi/2} & 0 & e^{-\rmi(\pi-\chi)/2} \\ e^{-\rmi \chi/2} & 1 & e^{\rmi(\pi-\chi)/2} & 0 \\ 0 & e^{-\rmi(\pi-\chi)/2} & 1 & e^{\rmi \chi/2} \\ e^{\rmi(\pi-\chi)/2} & 0 & e^{-\rmi \chi/2} & 1\end{pmatrix}.
\end{align}
Note that $\Phi_{11}=\Phi_{11}^\dagger$ is still valid. We have
\begin{align}
W(\vec{k}) = B_1(\vec{k})\Phi_{11}=\begin{pmatrix} e^{-\rmi k_1} & e^{-\rmi k_1+\rmi \chi/2} & 0 & e^{-\rmi k_1-\rmi(\pi-\chi)/2} \\ e^{-\rmi k_2-\rmi \chi/2} & e^{-\rmi k_2} & e^{-\rmi k_2+\rmi(\pi-\chi)/2} & 0 \\ 0 & e^{\rmi k_1-\rmi(\pi-\chi)/2} & e^{\rmi k_1} & e^{\rmi k_1+\rmi \chi/2} \\ e^{\rmi k_2+\rmi(\pi-\chi)/2} & 0 & e^{\rmi k_2-\rmi \chi/2} & e^{\rmi k_2}\end{pmatrix}
\end{align}
and find that (independently of $\chi$) that
\begin{align}
 \mbox{det}(\mathbb{1} - t W(\vec{k})) = (1+t^2)^2+t(1-t^2)\vare_\square(\vec{k}).
\end{align}

\subsubsection{Matrices $W_0$ and $\mathcal{B}$} 
We have
\begin{align}
 W_0=W_0^\dagger = \Phi_{11}
\end{align}
and
\begin{align}
 \mathcal{B}= W_0^\dagger W_0 = \Phi_{11}^2.
\end{align}
We find
\begin{align}
 \det(u\mathbb{1}-W_0^\dagger W_0) &=  \det(u\mathbb{1}-\Phi_{11}^2)=(u^2-6u+1)^2.
\end{align}

\subsubsection{Relation to reference $\Phi_0^2$} 
To make the connection to the untilted reference $\Phi_0$ clear, we note that $\vec{\theta}_1=U_1\vec{\theta}_0$ with
\begin{align}
 U_1 = \begin{pmatrix} 1 & & & \\ & e^{-\rmi (\frac{\pi}{2}-\chi)} & & \\ & & 1 & \\ & & & e^{-\rmi (\frac{\pi}{2}-\chi)} \end{pmatrix}.
\end{align}
and we have $\phi_{11}=U_1^\dagger \phi_0^{(q=0)}U_1$. Define
\begin{align}
 V_1 = \sqrt{U_1} = \begin{pmatrix} 1 & & & \\ & e^{-\rmi (\frac{\pi}{2}-\chi)/2} & & \\ & & 1 & \\ & & & e^{-\rmi (\frac{\pi}{2}-\chi)/2} \end{pmatrix}.
\end{align}
We find that 
\begin{align}
\mathcal{B}=W_0^\dagger W_0=\Phi_{11}^2 = V_1^\dagger \Phi_0^2V_1.
\end{align}
To show this, consider the matrix $M_1^{(1)} = \Phi_{11}^\dagger E_{1;1} =\Phi_{11} $  given by
\begin{align}
M_1^{(1)} = \begin{pmatrix} 1 & e^{\rmi \chi/2} & 0 & e^{-\rmi(\pi-\chi)/2} \\ e^{-\rmi \chi/2} & 1 & e^{\rmi(\pi-\chi)/2} & 0 \\ 0 & e^{-\rmi(\pi-\chi)/2} & 1 & e^{\rmi \chi/2} \\ e^{\rmi(\pi-\chi)/2} & 0 & e^{-\rmi \chi/2} & 1\end{pmatrix}.
\end{align}
The matrix contains $q=4$ nonvanishing columns $\vec{Y}_\mu$, which we order according to
\begin{align}
 \vec{Y}_1 &= \begin{pmatrix} 1 \\ e^{-\rmi \chi/2}  \\ 0  \\ e^{\rmi(\pi-\chi)/2} \end{pmatrix},\ \vec{Y}_2 = \begin{pmatrix}  e^{\rmi \chi/2}  \\  1  \\  e^{-\rmi(\pi-\chi)/2}  \\  0 \end{pmatrix},\ \vec{Y}_3 = \begin{pmatrix}  0  \\  e^{\rmi(\pi-\chi)/2}  \\  1 \\  e^{-\rmi \chi/2}\end{pmatrix},\ \vec{Y}_4 = \begin{pmatrix}  e^{-\rmi(\pi-\chi)/2} \\  0 \\  e^{\rmi \chi/2} \\  1\end{pmatrix}.
\end{align}
We have
\begin{align}
 \mathcal{B} = M_1^{(1)} [M_1^{(1)}]^\dagger = \sum_{\mu=1}^4 \vec{Y}_\mu \vec{Y}_\mu^\dagger.
\end{align}
The vectors $\vec{Y}_\mu$ are related to the reference vectors $\vec{X}_\mu^{(q=4)}$ from Eq. (\ref{ref4g}) by means of
\begin{align}
 \vec{Y}_\mu = S_\mu \vec{X}_\mu
\end{align}
with
\begin{align}
 S_1 &= S_3 = V_1^\dagger,\\
 S_2 &= S_4 = e^{\rmi(\chi-\frac{\pi}{2})}S_1=e^{\rmi(\chi-\frac{\pi}{2})}V_1^\dagger.
\end{align}
Hence
\begin{align}
 \mathcal{B} = S_1 \Bigl(\sum_{\mu=1}^q \vec{X}_\mu \vec{X}_\mu^\dagger\Bigr) S_1^\dagger = S_1 \Phi_0^2 S_1^\dagger = V_1^\dagger \Phi_0^2V_1.
\end{align}

\clearpage
\subsection{Ruby lattice ($q=4$)}

\subsubsection{Matrix $W(\vec{k})$} 

The Ruby lattice has coordination number $q=4$ and $N_{\rm u}=6$ sites in the unit cell. We label the sites in the unit cell by $a\in\{1,\dots,6\}$ and the local edges at each site by $\mu\in\{1,\dots,4\}$ according to the following schematic:
\begin{figure}[h!]
    \includegraphics[width=7cm]{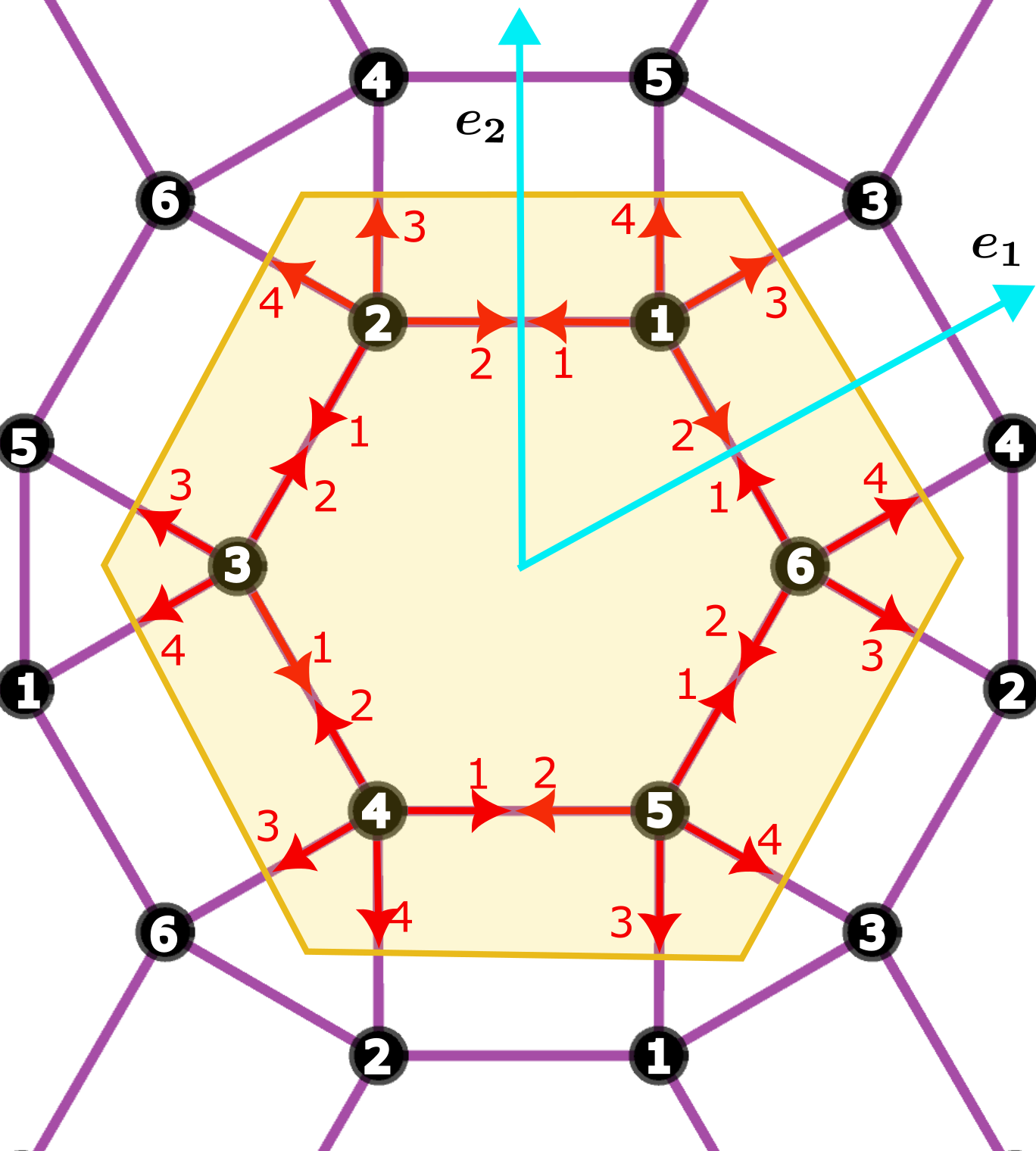}
     \label{FigRubyLabels}
\end{figure}\\
The nonzero blocks of the matrix $W(\vec{k})$ are distributed according to
\begin{align}
 W(\vec{k}) = \begin{pmatrix} 0 & * & * & 0 & * & * \\ * & 0 & * & * & 0 & * \\ * & * & 0 & * & * & 0 \\ 0 & * & * & 0 & * & * \\ * & 0 & * & * & 0 & * \\ * & * & 0 & * & * & 0 \end{pmatrix},
\end{align}
with
\begin{align}
 \label{rub2} W(\vec{k})_{ab} =  E_{a;b}B_a(\vec{k})\Phi_{ab}.
\end{align}
Each block has size $q\times q$. The Bloch matrices at the unit cell sites are
\begin{align}
 \nonumber B_1(\vec{k}) &= \text{diag}(1,1,e^{-\rmi k_1},e^{-\rmi k_2}),\\
 \nonumber B_2(\vec{k}) &= \text{diag}(1,1,e^{-\rmi k_2},e^{\rmi(k_1-k_2)}),\\
 \nonumber B_3(\vec{k}) &= \text{diag}(1,1,e^{\rmi(k_1-k_2)},e^{\rmi k_1}),\\
 \nonumber B_4(\vec{k}) &= \text{diag}(1,1,e^{\rmi k_1},e^{\rmi k_2}),\\
 \nonumber   B_5(\vec{k}) &= \text{diag}(1,1,e^{\rmi k_2},e^{-\rmi (k_1-k_2)}),\\
 \label{rub3}   B_6(\vec{k}) &= \text{diag}(1,1,e^{-\rmi (k_1-k_2)},e^{-\rmi k_1}).
\end{align}
The edge-connectivity matrices read
\begin{align}
 \label{rub9} E_{1;2} &= E_{2;3}=E_{3;4}=E_{4;5}=E_{5;6}=E_{6;1} = \vare_1,\\
 \label{rub10} E_{1;6} &= E_{2;1} = E_{3;2} = E_{4;3} = E_{5;4} = E_{6;5} = \vare_2,\\
  \label{rub11} E_{1;3} &= E_{2;4} = E_{3;5} = E_{4;6} = E_{5;1} = E_{6;2}  =\vare_3,\\
 \label{rub12} E_{1;5} &= E_{2;6} = E_{3;1} = E_{4;2}  = E_{5;3} = E_{6;4} = \vare_4,
\end{align}
with
\begin{align}
 \label{rub13} \vare_1 &= \text{diag}(1,0,0,0),\\
 \vare_2 &= \text{diag}(0,1,0,0),\\
 \vare_3 &= \text{diag}(0,0,1,0),\\
 \vare_4 &= \text{diag}(0,0,0,1).
\end{align}
Define $g=e^{\pi \rmi /3}$ to obtain
\begin{align}
 \label{rub15} \vec{\theta}_a &= g^{a-1}(-1,e^{-\pi\rmi/3},e^{\pi \rmi/6},\rmi)^T
\end{align}
for $a=1\dots,6$. We then have
\begin{align}
 \label{rub16} \Phi_{12} &= \Phi_{23} = \Phi_{34} = \Phi_{45} = \Phi_{56} = \Phi_{61} =: \Psi_1,\\
 \label{rub17} \Phi_{16} &= \Phi_{21} = \Phi_{32} = \Phi_{43} = \Phi_{54} = \Phi_{65} = \Psi_1^\dagger ,\\
 \label{rub18} \Phi_{13} &= \Phi_{24} = \Phi_{35} = \Phi_{46} = \Phi_{51} = \Phi_{62} =: \Psi_2,\\
 \label{rub19} \Phi_{15} &= \Phi_{26} = \Phi_{31} = \Phi_{42} = \Phi_{53} = \Phi_{64} = \Psi_2^\dagger.
\end{align}
The fact that the matrices $\vare_\mu$ are diagonal with only one non-vanishing entry simplifies the products $\vare_\mu B_a$ appearing in $W(\vec{k})$. For instance, $\vare_1B_1(\vec{k})=\vare_1$ and $\vare_3 B_1(\vec{k}) = e^{-\rmi k_1}\vare_3$. Thus we arrive at
\begin{align}
 \label{rub20} W(\vec{k}) &= \begin{pmatrix} 0 & \vare_1B_1\Psi_1 & \vare_3B_1\Psi_2 & 0 & \vare_4B_1\Psi_2^\dagger & \vare_2B_1\Psi_1^\dagger \\ \vare_2B_2\Psi_1^\dagger & 0 & \vare_1B_2\Psi_1 & \vare_3B_2\Psi_2 & 0 & \vare_4B_2\Psi_2^\dagger \\ \vare_4B_3\Psi_2^\dagger & \vare_2B_3\Psi_1^\dagger & 0 & \vare_1B_3\Psi_1 & \vare_3B_3\Psi_2 & 0 \\ 0 & \vare_4B_4\Psi_2^\dagger & \vare_2B_4\Psi_1^\dagger & 0 & \vare_1B_4\Psi_1 & \vare_3B_4\Psi_2 \\ \vare_3B_5\Psi_2 & 0 & \vare_4B_5\Psi_2^\dagger & \vare_2B_5\Psi_1^\dagger & 0 & \vare_1B_5\Psi_1 \\ \vare_1B_6\Psi_1 & \vare_3B_6\Psi_2 & 0 & \vare_4B_6\Psi_2^\dagger & \vare_2B_6\Psi_1^\dagger & 0 \end{pmatrix}\\
 \label{rub21} &= \begin{pmatrix} 0 & \vare_1\Psi_1 & e^{-\rmi k_1}\vare_3\Psi_2 & 0 & e^{-\rmi k_2}\vare_4\Psi_2^\dagger & \vare_2\Psi_1^\dagger \\ \vare_2\Psi_1^\dagger & 0 & \vare_1\Psi_1 & e^{-\rmi k_2}\vare_3\Psi_2 & 0 & e^{\rmi(k_1-k_2)}\vare_4\Psi_2^\dagger \\ e^{\rmi k_1}\vare_4\Psi_2^\dagger & \vare_2\Psi_1^\dagger & 0 & \vare_1\Psi_1 & e^{\rmi(k_1-k_2)}\vare_3\Psi_2 & 0 \\ 0 & e^{\rmi k_2}\vare_4\Psi_2^\dagger & \vare_2\Psi_1^\dagger & 0 & \vare_1\Psi_1 & e^{\rmi k_1}\vare_3\Psi_2 \\ e^{\rmi k_2}\vare_3\Psi_2 & 0 & e^{-\rmi(k_1-k_2)}\vare_4\Psi_2^\dagger & \vare_2\Psi_1^\dagger & 0 & \vare_1\Psi_1 \\ \vare_1\Psi_1 & e^{-\rmi(k_1-k_2)}\vare_3\Psi_2 & 0 & e^{-\rmi k_1}\vare_4\Psi_2^\dagger & \vare_2\Psi_1^\dagger & 0 \end{pmatrix}.
\end{align}
We have
\begin{align}
  \nonumber \mbox{det}(\mathbb{1}-tW(\vec{k}))={}& (1+t)^4 \Bigl[1-4 t+10 t^2-16 t^3+25 t^4-28 t^5+42 t^6-12 t^7+78 t^8-68 t^9+200 t^{10}-68 t^{11}\\
  \nonumber &+78 t^{12}-12 t^{13}+42 t^{14}-28 t^{15}+25 t^{16}-16 t^{17}+10 t^{18}-4 t^{19}+t^{20}\Bigr]\\
  \nonumber &+2(1-t)^2 t^3 (1+t)^4 (1+4 t^2-t^3+9 t^4+t^5+4 t^6+t^7+9 t^8-t^9+4 t^{10}+t^{12}) \vare_{\Delta}(\vec{k}')\\
  \label{rub22} &+2 (1-t)^6 t^6 (1+t)^6\vare_\Delta(\vec{k})-(1-t)^6 t^6 (1+t)^6 \vare_\Delta(2\vec{k})
\end{align}
with $\vec{k}'=(2k_1-k_2,k_1-2k_2)$. Using the relation
\begin{align}
 \label{rub23} \vare_\Delta(2\vec{k}) = 6-2\vare_\Delta(\vec{k})-\vare_\Delta(\vec{k})^2-2\vare_\Delta(\vec{k}')
\end{align}
we can eliminate $\vare_\Delta(\vec{k}')$ to arrive at
\begin{align}
 \nonumber \mbox{det}(\mathbb{1}-tW(\vec{k}))={}& (1+t)^4(1+t^2)^4\Bigl[1-4t+6t^2-5t^4-4t^5+28t^6-4t^7-5t^8+6t^{10}-4t^{11}+t^{12}\Bigr]\\
 \nonumber &+2(1-t)^2(1+t)^4t^3\Bigl[1+4t^2-2t^3+11t^4+2t^5+2t^7+11t^8-2t^9+4t^{10}+t^{12}\Bigr]\vare_\Delta(\vec{k})\\
 \label{rub24} &-(1-t^2)^6t^6[\vare_\Delta(\vec{k})^2+2\vare_\Delta(2\vec{k})].
\end{align}

\subsubsection{Matrices $W_0$ and $\mathcal{B}_a$} 

For $\vec{k}=0$ we have
\begin{align}
 \label{rub25} W_0 = \begin{pmatrix} 0 & \vare_1\Psi_1 & \vare_3\Psi_2 & 0 & \vare_4\Psi_2^\dagger & \vare_2\Psi_1^\dagger \\ \vare_2\Psi_1^\dagger & 0 & \vare_1\Psi_1 & \vare_3\Psi_2 & 0 & \vare_4\Psi_2^\dagger \\ \vare_4\Psi_2^\dagger & \vare_2\Psi_1^\dagger & 0 & \vare_1\Psi_1 & \vare_3\Psi_2 & 0 \\ 0 & \vare_4\Psi_2^\dagger & \vare_2\Psi_1^\dagger & 0 & \vare_1\Psi_1 & \vare_3\Psi_2 \\ \vare_3\Psi_2 & 0 & \vare_4\Psi_2^\dagger & \vare_2\Psi_1^\dagger & 0 & \vare_1\Psi_1 \\ \vare_1\Psi_1 & \vare_3\Psi_2 & 0 & \vare_4\Psi_2^\dagger & \vare_2\Psi_1^\dagger & 0 \end{pmatrix}.
\end{align}
Using $\vare_a^2=\vare_a$ and $\vare_a\vare_b=0$ for $a\neq b$ we find 
\begin{align}
 \label{rub26} W_0^\dagger W_0 = \begin{pmatrix} \mathcal{B} & & & & & \\ & \mathcal{B} & & & & \\ & & \mathcal{B} & & & \\ & & & \mathcal{B} & & \\ & & & & \mathcal{B} & \\ & & & & & \mathcal{B} \end{pmatrix}
\end{align}
with $q \times q$ block
\begin{align}
 \label{rub27} \mathcal{B} = \Psi_1^\dagger\vare_1\Psi_1+\Psi_1\vare_2\Psi_1^\dagger+\Psi_2^\dagger\vare_3\Psi_2+\Psi_2\vare_4\Psi_2^\dagger.
\end{align}
For the characteristic polynomial of $W_0^\dagger W_0$ we have
\begin{align}
 \label{rub28} \det(u \mathbb{1}-W_0^\dagger W_0) &= \Bigl[ \det(u \mathbb{1}-\mathcal{B})\Bigr]^{N_{\rm u}}=(u^2-6u+1)^{12}.
\end{align}

\subsubsection{Relation to reference $\Phi_0^2$} 
To make the connection to the reference matrix $\Phi_0^{(q=4)}$, we write $\vec{\theta}_a=U_a\vec{\theta}_0$ with $\vec{\theta}_0=(1,\rmi,-1,-\rmi)^T$ and diagonal unitary matrices
\begin{align}
 U_1 &= \text{diag}(-1,e^{-5\pi\rmi/6},e^{-5\pi\rmi/6},-1),\\
 U_2 &= gU_1,\ U_3 = g^2U_1,\ U_4 = g^3U_1,\\
 U_5 &= g^4U_1,\ U_6 = g^5U_1.
\end{align}
Define $V_a$ so that $V_a^2=U_a$ according to
\begin{align}
 V_1 &= \text{diag}(-\rmi,e^{-5\pi\rmi/12},e^{-5\pi\rmi/12},-\rmi),\\
 V_2 &= \text{diag}(e^{-\rmi \pi/3},e^{-\rmi \pi/4},e^{-\rmi \pi/4},e^{-\rmi \pi/3}),\\
 V_3 &= \text{diag}(e^{-\rmi \pi/6},e^{-\rmi \pi/12},e^{-\rmi \pi/12},e^{-\rmi \pi/6}),\\
 V_4 &= \text{diag}(1,e^{\rmi \pi/12},e^{\rmi \pi/12},1),\\
 V_5 &= \text{diag}(e^{\rmi \pi/6},e^{\rmi \pi/4},e^{\rmi \pi/4},e^{\rmi \pi/6}),\\
 V_6 &= \text{diag}(e^{\rmi \pi/3},e^{5\pi\rmi/12},e^{5\pi\rmi/12},e^{\rmi \pi/3}).
\end{align}
These matrices satisfy
\begin{align}
 V_2 &= e^{\rmi\pi/6}V_1,\ V_3=e^{\rmi \pi/3}V_1,\ V_4=\rmi V_1,\\
 V_5 &= e^{2\pi\rmi/3} V_1,\ V_6 = e^{5\pi\rmi/6}V_1.
\end{align}
Consider the matrices $M_k^{(a)}=\Phi_{ka}^\dagger E_{k;a}$ such that $\mathcal{B}_a = \sum_{k=1}^{N_{\rm u}} M_k^{(a)} [M_k^{(a)}]^\dagger$. These matrices vanish if $E_{k;a}=0$. The nonvanishing ones for $a=1,2,3$ are
\begin{align}
 a=1:\ M_2^{(a)} &= (\vec{0}\ \vec{Y}_1\ \vec{0}\ \vec{0}),\ M_3^{(a)} = (\vec{0}\ \vec{0}\ \vec{0}\ \vec{Y}_3),\ M_5^{(a)} = (\vec{0}\ \vec{0}\ \vec{Y}_4\ \vec{0}),\ M_6^{(a)} = (\vec{Y}_2\ \vec{0}\ \vec{0}\ \vec{0}),\\
 a=2:\ M_1^{(a)} &= (\vec{Y}_2\ \vec{0}\ \vec{0}\ \vec{0}),\ M_3^{(a)} = (\vec{0}\ \vec{Y}_1\ \vec{0}\ \vec{0}),\ M_4^{(a)} = (\vec{0}\ \vec{0}\ \vec{0}\ \vec{Y}_3),\ M_6^{(a)} = (\vec{0}\ \vec{0}\ \vec{Y}_4\ \vec{0}),\\
 a=3:\ M_1^{(a)} &= (\vec{0}\ \vec{0}\ \vec{Y}_4\ \vec{0}),\ M_2^{(a)} = (\vec{Y}_2\ \vec{0}\ \vec{0}\ \vec{0}),\ M_4^{(a)} = (\vec{0}\ \vec{Y}_1\ \vec{0}\ \vec{0}),\ M_5^{(a)} = (\vec{0}\ \vec{0}\ \vec{0}\ \vec{Y}_3),
\end{align}
and similarly for $a=4,5,6$. These matrices have only $q=4$ nonvanishing columns $\vec{Y}_\mu^{(a)}=\vec{Y}_\mu$, which are given by 
\begin{align}
 \vec{Y}_1 &= \begin{pmatrix} 0 \\ e^{\frac{\rmi \pi}{6}} \\ e^{\frac{-\rmi \pi}{12}} \\ e^{\frac{-\rmi \pi}{4}} \end{pmatrix},\ \vec{Y}_2 = \begin{pmatrix} e^{\frac{-\rmi \pi}{6}} \\ 0 \\ e^{\frac{\rmi \pi}{4}} \\ e^{\frac{\rmi \pi}{12}} \end{pmatrix},\
 \vec{Y}_3 = \begin{pmatrix} e^{\frac{\rmi \pi}{12}}\\ e^{\frac{-\rmi \pi}{4}} \\ 0 \\ e^{\frac{\rmi \pi}{3}} \end{pmatrix},\ \vec{Y}_4 = \begin{pmatrix} e^{\frac{\rmi \pi}{4}} \\ e^{\frac{-\rmi \pi}{12}} \\ e^{\frac{-\rmi \pi}{3}} \\ 0 \end{pmatrix}.
\end{align} 
We have
\begin{align}
a=1,\dots,6:\ \mathcal{B}_a = \sum_{k=1}^{N_{\rm u}} M_k^{(a)}[M_k^{(a)}]^\dagger=\sum_{\mu=1}^4 \vec{Y}_\mu \vec{Y}_\mu^\dagger.
\end{align}
The vectors $\vec{Y}_\mu$ satisfy
\begin{align}
 \nonumber a=1,\dots,6:\ Y_\mu^{(a)}(\alpha) &= \begin{cases} \sqrt{-\theta_a(\mu)\theta_a^*(\alpha)} & \alpha\neq \mu \\ 0 & \alpha = \mu\end{cases} \\
  &= \begin{cases} (V_a)_{\mu\mu}(V_a^*)_{\alpha\alpha} \sqrt{-\theta_0(\mu)\theta_0^*(\alpha)} & \alpha\neq \mu \\ 0 & \alpha = \mu\end{cases},
\end{align}
which is equivalent to
\begin{align}
 a=1,\dots,6:\ \vec{Y}_\mu = (V_a)_{\mu\mu} V_a^\dagger \vec{X}_\mu
\end{align}
with the $q=4$ reference vectors $\vec{X}_\mu=\vec{X}_\mu^{(q=4)}$. Consequently,
\begin{align}
 a=1,\dots,6:\ \mathcal{B}_a = V_a^\dagger \Bigl(\sum_{\mu=1}^q \vec{X}_\mu \vec{X}_\mu^\dagger\Bigr)V_a = V_a^\dagger \Phi_0^2 V_a.
\end{align}
Hence $\mathcal{B}_a$ is unitarily equivalent to $\Phi_0^2$ and 
\begin{align}
 \det(u \mathbb{1}-W_0^\dagger W_0) &= \Bigl[ \det(u \mathbb{1}-\Phi_0^2)\Bigr]^{N_{\rm u}}=[(u^2-6u+1)^2]^6 = (u^2-6u+1)^{12},
\end{align}
where we used Eq. (\ref{ref4f}).

\subsection{House lattice / t2.007 ($q=4$)}

\subsubsection{Matrix $W(\vec{k})$} The 2-uniform House lattice (labelled t2.007 above) has coordination number $q=4$ and $N_{\rm u}=5$ sites in the unit cell. We label the sites in the unit cell by $a\in\{1,\dots,5\}$ and the local edges at each site by $\mu\in\{1,\dots,4\}$ according to the following schematic:
\begin{figure}[h!]
    \includegraphics[width=8cm]{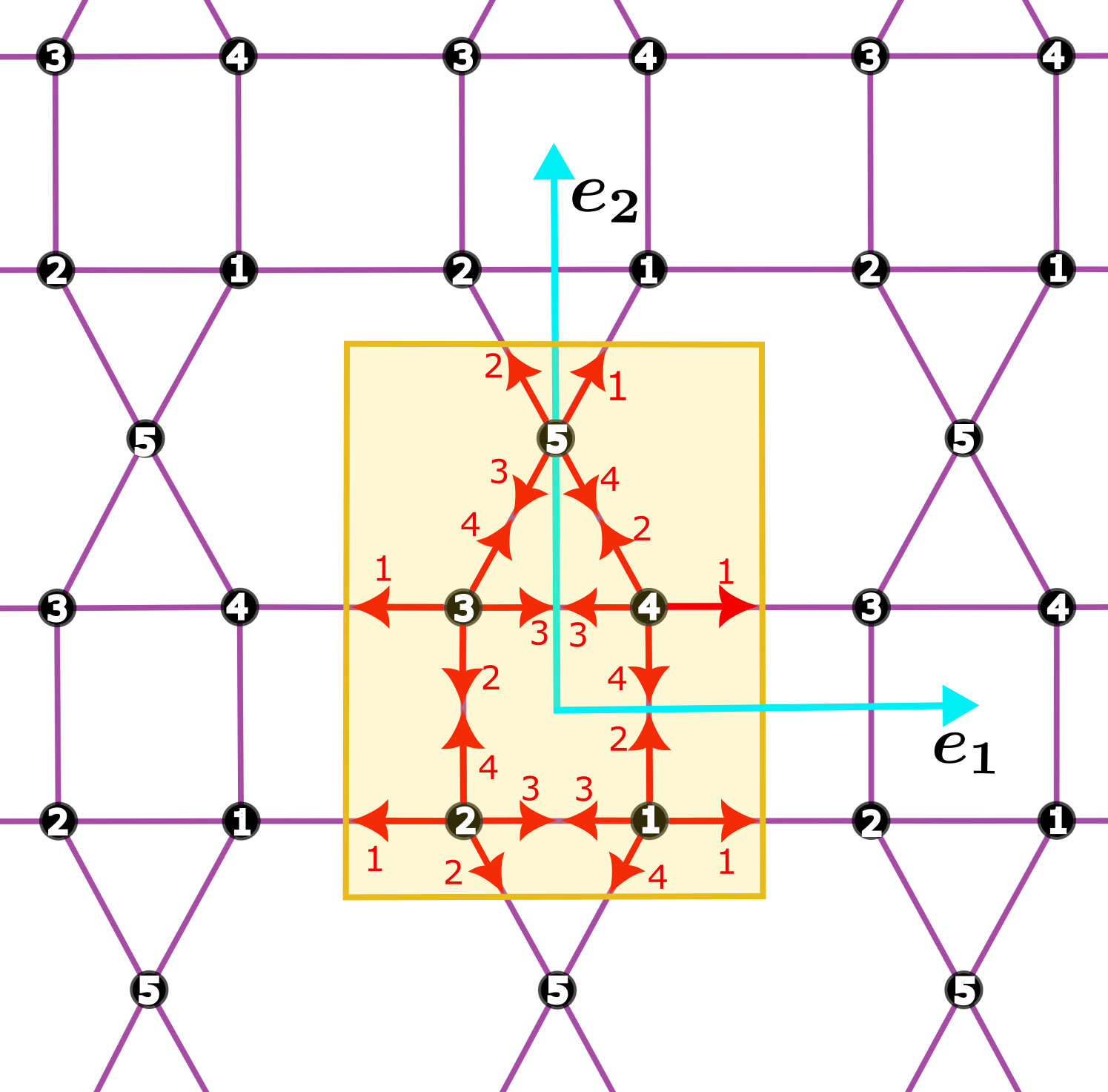}
     \label{FigHouseLabels}
\end{figure}\\
The lattice is non-Archimedean, since the local environment of unit-cell site $5$ is different from the one of unit-cell sites $1,2,3,4$. The nonzero blocks of $W(\vec{k})$ are distributed according to
\begin{align}
 W(\vec{k}) = \begin{pmatrix} 0 & * & 0 & * & * \\ * & 0 & * & 0 & * \\ 0 & * & 0 & * & * \\ * & 0 & * & 0 & *\\ * & * & * & *& 0 \end{pmatrix}.
\end{align}
with
\begin{align}
 W(\vec{k})_{ab} =  E_{a;b}B_a(\vec{k})\Phi_{ab}.
\end{align}
The Bloch matrices at the unit-cell sites are
\begin{align}
 B_1(\vec{k}) &= \begin{pmatrix} e^{-\rmi k_1} & & & \\ & 1 & & \\ & & 1 & \\ & & & e^{\rmi k_2}\end{pmatrix},\ B_2(\vec{k}) = \begin{pmatrix} e^{\rmi k_1} & & & \\ & e^{\rmi k_2} & & \\ & & 1 & \\ & & & 1\end{pmatrix},\ B_3(\vec{k}) = \begin{pmatrix} e^{\rmi k_1} & & & \\ & 1 & & \\ & & 1 & \\ & & & 1\end{pmatrix},\\
 B_4(\vec{k}) &= \begin{pmatrix} e^{-\rmi k_1} & & & \\ & 1 & & \\ & & 1 & \\ & & & 1\end{pmatrix},\  B_5(\vec{k}) = \begin{pmatrix} e^{-\rmi k_2} & & & \\ & e^{-\rmi k_2} & & \\ & & 1 & \\ & & & 1\end{pmatrix}.
\end{align}
The edge-connectivity matrices read
\begin{align}
 E_{5;1} &= \vare_1,\\
 E_{1;4} &= E_{2;5} =E_{3;2} =E_{4;5} = E_{5;2} =\vare_2,\\
 E_{5;3} &= \vare_3,\\
 E_{1;5} &= E_{2;3} = E_{3;5} = E_{4;1}=E_{5;4} =\vare_4,\\
 E_{1;2} &= E_{2;1}=E_{3;4} =E_{4;3} = \vare_5
\end{align}
with
\begin{align}
 \vare_1 = \begin{pmatrix} 1 & & &\\ & 0 & & \\  & & 0 & \\ & & & 0 \end{pmatrix},\ \vare_2= \begin{pmatrix} 0 & & &\\ & 1 & & \\  & & 0 & \\ & & & 0 \end{pmatrix},\ \vare_3=\begin{pmatrix} 0 & & &\\ & 0 & & \\  & & 1 & \\ & & & 0 \end{pmatrix},\ \vare_4 = \begin{pmatrix} 0 & & &\\ & 0 & & \\  & & 0 & \\ & & & 1 \end{pmatrix},\ \vare_5 = \begin{pmatrix} 1 & & &\\ & 0 & & \\  & & 1 & \\ & & & 0 \end{pmatrix}.
\end{align}
We have
\begin{align}
 \vec{\theta}_1 &=-\vec{\theta}_3 = (1,\rmi,-1,e^{-2\pi\rmi/3})^T,\\
 \vec{\theta}_2 &=-\vec{\theta}_4 = (-1,\rmi,1,e^{-\pi\rmi/3})^T,\\
 \vec{\theta}_5 &= (e^{\rmi \pi/3},e^{2\pi\rmi/3},e^{-2\pi\rmi,3},e^{-\pi\rmi/3})^T.
\end{align} 
We arrive at
\begin{align}
 \nonumber \mbox{det}(\mathbb{1}-t W(\vec{k})) ={}& 1+4t^3+8t^4+12t^5+30t^6+68t^7+103t^8+172t^9+228t^{10}\\
 \nonumber &+172 t^{11}+103 t^{12}+68t^{13}+30t^{14}+12t^{15}+8t^{16}+4t^{17}+t^{20}\\
 \nonumber &-4t^3(1-t)^2(1+t)^4(1+t^2)\Bigl[1+2t^2-2t^3+2t^4+t^6+2t(1-t)^2(1+t+t^2)\cos k_1\Bigr]\cos k_2\\
 \nonumber &-4t^2(1-t)^2(1+t)^4(1+t^2)(1+t+t^2)(1-2t+4t^2-2t^3+4t^4-2t^5+t^6)\cos k_1\\
 &+2t^4(1-t)^4(1+t)^6(1+t^2)\cos(2k_1).
\end{align}

\subsubsection{Matrices $W_0$ and $\mathcal{B}_a$} For $\vec{k}=0$ we have
\begin{align}
\mbox{det}(\mathbb{1}-tW_0) ={}& (1+t)^4(1-2t+t^2-2t^3-4t^4-2t^5+t^6-2t^7+t^8)^2.
\end{align}
The resulting critical value is
\begin{align}
 t_{\rm c} &=0.430464,\\
 \Rightarrow\ \frac{T_{\rm c}}{J} &= 2.17171.
\end{align}
We obtain the block-diagonal matrix structure
\begin{align}
 W_0^\dagger W_0 = \begin{pmatrix} \mathcal{B}_1 & & & & \\ & \mathcal{B}_2 & & & \\ & & \mathcal{B}_1 & & \\ & & & \mathcal{B}_2 & \\ & & & & \mathcal{B}_5\end{pmatrix}.
\end{align}
The matrices $\mathcal{B}_1$, $\mathcal{B}_2$, and $\mathcal{B}_5$ have identical characteristic polynomial $\det(u\mathbb{1}-\mathcal{B}_a)=(\mu^2-6\mu+1)^2$ and same eigenvalues. The characteristic polynomial of $W_0^\dagger W_0$ is
\begin{align}
 \det(u\mathbb{1}-W_0^\dagger W_0) &=P_4(u)^5 = (u^2-6u+1)^{10}.
\end{align}

\subsubsection{Relation to reference $\Phi_0^2$} The relation between the $\vec{\theta}_a$ and the reference $\vec{\theta}_0^{(q=4)}$ is $\vec{\theta}_a=U_a\vec{\theta}_0$ with
\begin{align}
 U_1 &= \text{diag}(1,1,1,e^{-\rmi \pi/6}),\\
 U_2 &= \text{diag}(-1,e^{-5\pi\rmi/6},-1,-1),\\
 U_3 &= \text{diag}(-1,-1,-1,e^{5\pi\rmi/6}),\\
 U_4 &= \text{diag}(1,e^{\rmi \pi/6},1,1),\\
 U_5 &= \text{diag}(e^{\rmi \pi/3},e^{\rmi\pi/6},e^{\rmi\pi/3},e^{\rmi \pi/6}).
\end{align}
The corresponding matrices $V_a=\sqrt{U_a}$ are
\begin{align}
 V_1 &= \text{diag}(1,1,1,e^{-\rmi\pi/12}),\\
 V_2 &= \text{diag}(-\rmi,e^{-5\pi\rmi/12},-\rmi,-\rmi),\\
 V_3 &= \text{diag}(\rmi,\rmi,\rmi,e^{5\pi\rmi/12}),\\
 V_4 &= \text{diag}(1,e^{\rmi\pi/12},1,1),\\
 V_5 &= \text{diag}(e^{\rmi \pi/6},e^{\rmi\pi/12},e^{\rmi\pi/6},e^{\rmi\pi/12}),
\end{align}
which satisfy
\begin{align}
 V_3 &= \rmi V_1,\  V_4 = \rmi V_2.
\end{align}
The remaining matrices $V_1$, $V_2$, and $V_5$, however, are not related by a simple global phase. Consider the matrices $M_k^{(a)}=\Phi_{ka}^\dagger E_{k;a}$ such that $\mathcal{B}_a = \sum_{k=1}^{N_{\rm u}} M_k^{(a)}[M_k^{(a)}]^\dagger$. The nonvanishing matrices are
\begin{align}
 a=1:\ M_2^{(a)} &= (\vec{Y}_1\ \vec{0}\ \vec{Y}_3\ \vec{0}),\ M_4^{(a)} = (\vec{0}\ \vec{0}\ \vec{0}\ \vec{Y}_2),\ M_5^{(a)} = (\vec{Y}_4\ \vec{0}\ \vec{0}\ \vec{0}),\\
 a=2:\ M_1^{(a)} &= (\vec{Y}_1\ \vec{0}\ \vec{Y}_3\ \vec{0}),\ M_3^{(a)} = (\vec{0}\ \vec{Y_4}\ \vec{0}\ \vec{0}),\ M_5^{(a)} = (\vec{0}\ \vec{Y}_2\ \vec{0}\ \vec{0}),\\
 a=3:\ M_2^{(a)} &= (\vec{0}\ \vec{0}\ \vec{0}\ \vec{Y}_2),\ M_4^{(a)} = (\vec{Y}_1\ \vec{0}\ \vec{Y}_3\ \vec{0}),\ M_5^{(a)} = (\vec{0}\ \vec{0}\ \vec{Y}_4\ \vec{0}),\\
 a=4:\ M_1{(a)} &= (\vec{0}\ \vec{Y}_4\ \vec{0}\ \vec{0}),\ M_3^{(a)} = (\vec{Y}_1\ \vec{0}\ \vec{Y}_3\ \vec{0}),\ M_5^{(a)} = (\vec{0}\ \vec{0}\ \vec{0}\ \vec{Y}_2),\\
 a=5:\ M_1^{(a)} &= (\vec{0}\ \vec{0}\ \vec{0}\ \vec{Y}_1),\ M_2^{(a)} = (\vec{0}\ \vec{Y}_2\ \vec{0}\ \vec{0}),\ M_3^{(a)} = (\vec{0}\ \vec{0}\ \vec{0}\ \vec{Y}_3),\ M_4^{(a)} = (\vec{0}\ \vec{Y}_4\ \vec{0}\ \vec{0}).
\end{align}
For a given $a$, each set of matrices $\{M_k^{(a)}\}_k$ has exactly $q=4$ nonvanishing columns $\vec{Y}_\mu^{(a)}$ given by
\begin{align}
 a=1,3:\ \vec{Y}_1=\begin{pmatrix} 0 \\ e^{\rmi\pi/4} \\ 1 \\ e^{-\rmi\pi/6}\end{pmatrix},\ \vec{Y}_2 = \begin{pmatrix} e^{-\rmi \pi/4} \\ 0 \\ e^{\rmi \pi/4} \\ e^{\rmi\pi/12} \end{pmatrix},\ \vec{Y}_3 &= \begin{pmatrix} 1 \\ e^{-\rmi\pi/4} \\ 0 \\ e^{\rmi\pi/3} \end{pmatrix},\ \vec{Y}_4 = \begin{pmatrix} e^{\rmi \pi/6} \\ e^{-\rmi\pi/12} \\ e^{-\rmi \pi/3} \\ 0 \end{pmatrix},\ ,\\
  a=2,4:\ \bar{\vec{Y}}_1 = \begin{pmatrix} 0 \\ e^{\rmi \pi/6} \\ 1 \\ e^{-\rmi \pi/4} \end{pmatrix},\ \bar{\vec{Y}}_2=\begin{pmatrix} e^{-\rmi\pi/6} \\ 0 \\ e^{\rmi \pi/3} \\ e^{\rmi \pi/12} \end{pmatrix}, \bar{\vec{Y}}_3 &= \begin{pmatrix} 1 \\ e^{-\rmi \pi/3} \\ 0 \\ e^{\rmi \pi/4} \end{pmatrix},\ \bar{\vec{Y}}_4 = \begin{pmatrix} e^{\rmi \pi/4} \\ e^{-\rmi \pi/12} \\ e^{-\rmi \pi/4} \\ 0 \end{pmatrix},\\
  a=5:\ \tilde{\vec{Y}}_1 = \begin{pmatrix} 0 \\ e^{\rmi\pi/3} \\ 1 \\ e^{-\rmi\pi/6} \end{pmatrix},\ \tilde{\vec{Y}}_2 = \begin{pmatrix} e^{-\rmi\pi/3} \\ 0 \\ e^{\rmi\pi/6} \\ 1 \end{pmatrix}, \tilde{\vec{Y}}_3 &= \begin{pmatrix} 1 \\ e^{-\rmi \pi/6} \\ 0 \\ e^{\rmi\pi/3}\end{pmatrix},\ \tilde{\vec{Y}}_4 = \begin{pmatrix} e^{\rmi\pi/6} \\ 1 \\ e^{-\rmi\pi/3} \\ 0\end{pmatrix}.
\end{align}
Thus we have
\begin{align}
 a=1,3:\ \mathcal{B}_1 &= \mathcal{B}_3 = \sum_{k=1}^{N_{\rm u}} M_k^{(a)}[M_k^{(a)}]^\dagger= \sum_{\mu=1}^4 \vec{Y}_\mu \vec{Y}_\mu^\dagger,\\
 a=2,4:\ \mathcal{B}_2&=\mathcal{B}_4 = \sum_{k=1}^{N_{\rm u}} M_k^{(a)}[M_k^{(a)}]^\dagger= \sum_{\mu=1}^4 \bar{\vec{Y}}_\mu \bar{\vec{Y}}_\mu^\dagger,\\
 a=5:\ \mathcal{B}_5 &=\sum_{k=1}^{N_{\rm u}} M_k^{(a)}[M_k^{(a)}]^\dagger= \sum_{\mu=1}^4 \tilde{\vec{Y}}_\mu \tilde{\vec{Y}}_\mu^\dagger.
\end{align}
The vectors $\vec{Y}_\mu^{(a)}$ satisfy
\begin{align}
 a=1,\dots,5:\ Y_\mu^{(a)}(\alpha) &= \begin{cases} \sqrt{-\theta_a(\mu)\theta_a^*(\alpha)} & \alpha\neq \mu \\ 0 & \alpha = \mu\end{cases} \\
  &= \begin{cases} (V_a)_{\mu\mu}(V_a^*)_{\alpha\alpha} \sqrt{-\theta_0(\mu)\theta_0^*(\alpha)} & \alpha\neq \mu \\ 0 & \alpha = \mu\end{cases},
\end{align}
which is equivalent to
\begin{align}
 a=1,\dots,5:\ \vec{Y}_\mu = (V_a)_{\mu\mu} V_a^\dagger \vec{X}_\mu
\end{align}
with the $q=4$ reference vectors $\vec{X}_\mu=\vec{X}_\mu^{(q=4)}$. Consequently,
\begin{align}
 a=1,\dots,5:\ \mathcal{B}_a = V_a^\dagger \Bigl(\sum_{\mu=1}^q \vec{X}_\mu \vec{X}_\mu^\dagger\Bigr)V_a = V_a^\dagger \Phi_0^2 V_a.
\end{align}
Hence all $\mathcal{B}_a$ are unitarily equivalent to $\Phi_0^2$ and we conclude that
\begin{align}
\det(u \mathbb{1}-W_0^\dagger W_0) = \prod_{a=1}^{N_{\rm u}}\det(u \mathbb{1}_q-\mathcal{B}_a) = \Bigl[ \det(u \mathbb{1}_q-\Phi_0^2)\Bigr]^{N_{\rm u}} = [(u^2-6u+1)^2]^5 = (u^2-6u+1)^{10}.
\end{align}

\clearpage
\subsection{Trellis lattice ($q=5$)}

\subsubsection{Matrix $W(\vec{k})$} The Trellis lattice consists of three regular triangles and two regular squares meeting at a vertex. Thus, it has coordination number $q=5$ with $N_{\rm u}=2$ sites in the unit cell. We label the sites in the unit cell by $a\in\{1,2\}$ and the local edges at each site by $\mu\in\{1,\dots,5\}$ according to the following schematic:
\begin{figure}[h!]
    \centering
    \includegraphics[width=0.4\linewidth]{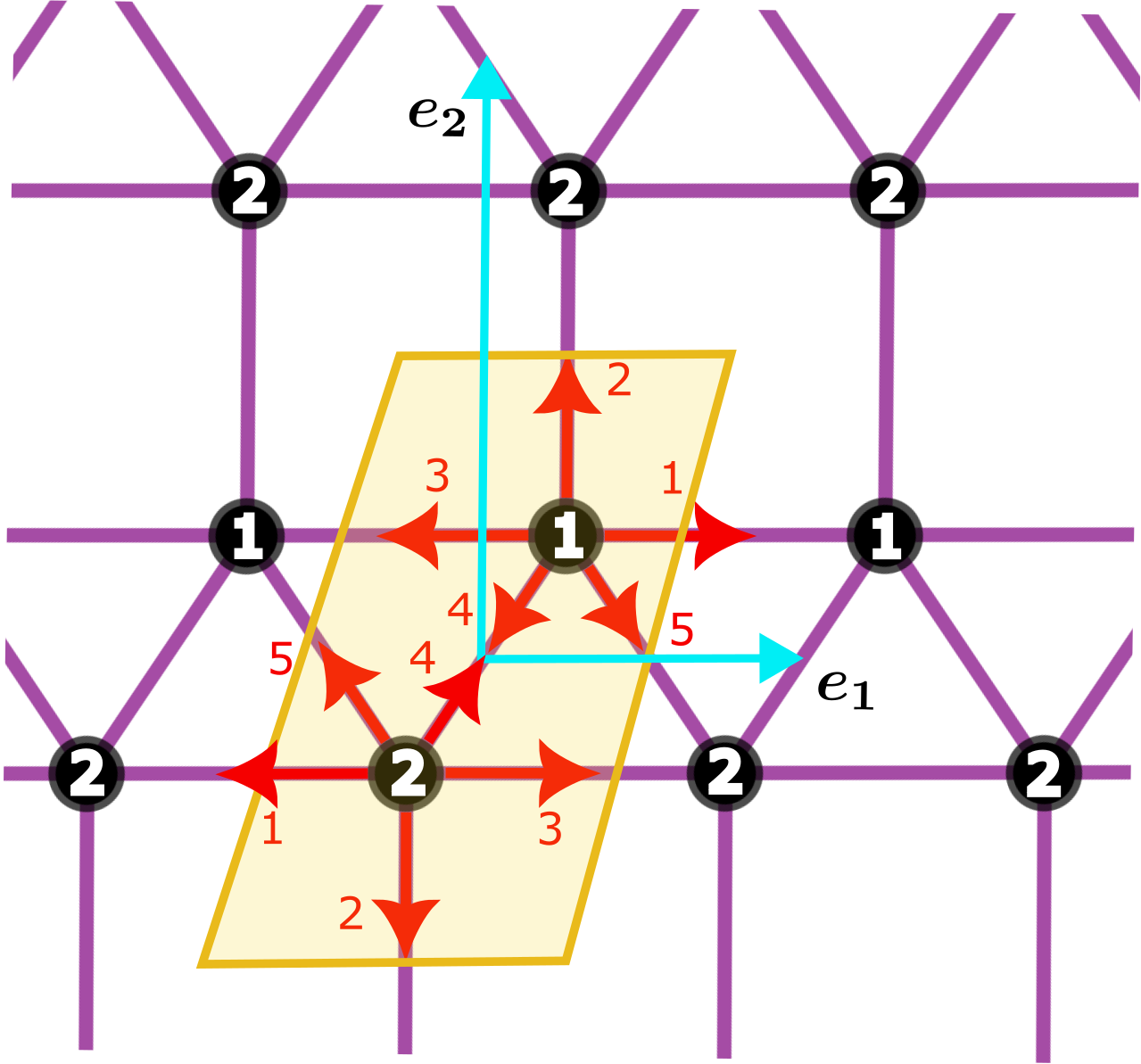}
\end{figure}\\
We have
\begin{align}
 W(\vec{k}) = \begin{pmatrix} E_{1;1} B_1(\vec{k}) \Phi_{11} & E_{1;2} B_1(\vec{k}) \Phi_{12} \\ E_{2;1} B_2(\vec{k}) \Phi_{21} & E_{2;2} B_2(\vec{k}) \Phi_{22} \end{pmatrix}.
\end{align}
The Bloch matrices are given by
\begin{align}
 B_1(\vec{k}) &= \text{diag}(e^{-\rmi k_1},e^{-\rmi k_2},e^{\rmi k_1},1,e^{-\rmi k_1}),\\
 B_2(\vec{k}) &= \text{diag}(e^{\rmi k_1},e^{\rmi k_2},e^{-\rmi k_1},1,e^{\rmi k_1})
\end{align}
and the edge-connectivity matrices read
\begin{align}
 E_{1;1} &= E_{2;2} = \vare_1 = \text{diag}(1,0,1,0,0),\\
 E_{1;2} &= E_{2;1} = \vare_2 = \text{diag}(0,1,0,1,1).
\end{align}
We have
\begin{align}
 \vec{\theta}_1 &= (1,\rmi,-1,e^{-2\pi\rmi/3},e^{-\rmi\pi/3}),\\
 \vec{\theta}_2 &= (-1,-\rmi,1,e^{\rmi\pi/3},e^{2\pi\rmi/3}).
\end{align}
From this we compute $\Phi_{ka}$ and arrive at
\begin{align}
 \nonumber \det(\mathbb{1}-t W(\vec{k}))={}&(1+t)^2(1+t^2)^2(1-2t+5t^2) -4t(1+t)^2(1-2t+4t^2-2t^3-t^4)\cos k_1 \\
 &+t^2 (1-t^2)^3\Bigl[2\cos(2k_1)+\vare_\Delta(\vec{k})\Bigr].
\end{align}

\subsubsection{Matrices $W_0$ and $\mathcal{B}$} For $\vec{k}=0$ we have
\begin{align}
   \det(\mathbb{1}-t W_0)=(1-3 t)^2 (1+t)^2 (1+t^2)^2,
\end{align}
which gives $t_c=\frac{1}{3}$. We obtain the block-diagonal structure
\begin{align}
 W_0^\dagger W_0 = \begin{pmatrix} \mathcal{B} & 0 \\ 0 & \mathcal{B} \end{pmatrix}
\end{align} 
with
\begin{align}
 \mathcal{B} &= \Phi_{11}^\dagger \vare_1 \Phi_{11} + \Phi_{21}^\dagger \vare_2 \Phi_{21}.
\end{align}
We have $\det(u \mathbb{1}_5 -\mathcal{B})=P_5(\mu)=\mu(\mu^2-10\mu+5)^2$ and thus
\begin{equation}
    \det(u\mathbbm{1}- W_0^\dagger W_0)=\Bigl[ \det(u \mathbb{1}_5-\mathcal{B})\Bigr]^{N_{\rm u}} =\Bigl[u(u^2-10u+5)^2\Bigr]^2 =  u ^2 (u^2-10u +5)^4.
\end{equation}

\subsubsection{Relation to reference $\Phi_0^2$} The vectors $\vec{\theta}_a$ are related to the $q=5$ reference $\vec{\theta}_0=(1,e^{2\pi\rmi/5},e^{4\pi\rmi/5},e^{-4\pi\rmi/5},e^{-2\pi\rmi/5})^T$ through $\vec{\theta}_a = U_a \vec{\theta}_0$ with
\begin{align}
 U_1 &= \text{diag}(1,e^{\rmi\pi/10},e^{\rmi\pi/5},e^{2\pi\rmi/15},e^{\rmi\pi/15}),\\
 U_2 &= \text{diag}(-1,e^{-9\pi\rmi/10},e^{-4\pi\rmi/5},e^{-13\pi\rmi/15},e^{-14\pi\rmi/15}) = -U_1.
\end{align}
Define $V_a=\sqrt{U_a}$ to find
\begin{align}
 V_1 &= \text{diag}(1,e^{\rmi\pi/20},e^{\rmi\pi/10},e^{\pi\rmi/15},e^{\rmi\pi/30}),\\
 V_2 &=\text{diag}(-\rmi, e^{-9\pi\rmi/20},e^{-2\pi\rmi/5},e^{-13\pi\rmi/30},e^{-7\pi\rmi/15})=-\rmi V_1.
\end{align}
Consider the matrices $M_k^{(a)}=\Phi_{ka}^\dagger E_{k;a}$. We have
\begin{align}
 a=1:\ M_1^{(a)} &= (\vec{Y}_3\ \vec{0}\ \vec{Y}_1\ \vec{0}\ \vec{0}),\ M_2^{(a)} = (\vec{0}\ \vec{Y}_2\ \vec{0}\ \vec{Y}_4\ \vec{Y}_5),\\
 a=2:\ M_1^{(a)} &= (\vec{0}\ \vec{Y}_2\ \vec{0}\ \vec{Y}_4\ \vec{Y}_5),\ M_2^{(a)} =  (\vec{Y}_3\ \vec{0}\ \vec{Y}_1\ \vec{0}\ \vec{0}).
\end{align}
For each $a$, the matrices have exactly $q=5$ nonvanishing columns given by
\begin{align}
 \vec{Y}_1 = \begin{pmatrix}  0  \\  e^{\rmi\pi/4} \\  1  \\  e^{-\rmi\pi/6}  \\  e^{-\rmi\pi/3}  \end{pmatrix},\ \vec{Y}_2 = \begin{pmatrix}  e^{-\rmi\pi/4}  \\ 0 \\  e^{\rmi\pi/4}  \\  e^{\rmi \pi/12}  \\  e^{-\rmi \pi/12} \end{pmatrix}, \vec{Y}_3=\begin{pmatrix} 1  \\ e^{-\rmi \pi/4} \\ 0  \\ e^{\rmi\pi/3}  \\ e^{\rmi\pi/6}  \end{pmatrix},\ \vec{Y}_4 = \begin{pmatrix}  e^{\rmi\pi/6}  \\ e^{-\rmi\pi/12}  \\  e^{-\rmi\pi/3}  \\  0  \\ e^{\rmi \pi/3}  \end{pmatrix},\ \vec{Y}_5 = \begin{pmatrix}  e^{\rmi\pi/3} \\  e^{\rmi\pi/12} \\  e^{-\rmi\pi/6} \\  e^{-\rmi \pi/3} \\ 0 \end{pmatrix}.
\end{align}
We have
\begin{align}
 a=1,2:\ \mathcal{B} = \mathcal{B}_a = \sum_{k=1}^{N_{\rm u}} M_k^{(a)} [M_k^{(a)}]^\dagger = \sum_{\mu=1}^5 \vec{Y}_\mu\vec{Y}_\mu^\dagger.
\end{align}
The vectors $\vec{Y}_\mu^{(a)}$ satisfy
\begin{align}
 a=1,2:\ Y_\mu^{(a)}(\alpha) &= \begin{cases} \sqrt{-\theta_a(\mu)\theta_a^*(\alpha)} & \alpha\neq \mu \\ 0 & \alpha = \mu\end{cases} \\
  &= \begin{cases} (V_a)_{\mu\mu}(V_a^*)_{\alpha\alpha} \sqrt{-\theta_0(\mu)\theta_0^*(\alpha)} & \alpha\neq \mu \\ 0 & \alpha = \mu\end{cases},
\end{align}
which is equivalent to
\begin{align}
 a=1,2:\ \vec{Y}_\mu = (V_a)_{\mu\mu} V_a^\dagger \vec{X}_\mu
\end{align}
with the $q=5$ reference vectors $\vec{X}_\mu=\vec{X}_\mu^{(q=5)}$. Consequently,
\begin{align}
 a=1,2:\ \mathcal{B}_a = V_a^\dagger \Bigl(\sum_{\mu=1}^q \vec{X}_\mu \vec{X}_\mu^\dagger\Bigr)V_a = V_a^\dagger \Phi_0^2 V_a.
\end{align}
This shows that $\mathcal{B}$ is unitarily equivalent to $\Phi_0^2$. We conclude that
\begin{align}
 \det(u \mathbb{1}-W_0^\dagger W_0) = \Bigl[ \det(u \mathbb{1}_5 - \Phi_0^2)\Bigr]^2 = [P_5(u)]^2 = u^2(u^2-10u+5)^4.
\end{align}

\clearpage
\subsection{SrCuBO lattice ($q=5$)}

\subsubsection{Matrix $W(\vec{k})$} The SrCuBO lattice consists of three regular triangles and two regular polygons meeting at a vertex. Thus, it has coordination number $q=5$ with $N_{\rm u}=4$ sites in the unit cell. We label the sites in the unit cell by $a\in\{1,2,3,4\}$ and the local edges at each site by $\mu\in\{1,\dots,5\}$ according to the following schematic:
\begin{figure}[h!]
    \includegraphics[width=9cm]{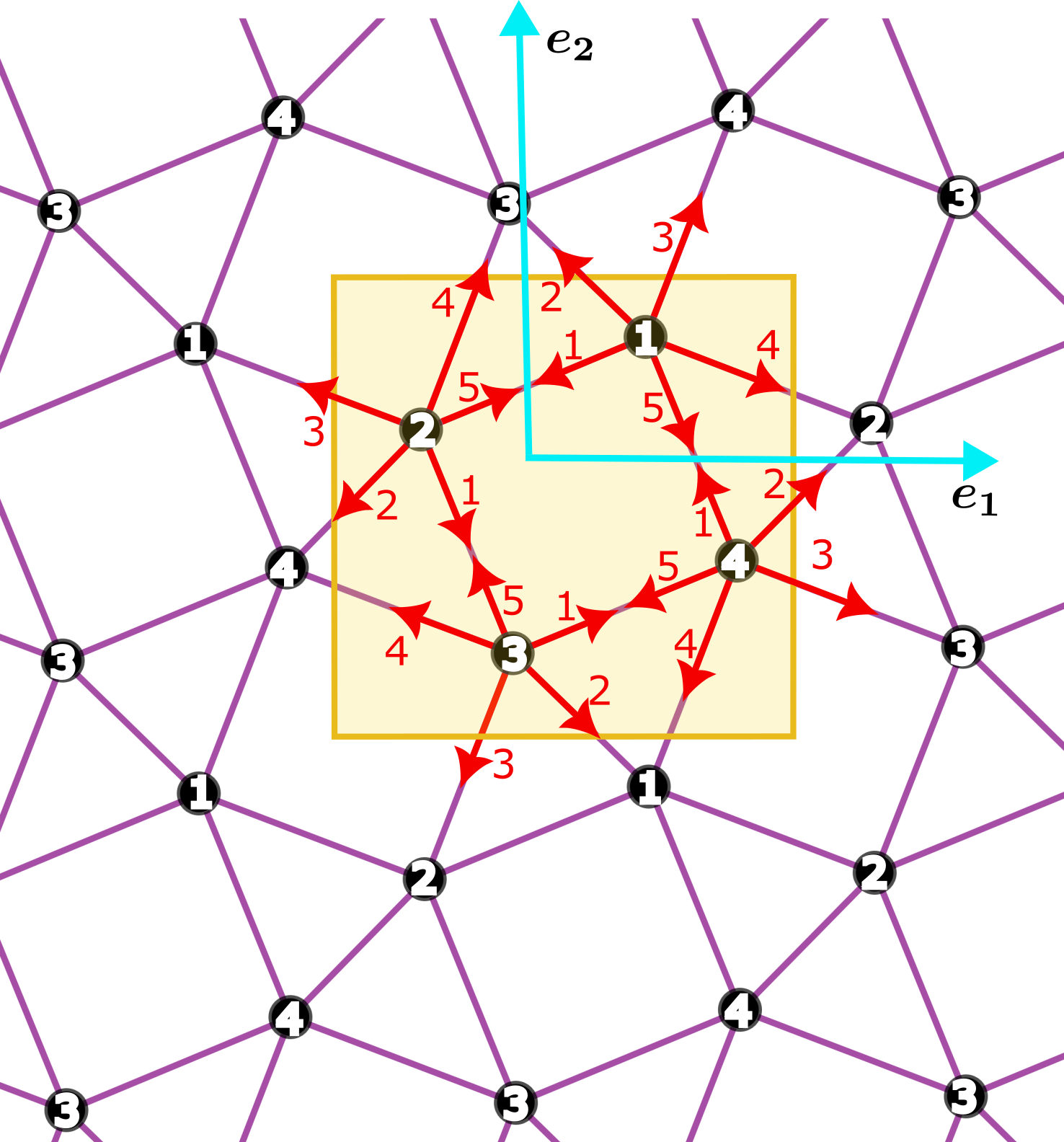}
     \label{FigSrCuBOLabels}
\end{figure}\\ 
The nonvanishing blocks of $W(\vec{k})$ are 
\begin{equation}
W(\vec{k})=\left(
\begin{array}{cccc}
 0 & * & * & * \\
 * & 0 & * & * \\
 * & * & 0 & * \\
 * & * & * & 0 \\
\end{array}
\right).
\end{equation}
The Bloch matrices read
\begin{align}
B_1(\boldsymbol{k})&=\left(
\begin{array}{ccccc}
 1 & 0 & 0 & 0 & 0 \\
 0 & e^{-\rmi k_2} & 0 & 0 & 0 \\
 0 & 0 & e^{-\rmi k_2} & 0 & 0 \\
 0 & 0 & 0 & e^{-\rmi k_1} & 0 \\
 0 & 0 & 0 & 0 & 1 \\
\end{array},\right)\ B_2(\boldsymbol{k})=\left(
\begin{array}{ccccc}
 1 & 0 & 0 & 0 & 0 \\
 0 & e^{\rmi k_1} & 0 & 0 & 0 \\
 0 & 0 & e^{\rmi k_1} & 0 & 0 \\
 0 & 0 & 0 & e^{-\rmi k_2} & 0 \\
 0 & 0 & 0 & 0 & 1 \\
\end{array}
\right),\\
 B_3(\boldsymbol{k})&=\left(
\begin{array}{ccccc}
 1 & 0 & 0 & 0 & 0 \\
 0 & e^{\rmi k_2} & 0 & 0 & 0 \\
 0 & 0 & e^{\rmi k_2} & 0 & 0 \\
 0 & 0 & 0 & e^{\rmi k_1} & 0 \\
 0 & 0 & 0 & 0 & 1 \\
\end{array}
\right),\ B_4(\boldsymbol{k})=\left(
\begin{array}{ccccc}
 1 & 0 & 0 & 0 & 0 \\
 0 & e^{-\rmi k_1} & 0 & 0 & 0 \\
 0 & 0 & e^{-\rmi k_1} & 0 & 0 \\
 0 & 0 & 0 & e^{\rmi k_2} & 0 \\
 0 & 0 & 0 & 0 & 1 \\
\end{array}
\right).
\end{align}
Since the tiling consists of regular triangles and squares, the angles between edges are uniquely defined. Define $\alpha=e^{\rmi\frac{\pi}{2}}, \sigma=e^{\rmi\frac{\pi}{3}}$ and $g=e^{\rmi\frac{\pi}{2}}$, then 
\begin{equation}\boldsymbol{\theta}_a= g^{a-1}(-1,-\sigma^{-1},-\sigma^{-2},-\sigma^{-2}\alpha^{-1},-\sigma^{-3}\alpha^{-1})^{T}\end{equation}
for $a=1,\dots,4.$
The edge-connectivity matrices read \begin{equation}E_{1;2}=E_{2;3}=E_{3;4}=E_{4;1}=\left(
\begin{array}{ccccc}
 1 & 0 & 0 & 0 & 0 \\
 0 & 0 & 0 & 0 & 0 \\
 0 & 0 & 0 & 0 & 0 \\
 0 & 0 & 0 & 1 & 0 \\
 0 & 0 & 0 & 0 & 0 \\
\end{array}
\right)=\varepsilon_1, \end{equation}
    \begin{equation}E_{2;1}=E_{3;2}=E_{4;3}=E_{1;4
    }=\left(
\begin{array}{ccccc}
 0 & 0 & 0 & 0 & 0 \\
 0 & 0 & 0 & 0 & 0 \\
 0 & 0 & 1 & 0 & 0 \\
 0 & 0 & 0 & 0 & 0 \\
 0 & 0 & 0 & 0 & 1 \\
\end{array}
\right)=\varepsilon_2 ,\end{equation}
\begin{equation}E_{1;3}=E_{2;4}=E_{3;1}=E_{4;2}=\left(
\begin{array}{ccccc}
 0 & 0 & 0 & 0 & 0 \\
 0 & 1 & 0 & 0 & 0 \\
 0 & 0 & 0 & 0 & 0 \\
 0 & 0 & 0 & 0 & 0 \\
 0 & 0 & 0 & 0 & 0 \\
\end{array}
\right)=\varepsilon_3 .\end{equation}
We then have
\begin{align}
\det(\mathbbm{1}-t W(\boldsymbol{k}))&=(1+t)^4(1+t^2)^2(1-4t+8t^2-4t^3+2t^4+20t^5+32t^6-12t^7+21t^8) - t^4(1-t^2)^6\varepsilon_{\square}(2\bk) \nonumber \\
&+ 2(1-t)^3(1+t)^5(1+3t^2)(t+t^3)^2\varepsilon_{\square}(\bk) \nonumber \\
&+2t(1-t)^3t^2(1+t)^5(1+t^2)(1-t^2)\varepsilon_{\square}(k_1+k_2,k_1-k_2).
\end{align}
The critical temperature is determined by the roots of
\begin{align}
    \det(\mathbbm{1}-t W_0)&=(1+t)^4 \left(-1+2 t+t^2+4 t^3+9 t^4-6 t^5+7 t^6\right)^2,
\end{align}
which yields
\begin{equation}
t_c = 0.329024.
\end{equation}
The characteristic polynomial of $W_0^\dagger W_0$ is given by
\begin{equation}
    \det(u \mathbbm{1}- W_0^\dagger W_0)=u^4 (u^2-10u+5)^8.
\end{equation}

\clearpage
\subsection{Maple-Leaf lattice ($q=5$)}

\subsubsection{Matrix $W(\vec{k})$} The Maple-Leaf lattice consists of four regular triangles and a regular hexagon meeting at a vertex. It has coordination number $q=5$ and contains $N_{\rm u}=6$ sites in the unit cell. We label the sites in the unit cell by $a\in\{1,\dots,6\}$ and the local edges at each site by $\mu\in\{1,\dots,5\}$ according to the following schematic:
\begin{figure}[h!]
    \centering
    \includegraphics[width=0.5\linewidth]{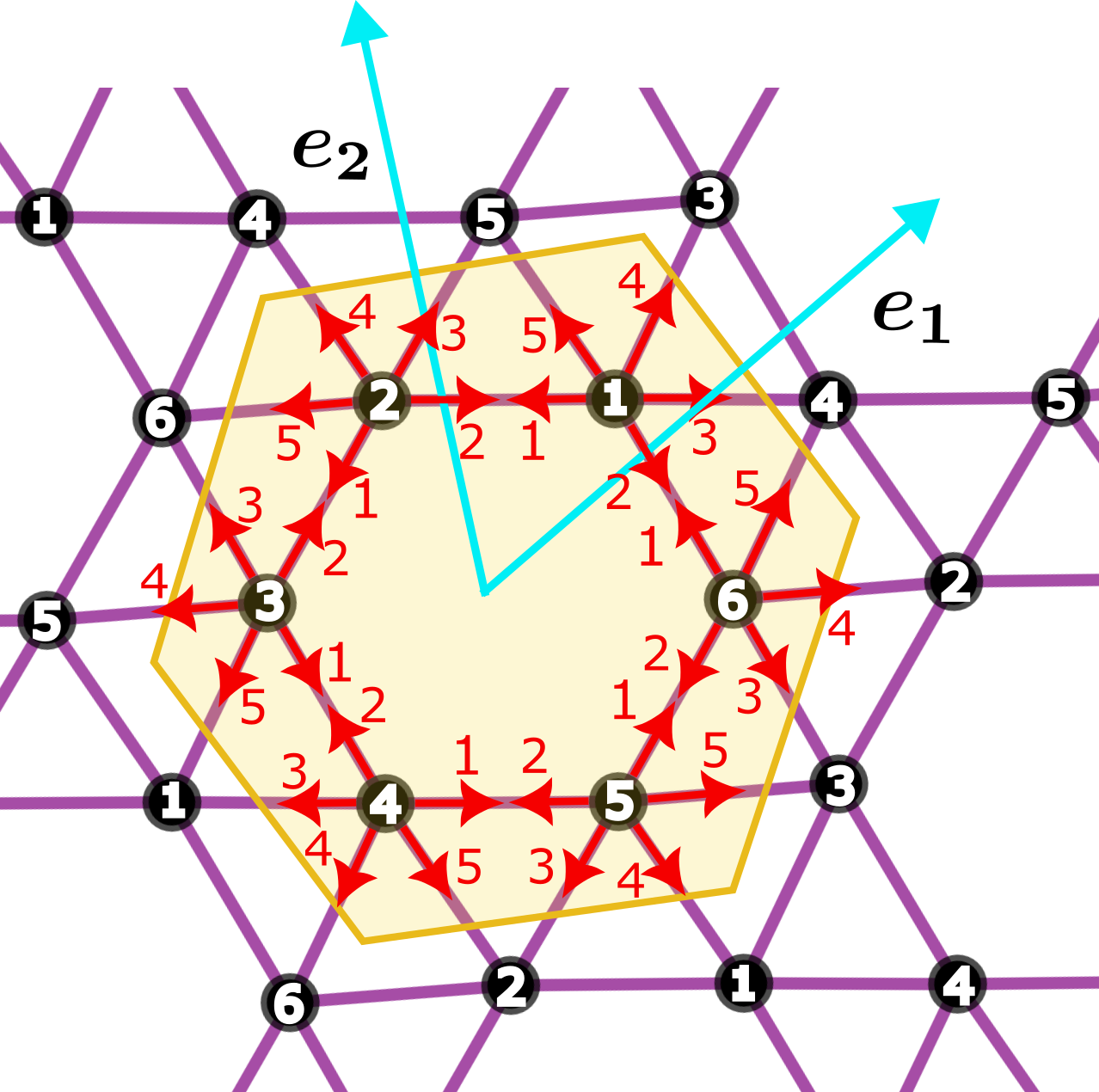}
\end{figure}\\
The nonvanishing entries of $W(\vec{k})$ are given by
\begin{equation}W(\vec{k})=\left(
\begin{array}{cccccc}
 0 & * & * & * & * & * \\
 * & 0 & * & * & * & * \\
 * & * & 0 & * & * & * \\
 * & * & * & 0 & * & * \\
 * & * & * & * & 0 & * \\
 * & * & * & * & * & 0 \\
\end{array}
\right).\end{equation}
The Bloch matrices read
\begin{align}
B_1(\boldsymbol{k})&=\left(
\begin{array}{ccccc}
 1 & 0 & 0 & 0 & 0 \\
 0 & 1 & 0 & 0 & 0 \\
 0 & 0 & e^{-\rmi k_1} & 0 & 0 \\
 0 & 0 & 0 & e^{-\rmi k_2} & 0 \\
 0 & 0 & 0 & 0 & e^{-\rmi k_2} \\
\end{array}
\right),\ B_2(\boldsymbol{k})=\left(
\begin{array}{ccccc}
 1 & 0 & 0 & 0 & 0 \\
 0 & 1 & 0 & 0 & 0 \\
 0 & 0 & e^{-\rmi k_2} & 0 & 0 \\
 0 & 0 & 0 & e^{-\rmi \left(k_2-k_1\right)} & 0 \\
 0 & 0 & 0 & 0 & e^{-\rmi \left(k_2-k_1\right)} \\
\end{array}
\right),\\
B_3(\boldsymbol{k})&=\left(
\begin{array}{ccccc}
 1 & 0 & 0 & 0 & 0 \\
 0 & 1 & 0 & 0 & 0 \\
 0 & 0 & e^{-\rmi \left(k_2-k_1\right)} & 0 & 0 \\
 0 & 0 & 0 & e^{\rmi k_1} & 0 \\
 0 & 0 & 0 & 0 & e^{\rmi k_1} \\
\end{array}
\right),\ B_4(\boldsymbol{k})=\left(
\begin{array}{ccccc}
 1 & 0 & 0 & 0 & 0 \\
 0 & 1 & 0 & 0 & 0 \\
 0 & 0 & e^{\rmi k_1} & 0 & 0 \\
 0 & 0 & 0 & e^{\rmi k_2} & 0 \\
 0 & 0 & 0 & 0 & e^{\rmi k_2} \\
\end{array}
\right),\\
B_5(\boldsymbol{k})&=\left(
\begin{array}{ccccc}
 1 & 0 & 0 & 0 & 0 \\
 0 & 1 & 0 & 0 & 0 \\
 0 & 0 & e^{\rmi k_2} & 0 & 0 \\
 0 & 0 & 0 & e^{-\rmi \left(k_1-k_2\right)} & 0 \\
 0 & 0 & 0 & 0 & e^{-\rmi \left(k_1-k_2\right)} \\
\end{array}
\right),\ B_6(\boldsymbol{k})=\left(
\begin{array}{ccccc}
 1 & 0 & 0 & 0 & 0 \\
 0 & 1 & 0 & 0 & 0 \\
 0 & 0 & e^{-\rmi \left(k_1-k_2\right)} & 0 & 0 \\
 0 & 0 & 0 & e^{-\rmi k_1} & 0 \\
 0 & 0 & 0 & 0 & e^{-\rmi k_1} \\
\end{array}
\right).
\end{align}
Let $\alpha=e^{\rmi\frac{\pi}{3}}$ and $g= e^{\rmi \frac{\pi}{3}}$, then we identify the phases through the fixed angles as 
\begin{equation}\boldsymbol{\theta}_a=g^{a-1}(-1,-\alpha^2,-\alpha^3,-\alpha^4,-\alpha^5)^T \end{equation}
for $a=1,\dots,6$.
The edge-connectivity matrices are given by
\begin{equation}E_{1;2}=E_{2;3}=E_{3;4}=E_{4;5}=E_{5;6}=E_{6;1}=\left(
\begin{array}{ccccc}
 1 & 0 & 0 & 0 & 0 \\
 0 & 0 & 0 & 0 & 0 \\
 0 & 0 & 0 & 0 & 0\\
 0 & 0 & 0 & 0 & 0\\
 0 & 0 & 0 & 0 & 0\\
\end{array}
\right)=\varepsilon_1,\end{equation}
\begin{equation}E_{2;1}=E_{3;2}=E_{4;3}=E_{5;4}=E_{6;5}=E_{1;6}=\left(
\begin{array}{ccccc}
 0 & 0 & 0 & 0 & 0 \\
 0 & 1 & 0 & 0 & 0 \\
 0 & 0 & 0 & 0 & 0\\
 0 & 0 & 0 & 0 & 0\\
 0 & 0 & 0 & 0 & 0\\
\end{array}
\right)=\varepsilon_2,\end{equation}
\begin{equation}E_{1;4}=E_{2;5}=E_{3;6}=E_{4;1}=E_{5;2}=E_{6;3}=\left(
\begin{array}{ccccc}
 0 & 0 & 0 & 0 & 0 \\
 0 & 0 & 0 & 0 & 0 \\
 0 & 0 & 1 & 0 & 0\\
 0 & 0 & 0 & 0 & 0\\
 0 & 0 & 0 & 0 & 0\\
\end{array}
\right)=\varepsilon_3,\end{equation}
\begin{equation}E_{5;1}=E_{6;2}=E_{1;3}=E_{2;4}=E_{3;5}=E_{4;6}=\left(
\begin{array}{ccccc}
 0 & 0 & 0 & 0 & 0 \\
 0 & 0 & 0 & 0 & 0 \\
 0 & 0 & 0 & 0 & 0\\
 0 & 0 & 0 & 1 & 0\\
 0 & 0 & 0 & 0 & 0\\
\end{array}
\right)=\varepsilon_4,\end{equation}
\begin{equation}E_{1;5}=E_{2;6}=E_{3;1}=E_{4;2}=E_{5;3}=E_{6;4}=\left(
\begin{array}{ccccc}
 0 & 0 & 0 & 0 & 0 \\
 0 & 0 & 0 & 0 & 0 \\
 0 & 0 & 0 & 0 & 0\\
 0 & 0 & 0 & 0 & 0\\
 0 & 0 & 0 & 0 & 1\\
\end{array}
\right)=\varepsilon_5,\end{equation}
which yields
\begin{align}
   \nonumber \det(\mathbbm{1}-tW(\boldsymbol{k}))={}& (1 + t)^8\Bigl(1 - 8t + 36t^2 - 104t^3 + 220t^4 - 336t^5 + 438t^6 - 384t^7 + 444t^8 - 168t^9 \\
\nonumber & \quad + 384t^{10} + 24t^{11} + 264t^{12} + 198t^{14} - 48t^{15} + 63t^{16}\Bigr) \\
\nonumber & +2(1-t)^7t^5(1+t)^9(1+t^2)\varepsilon_\Delta(2k_1-k_2,k_1-2k_2) \\
\nonumber & +2(1-t)^3t^3(1+t)^8(1+t^2)\Bigl(2-4t+13t^2-7t^3+14t^4 +6t^5+t^6+5t^7+2t^8\Bigr)\varepsilon_\Delta(\bk) \\
&- (1-t)^9t^6(1+t)^9\varepsilon_\Delta(2\bk) .
\end{align}
The critical temperature is given by the root of the equation
\begin{align}
   \nonumber \det(\mathbbm{1}-tW_0)&=(1+t)^8 \left(1+3 t^2\right)^2 \left(1-4 t+7 t^2-12 t^3+3 t^4-3 t^6\right)^2=0\\ \implies t_c &= 0.344296.
\end{align}
The characteristic polynomial of $W_0^\dagger W_0$ is then given by
\begin{equation}
    \det(u \mathbbm{1}- W_0^\dagger W_0)=u^6 (u^2-10u +5)^{12}
\end{equation}

\clearpage
\subsection{Triangular lattice ($q=6$)}

\subsubsection{Matrix $W(\vec{k})$} The Triangular lattice is a Bravais lattice with only one site in the unit cell, labelled $a=1$, with coordination number $q=6$. The Bloch matrix is
\begin{align}
 B_1(\vec{k}) = \begin{pmatrix} e^{-\rmi k_1} & & & & & \\ & e^{-\rmi k_2} & & & &  \\ & & e^{\rmi (k_1-k_2)} & & &  \\ & & & e^{\rmi k_1} & &  \\ & & &  & e^{\rmi k_2} &  \\ & & & & & e^{-\rmi(k_1-k_2)} \end{pmatrix}
\end{align}
and we have $E_{1;1}=\mathbb{1}_6$. We have
\begin{align}
 \vec{\theta}_1 = (e^{\rmi \pi/6},\rmi,e^{5\pi\rmi/6},e^{-5\pi\rmi/6},-\rmi,e^{-\rmi \pi/6})^T = e^{\rmi \pi/6}\vec{\theta}_0^{(q=6)}.
\end{align}
From this we obtain
\begin{align}
 \Phi_{11} = \begin{pmatrix} 1 & e^{\rmi \pi/6} & e^{\rmi \pi/3} & 0 & e^{-\rmi \pi/3} & e^{-\rmi \pi/6} \\ e^{-\rmi \pi/6} & 1 & e^{\rmi \pi/6} & e^{\rmi \pi/3} & 0 & e^{-\rmi \pi/3}  \\ e^{-\rmi \pi/3} & e^{-\rmi \pi/6} & 1 & e^{\rmi \pi/6} & e^{\rmi \pi/3} & 0   \\  0 & e^{-\rmi \pi/3} & e^{-\rmi \pi/6} & 1 & e^{\rmi \pi/6} & e^{\rmi \pi/3} \\ e^{\rmi \pi/3} &  0 & e^{-\rmi \pi/3} & e^{-\rmi \pi/6} & 1 & e^{\rmi \pi/6}  \\ e^{\rmi \pi/6} & e^{\rmi \pi/3} &  0 & e^{-\rmi \pi/3} & e^{-\rmi \pi/6} & 1 \end{pmatrix} = \Phi_0^{(q=6)}
\end{align}
and 
\begin{align}
 \nonumber W(\vec{k}) &= B_1(\vec{k}) \Phi_{11} \\
 &= \begin{pmatrix} e^{-\rmi k_1} & e^{-\rmi k_1}e^{\rmi \pi/6} & e^{-\rmi k_1}e^{\rmi \pi/3} & 0 & e^{-\rmi k_1}e^{-\rmi \pi/3} & e^{-\rmi k_1}e^{-\rmi \pi/6} \\ e^{-\rmi k_2}e^{-\rmi \pi/6} & e^{-\rmi k_2} & e^{-\rmi k_2}e^{\rmi \pi/6} & e^{-\rmi k_2}e^{\rmi \pi/3} & 0 & e^{-\rmi k_2}e^{-\rmi \pi/3}  \\ e^{\rmi(k_1-k_2)} e^{-\rmi \pi/3} & e^{\rmi(k_1-k_2)}e^{-\rmi \pi/6} & e^{\rmi(k_1-k_2)} & e^{\rmi(k_1-k_2)}e^{\rmi \pi/6} & e^{\rmi(k_1-k_2)}e^{\rmi \pi/3} & 0   \\  0 & e^{\rmi k_1}e^{-\rmi \pi/3} & e^{\rmi k_1}e^{-\rmi \pi/6} & e^{\rmi k_1} & e^{\rmi k_1}e^{\rmi \pi/6} & e^{\rmi k_1}e^{\rmi \pi/3} \\ e^{\rmi k_2}e^{\rmi \pi/3} &  0 & e^{\rmi k_2}e^{-\rmi \pi/3} & e^{\rmi k_2}e^{-\rmi \pi/6} & e^{\rmi k_2} & e^{\rmi k_2}e^{\rmi \pi/6}  \\ e^{-\rmi(k_1-k_2)}e^{\rmi \pi/6} & e^{-\rmi(k_1-k_2)}e^{\rmi \pi/3} &  0 & e^{-\rmi(k_1-k_2)}e^{-\rmi \pi/3} & e^{-\rmi(k_1-k_2)}e^{-\rmi \pi/6} & e^{-\rmi(k_1-k_2)} \end{pmatrix}.
\end{align}
We arrive at
\begin{align}
 \mbox{det}(\mathbb{1}-t W(\vec{k})) = 1 + (3 + 8t + 3t^2 + t^4)t^2 + t(1-t^2)^2\vare_\Delta(\vec{k}).
\end{align}

\subsubsection{Matrices $W_0$ and $\mathcal{B}$} For $\vec{k}=0$ we have
\begin{align}
 W_0 = \Phi_{11} = \Phi_0^{(q=6)}
\end{align}
and
\begin{align}
 \mbox{det}(\mathbb{1}-t W_0) =(1-3t-3t^2+t^3)^2.
\end{align}
The critical values $t_{\rm c}$ and $T_{\rm c}$ saturate the bound for $q=6$ according to
\begin{align}
 t_{\rm c} &= \tan\Bigl(\frac{\pi}{12}\Bigr) = 0.267949,\\
 \frac{T_{\rm c}}{J} &= \frac{2}{\log \sqrt{3}} = 3.64096.
\end{align}
We confirm that
\begin{align}
 \mbox{det}(u\mathbb{1}-W_0^\dagger W_0) = P_6(u) = (u^3-15u^2+15u-1)^2.
\end{align}

\clearpage
\subsection{Laves-CaVO lattice $(q_a=4,8$)}

\subsubsection{Matrix $W(\vec{k})$} The Laves-CaVO lattice has $N_{\rm u}=2$ sites in the unit cell with coordination numbers $q_1=8$ and $q_2=4$, respectively. We label the sites in the unit cell by $a\in\{1,2\}$ and the local edges at each site by $\mu_1\in\{1,\dots,8\}$ and $\mu_2\in\{1,\dots,4\}$ according to the following schematic:
\begin{figure}[htpb!]
    \includegraphics[width=7cm]{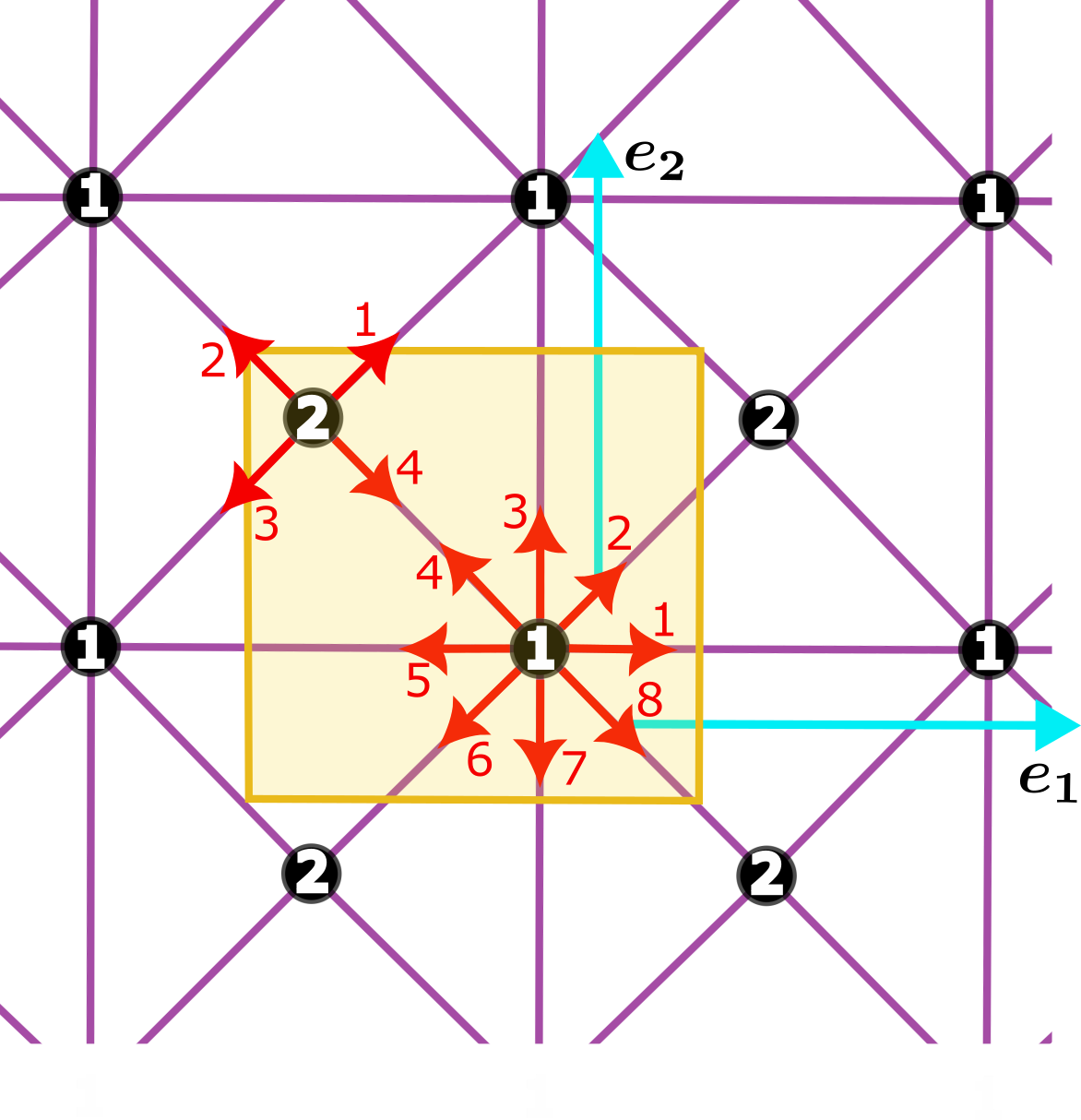}
     \label{FigLavesCaVOLabels}
\end{figure}\\
We have
\begin{align}
 W(\vec{k}) = \begin{pmatrix} E_{1;1} B_1(\vec{k})\Phi_{11} & E_{1;2} B_1(\vec{k})\Phi_{12} \\ E_{2;1} B_2(\vec{k})\Phi_{21} & E_{2;2} B_2(\vec{k})\Phi_{22}\end{pmatrix}
\end{align}
We have
\begin{align}
 B_1(\vec{k}) &= \text{diag}(e^{-\rmi k_1},e^{-\rmi k_1},e^{-\rmi k_2},1,e^{\rmi k_1},e^{\rmi k_2},e^{\rmi k_2},e^{-\rmi (k_1-k_2)}),\\
 B_2(\vec{k}) &= \text{diag}(e^{-\rmi k_2},e^{-\rmi (k_2-k_1)},e^{\rmi k_1},1).
\end{align}
The edge-connectivity matrices are
\begin{align}
 E_{1;1} &= \text{diag}(1,0,1,0,1,0,1,0),\\
 E_{1;2} &= \text{diag}(0,1,0,1,0,1,0,1),\\
 E_{2;1} &= \mathbb{1}_4,\ E_{2;2} = 0.
\end{align}
We have ($\eta=e^{\pi\rmi/4}$)
\begin{align}
 \vec{\theta}_1 &= (1,\eta,\eta^2,\eta^3,\eta^4,\eta^5,\eta^6,\eta^7)^T,\\
 \vec{\theta}_2 &= (\eta,\eta^3,\eta^5,\eta^7)^T.
\end{align}
Thus $\Phi_{11}$, $\Phi_{12}$, $\Phi_{21}$, and $\Phi_{22}$ are $8\times 8$, $8\times 4$, $4\times 8$, and $4\times 4$ matrices, respectively. We have
\begin{align}
 \Phi_{11}^\dagger &=\Phi_{11},\ \Phi_{12}^\dagger = \Phi_{21}.
\end{align}
The resulting matrix Kac--Ward matrix is $12\times 12$ because $Q=q_1+q_2=12$. We find
\begin{align}
 W(\vec{k}) = \begin{pmatrix} E_{1;1} B_1(\vec{k})\Phi_{11} & E_{1;2} B_1(\vec{k})\Phi_{12} \\ B_2(\vec{k})\Phi_{21} & 0\end{pmatrix}.
\end{align}
The determinant $\mbox{det}(\mathbb{1}-tW(\vec{k}))$ can be computed from these expressions.

\subsubsection{Matrices $W_0$ and $\mathcal{B}_a$} For $\vec{k}=0$ we have
\begin{align}
 \mbox{det}(\mathbb{1}-tW_0) = (1+t)^4(1-4t-t^4)^2.
\end{align}
The resulting critical temperature is
\begin{align}
 t_{\rm c} &=0.249038,\\
 \frac{T_{\rm c}}{J} &= 3.93101.
\end{align}
We find
\begin{align}
 W_0^\dagger W_0 &= \begin{pmatrix} \mathcal{B}_1 & 0 \\ 0 & \mathcal{B}_2 \end{pmatrix}
\end{align}
with, respectively, $8\times 8$ and $4\times 4$ blocks
\begin{align}
  \mathcal{B}_1 &= \Phi_{11}E_{1;1} \Phi_{11}+\Phi_{21}^\dagger \Phi_{21},\\
 \mathcal{B}_2 &= \Phi_{12}^\dagger E_{1;2} \Phi_{12}.
\end{align}
We have
\begin{align}
 \det(u\mathbb{1}-W_0^\dagger W_0)= \det(u\mathbb{1}-\mathcal{B}_1)\det(u\mathbb{1}-\mathcal{B}_2)
\end{align}
with
\begin{align}
 \det(u\mathbb{1}-\mathcal{B}_1) &= P_8(u) = (u^4-28u^3+70u^2-28u+1)^2,\\
 \det(u\mathbb{1}-\mathcal{B}_2) &= P_4(u) = (u^2-6u+1)^2.
\end{align}

\subsubsection{Relation to reference $\Phi_0^2$} We have $\vec{\theta}_1=U_1\vec{\theta}_0^{(q=8)}$ and $\vec{\theta}_2=U_2\vec{\theta}_0^{(q=4)}$ with
\begin{align}
 U_1 &= \mathbb{1}_8,\\
 U_2 &=e^{\rmi \pi/4} \mathbb{1}_4.
\end{align}
Define $V_a=\sqrt{U_a}$ such that
\begin{align}
 V_1 &=\mathbb{1}_8,\\
 V_2 &= e^{\rmi \pi/8} \mathbb{1}_4.
\end{align} 
Consider then the matrices $M_k^{(a)}=\Phi_{ka}^\dagger E_{k;a}$. We have
\begin{align}
 a=1:\ M_1^{(a)}&= (\vec{Y}_5^{(1)}\ \vec{0}\ \vec{Y}_7^{(1)}\ \vec{0}\ \vec{Y}_1^{(1)}\ \vec{0}\ \vec{Y}_3^{(1)}\ \vec{0}),\ M_2^{(a)} = (\vec{Y}_6^{(1)}\ \vec{Y}_8^{(1)}\ \vec{Y}_2^{(1)}\ \vec{Y}_4^{(1)}),\\
a=2:\ M_1^{(a)} &= (\vec{0}\ \vec{Y}_3^{(2)}\ \vec{0}\ \vec{Y}_4^{(2)}\ \vec{0}\ \vec{Y}_1^{(2)}\ \vec{0}\ \vec{Y}_2^{(2)}),\ M_2^{(a)} = 0.
\end{align}
 For fixed $a$, the set of matrices $\{M_k^{(a)}\}_k$ contains $q_a$ nonvanishing columns $\vec{Y}_\mu^{(a)}$ that are related to the vectors $\vec{X}_\mu^{(q=8)}$ and $\vec{X}_\mu^{(q=4)}$ according to
 \begin{align}
 a=1:\ \vec{Y}_\mu^{(a)} &= \vec{X}_\mu^{(q=8)},\\
 a=2:\ \vec{Y}_\mu^{(a)} &= \vec{X}_\mu^{(a=4)}.
 \end{align}
Hence we arrive at
\begin{align}
 a=1:\ \mathcal{B}_1 &= \sum_{\mu=1}^8 \vec{Y}_\mu \vec{Y}_\mu^\dagger = \sum_{\mu=1}^8 \vec{X}_\mu \vec{X}_\mu^\dagger = [\Phi_0^{(q=8)}]^2,\\
 a=2:\ \mathcal{B}_2 &= \sum_{\mu=1}^4 \vec{Y}_\mu \vec{Y}_\mu^\dagger = \sum_{\mu=1}^4 \vec{X}_\mu \vec{X}_\mu^\dagger = [\Phi_0^{(q=4)}]^2.
\end{align} 
Indeed, $\mathcal{B}_1$ and $\mathcal{B}_2$ simply coincide with $\Phi_0^2$ for $q=8$ and $q=4$, respectively. Hence we have
\begin{align}
 \det(u\mathbb{1}-W_0^\dagger W_0) &= \det(u\mathbb{1}_8-\mathcal{B}_8)\det(u\mathbb{1}_4-\mathcal{B}_4) = P_8(u)P_4(u) \\
 &= (u^4-28u^3+70u^2-28u+1)^2(u^2-6u+1)^2.
\end{align}

\clearpage
\subsection{Laves-Maple-Leaf lattice $(q_a=3,6)$}

\subsubsection{Matrix $W(\vec{k})$} The Laves-Maple-Leaf lattice has $N_{\rm u}=9$ sites in the unit cell with coordination numbers $q_2=6$ and $q_a=3$ for $a=1,3,4,\dots,9$. We label the sites in the unit cell by $a\in\{1,\dots,9\}$ and the local edges at each site by $\mu_2\in\{1,\dots,6\}$ and the remaining edges $\mu_a\in\{1,2,3\}$ for $a=1,3,\dots,9$ according to the following schematic:
\begin{figure}[htpb!]
    \includegraphics[width=0.6\linewidth]{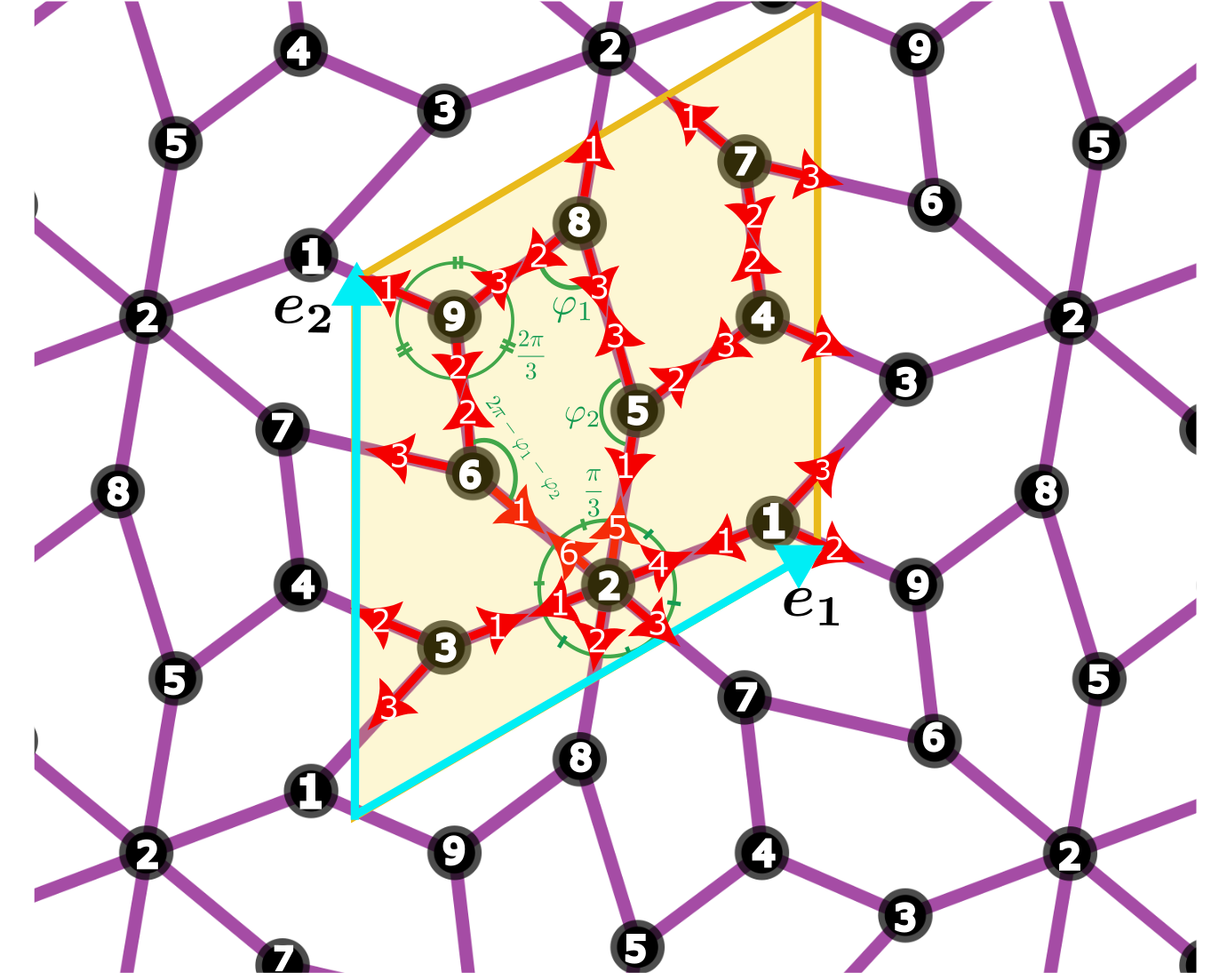}
     \label{FigLavesMLLabels}
\end{figure}\\
Since the Maple-Leaf lattice consists of a single type of vertex, the Laves-Maple-Leaf lattice consists of the same type of tile which has five corners. The interior angles meeting at site $a=2$ are fixed to be $\frac{\pi}{3}$. Similarly,  angles at site $a=9$ are fixed to be $\frac{2\pi}{3}$. However,, there are two free angles $\varphi_1,\varphi_2\in (0,\pi)$ that parameterize a whole family of topologically equivalent Laves Maple-Leaf lattices that produces the same critical temperature $t_{\rm c}$. The matrix $W(\vec{k})$ is of dimension $Q\times Q$ with $Q=8\cdot 3+6 =30$. The Bloch matrices are given by
\begin{align}
    B_1(\bk)&=\diag(1,e^{-\rmi(k_1-k_2)},e^{-\rmi k_1}),\ B_2(\bk)=\diag(1,e^{+\rmi k_2},e^{+\rmi k_2}),\\
    B_3(\bk)&=\diag(1,e^{+\rmi k_1},e^{+\rmi k_1}),\ B_4(\bk)=\diag(e^{-\rmi k_1},1,1),\\
    B_5(\bk)&=\diag(1,1,1),\ B_6(\bk)=\diag(1,1,e^{+\rmi k_1}),\\
    B_7(\bk)&=\diag(e^{-\rmi k_2},1,e^{-\rmi k_1}),\ B_8(\bk)=\diag(e^{-\rmi k_2},1,1),\\
    B_9(\bk)&=\diag(e^{-\rmi (k_2-k_1)},1,1).
\end{align}
The diagonal edge-connectivity matrices for the eight sites with $q_a=3$ are given by the $3\times 3$ diagonal matrices
\begin{align}
E_{1;2}&=E_{3;2}=E_{6;2}=E_{5;2}=E_{8;2}=E_{7;2}=E_{9;1}=E_{4;3}=\varepsilon_1,\\
    E_{3;4}&=E_{6;9}=E_{5;4}=E_{1;9}=E_{7;4}=E_{8;9}=E_{9;6}=E_{4;7}=\varepsilon_2,\\
    E_{1;3}&=E_{3;1}=E_{6;7}=E_{5;8}=E_{4;5}=E_{9;8}=E_{8;5}=E_{7;6}=\varepsilon_3,
\end{align}
where
\begin{equation}
    \varepsilon_1=\begin{pmatrix}
        1 & & \\ & 0 & \\ & & 0
    \end{pmatrix},\
    \varepsilon_2=\begin{pmatrix}
        0 & & \\ & 1 & \\ & & 0
    \end{pmatrix},\
    \varepsilon_3=\begin{pmatrix}
        0 & & \\ & 0 & \\ & & 1
    \end{pmatrix}.
\end{equation}
The edge-connectivity matrices emanating from site $a=2$, however, are $6\times 6$ diagonal matrices given by
\begin{align}
    E_{2;3} &= \begin{pmatrix}
        1 & & & & & \\
        & 0 & & & & \\
        & & 0 & & & \\
        & & & 0 & & \\
        & & & & 0 & \\
        & & & & & 0
    \end{pmatrix},\ E_{2;8}= \begin{pmatrix}
        0 & & & & & \\
        & 1 & & & & \\
        & & 0 & & & \\
        & & & 0 & & \\
        & & & & 0 & \\
        & & & & & 0
    \end{pmatrix},\ E_{2;7}= \begin{pmatrix}
        0 & & & & & \\
        & 0 & & & & \\
        & & 1 & & & \\
        & & & 0 & & \\
        & & & & 0 & \\
        & & & & & 0
    \end{pmatrix},\\
    E_{2;1} &= \begin{pmatrix}
        0 & & & & & \\
        & 0 & & & & \\
        & & 0 & & & \\
        & & & 1 & & \\
        & & & & 0 & \\
        & & & & & 0
    \end{pmatrix},\
    E_{2;5}= \begin{pmatrix}
        0 & & & & & \\
        & 0 & & & & \\
        & & 0 & & & \\
        & & & 0 & & \\
        & & & & 1 & \\
        & & & & & 0
    \end{pmatrix},\
    E_{2;6}= \begin{pmatrix}
        0 & & & & & \\
        & 0 & & & & \\
        & & 0 & & & \\
        & & & 0 & & \\
        & & & & 0 & \\
        & & & & & 1
    \end{pmatrix}.
\end{align}
    We define $\alpha=e^{\rmi \varphi_1},\beta=e^{\rmi \varphi_2}$, $\omega=e^{\rmi \frac{\pi}{3}}, \sigma=\omega^2=e^{\rmi \frac{2\pi}{3}}$. Then the phases at each of the sites $\bd{\theta}_a$ are given by
\begin{align}
    \bd{\theta}_1&= -\left(1,\alpha^{-1}\beta^{-1},\beta^{-1}\right), \\
    \bd{\theta}_2&= -\left(1,\omega,\omega^2,\omega^3,\omega^4,\omega^5\right) ,\\
    \bd{\theta}_3&= \left(1,\alpha^{-1}\beta^{-1},\beta^{-1}\right) ,\\
    \bd{\theta}_4&= -\alpha^{-1} \beta^{-1}\left(1,\sigma,\sigma^2 \right), \\
    \bd{\theta}_5&= \sigma^2\left(1,\alpha^{-1}\beta^{-1},\beta^{-1}\right), \\
    \bd{\theta}_6&= \omega^{-1}\left(1,\alpha^{-1}\beta^{-1},\beta^{-1}\right), \\
    \bd{\theta}_7&= \sigma\left(1, \alpha^{-1}\beta^{-1},\beta^{-1}\right) ,\\
    \bd{\theta}_8&=-\sigma^2 \left(1,\alpha^{-1}\beta^{-1},\beta^{-1}\right) ,\\
    \bd{\theta}_9&= \alpha^{-1} \beta^{-1}\left(1,\sigma ,\sigma^2 \right).
\end{align}
These phases satisfy 
\begin{align}
    (1,\alpha^{-1}
\beta^{-1},\beta^{-1})&=-\bd{\theta}_1= \bd{\theta}_3=\frac{1}{\sigma^2}\bd{\theta}_5=\omega \bd{\theta}_6= \frac{1}{\sigma} \bd{\theta}_7=-\frac{1}{\sigma^2}\bd{\theta}_8,\\
  \alpha^{-1}\beta^{-1}(1,\sigma,\sigma^2)&=\bd{\theta}_9= -\bd{\theta}_4.
\end{align}
We arrive at
\begin{align} 
 \nonumber \det(\mathbbm{1}-t W(\bk))&=- \left(t^2-1\right)^6 t^9 \varepsilon_\Delta(2\bk)+ (1-t)^5 (1+t)^6 \left(t^3+t^2+t+1\right)
   t^7 \varepsilon_\Delta(2k_1-k_2,k_1-2k_1)\\\nonumber &+(1-t)^3 (1+t)^6 \left(t^{10}+4 t^9+4 t^8+4
   t^7+10 t^6-3 t^5+11 t^4-4 t^3+5 t^2-t+1\right) t^3 \varepsilon_\Delta(\bk)\\\nonumber &+(t+1)^6 \Bigl(3
   t^{16}+39 t^{14}-78 t^{13}+226 t^{12}-390 t^{11}+579 t^{10}-668 t^9+660 t^8-546 t^7+393 t^6-240 t^5\\ &+126 t^4-56 t^3+21 t^2-6 t+1\Bigr).\label{LavesMLf}
\end{align}
We obtain the critical temperature as the root of
\begin{align}
    \det(\mathbbm{1}- t W_0)&=(t+1)^6\left(t^4-2 t^3+3 t^2-2 t+1\right)\nonumber \\ &\times \Bigl(9 t^{12}+36 t^{11}-6 t^{10}-36 t^9+127 t^8-136 t^7+124 t^6-88 t^5+55 t^4-28 t^3+10 t^2-4
   t+1\Bigr)
\end{align}
to be
\begin{align}
 t_{\rm c} &=0.487768,\\
 \frac{T_{\rm c}}{J} &= 1.87572.
\end{align}
The characteristic polynomial for $W_{0}^\dagger W_0$ is given by
\begin{equation}
    \det(u\mathbbm{1}- W_0^\dagger W_0)=[u (u-3)^2]^8 (u^3-15 u^2+15 u -1)^2.
\end{equation}

\clearpage
\subsection{Laves-Kagome lattice $(q_a=6,3)$}

\subsubsection{Matrix $W(\vec{k})$} The Laves-Kagome lattice has $N_{\rm u}=3$ sites in the unit cell with coordination numbers $q_1=6$ and $q_2=q_3=3$, hence $Q=12$. We label the sites in the unit cell by $a\in\{1,2,3\}$ and the local edges at each site by $\mu_1\in\{1,\dots,6\}$ and the remaining edges $\mu_a\in\{1,2,3\}$ for $a=2,3$ according to the following schematic:
\begin{figure}[h!]
    \includegraphics[width=0.4\linewidth]{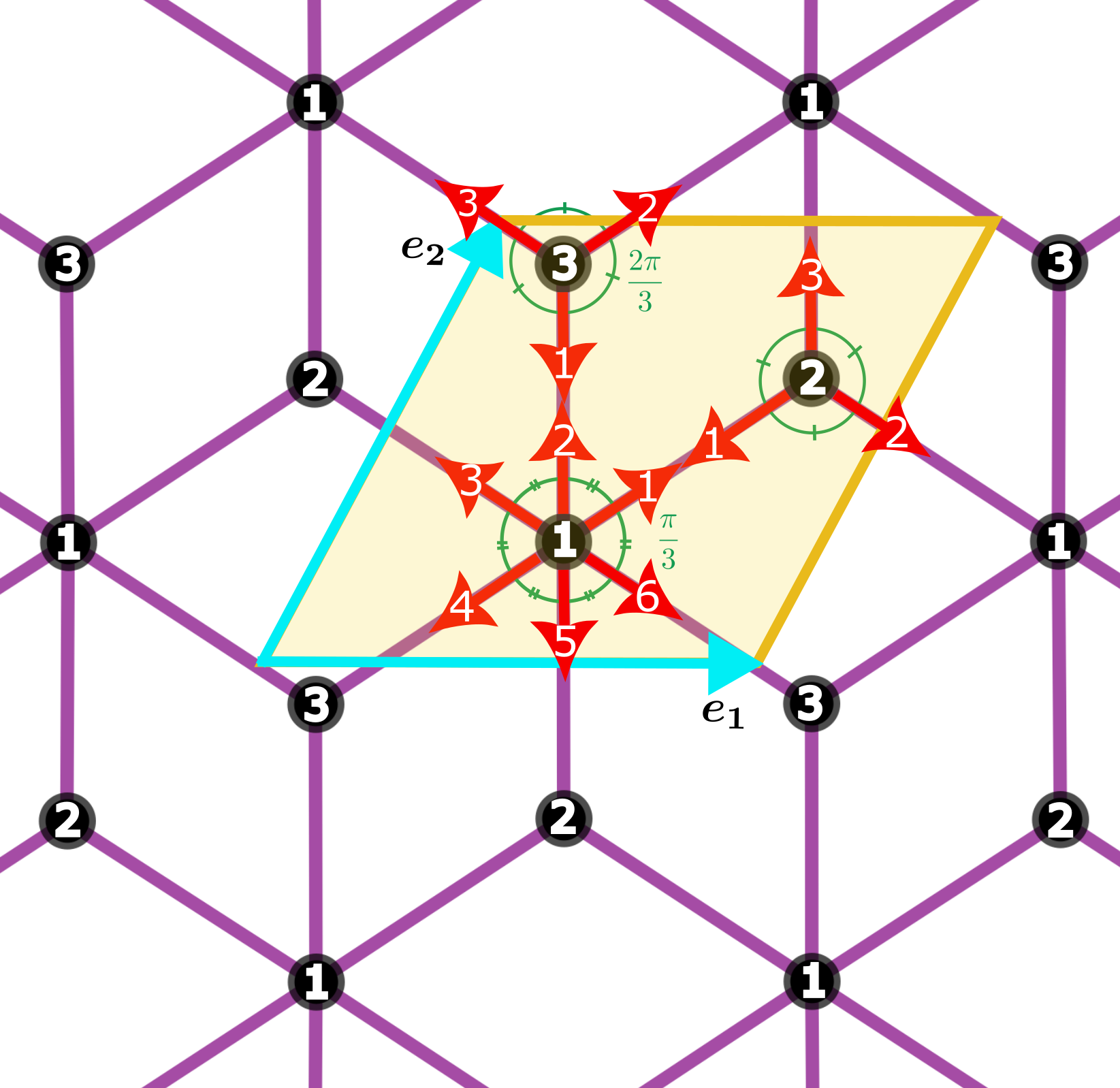}
     \label{FigLavesKagomeLabels}
\end{figure}
The matrix $W(\vec{k})$ is given by
\begin{align}
 W(\vec{k}) =\begin{pmatrix} 0 & E_{1;2}B_1(\vec{k})\Phi_{12} & E_{1;3}B_1(\vec{k})\Phi_{13} \\ E_{2;1} B_2(\vec{k}) \Phi_{21} & 0 & 0 \\ E_{3;1} B_3(\vec{k}) \Phi_{31} & 0 & 0 \end{pmatrix}.
\end{align}
The Bloch matrices are
\begin{align}
 B_1(\vec{k}) &= \text{diag}(1,1,e^{\rmi k_1},e^{\rmi k_2},e^{\rmi k_2},e^{-\rmi(k_1-k_2)}),\\
 B_2(\vec{k}) &= \text{diag}(1,e^{-\rmi k_1},e^{-\rmi k_2}),\\
 B_3(\vec{k}) &=\text{diag}(1,e^{-\rmi k_2},e^{\rmi(k_1-k_2)}),
\end{align}
and the nonvanishing edge-connectivity matrices are
\begin{align}
 E_{1;2} &=\text{diag}(1,0,1,0,1,0),\\
 E_{1;3} &=\text{diag}(0,1,0,1,0,1),\\
 E_{2;1} &= E_{3;1} = \mathbb{1}_3.
\end{align}
We move the coordinates of the sites such that we have
\begin{align}
 \vec{\theta}_1 &= (e^{\rmi \pi/6},\rmi,e^{5\pi\rmi/6},e^{-5\pi\rmi/6},-\rmi,e^{-\rmi\pi/6})^T,\\
 \vec{\theta}_2 &= (e^{-\rmi \pi/6},\rmi,e^{-5\pi\rmi/6})^T,\\
 \vec{\theta}_3 &= (-\rmi, e^{\rmi \pi/6},e^{5\pi\rmi/6})^T.
\end{align}
The determinant $\mbox{det}(\mathbb{1}-tW(\vec{k}))$ can be computed from these expressions.

\subsubsection{Matrices $W_0$ and $\mathcal{B}_a$} For $\vec{k}=0$ we have
\begin{align}
 \mbox{det}(\mathbb{1}-tW_0) = (1-6t^2-3t^4)^2.
\end{align}
The corresponding critical temperature is
\begin{align} 
 t_{\rm c}&=\sqrt{\frac{2}{\sqrt{3}}-1} = 0.39332,\\
 \frac{T_{\rm c}}{J} &= 2.40546.
\end{align}
We confirm that
\begin{align}
 W_0^\dagger W_0 &= \begin{pmatrix} \mathcal{B}_1 & 0 & 0\\ 0 & \mathcal{B}_2 &0 \\ 0 & 0 & \mathcal{B}_3 \end{pmatrix}.
\end{align}
We have
\begin{align}
 \mbox{det}(u\mathbb{1}-W_0^\dagger W_0)&= P_6(u) P_3(u)^2 = (u^3-15u^2+15u-1)^2u^2(u-3)^4.
\end{align}

\clearpage
\subsection{Laves-Trellis lattice $(q_a=4,3)$}

\subsubsection{Matrix $W(\vec{k})$}  The Laves-Trellis lattice has $N_{\rm u}=3$ sites in the unit cell with coordination numbers $q_1=4$ and $q_2=q_3=3$, hence $Q=10$. We label the sites in the unit cell by $a\in\{1,2,3\}$ and the local edges at each site by $\mu_1\in\{1,\dots,4\}$ and the remaining edges $\mu_a\in\{1,2,3\}$ for $a=2,3$ according to the following schematic:
\begin{figure}[h!]
\includegraphics[width=0.35\linewidth]{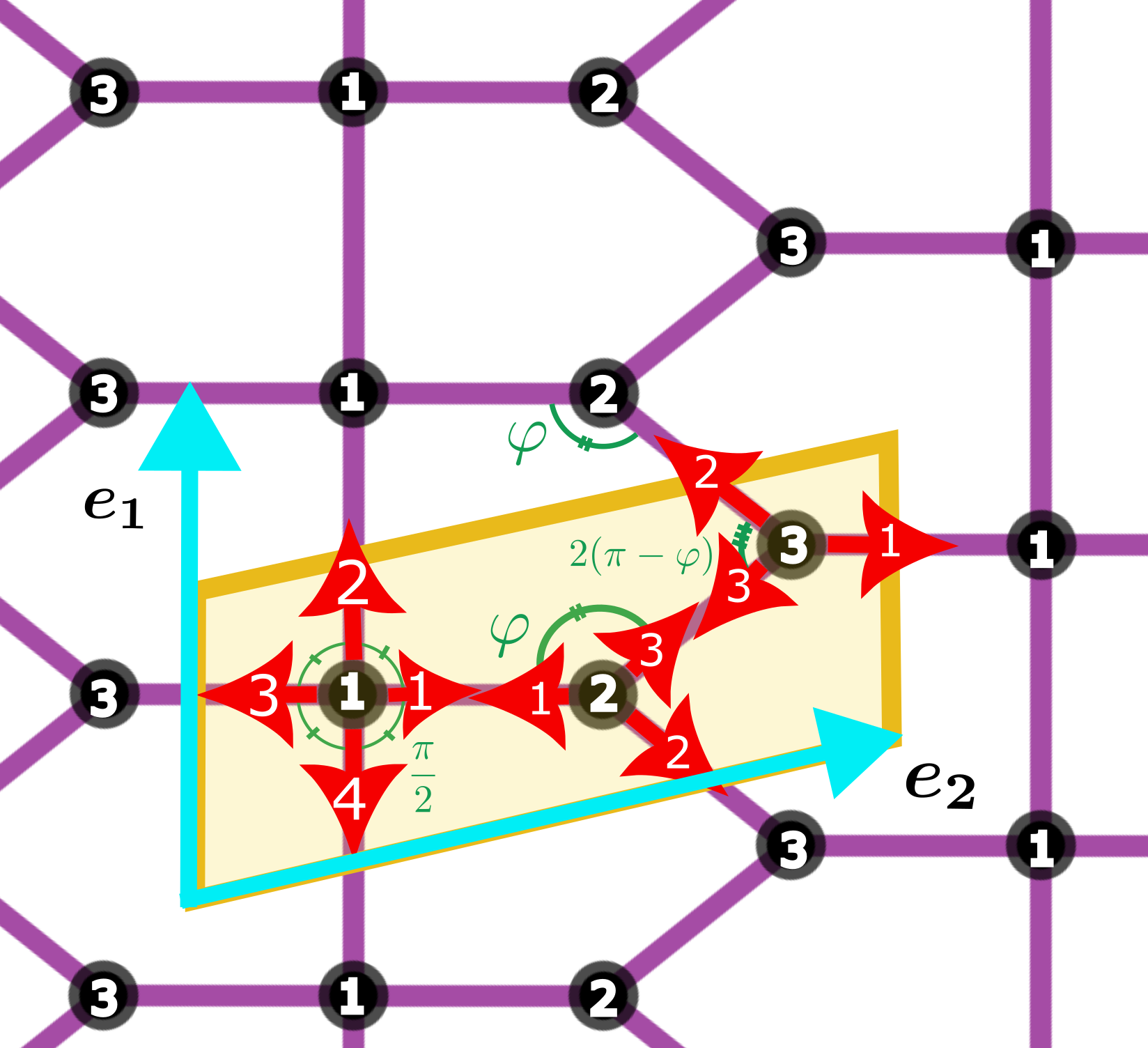}
\label{FigLavesTrellisLabels}
\end{figure}\\
The matrix $W(\vec{k})$ has the structure
\begin{align}
 W(\vec{k}) &= \begin{pmatrix} E_{1;1} B_1(\vec{k})\Phi_{11} & E_{1;2} B_1(\vec{k})\Phi_{12} & E_{1;3} B_1(\vec{k})\Phi_{13} \\ E_{2;1} B_2(\vec{k}) \Phi_{21} & 0 & E_{2;3} B_2(\vec{k}) \Phi_{23} \\ E_{3;1}B_3(\vec{k})\Phi_{31} & E_{3;2}B_3(\vec{k})\Phi_{32} & 0 \end{pmatrix}.
\end{align}
The Bloch adjacency matrices read
\begin{align}
 B_1(\vec{k}) &= \begin{pmatrix} 1 & & & \\ & e^{-\rmi k_1} & & \\ & & e^{\rmi k_2} & \\ & & & e^{\rmi k_1}\end{pmatrix},\ B_2(\vec{k}) = \begin{pmatrix} 1 & & \\ & e^{\rmi k_1} & \\ & & 1 \end{pmatrix},\ B_3(\vec{k}) = \begin{pmatrix} e^{-\rmi k_2} & & \\ & e^{-\rmi k_1} & \\ & & 1 \end{pmatrix}
\end{align} 
and the nonvanishing edge-connectivity matrices are given by 
\begin{align}
 E_{1;1} &=\text{diag}(0,1,0,1),\ E_{1;2}=\text{diag}(1,0,0,0),\ E_{1;3} = \text{diag}(0,0,1,0),\\
 E_{2;1} &=E_{3;1} = \text{diag}(1,0,0),\ E_{2;3}=E_{3;2} = \text{diag}(0,1,1).
\end{align}
We move the coordinates of the sites such that we have
\begin{align}
 \vec{\theta}_1 &= (1,\rmi,-1,-\rmi)^T,\\
 \vec{\theta}_2 &=  (-1,e^{-\rmi \pi/3},e^{\rmi \pi/3})^T,\\
 \vec{\theta}_3 &=(1,e^{2\pi\rmi/3},e^{-2\pi\rmi/3})^T.
\end{align}
This yields
\begin{align}
 \mbox{det}(\mathbb{1} - t W(\vec{k})) ={}& (1+t)^2\Bigl[(1+t^2)^2(1-2t+2t^2)-2t(1-t+2t^3-t^4-t^5)\cos k_1\Bigr]\\
 &+t^3(1-t^2)^2\Bigl[\vare_\Delta(\vec{k})+2\cos (2k_1)\Bigr].
\end{align}

\subsubsection{Matrices $W_0$ and $\mathcal{B}_a$} For $\vec{k}=0$ we have
\begin{align}
 \mbox{det}(\mathbb{1}-tW_0) = (1-t-t^2-t^3-2t^4)^2.
\end{align}
The corresponding critical temperature is
\begin{align}
 t_{\rm c}&=\frac{1}{2},\\
 \frac{T_{\rm c}}{J} &= \frac{2}{\log 3} = 1.82048.
\end{align}
We confirm that
\begin{align}
 W_0^\dagger W_0 &= \begin{pmatrix} \mathcal{B}_1 & 0 & 0\\ 0 & \mathcal{B}_2 &0 \\ 0 & 0 & \mathcal{B}_3 \end{pmatrix}.
\end{align}
We have
\begin{align}
 \mbox{det}(u\mathbb{1}-W_0^\dagger W_0)&=P_4(u) P_3(u)^2 = (u^2-6u+1)^2u^2(u-3)^4.
\end{align}

\clearpage
\subsection{Laves-SrCuBO lattice $(q_a=4,3)$}

\subsubsection{Matrix $W(\vec{k})$}  The Laves-SrCuBO lattice has $N_{\rm u}=6$ sites in the unit cell with coordination numbers $q_a=3$ for $a=1,\dots,4$ and $q_5=q_6=4$, hence $Q=20$. We label the sites in the unit cell by $a\in\{1,\dots,6\}$ and the local edges at each site by $\mu_a\in\{1,2,3\}$ for $a=1,\dots,4$ and the remaining edges $\mu_a\in\{1,2,3\}$ for $a=5,6$ according to the following schematic:
\begin{figure}[h!]
\includegraphics[width=0.55\linewidth]{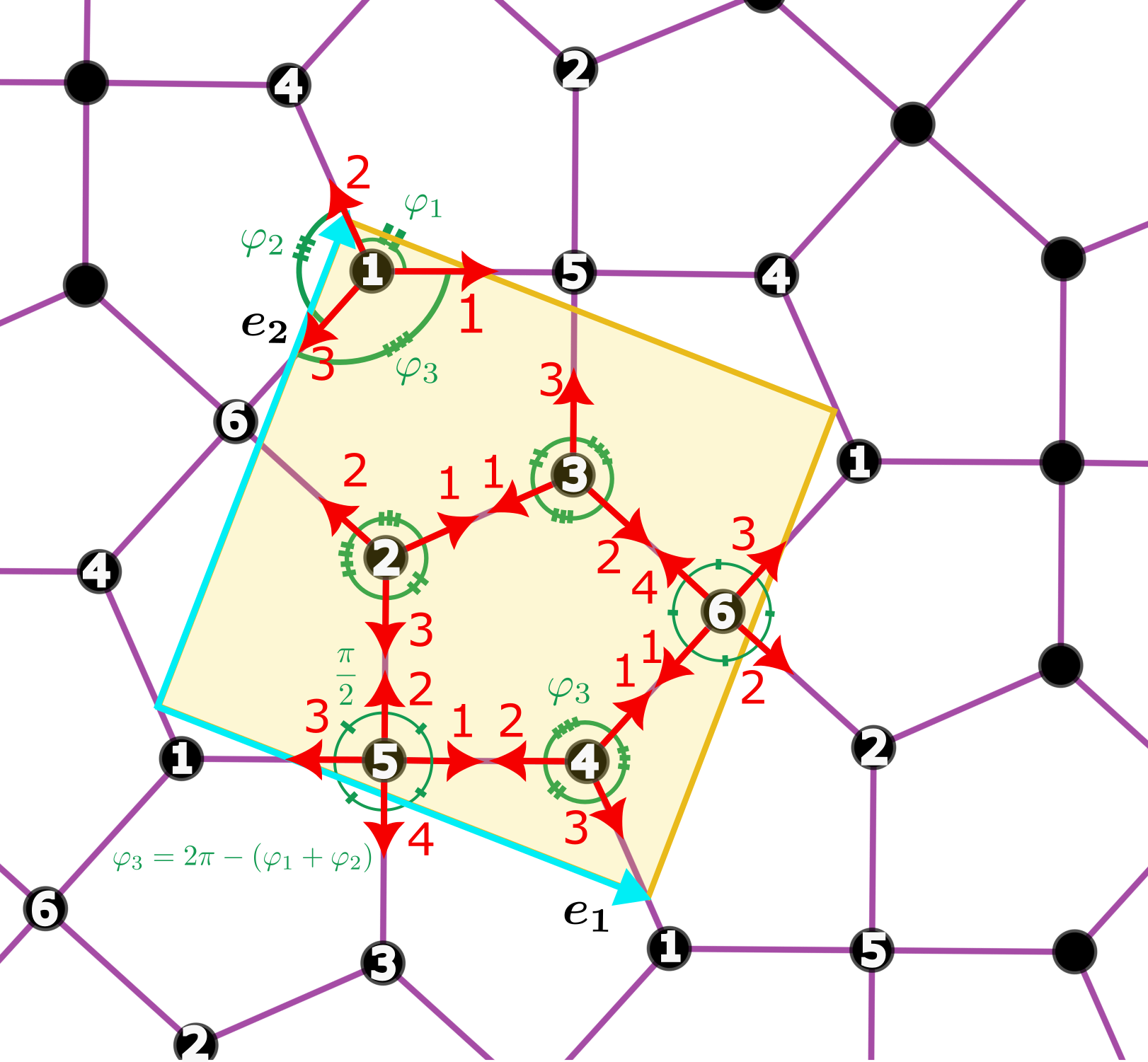}
\label{FigLavesSrCuBOLabels}
\end{figure}\\
The matrix $W(\vec{k)}$ has the structure
\begin{align}
 W(\vec{k}) = \begin{pmatrix} 0 & 0 & 0 & * & * & * \\ 0 & 0 & * & 0 & * & * \\ 0 & * & 0 & 0 & * & * \\ * & 0 & 0 & 0 & * & * \\ * & * & * & * & 0 & 0 \\ * & * & * & * & 0 & 0 \end{pmatrix}.
\end{align}
The Bloch adjacency matrices are
\begin{align}
 B_1(\vec{k}) &= \text{diag}(e^{-\rmi k_2},e^{\rmi(k_1-k_2)},e^{\rmi k_1}),\\
 B_2(\vec{k}) &= B_3(\vec{k}) = B_4(\vec{k}) = \mathbb{1}_3,\\
 B_5(\vec{k}) &= \text{diag}(1,1,e^{\rmi k_2},e^{\rmi k_2}),\\
 B_6(\vec{k}) &= \text{diag}(1,e^{-\rmi k_1},e^{-\rmi k_1},1),
\end{align}
and the nonvanishing edge-connectivity matrices are
\begin{align}
 E_{1;4} &= \text{diag}(0,1,0),\ E_{1;5} =\text{diag}(1,0,0),\ E_{1;6}=\text{diag}(0,0,1),\\
 E_{2;3} &=\text{diag}(1,0,0),\ E_{2;6}=\text{diag}(0,1,0),\ E_{2;5}=\text{diag}(0,0,1),\\
 E_{3;2} &=\text{diag}(1,0,0),\ E_{3;5}=\text{diag}(0,0,1),\ E_{3;6}=\text{diag}(0,1,0),\\
 E_{4;1} &= \text{diag}(0,0,1),\ E_{4;5}=\text{diag}(0,1,0),\ E_{4;6}=\text{diag}(1,0,0),\\
 E_{5;1} &= \text{diag}(0,0,1,0),\ E_{5;2}=\text{diag}(0,1,0,0),\ E_{5;3}=\text{diag}(0,0,0,1),\ E_{5;4}=\text{diag}(1,0,0,0),\\
 E_{6;1} &=\text{diag}(0,0,1,0),\ E_{6;2}=\text{diag}(0,1,0,0),\ E_{6;3}=\text{diag}(0,0,0,1),\ E_{6;4}=\text{diag}(1,0,0,0).
\end{align}
We move the coordinates of the sites such that we have
\begin{align}
 \vec{\theta}_1 &= (1,e^{2\pi\rmi/3},e^{-2\pi\rmi/3})^T,\\
 \vec{\theta}_2 &= (e^{\rmi\pi/3},e^{2\pi\rmi/3},-\rmi)^T,\\
 \vec{\theta}_3 &= (e^{-2\pi\rmi/3},e^{-\rmi\pi/3},\rmi)^T,\\
 \vec{\theta}_4 &= (e^{\rmi\pi/3},-1,e^{-\rmi\pi/3})^T,\\
 \vec{\theta}_5 &= (1,\rmi,-1,-\rmi)^T,\\
 \vec{\theta}_6 &= (e^{-2\pi\rmi/3},e^{-\rmi \pi/3},e^{\rmi\pi/3},e^{2\pi\rmi/3})^T.
\end{align}
The determinant $\mbox{det}(\mathbb{1}-tW(\vec{k}))$ can be computed from these expressions.

\subsubsection{Matrices $W_0$ and $\mathcal{B}_a$} For $\vec{k}=0$ we have
\begin{align}
 \mbox{det}(\mathbb{1}-tW_0) =(1+t)^4(1-2t+3t^2-8t^3+7t^4-6t^5+t^6)^2.
\end{align}
The corresponding critical temperature is
\begin{align}
 t_{\rm c}&=0.504864,\\
 \frac{T_{\rm c}}{J} &= 1.79917.
\end{align}
We confirm that
\begin{align}
 W_0^\dagger W_0 &= \text{diag}(\mathcal{B}_1,\mathcal{B}_2,\mathcal{B}_3,\mathcal{B}_4,\mathcal{B}_5,\mathcal{B}_6).
\end{align}
We have
\begin{align}
 \mbox{det}(u\mathbb{1}-W_0^\dagger W_0)&= P_3(u)^4P_4(u)^2= u^4(u-3)^8(u^2-6u+1)^4.
\end{align}

\clearpage
\subsection{Laves-Ruby lattice $(q_a=6,4,3)$}

\subsubsection{Matrix $W(\vec{k})$}  The Laves-Ruby lattice has $N_{\rm u}=6$ sites in the unit cell with coordination numbers $q_1=6,q_2=q_3=q_5=4$, and $q_4=q_6=3$, hence $Q=24$. We label the sites in the unit cell by $a\in\{1,\dots,6\}$ and the local edges at each site starting at site $a=1$ by $\mu_1\in\{1,2,\dots,6\}$, $\mu_a\in\{1,2,3,4\}$ for $a=2,3,5$ and the remaining edges $\mu_a\in\{1,2,3\}$ for $a=4,6$ according to the following schematic:
\begin{figure}[h!]
\includegraphics[width=0.55\linewidth]{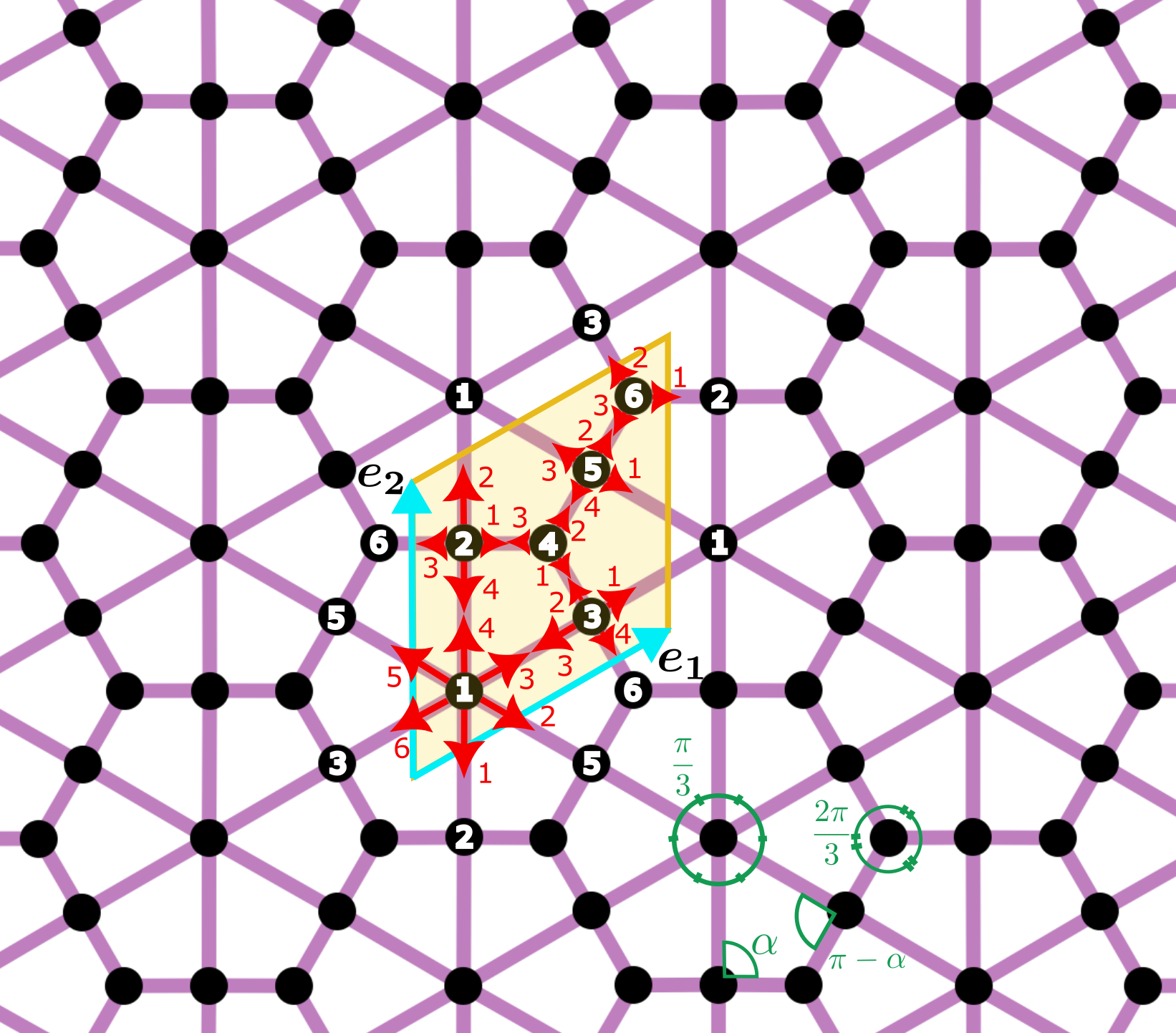}
\label{FigLavesRubyLabels}
\end{figure}\\
The matrix $W(\vec{k})$ has the structure
\begin{align}
 W(\vec{k}) = \begin{pmatrix} 0 & * & * & 0 & * & 0 \\ * & 0 & 0 & * & 0 & * \\ * & 0 & 0 & * & 0 & * \\ 0 & * & * & 0 & * & 0 \\ * & 0 & 0 & * & 0 & *  \\ 0 & * & * & 0 & * & 0 \end{pmatrix}.
\end{align}
The Bloch adjacency matrices read
\begin{align}
 B_1(\vec{k}) &= \text{diag}(e^{\rmi k_2},e^{\rm ik_2},1,1,e^{\rmi k_1},e^{\rmi k_1}),\\
 B_2(\vec{k}) &= \text{diag}(1,e^{-\rmi k_2},e^{\rmi k_1},1),\\
 B_3(\vec{k}) &= \text{diag}(e^{-\rmi k_1},1,1,e^{\rmi k_2}),\\
 B_4(\vec{k}) &= \text{diag}(1,1,1),\\
 B_5(\vec{k}) &= \text{diag}(e^{-\rmi k_1},1,e^{-\rmi k_2},1),\\
 B_6(\vec{k}) &= \text{diag}(e^{-\rmi k_1},e^{-\rmi k_2},1),
\end{align}
and the nonvanishing edge-connectivity matrices are given by
\begin{align}
 E_{1;2} &=\text{diag}(1,0,0,1,0,0),\ E_{1;3}=\text{diag}(0,0,1,0,0,1),\ E_{1;5}=\text{diag}(0,1,0,0,1,0),\\
 E_{2;1} &=\text{diag}(0,1,0,1),\ E_{2;4}=\text{diag}(1,0,0,0),\ E_{2;6}=\text{diag}(0,0,1,0),\\
 E_{3;1} &= \text{diag}(1,0,1,0),\ E_{3;4}=\text{diag}(0,1,0,0),\ E_{3;6}=\text{diag}(0,0,0,1),\\
 E_{4;2} &= \text{diag}(0,0,1),\ E_{4;3}=\text{diag}(1,0,0),\ E_{4;5}=\text{diag}(0,1,0),\\
 E_{5;1} &= \text{diag}(1,0,1,0),\ E_{5;4}=\text{diag}(0,0,0,1),\ E_{5;6}=\text{diag}(0,1,0,0),\\
 E_{6;2} &=\text{diag}(1,0,0),\ E_{6;3}=\text{diag}(0,1,0),\ E_{6;5}=\text{diag}(0,0,1).
\end{align}
We move the coordinates of the sites such that we have
\begin{align}
 \vec{\theta}_1 &= (-\rmi,e^{-\rmi \pi/3},e^{\rmi \pi/3},\rmi,e^{2\pi\rmi/3},e^{-2\pi\rmi/3})^T,\\
 \vec{\theta}_2 &= (1,\rmi,-1,-\rmi),\\
 \vec{\theta}_3 &= (e^{\rmi \pi/3},e^{2\pi\rmi/3},e^{-2\pi\rmi/3},e^{-\rmi\pi/3})^T,\\
 \vec{\theta}_4 &= (e^{-\rmi \pi/3},e^{\rmi \pi/3},-1)^T,\\
 \vec{\theta}_5 &= (e^{-\rmi \pi/3},e^{\rmi \pi/3},e^{2\pi\rmi/3},e^{-2\pi\rmi/3})^T,\\
 \vec{\theta}_6 &= (1,e^{2\pi\rmi/3},e^{-2\pi\rmi/3})^T.
 \end{align}
The determinant $\mbox{det}(\mathbb{1}-tW(\vec{k}))$ can be computed from these expressions.

\subsubsection{Matrices $W_0$ and $\mathcal{B}_a$} For $\vec{k}=0$ we have
\begin{align}
 \mbox{det}(\mathbb{1}-tW_0) = (1+t^2)^6(1-6t^2-3t^4)^2.
\end{align}
The corresponding critical temperature is
\begin{align} 
 t_{\rm c}&=\sqrt{\frac{2}{\sqrt{3}}-1} = 0.39332,\\
 \frac{T_{\rm c}}{J} &= 2.40546,
\end{align}
which coincides with the result of the Laves-Kagome lattice. We confirm that
\begin{align}
 W_0^\dagger W_0 &= \text{diag}(\mathcal{B}_1,\mathcal{B}_2,\mathcal{B}_3,\mathcal{B}_4,\mathcal{B}_5,\mathcal{B}_6).
\end{align}
We have
\begin{align}
 \mbox{det}(u\mathbb{1}-W_0^\dagger W_0)&= P_6(u)P_4(u)^3P_3(u)^2 =(u^3-15u^2+15u-1)^2(u^2-6u+1)^6u^2(u-3)^4.
\end{align}

\clearpage
\subsection{Laves-Star lattice $(q_a=12,3)$}

\subsubsection{Matrix $W(\vec{k})$} The Star lattice has 6 sites in the unit cell with coordination number 3, hence $Q=18$. Thus the Laves-Star lattice, which is the dual lattice, has $\frac{Q}{2}-6=3$ sites in the unit cell. The unit-cell site $a=1$ has $q_1=12$ and sites $a=2,3$ have $q_a=3$. We assign labels according to the following schematic:
\begin{figure}[h!]
    \includegraphics[width=0.4\linewidth]{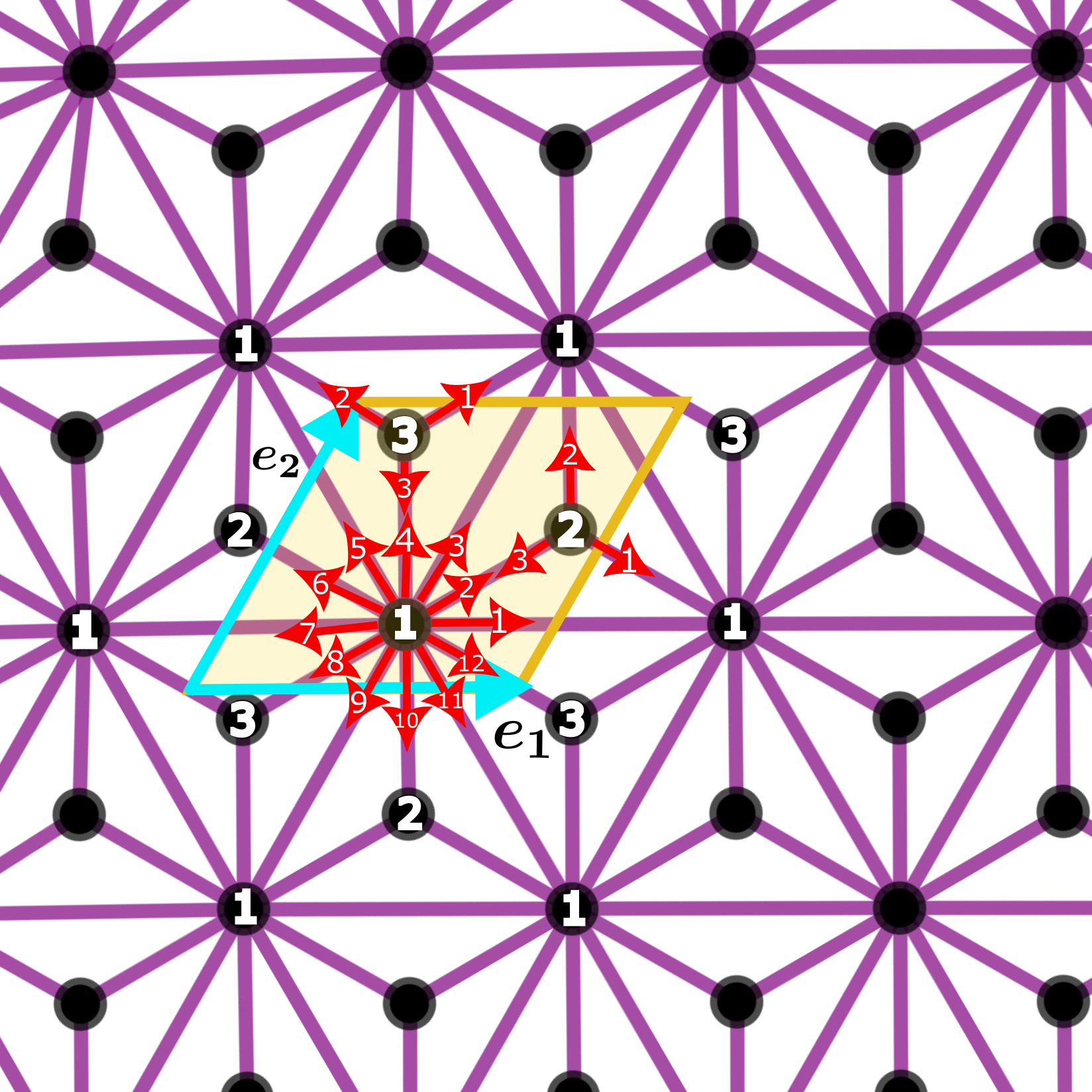}
     \label{FigLavesStarLabels2}
\end{figure}\\
The Bloch matrices are
\begin{align}
 B_1(\vec{k}) &=\text{diag}(e^{-\rmi k_1},1,e^{-\rmi k_2},1,e^{\rmi k_1},e^{\rmi k_1},e^{\rmi k_1},e^{\rmi k_1},e^{\rmi k_2},e^{\rmi k_2},e^{\rmi k_2},e^{\rmi k_2}),\\
 B_2(\vec{k}) &=\text{diag}(e^{-\rmi k_1},e^{-\rmi k_2},1),\\
 B_3(\vec{k}) &=\text{diag}(e^{-\rmi k_2},e^{\rmi k_1},1)
\end{align}
and the edge-connectivity matrices read
\begin{align}
 E_{1;1} &= \text{diag}(1,0,1,0,1,0,1,0,1,0,1,0),\\
 E_{1;2} &= \text{diag}(0,1,0,0,0,1,0,0,0,1,0,0),\\
 E_{1;3} &= \text{diag}(0,0,0,1,0,0,0,1,0,0,0,1),\\
 E_{2;1} &= E_{3;1} = \mathbb{1}_3.
\end{align}
With $\eta=e^{\rmi \pi/6}$ we have
\begin{align}
 \vec{\theta}_1 &=(1,\eta,\eta^2,\eta^3,\eta^4,\eta^5,\eta^6,\eta^7,\eta^8,\eta^9,\eta^{10},\eta^{11})^T,\\
 \vec{\theta}_2 &= (e^{-\rmi \pi/6},\rmi,e^{-5\pi\rmi/6})^T,\\
 \vec{\theta}_3 &= (e^{\rmi \pi/6},e^{5\pi\rmi/6},-\rmi)^T.
\end{align}
We arrive at
\begin{align}
 W(\vec{k}) = \begin{pmatrix} E_{1;1} B_1(\vec{k})\Phi_{11} & E_{1;2} B_1(\vec{k})\Phi_{12}  & E_{1;3} B_1(\vec{k})\Phi_{13} \\ E_{2;1} B_2(\vec{k})\Phi_{21}  & 0 & 0 \\ E_{3;1} B_3(\vec{k})\Phi_{21}  & 0 & 0 \end{pmatrix}.
\end{align}
The determinant $\mbox{det}(\mathbb{1}-tW(\vec{k}))$ can be computed from these expressions.

\subsubsection{Matrices $W_0$ and $\mathcal{B}_a$} For $\vec{k}=0$ we find
\begin{align}
 \mbox{det}(\mathbb{1}_{18}- t W_0) = (1+t)^6(1-6t+6t^2-6t^3-3t^4)^2.
\end{align} 
The critical temperature is
\begin{align}
 t_{\rm c}&= 0.197105,\\
 \frac{T_{\rm c}}{J} &= 5.00705.
\end{align}
We confirm that
\begin{align}
 W_0^\dagger W_0 = \begin{pmatrix} \mathcal{B}_1 & & \\ & \mathcal{B}_2 & \\ & & \mathcal{B}_3 \end{pmatrix}
\end{align}
and
\begin{align}
\det(u\mathbb{1}_{18} - W_0^\dagger W_0) = P_{12}(u)P_3(u)^2 = (u^6-66u^5+495u^4-924u^3+495u^2-66u+1)^2\Bigl[u(u-3)^2\Bigr]^2.
\end{align}

\clearpage
\subsection{Laves-SHD lattice $(q_a=12,6,4)$}

\subsubsection{Matrix $W(\vec{k})$} The Laves-SHD lattice has $N_{\rm u}=6$ sites in the unit cell with coordination numbers $q_1=q_2=q_3=4$, $q_4=q_5=6$ and $q_6=12$, thus $Q=36$. We label the sites in the unit cell by $a\in\{1,\dots,6\}$ and the local edges at each site starting by $\mu_a\in\{1,\dots,4\}$, for sites $a=1,2,3$, $\mu_a\in\{1,2,\dots,6\}$ for sites $a=4,5$, and the remaining edges $\mu_6 \in\{1,2,\dots,12\}$ for the twelve coordinated site according to the following schematic:
\begin{figure}[h!]
\includegraphics[width=0.5\linewidth]{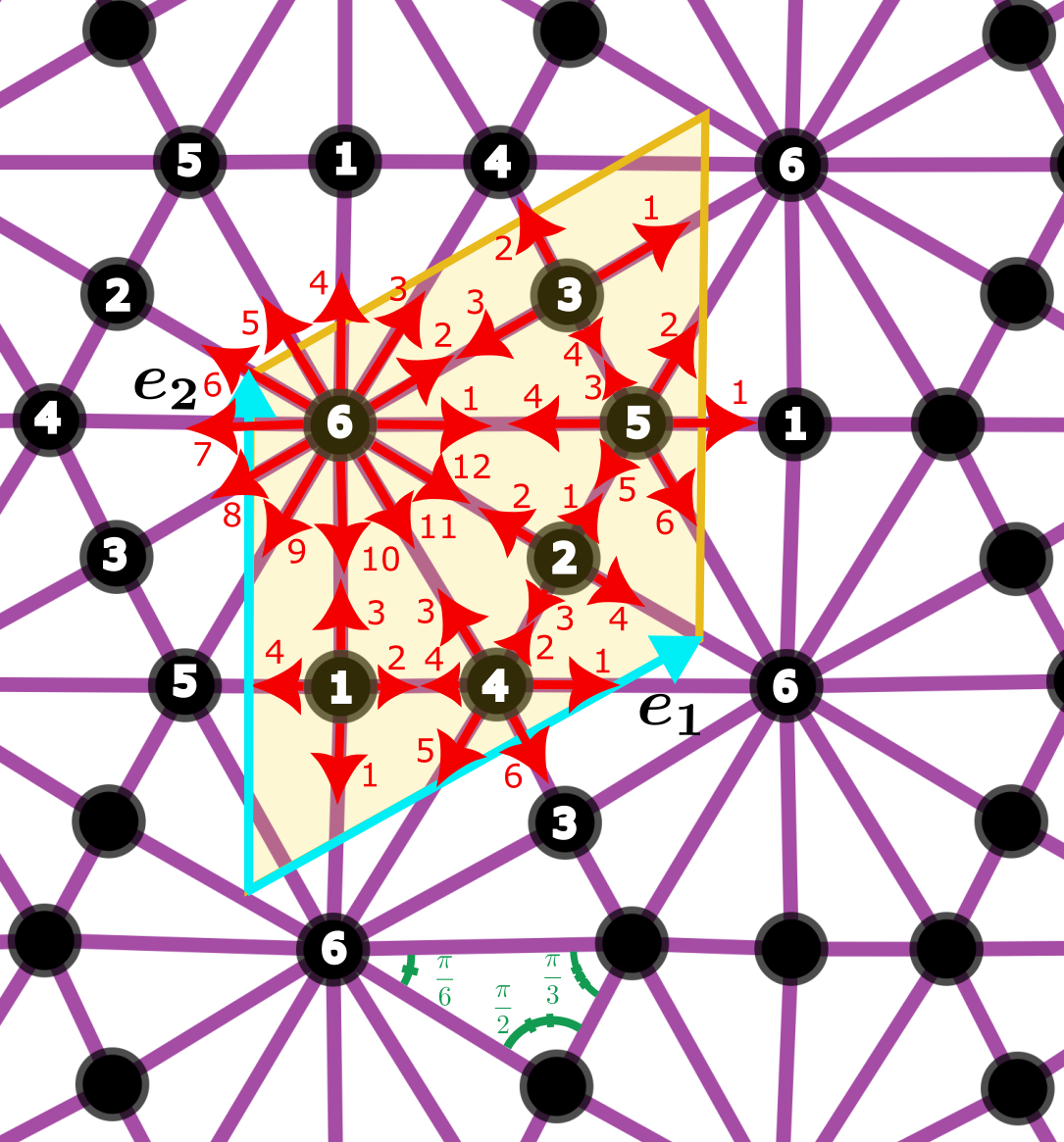}
\label{FigLavesSHDLabels}
\end{figure}\\
The Bloch matrices are given by
\begin{align}
    B_1(\bk)&= \diag(e^{\rmi k_2},1,1,e^{\rmi k_1}),\\
    B_2(\bk)&= \diag(1,1,1,e^{- \rmi(k_1-k_2)}),\\
    B_3(\bk)&= \diag(e^{-\rmi k_1},e^{-\rmi k_2},1,1),\\
    B_4(\bk)&= \diag(e^{-\rmi (k_1-k_2)},1,1,1,e^{\rmi k_2},e^{\rmi k_2}),\\
    B_5(\bk)&= \diag(e^{-\rmi k_2},e^{-\rmi k_2},1,1,1,e^{-\rmi (k_1-
    k_2)})\\
    B_6(\bk)&= \diag(1,1,e^{-\rmi k_1},e^{-\rmi k_1},e^{-\rmi( k_2-k_1)},e^{-\rmi( k_2-k_1)},e^{-\rmi( k_2-k_1)},e^{\rmi k_1},e^{\rmi k_1},1,1,1).
\end{align}
The edge-connectivity matrices are given by 
\begin{align}
     E_{1;4}&=E_{3;4}=\varepsilon_1= \begin{pmatrix}
        0 & & &  \\ 
        & 1 & & \\
        & & 0 & \\ 
        & & & 0
    \end{pmatrix},\
     E_{1;6}=E_{3;6}=\varepsilon_2= \begin{pmatrix}
        1 & & &  \\ 
        & 0 & & \\
        & & 1 & \\ 
        & & & 0
    \end{pmatrix},\
     E_{1;5}=E_{3;5}=\varepsilon_3= \begin{pmatrix}
        0 & & &  \\ 
        & 0 & & \\
        & & 0 & \\ 
        & & & 1
    \end{pmatrix},\\
    E_{2;6}&= \begin{pmatrix}
        0 & & &  \\ 
        & 1 & & \\
        & & 0 & \\ 
        & & & 0
    \end{pmatrix},\ E_{2;4}= \begin{pmatrix}
        0 & & &  \\ 
        & 0 & & \\
        & & 1 & \\ 
        & & & 0
    \end{pmatrix},\ E_{2;5}= \begin{pmatrix}
        1 & & &  \\ 
        & 0 & & \\
        & & 0 & \\ 
        & & & 0
    \end{pmatrix},
\end{align}
\begin{align}
    E_{4;1} &= \begin{pmatrix}
        0 & & & & & \\ 
        & 0 & & & & \\
        & & 0 & & & \\ 
        & & & 1 & & \\
        & & & & 0 & \\
        & & & & & 0 
    \end{pmatrix}, E_{4;2}= \begin{pmatrix}
        0 & & & & & \\ 
        & 1 & & & & \\
        & & 0 & & & \\ 
        & & & 0 & & \\
        & & & & 0 & \\
        & & & & & 0 
    \end{pmatrix}, E_{4;3}= \begin{pmatrix}
        0 & & & & & \\ 
        & 0 & & & & \\
        & & 0 & & & \\ 
        & & & 0 & & \\
        & & & & 0 & \\
        & & & & & 1 
    \end{pmatrix},  E_{4;6}= \begin{pmatrix}
        1 & & & & & \\ 
        & 0 & & & & \\
        & & 1 & & & \\ 
        & & & 0 & & \\
        & & & & 1 & \\
        & & & & & 0 
    \end{pmatrix},\\
    E_{5;1} &= \begin{pmatrix}
        1 & & & & & \\ 
        & 0 & & & & \\
        & & 0 & & & \\ 
        & & & 0 & & \\
        & & & & 0 & \\
        & & & & & 0 
    \end{pmatrix}, E_{5;2}= \begin{pmatrix}
        0 & & & & & \\ 
        & 0 & & & & \\
        & & 0 & & & \\ 
        & & & 0 & & \\
        & & & & 1 & \\
        & & & & & 0 
    \end{pmatrix}, E_{5;3}= \begin{pmatrix}
        0 & & & & & \\ 
        & 0 & & & & \\
        & & 1 & & & \\ 
        & & & 0 & & \\
        & & & & 0 & \\
        & & & & & 0 
    \end{pmatrix},  E_{5;6}= \begin{pmatrix}
        0 & & & & & \\ 
        & 1 & & & & \\
        & & 0 & & & \\ 
        & & & 1 & & \\
        & & & & 0 & \\
        & & & & & 1 
    \end{pmatrix},
\end{align}
and 
\begin{align}
E_{6;1}&=\Delta_4+\Delta_6,\\  
E_{6;2}&=\Delta_6+\Delta_{12}\\
E_{6;3}&=\Delta_2+\Delta_8\\
E_{6;4}&=\Delta_3+\Delta_7+\Delta_{11}\\
 E_{6;5}&=\Delta_1+\Delta_5+\Delta_9.
\end{align}
Here $\Delta_i$ are $12\times 12$ matrices defined by $(\Delta_i)_{jk}= \delta_{ij}\delta_{ik}$ $i=1,j,k,\dots,12$. The tiling consists of a triangle whose angles are uniquely determined due to the fact that the sum of the angles around around vertices $a=6$, $a=4,5$ and $a=1,2,3$ must sum to $2\pi$, giving uniform angles $\frac{\pi}{6}$, $\frac{\pi}{3}$ and $\frac{\pi}{2}$ around these vertices, respectively.
To derive the above, let $\omega=e^{\rmi \frac{\pi}{2}}, \sigma=e^{\rmi \frac{\pi}{3}}, \zeta=e^{\rmi \frac{\pi}{6}}$ and $g= e^{\rmi \frac{5\pi}{6}},$ then the phases at each site in the unit cell are given by
\begin{align}
    \bd{\theta}_1&=(\omega^{-1},1,\omega,\omega^2),\\
    \bd{\theta}_2&=g\bd{\theta_1},\\
    \bd{\theta}_3&=\zeta^{-1} \bd{\theta}_2,\\
    \bd{\theta}_4&=(1,\sigma,\sigma^2,\sigma^3,\sigma^4,\sigma^5),\\
    \bd{\theta}_5&=\bd{\theta}_4,\\
    \bd{\theta}_6&=(1,\zeta,\zeta^2,\zeta^3,\zeta^4,\zeta^5,\zeta^6,\zeta^7,\zeta^8,\zeta^9,\zeta^{10},\zeta^{11}).
\end{align}
We find
\begin{align}
    \nonumber \det(\mathbbm{1}_Q- t W(\bk))&=(1+t)^{12}\Bigl(t^{24}-12 t^{23}+78 t^{22}-340
   t^{21}+1146 t^{20}-3084 t^{19}+7152 t^{18}-14220 t^{17}
   \\ \nonumber & +25731 t^{16}-40472 t^{15}+58146 t^{14}-70896 t^{13}+77636 t^{12}-70896 t^{11}+58146 t^{10}\\
   \nonumber & -40472 t^9+25731 t^8-14220
   t^7+7152 t^6-3084 t^5+1146 t^4-340 t^3+78 t^2-12 t+1\Bigr) 
   \\ 
   \nonumber &+t^2 (1-t)^6 (1+t)^{12}  \Bigl(3 t^{14}-10 t^{13}+44 t^{12}-70 t^{11}+160 t^{10}-114 t^9+209 t^8-60
   t^7\\
   \nonumber & +209 t^6-114 t^5 +160 t^4-70 t^3+44 t^2-10 t+3\Bigr) \vare_{\Delta}(\bk)-t^6 \left(1-t^2\right)^{12} \vare_{\Delta}(2\bk)\\ 
   &+t^4 (1-t)^{10}(1+t)^{12}  \left(t^6+2 t^5+t^4+8 t^3+t^2+2 t+1\right) \vare_{\Delta}(2k_1-k_2,k_1-2k_2).
\end{align}
The critical temperature $t_{\rm c}=0.2371$ is the root of the equation 
\begin{equation}
    \det(\mathbbm{1}_Q- t W_0)=(t+1)^{12} \left(t^4-2 t^3+4 t^2-2 t+1\right)^2 \left(t^8-4 t^7+4 t^5-34 t^4+4 t^3-4 t+1\right)^2.
\end{equation}
We confirm that
\begin{align}
    \nonumber \det(u \mathbbm{1}_Q- W_0^{\dagger}W_0)&= P_4(u)^3 P_6(u)^2 P_{12}(u)\\
   &= (u^2-6u+1)^6(u^3-15u^2+15u-1)^4(u^6-66u^5+495u^4-924u^3+495u^2-66u+1)^2.
\end{align}

\clearpage
\subsection{Lieb lattice ($q_a=4,2$)}

\subsubsection{Matrix $W(\vec{k})$} 

\noindent The Lieb lattice has $N_{\rm u}=3$ sites in the unit cell with coordination numbers $q_1=4$ and $q_2=q_3=2$, respectively. We label the sites in the unit cell by $a\in\{1,2,3\}$ and the local edges at each site by $\mu_1\in\{1,\dots,4\}$ and $\mu_{2,3}\in\{1,1\}$ according to the following schematic:
\begin{figure}[h!]
    \includegraphics[width=9cm]{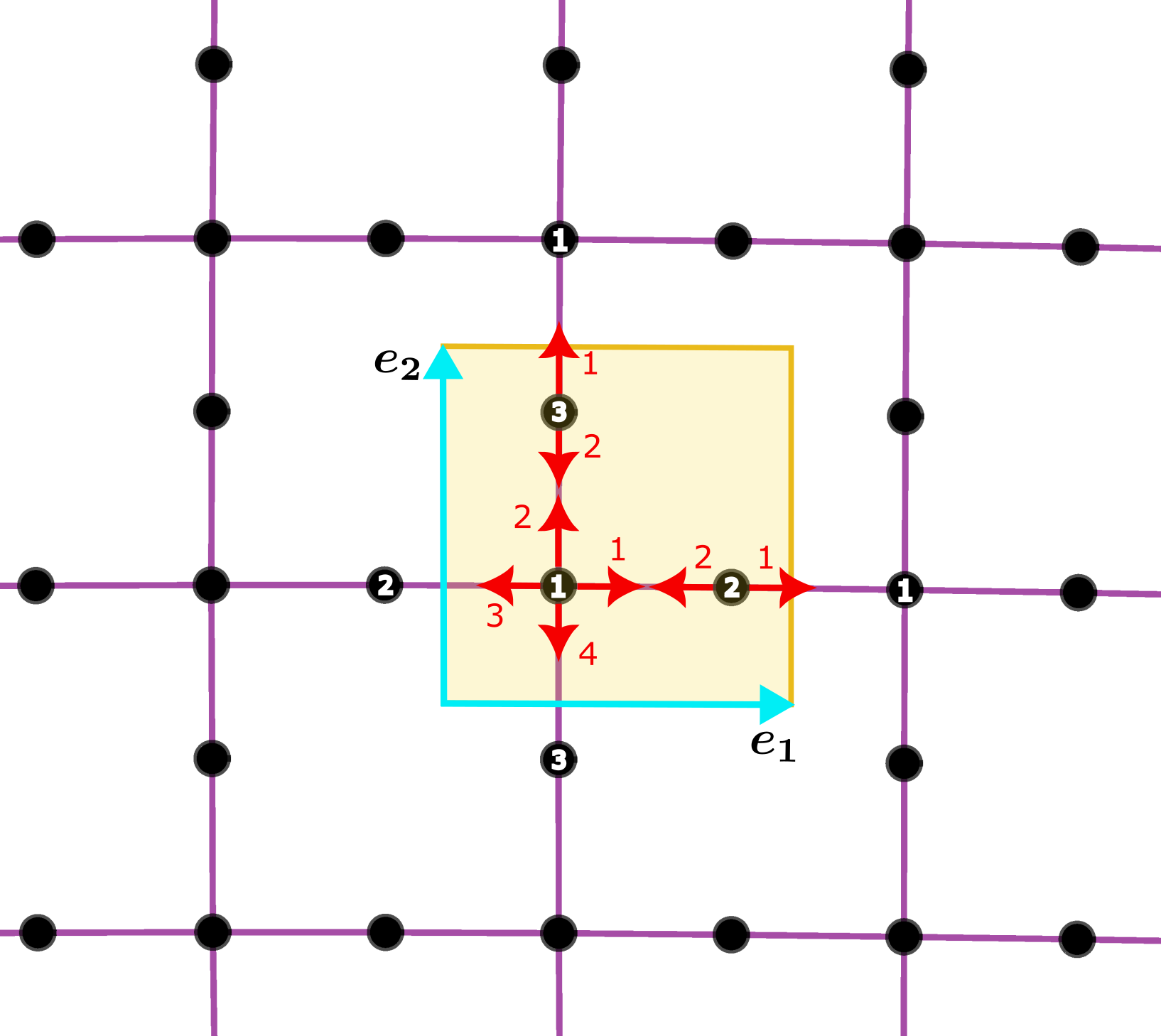}
     \label{FigLiebLabels}
\end{figure}\\
The Bloch matrices read
\begin{align}
 \nonumber B_1(\vec{k}) &=\begin{pmatrix} 1 & & & \\ & 1 & & \\ & & e^{\rmi k_1} & \\ & & & e^{\rmi k_2}\end{pmatrix},\\ B_2(\vec{k}) &=\begin{pmatrix} e^{-\rmi k_1} & \\ & 1\end{pmatrix},\ B_3(\vec{k}) = \begin{pmatrix} e^{-\rmi k_2} & \\ & 1 \end{pmatrix}
\end{align}
and the edge-connectivity matrices are given by
\begin{align}
 E_{1;1} &= E_{2;2} = E_{3;3} = E_{2;3} = E_{3;2} = 0,\\
 E_{1;2} &= \vare_1,\ E_{1;3} = \vare_2,\ E_{2;1}=E_{3;1} = \mathbb{1}_2
\end{align}
with
\begin{align} 
 \vare_1 = \text{diag}(1, 0,1, 0),\ \vare_2 = \text{diag}(0, 1, 0, 1).
\end{align}
The angles at the vertices are
\begin{align}
 \vec{\theta}_1 &= (1,\rmi,-1,-\rmi)^T\\
 \vec{\theta}_2 &= (1,-1)^T,\ \vec{\theta}_3= (\rmi,-\rmi)^T.
\end{align}
We obtain
\begin{align}
 W(\vec{k}) = \begin{pmatrix} 0 & E_{1;2} B_1(\vec{k}) \Phi_{12} & E_{1;3} B_1(\vec{k}) \Phi_{13} \\ E_{2;1} B_2(\vec{k}) \Phi_{21} & 0 & 0 \\ E_{3;1} B_3(\vec{k}) \Phi_{31} & 0 & 0 \end{pmatrix}
\end{align}
with
\begin{align}
  \Phi_{21} &= \Phi_{12}^\dagger = \begin{pmatrix} 1 & e^{\rmi \pi/4} & 0 & e^{-\rmi \pi/4} \\ 0 & e^{-\rmi \pi/4} &1 & e^{\rmi \pi/4} \end{pmatrix},\\
  \Phi_{31} &= \Phi_{13}^\dagger = \begin{pmatrix} e^{-\rmi \pi/4} & 1 & e^{\rmi \pi/4} & 0 \\ e^{\rmi \pi/4} & 0 & e^{-\rmi \pi/4} & 1 \end{pmatrix}.
\end{align}
We find
\begin{align}
 \mbox{det}(\mathbb{1} - t W(\vec{k})) = (1+t^4)^2+t^2(1-t^4)\vare_\square(\vec{k}).
\end{align}
The critical temperature follows from solving
\begin{align}
 \mbox{det}(\mathbb{1}-t W_0) = (1-2t^2-t^4)^2=0
\end{align}
and is given by
\begin{align}
 t_{\rm c} &= \sqrt{\sqrt{2}-1} = 0.643594,\\
 \frac{T_{\rm c}}{J} &= \frac{2}{\log\Bigl(1+\sqrt{2}+[2(\sqrt{2}+1)]^{1/2}\Bigr)} = 1.30841.
\end{align}

\subsubsection{Matrices $W_0$ and $\mathcal{B}$} 

\noindent For $\vec{k}=0$ we have
\begin{align}
 W_0^\dagger W_0 = \begin{pmatrix} \mathcal{B}_1 &  &  \\  & \mathcal{B}_2 &  \\  &  & \mathcal{B}_3 \end{pmatrix}
\end{align}
with
\begin{align}
\mathcal{B}_1 &= [\Phi_0^{(q=4)}]^2,\ \mathcal{B}_2 = \mathcal{B}_3 = \mathbb{1}_2.
\end{align} 
The corresponding characteristic polynomial is 
\begin{align}
 \nonumber \det(u \mathbb{1}_8 - W_0^\dagger W_0) &= P_4(u) P_2(u)^2 = (u^2-6u+1)^2 (u-1)^4.
\end{align}

\clearpage

\subsection{Lieb-$2$ and Lieb-$n$  lattices ($q_a=4,2$)}

\subsubsection{Matrix $W(\vec{k})$}  Consider the Lieb-$n$ lattice with $n$ insertions of 2-coordinated sites on each edge of a Square lattice according to the following schematic:
\begin{figure}[h!]
    \includegraphics[width=8cm]{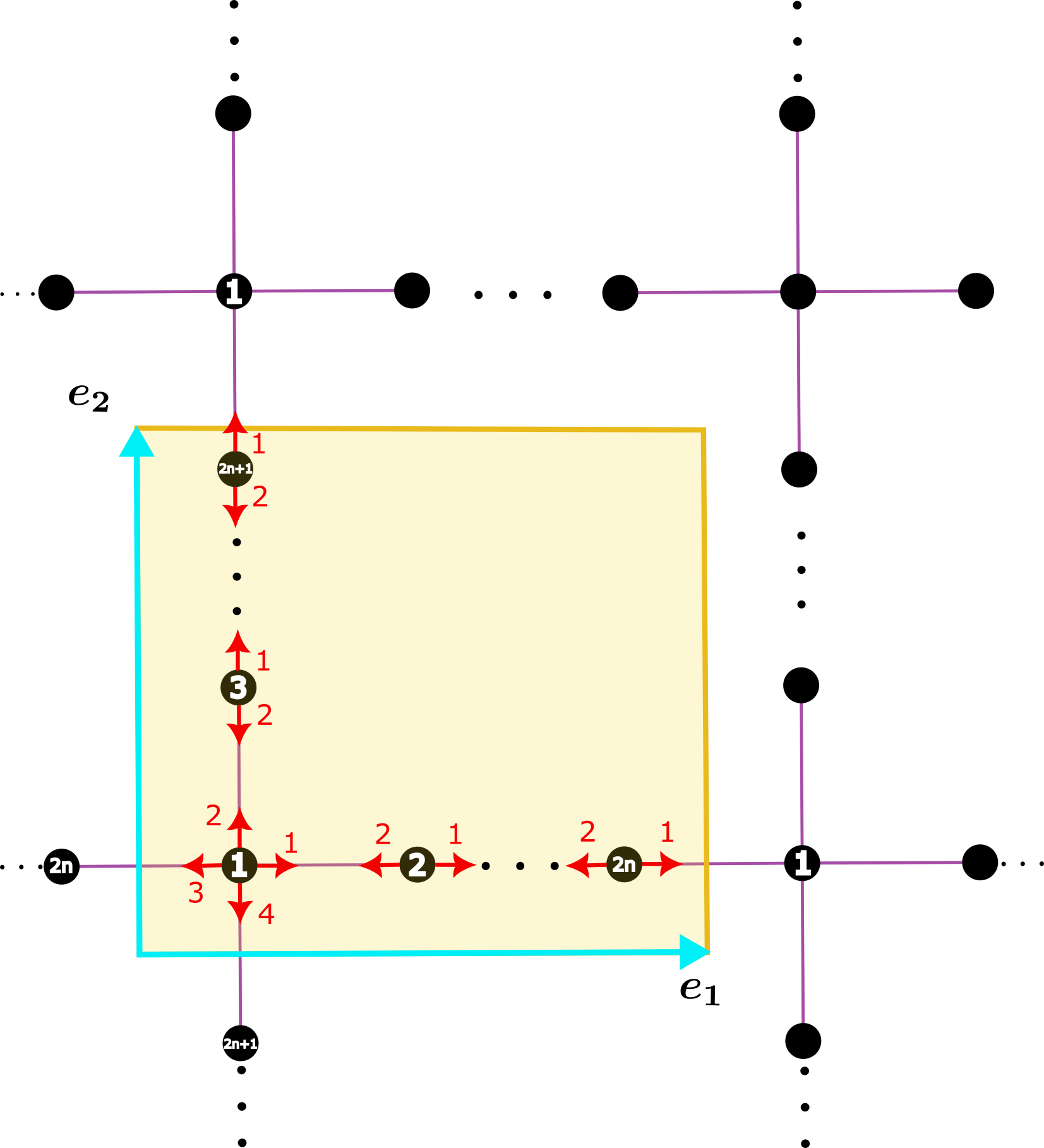}
     \label{FigLieb-nLabels}
\end{figure}\\
For $n=2$, there are $N_{\rm u}=2n+1=5$ sites in the unit cell that we label by $a\in\{1,\dots,5\}$, with $\mu_1\in\{1,\dots,4\}$ and $\mu_{a>1}\in\{1,2\}$. The Bloch matrices are
\begin{align}
 B_1(\vec{k}) &=\text{diag}(1,1,e^{-\rmi k_1},e^{-\rmi k_2}),\\
 B_2(\vec{k}) &=B_{3}(\vec{k}) = \mathbb{1}_2,\\
 B_{4}(\vec{k}) &=\text{diag}(e^{\rmi k_1},1),\ B_{5}(\vec{k})=\text{diag}(e^{\rmi k_2},1).
\end{align}
The nonvanishing edge-connectivity matrices are given by 
\begin{align}
 E_{1;2} &=\text{diag}(1,0,0,0),\ E_{1;3} =\text{diag}(0,1,0,0),\ E_{1;4} =\text{diag}(0,0,1,0),\ E_{1;5} =\text{diag}(0,0,0,1),\\
 E_{2;4} &= E_{3;5} = E_{4;1} = E_{5;1} = \begin{pmatrix} 1 & \\ & 0\end{pmatrix},\ E_{4;2} = E_{5;3} = E_{2;1} = E_{3;1} = \begin{pmatrix} 0 & \\ & 1\end{pmatrix}.
\end{align}
The angle vectors are
\begin{align}
 \vec{\theta}_1 &=(1,\rmi,-1,-\rmi)^T,\ \vec{\theta}_2 =\vec{\theta}_4 =(1,-1)^T,\ \vec{\theta}_3=\vec{\theta}_5=(\rmi,-\rmi)^T.
\end{align}
We arrive at
\begin{align}
 W(\vec{k}) = \begin{pmatrix} 0 & E_{1;2}B_1 \Phi_{12} & E_{1;3}B_1 \Phi_{13}  & E_{1;4}B_1 \Phi_{14}  & E_{1;5}B_1 \Phi_{15}  \\ E_{2;1} B_2 \Phi_{21} & 0 & 0 & E_{2;4}B_2\Phi_{24} & 0  \\ E_{3;1} B_3 \Phi_{31} & 0 & 0 & 0 & E_{3;5}B_3\Phi_{35}  \\ E_{4;1} B_4 \Phi_{41} & E_{4;2}B_4\Phi_{42} & 0 & 0 & 0  \\ E_{5;1} B_5 \Phi_{51} & 0 & E_{5;3}B_5\Phi_{53} & 0 & 0  \end{pmatrix}
\end{align}
and
\begin{align}
 \mbox{det}(\mathbb{1} - t W(\vec{k})) = 1+2t^6+t^{12}+t^3(1-t^6)\vare_\square(\vec{k}).
\end{align}
The critical temperature is found from
\begin{align}
 \mbox{det}(\mathbb{1}-t W_0) = (1-2t^3-t^6)^2=0.
\end{align}
We obtain
\begin{align}
 t_{\rm c} &= (\sqrt{2}-1)^{1/3} = 0.745432,\\
 \frac{T_{\rm c}}{J} &= 1.03886.
\end{align}
We have
\begin{align}
 W_0^\dagger W_0 = \text{diag}(\mathcal{B}_1,\dots,\mathcal{B}_5)
\end{align}
with
\begin{align}
 \mathcal{B}_1= [\Phi_0^{(q=4)}]^2,\ \mathcal{B}_2=\dots=\mathcal{B}_5 = \mathbb{1}_2.
\end{align}
This eventually yields
\begin{align}
\det(u \mathbb{1}_{12} - W_0^\dagger W_0) = P_4(u)P_2(u)^4= (u^2-6u+1)^2(u-1)^8.
\end{align}

\subsubsection{Lieb-$n$ lattice sequence ($q_a=4,2$)}
Following the arguments presented in the EM, the critical temperature of the the Lieb-$n$ lattice is $t_{\rm c}^{(n)}= (\sqrt{2}-1)^{1/(n+1)}$. For large $n$, we have
\begin{align}
 (\sqrt{2}-1)^{1/(n+1)} = 1 + \frac{\log(\sqrt{2}-1)}{n}+\mathcal{O}\Bigl(\frac{1}{n^2}\Bigr),
\end{align}
and hence
\begin{align}
 \frac{T_{\rm c}}{J} &= \frac{2}{\log\Big(\frac{1+t_{\rm c}^{(n)}}{1-t_{\rm c}^{(n)}}\Bigr)}= \frac{2}{\log n+0.819421}+\mathcal{O}(\frac{1}{n}\Bigr). 
\end{align}
Indeed, we have
\begin{align}
 t_{\rm c}^{(n)} = 1+\frac{\log t_\square}{n} + \mathcal{O}(n^{-2}),
\end{align}
and so
\begin{align}
\frac{T_{\rm c}}{J} &= \frac{2}{\log\Big(\frac{1+t_{\rm c}^{(n)}}{1-t_{\rm c}^{(n)}}\Bigr)} \simeq \frac{2}{\log\Biggl( \frac{2+\frac{\log t_\square}{n}}{-\frac{\log t_\square}{n}}\Biggr)} \simeq \frac{2}{\log n + \log\Bigl( \frac{2}{-\log t_\square}\Bigr)}
\end{align} 
with
\begin{align}
\log\Bigl( \frac{2}{-\log t_\square}\Bigr) = 0.819421.
\end{align}

\clearpage
\subsection{Lattice with $q_{\rm max}=7$ ($q_a=3,7)$}

\subsubsection{Matrix $W(\vec{k})$}  To obtain a lattice with $q_{\rm max}=7$, we consider a Laves-CaVO lattice where one edge of each unit cell is deleted. The lattice has $N_{\rm u}=2$ sites in the unit cell, which we label $a=1,2$. We have $q_1=7$ and $q_2=3$ according to the following schematic:
\begin{figure}[h!]
    \includegraphics[width=8cm]{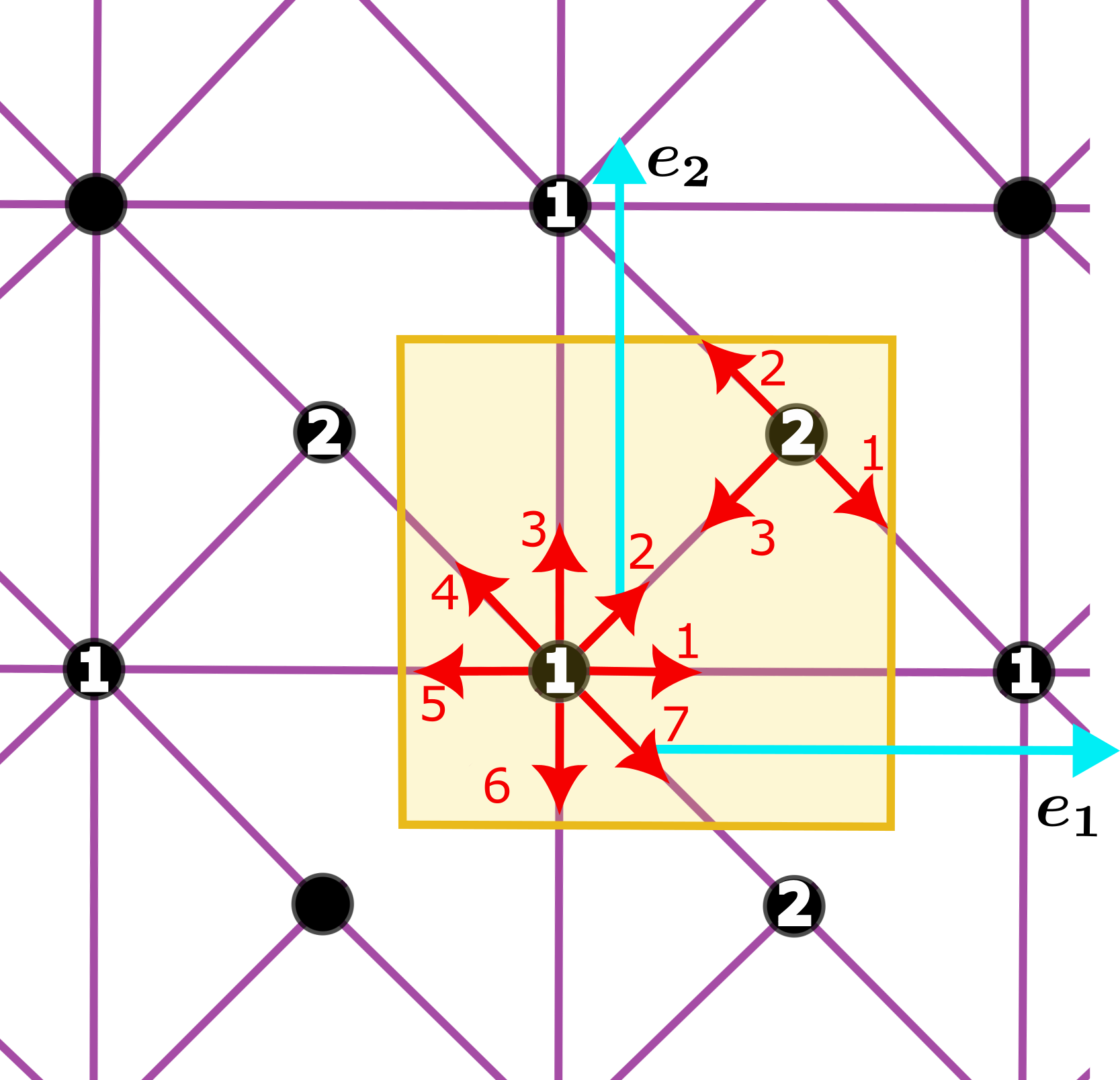}
     \label{Figqmax7}
\end{figure}\\
The average coordination number is $\bar{q}=5$. The angle vectors are
\begin{align}
 \vec{\theta}_1 &= (1,e^{\rmi \pi/4},\rmi,e^{3\pi\rmi/4},-1,-\rmi,e^{-\rmi\pi/4})^T,\\
 \vec{\theta}_2 &= (e^{-\rmi \pi/4},e^{3\pi\rmi/4},e^{-3\pi\rmi/4})^T
\end{align}
and the edge-connectivity matrices read
\begin{align}
 E_{1;1} &=\text{diag}(1,0,1,0,1,1,0),\\
 E_{1;2} &=\text{diag}(0,1,0,1,0,0,1),\\
 E_{2;1} &=\mathbb{1}_3,\ E_{2;2}=0.
\end{align}
Thus
\begin{align}
 W_0 &= \begin{pmatrix} E_{1;1} \Phi_{11} & E_{1;2} \Phi_{12} \\ E_{2;1} \Phi_{21} & 0 \end{pmatrix}.
\end{align}
We have
\begin{align}
\mbox{det}(\mathbb{1}_{10} - t W_0) = (1+t)^2(1-3t-t^2-t^3)^2,
\end{align}
so that
\begin{align}
 t_{\rm c} &=0.295598,\\
 \frac{T_{\rm c}}{J} &=3.28204.
\end{align}
We further have
\begin{align}
 W_0^\dagger W_0 = \begin{pmatrix} \mathcal{B}_1 & \\ & \mathcal{B}_2 \end{pmatrix}
\end{align}
with
\begin{align}
 \det(u\mathbb{1}_{10} -W_0^\dagger W_0) = P_7(u)P_3(u) = u^2(u^3-21u^2+35u-7)^2(u-3)^2.
\end{align}

\clearpage
\subsection{Lattice with $q_{\rm max}=9$ ($q_a=4,7,9$)}

\subsubsection{Matrix $W(\vec{k})$}  The lattice has $N_{\rm u}=5$ sites in the unit cell, which we label $a=1,\dots,5$. We have $q_1=9$, $q_2=q_4=q_5=4$ and $q_3=7$. The lattice is obtained from the $q_{\rm max}=10$ lattice by erasing an edge in the unit cell. We have the following schematic:
\begin{figure}[h!]
    \includegraphics[width=8cm]{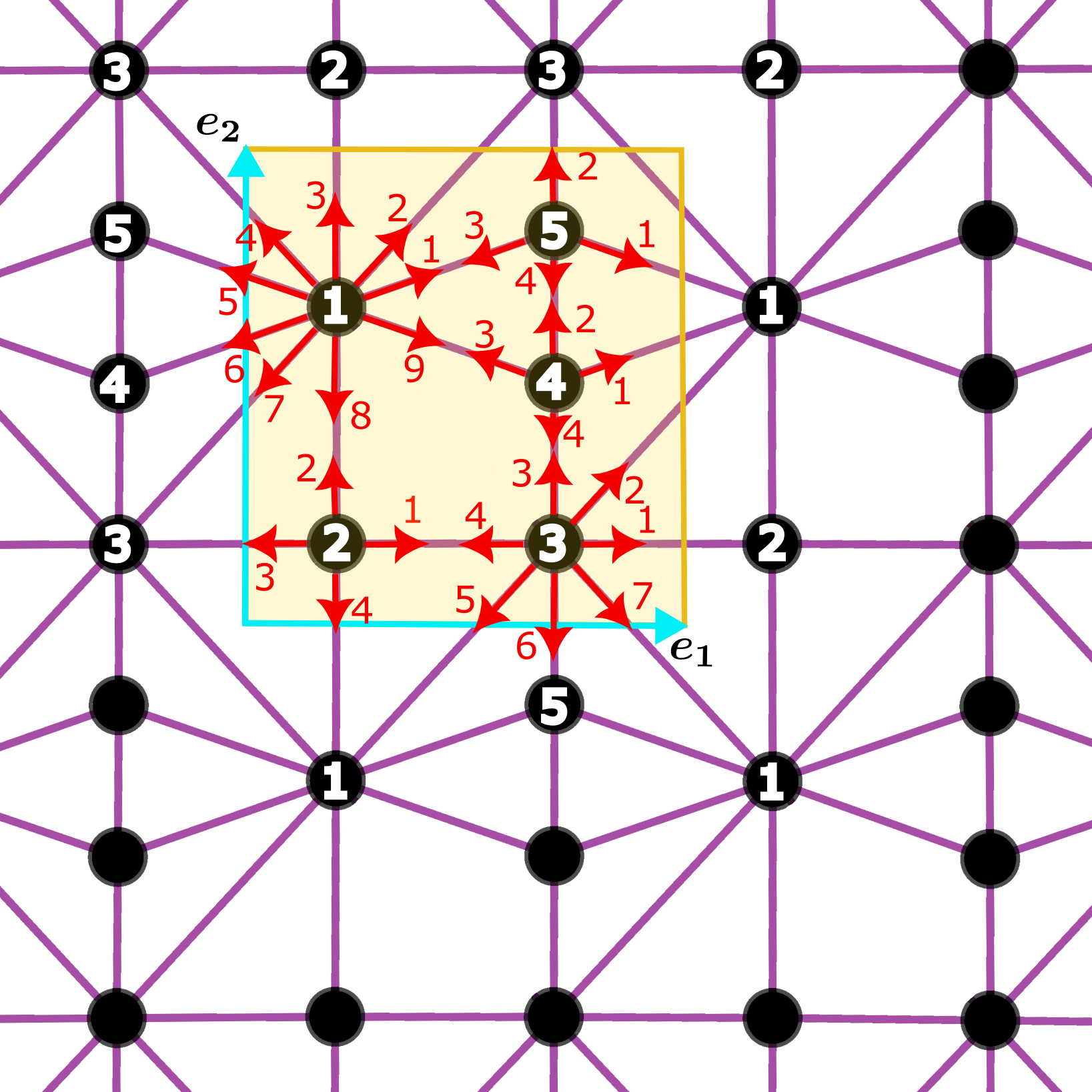}
     \label{Figqmax9}
\end{figure}\\
The matrix $W(\vec{k})$ has size $Q \times Q$ with $Q=28$ and the average coordination number is $\bar{q}=5.6$. The angle vectors are
\begin{align}
 \vec{\theta}_1 &= (e^{\rmi \pi/12}, e^{\rmi \pi/4},\rmi,e^{3\pi\rmi/4},e^{11\rmi\pi/12},e^{-11\pi\rmi/12}e^{-3\rmi\pi/4},-\rmi,e^{-\rmi \pi/12})^T,\\
 \vec{\theta}_2 &= (1,\rmi,-1,-\rmi)^T,\\
 \vec{\theta}_3 &= (1,e^{\rmi \pi/4},\rmi,-1,e^{-3\rmi\pi/4},-\rmi,e^{-\rmi \pi/4})^T,\\
 \vec{\theta}_4 &=(e^{\rmi \pi/12},\rmi,e^{\rmi 11\pi/12},-\rmi)^T,\\
 \vec{\theta}_5 &= (e^{-\rmi \pi/12},\rmi,e^{-\rmi11\pi/12},-\rmi)^T.
\end{align}
The nonvanishing edge-connectivity matrices are
\begin{align}
 E_{1;2} &= \text{diag}(0,0,1,0,0,0,0,1,0),\\
 E_{1;3} &= \text{diag}(0,1,0,1,0,0,1,0,0),\\
 E_{1;4} &= \text{diag}(0,0,0,0,0,1,0,0,1),\\
 E_{1;5} &= \text{diag}(1,0,0,0,1,0,0,0,0),\\
 E_{2;1} &= \text{diag}(0,1,0,1),\ E_{2;3} = \text{diag}(1,0,1,0),\\
 E_{3;1} &=\text{diag}(0,1,0,0,1,0,1),\\
 E_{3;2} &=\text{diag}(1,0,0,1,0,0,0),\\
 E_{3;4} &=\text{diag}(0,0,1,0,0,0,0),\\
 E_{3;5} &=\text{diag}(0,0,0,0,0,1,0),\\
 E_{4;1} &= \text{diag}(1,0,1,0),\ E_{4;3} = \text{diag}(0,0,0,1),\ E_{4;5} = \text{diag}(0,1,0,0),\\
 E_{5;1} &= \text{diag}(1,0,1,0),\ E_{5;3} = \text{diag}(0,1,0,0),\ E_{5;4} = \text{diag}(0,0,0,1).
\end{align}
We have
\begin{align}
 W_0 &= \begin{pmatrix} 0 & E_{1;2} \Phi_{12} & E_{1;3} \Phi_{13} & E_{1;4} \Phi_{14} & E_{1;5} \Phi_{15} \\  E_{2;1} \Phi_{21} & 0 & E_{2;3} \Phi_{23} & 0 & 0 \\ E_{3;1} \Phi_{31} & E_{3;2} \Phi_{32} & 0 & E_{3;4} \Phi_{34} & E_{3;5} \Phi_{35} \\ E_{4;1} \Phi_{41} & 0 & E_{4;3} \Phi_{43} & 0 & E_{4;5} \Phi_{45} \\ E_{5;1} \Phi_{51} & 0 & E_{5;3} \Phi_{53} & E_{5;4} \Phi_{54} & 0 \end{pmatrix}.
\end{align}
We have
\begin{align}
 \mbox{det}(\mathbb{1}_{28}-t W_0) = (1+t)^8(1-4t+3t^2-5t^3-5t^4-15t^5+t^6-7t^7-t^9)^2,
\end{align}
which yields the critical temperature
\begin{align}
 t_{\rm c} &=0.268086,\\
 \frac{T_{\rm c}}{J} &= 3.63901.
\end{align}
We confirm that
\begin{align}
 \det(u\mathbb{1}_{28}-W_0^\dagger W_0) &= P_{9}(u) P_7(u) P_4(u)^3 \\
 &= u^2(u^4-36u^3+126u^2-84u+9)^2(u^3-21u^2+35u-7)^2(u^2-6u+1)^6.
\end{align}

\clearpage
\subsection{Lattice with $q_{\rm max}=10$ ($q_a=4,8,10$)}

\subsubsection{Matrix $W(\vec{k})$}  The lattice has $N_{\rm u}=5$ sites in the unit cell, which we label $a=1,\dots,5$. We have $q_1=10$, $q_2=q_4=q_5=4$ and $q_3=8$. We assign coordinates according to the following schematic:
\begin{figure}[h!]
    \includegraphics[width=10cm]{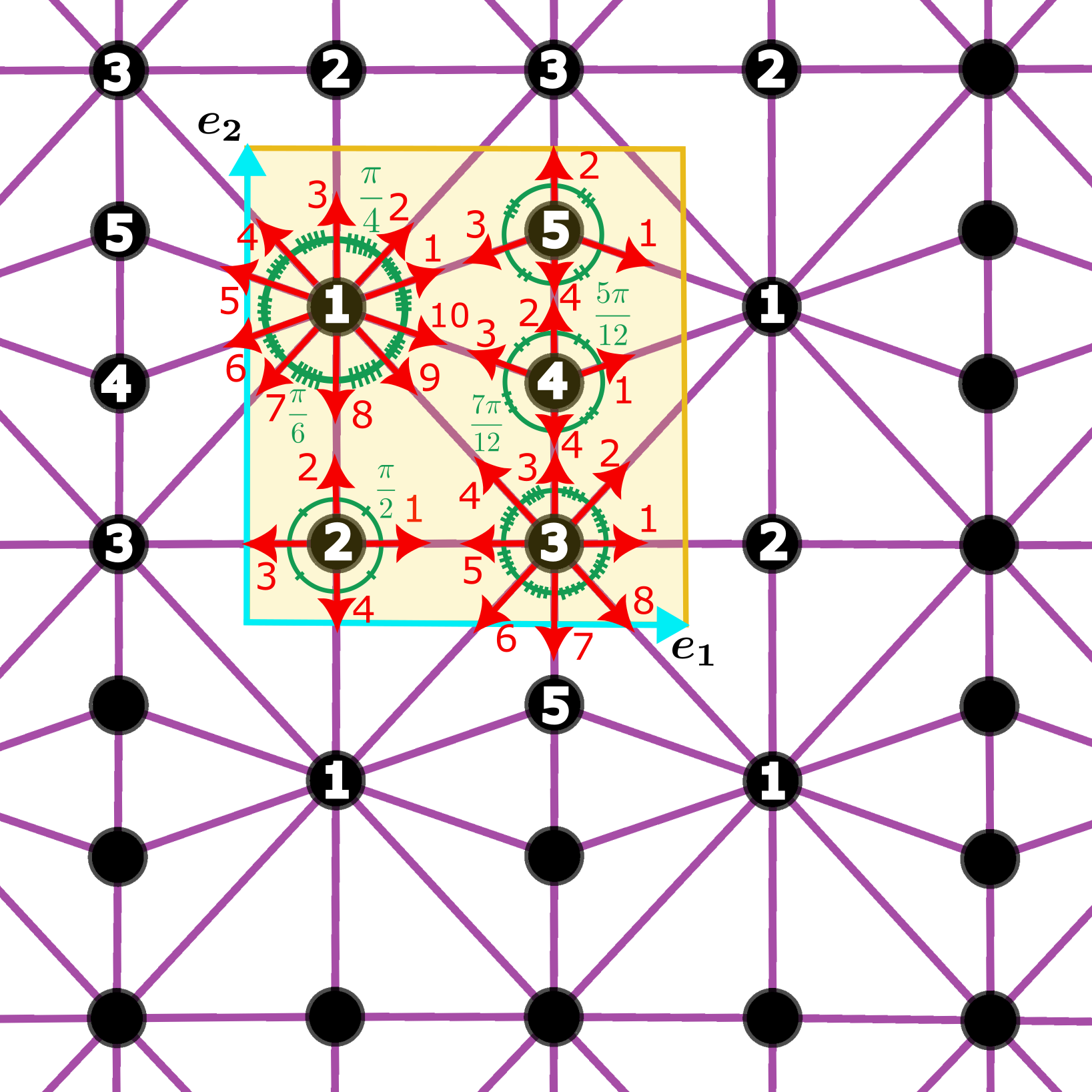}
     \label{Figqmax10}
\end{figure}\\
The matrix $W(\vec{k})$ has size $Q \times Q$ with $Q=30$ and the average coordination number is $\bar{q}=6$. The non-vanishing edge-connectivity matrices are
\begin{align}
 E_{1;2} &= \text{diag}(0,0,1,0,0,0,0,1,0,0),\\
 E_{1;3} &= \text{diag}(0,1,0,1,0,0,1,0,1,0),\\
 E_{1;4} &= \text{diag}(0,0,0,0,0,1,0,0,0,1),\\
 E_{1;5} &= \text{diag}(1,0,0,0,1,0,0,0,0,0),\\
 E_{2;1} &= \text{diag}(0,1,0,1),\ E_{2;3} = \text{diag}(1,0,1,0),\\
 E_{3;1} &=\text{diag}(0,1,0,1,0,1,0,1),\\
 E_{3;2} &=\text{diag}(1,0,0,0,1,0,0,0),\\
 E_{3;4} &=\text{diag}(0,0,1,0,0,0,0,0),\\
 E_{3;5} &=\text{diag}(0,0,0,0,0,0,1,0),\\
 E_{4;1} &= \text{diag}(1,0,1,0),\ E_{4;3} = \text{diag}(0,0,0,1),\ E_{4;5} = \text{diag}(0,1,0,0),\\
 E_{5;1} &= \text{diag}(1,0,1,0),\ E_{5;3} = \text{diag}(0,1,0,0),\ E_{5;4} = \text{diag}(0,0,0,1).
\end{align}
The lattice is a periodic tiling that uses three different triangles each with three free angles totaling $3\times 3=9$ free angles. However, the fact that the angles around each vertices must sum to $2\pi$ gives five equations, one of which fixes one angle to $\pi/2$. Furthermore, the sum of the angles inside the triangle should add to $\pi$, which gives three more equations. However, the rank of the matrix that hosts these equations (except the one that fixes $\pi/2$) is $6$. This means that the whole lattice has $9-(6+1)=2$ free angles (say $\alpha_1,\beta_1)$ that can be set to anything between $(0,\pi)$. The set of angles illustrated here are one of many such choices of these free angles when they are fixed to $\alpha_1=\frac{\pi}{4},\beta_1=\frac{\pi}{6}$.
The angle vectors are
\begin{align}
 \vec{\theta}_1 &= (e^{\rmi \pi/12}, e^{\rmi \pi/4},\rmi,e^{3\pi\rmi/4},e^{11\rmi\pi/12},e^{-11\pi\rmi/12}e^{-3\rmi\pi/4},-\rmi,e^{-\rmi \pi/4},e^{-\rmi \pi/12})^T,\\
 \vec{\theta}_2 &= (1,\rmi,-1,-\rmi)^T,\\
 \vec{\theta}_3 &= (1,e^{\rmi \pi/4},\rmi,e^{3\rmi\pi/4},-1,e^{-3\rmi\pi/4},-\rmi,e^{-\rmi \pi/4})^T,\\
 \vec{\theta}_4 &=(e^{\rmi \pi/12},\rmi,e^{\rmi 11\pi/12},-\rmi)^T,\\
 \vec{\theta}_5 &= (e^{-\rmi \pi/12},\rmi,e^{-\rmi11\pi/12},-\rmi)^T.
\end{align}

We have
\begin{align}
 W_0 &= \begin{pmatrix} 0 & E_{1;2} \Phi_{12} & E_{1;3} \Phi_{13} & E_{1;4} \Phi_{14} & E_{1;5} \Phi_{15} \\  E_{2;1} \Phi_{21} & 0 & E_{2;3} \Phi_{23} & 0 & 0 \\ E_{3;1} \Phi_{31} & E_{3;2} \Phi_{32} & 0 & E_{3;4} \Phi_{34} & E_{3;5} \Phi_{35} \\ E_{4;1} \Phi_{41} & 0 & E_{4;3} \Phi_{43} & 0 & E_{4;5} \Phi_{45} \\ E_{5;1} \Phi_{51} & 0 & E_{5;3} \Phi_{53} & E_{5;4} \Phi_{54} & 0 \end{pmatrix}.
\end{align}
We have
\begin{align}
 \mbox{det}(\mathbb{1}_{30}-t W_0) = (1+t)^{10}(1-5t+5t^2-2t^3-22t^4+16t^5-32t^6+18t^7-15t^8+5t^9-t^{10})^2,
\end{align}
which yields the critical temperature
\begin{align}
 t_{\rm c} &=0.238674,\\
 \frac{T_{\rm c}}{J} &= 4.10901.
\end{align}
We confirm that
\begin{align}
 \det(u\mathbb{1}_{30}-W_0^\dagger W_0) &= P_{10}(u) P_8(u) P_4(u)^3 \\
 &= (u^5-45u^4+210u^3-210u^2+45u-1)^2(u^4-28u^3+70u^2-28u+1)^2(u^2-6u+1)^6.
\end{align}

\clearpage
\subsection{Lattice with $q_{\rm max}=11$ ($q_a=3,11$)}

\subsubsection{Matrix $W(\vec{k})$} We consider the lattice where one of the edges of the Laves-Star lattice is removed. The lattice has $N_{\rm u}=6$ sites in the unit cell, which we label $a=1,\dots,6$. We have $q_1=q_2=11$ and $q_3=q_4=q_5=q_6=3$. We have the following schematic:\begin{figure}[h!]
    \includegraphics[width=10cm]{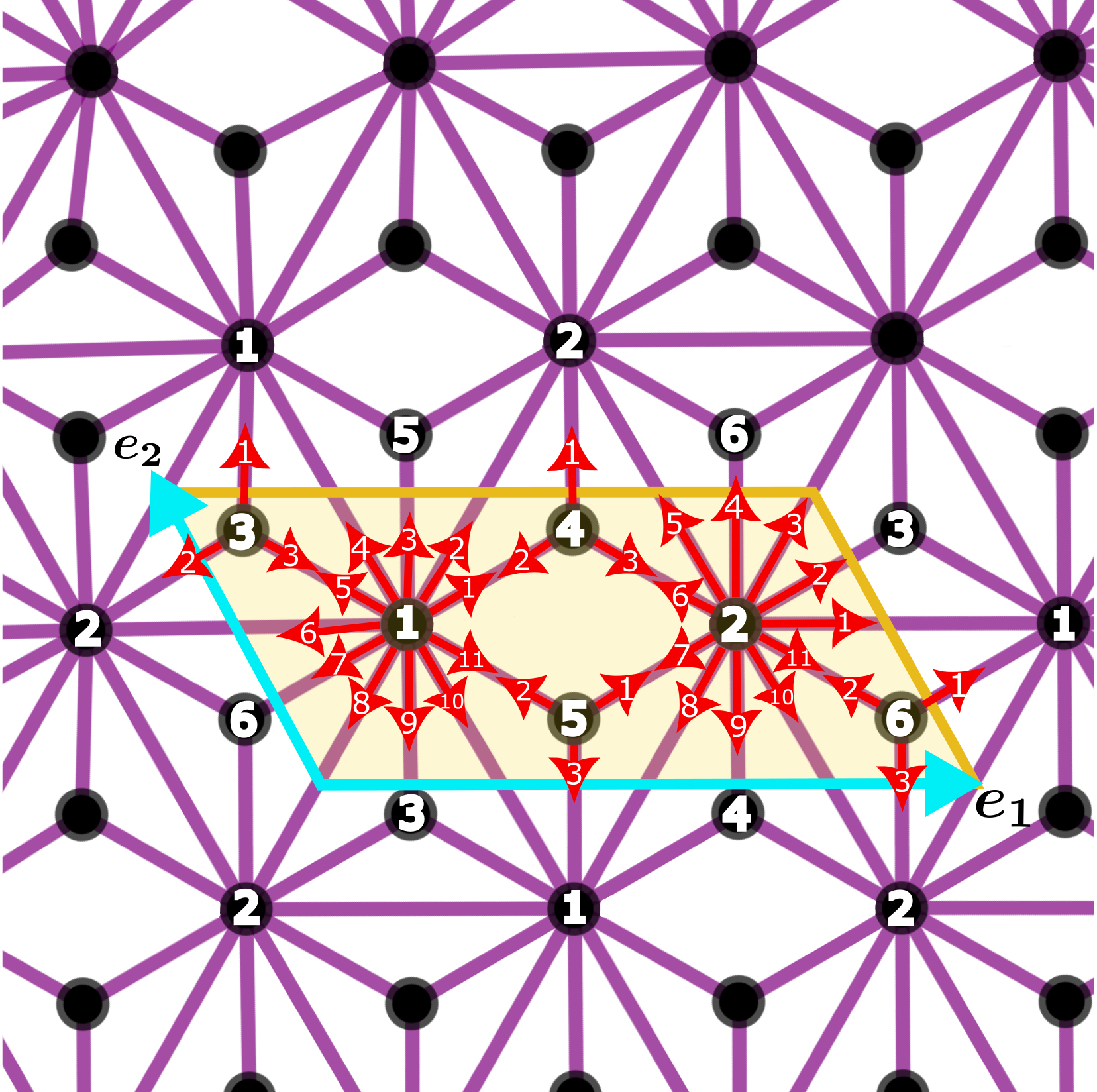}
     \label{Figqmax11}
\end{figure}\\
The matrix $W(\vec{k})$ has size $Q \times Q$ with $Q=34$ and the average coordination number is $\bar{q}=5.66$. The angle vectors with $\eta=e^{\rmi\pi/6}$ are
\begin{align}
 \vec{\theta}_1 &= (\eta,\eta^2,\eta^3,\eta^4,\eta^5,\eta^6,\eta^7,\eta^8,\eta^9,\eta^{10},\eta^{11})^T,\\
 \vec{\theta}_2 &= (1,\eta,\eta^2,\eta^3,\eta^4,\eta^5,\eta^7,\eta^8,\eta^9,\eta^{10},\eta^{11})^T,\\
 \vec{\theta}_3 &=\vec{\theta}_4 = (\rmi,\eta^7,\eta^{-1})^T,\\
 \vec{\theta}_5 &=\vec{\theta}_6 = (\eta,\eta^5,-\rmi)^T.
\end{align}
The nonvanishing edge-connectivity matrices are
\begin{align}
 E_{1;1} &= \text{diag}(0, 1, 0, 0, 0, 0, 0, 0, 0, 1, 0),\\
 E_{1;2} &= \text{diag}(0, 0, 0, 1, 0, 1, 0, 1, 0, 0, 0),\\
 E_{1;3} &= \text{diag}(0, 0, 0, 0, 1, 0, 0, 0, 1, 0, 0),\\
 E_{1;4} &= \text{diag}(1, 0, 0, 0, 0, 0, 0, 0, 0, 0, 0),\\
 E_{1;5} &= \text{diag}(0, 0, 0, 0, 0, 0, 0, 0, 0, 0, 1),\\
 E_{1;6} &= \text{diag}(0, 0, 1, 0, 0, 0, 1, 0, 0, 0, 0),\\
 E_{2;1} &= \text{diag}(1, 0, 0, 0, 1, 0, 0, 1, 0, 0, 0),\\
 E_{2;2} &= \text{diag}(0, 0, 1, 0, 0, 0, 0, 0, 0, 1, 0),\\
 E_{2;3} &= \text{diag}(0, 1, 0, 0, 0, 0, 0, 0, 0, 0, 0),\\
 E_{2;4} &= \text{diag}(0, 0, 0, 0, 0, 1, 0, 0, 1, 0, 0),\\
 E_{2;5} &= \text{diag}(0, 0, 0, 1, 0, 0, 1, 0, 0, 0, 0),\\
 E_{2;6} &= \text{diag}(0, 0, 0, 0, 0, 0, 0, 0, 0, 0, 1)
\end{align}
and
\begin{align}
 E_{3;1} &=\text{diag}(0,0,1),\ E_{3;2} =\text{diag}(1,1,0),\\
 E_{4;1} &=\text{diag}(1,1,0),\ E_{4;2} =\text{diag}(0,0,1),\\
 E_{5;1} &=\text{diag}(0,1,1),\ E_{5;2} =\text{diag}(1,0,0),\\
 E_{6;1} &=\text{diag}(1,0,0),\ E_{6;2} =\text{diag}(0,0,1).
\end{align} 
We have
\begin{align}
 W_0 &= \begin{pmatrix} E_{1;1}\Phi_{11} & E_{1;2} \Phi_{12} & E_{1;3} \Phi_{13} & E_{1;4} \Phi_{14}  & E_{1;5} \Phi_{15}  & E_{1;6} \Phi_{16} \\ E_{2;1}\Phi_{21} & E_{2;2} \Phi_{22} & E_{2;3} \Phi_{23} & E_{2;4} \Phi_{24}  & E_{2;5} \Phi_{25}  & E_{2;6} \Phi_{26} \\ E_{3;1}\Phi_{31} & E_{3;2} \Phi_{32} & 0 & 0  & 0  & 0 \\ E_{4;1}\Phi_{41} & E_{4;2} \Phi_{42} & 0 & 0  & 0  & 0 \\ E_{5;1}\Phi_{51} & E_{5;2} \Phi_{52} & 0 & 0  & 0  & 0 \\ E_{6;1}\Phi_{61} & E_{6;2} \Phi_{62} & 0 & 0  & 0  & 0 \end{pmatrix}.
\end{align}
We have
\begin{align}
 \mbox{det}(\mathbb{1}_{34}-t W_0) = (1+t)^6(1-5t+4t^2-16t^3+34t^4-10t^5+48t^6+8t^7+53t^8+23t^9-12t^{10})^2,
\end{align}
which yields the critical temperature
\begin{align}
 t_{\rm c} &=0.220907,\\
 \frac{T_{\rm c}}{J} &= 4.45218.
\end{align}
We confirm that
\begin{align}
 W_0^\dagger W_0 &= \text{diag}(\mathcal{B}_1,\mathcal{B}_2,\mathcal{B}_3,\mathcal{B}_4,\mathcal{B}_5,\mathcal{B}_6)
\end{align}
and
\begin{align}
 \mbox{det}(u\mathbb{1}_{34}-W_0^\dagger W_0) &= P_{11}(u)^2 P_3(u)^4 \\
 &= \Bigl[u(u^5-55u^4+330u^3-462u^2+165u-11)^2\Bigr]^2\Bigl[u(u-3)^2\Bigr]^4.
\end{align}

\clearpage
\subsection{Compass-Rose lattice ($q_a=3,6,24$)}\label{Spec}

\noindent The Compass-Rose lattice has $N_{\rm u}=9$ sites in the unit cell with coordination numbers $q_1=24,q_2=q_3=6$ and $q_4=q_5=q_6=q_7=q_8=q_9=3$, thus $Q=54$. We label the sites in the unit cell by $a\in\{1,\dots,9\}$ and the local edges at each site starting by $\mu_a\in\{1,\dots,24\}$, for site $a=1$, $\mu_a\in\{1,2,\dots,6\}$ for sites $a=2,6$, and the remaining edges $\mu_a \in\{1,2,\dots,3\}$ for the remaining three-coordinated sites according to the following schematic:

\begin{figure}[h!]
\centering
\includegraphics[width=0.75\linewidth]{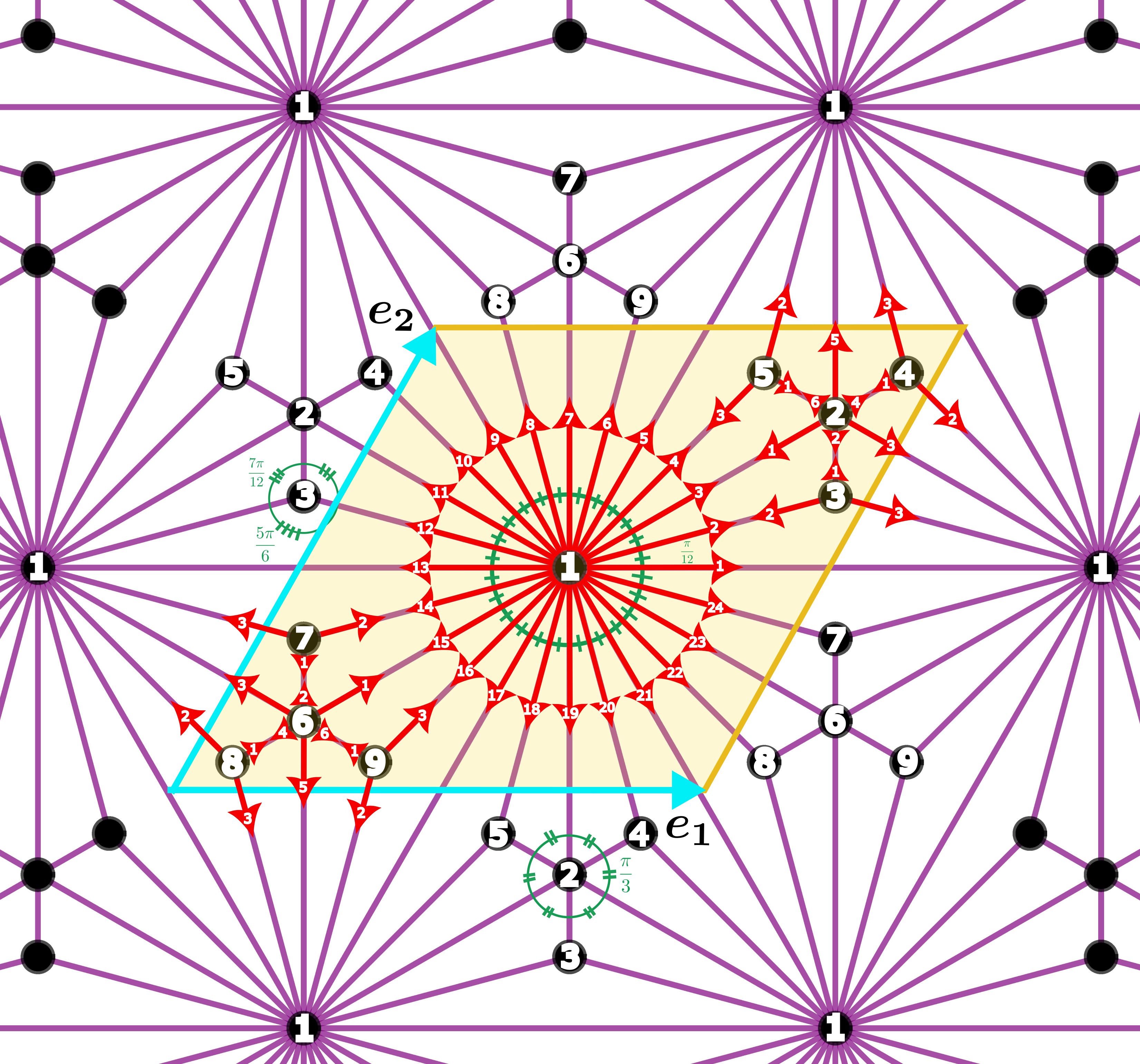}
\caption{The Compass-Rose lattice consists of $9$ sites in the unit cell with $q_{\rm max}=24$.}
\label{FigSpecLabels}
\end{figure}

\subsubsection{Matrix $W(\vec{k})$}

\noindent The matrix $W(\vec{k})$ has the block structure
\begin{align}
 W(\vec{k}) = \begin{pmatrix} * & * & * & * & * & * & * & * & * \\ * & 0 & * & * & * & 0 & 0 & 0 & 0 \\ * & * & 0 & 0 & 0 & 0 & 0 & 0 & 0 \\ * & * & 0 & 0 & 0 & 0 & 0 & 0 & 0 \\ * & * & 0 & 0 & 0 & 0 & 0 & 0 & 0 \\ * & 0 & 0 & 0 & 0 & 0 & * & * & * \\ * & 0 & 0 & 0 & 0 & * & 0 & 0 & 0 \\ * & 0 & 0 & 0 & 0 & * & 0 & 0 & 0 \\ * & 0 & 0 & 0 & 0 & * & 0 & 0 & 0 \end{pmatrix}.
\end{align}
The Bloch matrices read
\begin{align}
        B_1(\bk)&=\diag\Bigl(e^{-\rmi k_1},1,1,1,e^{-\rmi k_2},e^{-\rmi k_2},e^{-\rmi k_2},e^{-\rmi k_2},e^{-\rmi (k_2-k_1)},e^{\rmi k_1},e^{\rmi k_1},e^{\rmi k_1},\\
\phantom{B_1(\bk)}&\phantom{=\diag\Bigl(ss}e^{\rmi k_1},1,1,1,e^{\rmi k_2},e^{\rmi k_2},e^{\rmi k_2},e^{\rmi k_2},e^{-\rmi (k_1-k_2)},e^{-\rmi k_1},e^{-\rmi k_1},e^{-\rmi k_1}\Bigr)
\end{align}
for the $q_{\rm max}=24$ site, and
\begin{align}
B_2(\bk)&=\diag(1,1,e^{-\rmi k_1},1,e^{-\rmi k_2},1),\\
    B_3(\bk)&=\diag(1,1,e^{-\rmi k_1}),\\
    B_4(\bk)&=\diag(1,e^{-\rmi k_1},e^{-\rmi k_2}),\\
    B_5(\bk)&=\diag(1,e^{-\rmi k_2},1),\\
    B_6(\bk)&=\diag(1,1,e^{\rmi k_1},1,e^{\rmi k_2},1),\\
    B_7(\bk)&=\diag(1,1,e^{\rmi k_1}),\\
    B_8(\bk)&=\diag(1,e^{\rmi k_1},e^{\rmi k_2}),\\
    B_9(\bk)&=\diag(1,e^{\rmi k_2},1)
\end{align}
for the remaining ones. The edge-connectivity matrices read
\begin{align}
    &E_{2;1}=E_{6;1}=\begin{pmatrix}
        1 & & & & & \\
          & 0 & & & & \\
          & & 1& & & \\
          & & & 0& & \\
          & & & & 1& \\
          & & & & & 0
    \end{pmatrix},\ E_{2;3}=E_{6;7}=\begin{pmatrix}
        0& & & & & \\
        & 1& & & & \\
        & & 0& & & \\
        & & & 0& & \\
        & & & & 0& \\
        & & & & & 0\\
    \end{pmatrix},\\
    &E_{2;4}=E_{6;8}=\begin{pmatrix}
        0& & & & & \\
        & 0& & & & \\
        & & 0& & & \\
        & & & 1& & \\
        & & & & 0& \\
        & & & & & 0\\
    \end{pmatrix},\
    E_{2;5}=E_{6;9}=\begin{pmatrix}
        0& & & & & \\
        & 0& & & & \\
        & & 0& & & \\
        & & & 0& & \\
        & & & & 0& \\
        & & & & & 1\\
    \end{pmatrix},\\
     E_{3;2}&=E_{4;2}=E_{5;2}=E_{7;6}=E_{8;6}=E_{9;6}=\begin{pmatrix}        1 & & \\         & 0& \\         & & 0\\        \end{pmatrix},\\
    E_{3;1}&=E_{4;1}=E_{5;1}=E_{7;1}=E_{8;1}=E_{9;1}=\begin{pmatrix}        0 & & \\         & 1& \\         & & 1\end{pmatrix},\\ 
\end{align}
and 
\clearpage
\begin{align}
    E_{1;1}&=\Delta_1+\Delta_5+\Delta_9+\Delta_{13}+\Delta_{17}+\Delta_{21},\\
    E_{1;2}&=\Delta_{3}+\Delta_{11}+\Delta_{19},\\
    E_{1;3}&=\Delta_2+\Delta_{12},\\
    E_{1;4}&=\Delta_{10}+\Delta_{20},\\
    E_{1;5}&=\Delta_{4}+\Delta_{18},\\
    E_{1;6}&=\Delta_{7}+\Delta_{15}+\Delta_{23},\\
    E_{1;7}&=\Delta_{14}+\Delta_{24},\\
    E_{1;8}&=\Delta_{8}+\Delta_{22},\\
     E_{1;9}&=\Delta_{6}+\Delta_{16},\\
\end{align}
where $\Delta_i$ are $24\times 24$ matrices defined by $(\Delta_i)_{jk}= \delta_{ij}\delta_{ik}$. The orientation vectors are ($\sigma=e^{\rmi \pi/12}$)
\begin{align}
 \vec{\theta}_1 &= (1,\sigma,\sigma^2,\dots,\sigma^{23})^T,\\
 \vec{\theta}_2 &= -\vec{\theta}_6 = (e^{-5\pi \rmi/6},-\rmi,e^{-\rmi \pi/6},e^{\rmi \pi/6},\rmi,e^{5\pi\rmi/6})^T,\\
 \vec{\theta}_3 &= -\vec{\theta}_7 = \rmi(1,e^{7\pi \rmi/12},e^{-7\pi\rmi/12})^T,\\
 \vec{\theta}_4 &= -\vec{\theta}_8 = (e^{-5\pi\rmi/6},e^{-\rmi \pi/4},e^{7\pi\rmi/12})^T,\\
 \vec{\theta}_5 &= -\vec{\theta}_9 = (e^{-\rmi \pi/6},e^{5\pi\rmi/12},e^{-3\pi\rmi/4})^T.
\end{align}
The critical temperature is the root of the equation
\begin{align}
    \nonumber \det(\mathbbm{1}-t W_0)&=(1+t)^{18} \left(1-t+2 t^2\right)^4 \left(23 t^8+26 t^7+16 t^6+86 t^5-70 t^4+70 t^3-32 t^2+10 t-1\right)^2,
\end{align}
which gives 
\begin{align}
    t_{\rm c}&=0.152831,\\
    \frac{T_{\rm c}}{J}&=6.49190.
\end{align}
We confirm that
\begin{align}
    \nonumber\det(u\mathbbm{1}-W_0^\dagger W_0) = {}& P_{24}(u) P_6(u)^2 P_3(u)^6\\
    =\nonumber{}&\Bigl[(u^{12}-276 u^{11}+10626 u^{10}-134596 u^9+735471 u^8-1961256 u^7\\
    \nonumber&+2704156 u^6
    -1961256 u^5+735471 u^4-134596
   u^3+10626 u^2-276 u+1)^2\Bigr]\\
   &\times \Bigl[\left(u^3-15 u^2+15 u-1\right)^2\Bigr]^2 [(u-3)^2 u]^6.
\end{align}

\clearpage

\subsection{Critical temperature data for $k$-uniform lattices and duals ($k\leq 3$)}
\renewcommand{\arraystretch}{1.5}
\begin{table*}[h!]
\begin{tabular}{|c|c|c|c|c||c|c|c|c|c|}
\hline
 \multicolumn{10}{|c|}{1-uniform lattices}\\
 \hline\hline
 \ Lattice \ & \ $q_{\rm max}$ \ & \ $\bar{q}$ \ & \ $t_{\rm c}$ \ & \ $T_{\rm c}/J$ \ & \ Dual Lattice \ & \ $q_{\rm max}^{(\rm d)}$ \ & \ $\bar{q}^{(\rm d)}$ \ & \ $t_{\rm c}^{(\rm d)}$ \ & \ $T_{\rm c}^{(\rm d)}/J$ \ \\
\hline\hline
 \ Triangular \ & 6  & 6  &  0.26795  &  3.64095 & \ Honeycomb \ & 3  & 3  & 0.57735   &  1.51865 	\\
\hline
 \ SrCuBO \ & 5  & 5  & \ 0.32902 \  &  2.9263 & \ Laves-SrCuBO \ & 4  & 3.33  & \ 0.50486  \ &  1.79917	\\
\hline
 \ Trellis \ & 5  & 5  & $1/3$   &  2.88542 & \ Laves-Trellis \ & 4  & \ 3.33 \  & $1/2$   &  1.82048 	\\
\hline
 \ Maple-Leaf \ & 5  & 5  & 0.34430   &  2.7858 & \ Laves-Maple-Leaf \ & 6  & 3.33  & 0.48777   &  1.87572 	\\
\hline
\ Square \ & 4  & 4  & 0.41421   &  2.26921 & \ Square \ & 4  & 4  & 0.41421   &  2.26921 	\\
\hline
\ Kagome \ & 4  & 4  & 0.43542   &   2.14332 & \ Laves-Kagome \ & 6  & 4  & 0.39332   &  2.40546 	\\
\hline
 \ Ruby \ & 4  & 4  & 0.43542   &   2.14332 & \ Laves-Ruby \ & 6  & 4  & 0.39332  &  2.40546 	\\
\hline
 \ Honeycomb \ & 3  & 3  & \ 0.57735 \  &  \ 1.51865 \  & \ Triangular \ & 6  & \ 6 \  & \ 0.26795 \  & \ 3.64095 \	\\
\hline
 \ CaVO \ & 3  & 3  & 0.60123   & 1.4387 & \ Laves-CaVO \ & 8  & 6  & 0.24904   &  3.93102 	\\
\hline
 \ SHD \ & 3  & 3  & 0.61661   &  1.38982 & \ Laves-SHD \ & 12  & 6  & 0.23716   &  4.13629 	\\
\hline
 \ Star \ & 3  & 3  & 0.67070   &  1.23151 & \ Laves-Star \ & 12  & 6  & 0.19711   &  5.00704 	\\
\hline\hline
 \multicolumn{10}{|c|}{2-uniform lattices}\\
 \hline\hline
 \ Lattice \ & \ $q_{\rm max}$ \ & \ $\bar{q}$ \ & \ $t_{\rm c}$ \ & \ $T_{\rm c}/J$ \ & \ Dual Lattice \ & \ $q_{\rm max}^{(\rm d)}$ \ & \ $\bar{q}^{(\rm d)}$ \ & \ $t_{\rm c}^{(\rm d)}$ \ & \ $T_{\rm c}^{(\rm d)}/J$ \ \\
\hline\hline
 \ t2.001  \ & 4  & 3.33  & \ 0.56001 \  &  \ 1.58017 \  & \ t2.001dual \ &  12 & \ 5 \  & \ 0.28205 \  &  \ 3.44942 \	\\
\hline
 \ t2.002  \ & 4  &  3.5 & \ 0.58985 \  &  \ 1.47614 \  & \ t2.002dual \ &  12 & \ 4.67 \  & \ 0.25798 \  &  \ 3.78874 \	\\
\hline
 \ t2.003  \ & 5  & 4.5 & \ 0.37140 \  &  \ 2.56381 \  & \ t2.003dual \ &  4 & \ 3.6 \  & \ 0.45837 \  &  \ 2.01922 \	\\
\hline
 \ t2.004  \ &  5 & 4.67  & \ 0.35824 \  &  \ 2.66764 \  & \ t2.004dual \ &  4 & \ 3.5 \  & \ 0.47250 \  &  \ 1.94826 \	\\
\hline
 \ t2.005  \ &  4 & 4  & \ 0.43111 \  &  \ 2.16795 \  & \ t2.005dual \ &  6 & \ 4 \  & \ 0.39751 \  &  \ 2.37704 \	\\
\hline
 \ t2.006  \ &  4 &  4 & \ 0.42940 \  &  \ 2.17788 \  & \ t2.006dual \ &  6 & \ 4 \  & \ 0.39919 \  &  \ 2.36585 \	\\
\hline
 \ t2.007  \ &  4 &  4 & \ 0.43046 \  &  \ 2.17171 \  & \ t2.007dual \ & 6  & \ 4 \  & \ 0.39815 \  &  \ 2.37279 \	\\
\hline
 \ t2.008  \ &  5 &  4.5 & \ 0.38234 \  &  \ 2.48264 \  & \ t2.008dual \ &  6 & \ 3.6 \  & \ 0.44682 \  &  \ 2.0802 \	\\
\hline
 \ t2.009  \ &  5 &  4.5 & \ 0.37673 \  &  \ 2.52372 \  & \ t2.009dual \ &  6 & \ 3.6 \  & \ 0.45272 \  &  \ 2.04869 \	\\
\hline
 \ t2.010  \ &  6 & \ 4.29 \ & \ 0.42529 \  &  \ 2.20203 \  & \ t2.010dual \ &  6 & \ 3.75 \  & \ 0.40323 \  &  \ 2.3392 \	\\
\hline
 \ t2.011  \ &  4 &  4 & \ 0.44617 \  &  \ 2.08372 \  & \ t2.011dual \ &  6 & \  4 \  & \ 0.38296 \  &  \ 2.47816 \	\\
\hline
 \ t2.012  \ &  5 &  4.5 & \ 0.39661 \  &  \ 2.38312 \  & \ t2.012dual \ &  6 & \ 3.6 \  & \ 0.43204 \  &  \ 2.16262 \	\\
\hline
 \ t2.013  \ &  6 & 4.29  & \ 0.45780 \  &  \  2.02213 \  & \ t2.013dual \ &  12 & \ 3.75 \  & \ 0.37193 \  &  \ 2.55978 \	\\
\hline
 \ t2.014  \ & 6  &  5.33 & \ 0.31011 \  &  \ 3.11852 \  & \ t2.014dual \ &  4 & \ 3.2 \  & \ 0.52659 \  &  \ 1.70818 \	\\
\hline
 \ t2.015  \ & 6  & 5.5  & \ 0.29906 \  &  \ 3.24161 \  & \ t2.015dual \ &  4 & \ 3.14 \  & \ 0.53957 \  &   \ 1.65685 \	\\
\hline
 \ t2.016  \ &  5 &  5 & \ 0.33008 \  &  \ 2.91611 \  & \ t2.016dual \ & 4  & \ 3.33 \  & \ 0.50367 \  &  \ 1.80439 \	\\
\hline
 \ t2.017  \ & 5  & 5  & \ 0.33071 \  &  \ 2.91019 \  & \ t2.017dual \ &  4 & \ \ 3.33 \ \  & \ 0.50296 \  &  \ 1.80746 \	\\
\hline
 \ t2.018  \ &  6 &  5.14 & \ 0.31927 \  &  \ 3.02267 \  & \ t2.018dual \ & 4  & \ 3.27 \  & \ 0.51599 \  &  \ 1.75174 \	\\
\hline
 \ t2.019  \ &  6 & 5.25  & \ 0.32294 \  &  \ 2.98574 \  & \ t2.019dual \ & 6  & \ 3.23 \  & \ 0.51179 \  &  \ 1.76945 \	\\
\hline
 \ t2.020  \ & 6  &  5.5 & \ 0.30329 \  &  \ 3.19352 \  & \ t2.020dual \ & 6  & \ 3.14 \  & \ 0.53458 \  &  \ 1.67633 \	\\
\hline
\end{tabular}
    \caption{Critical temperatures for 1- and 2-uniform lattices and their duals (``d"). For the 2-uniform lattices, the labeling follows \cite{GalebachWebpage,Portillo} and the critical temperatures are taken from \cite{Portillo}. The maximal coordination numbers are $q_{\rm max}$ and the average coordination numbers are $\bar{q}$. The dual values $t_{\rm c}^{(d)}$ and $\bar{q}^{(\rm d)}$ are determined from Eqs. (\ref{data1}) and (\ref{data2}).}
\label{tcall1}
\end{table*}
\renewcommand{\arraystretch}{1}

\renewcommand{\arraystretch}{1.5}
\begin{table*}[t!]
\begin{tabular}{|c|c|c|c|c||c|c|c|c|c|}
\hline
 \multicolumn{10}{|c|}{3-uniform lattices I/II}\\
 \hline\hline
 \ Lattice \ & \ $q_{\rm max}$ \ & \ $\bar{q}$ \ & \ $t_{\rm c}$ \ & \ $T_{\rm c}/J$ \ & \ Dual Lattice \ & \ $q_{\rm max}^{(\rm d)}$ \ & \ $\bar{q}^{(\rm d)}$ \ & \ $t_{\rm c}^{(\rm d)}$ \ & \ $T_{\rm c}^{(\rm d)}/J$ \ \\
\hline\hline
 \ t3.001  \ & 4 &  3.67 & \ 0.484193 \  &  \ 1.89234 \  & \ t3.001dual \ &  6 &  4.4 & \  0.347534 \  &  \ 2.75758 \	\\
\hline
 \ t3.002  \ & 4 &  3.6 & \ 0.496091 \  &  \ 1.83787 \  & \ t3.002dual \ &  6 &  4.5 & \  0.336817 \  &  \ 2.85309 \	\\
\hline
 \ t3.003  \ & 5 & 4  & \ 0.451442 \  &  \ 2.05546 \  & \ t3.003dual \ &  6 &  4 & \  0.37794 \  &  \ 2.51475\	\\
\hline
 \ t3.004  \ & 6 & 4  & \ 0.465169 \  &  \ 1.98457 \  & \ t3.004dual \ &  6 &  4 & \  0.365031 \  &  \ 2.61316 \	\\
\hline
 \ t3.005  \ & 4 & 3.6  & \   0.500884 \  &  \ 1.81658 \  & \ t3.005dual \ & 12 &  4.5 & \    0.332548 \  &  \ 2.89276 \	\\
\hline
 \ t3.006  \ & 6 & 3.69  & \ 0.54617 \  &  \ 1.63157 \  & \ t3.006dual \ &  12 & 4.36  & \  0.293519 \  &  \ 3.30674 \	\\
\hline
 \ t3.007  \ & 4 &  3.75 & \ 0.537527 \  &  \ 1.66481 \  & \ t3.007dual \ &  12 & 4.29  & \  0.30079 \  &  \ 3.22177 \	\\
\hline
 \ t3.008  \ & 4 & 3.75 & \ 0.529379 \  &  \ 1.69698 \  & \ t3.008dual \ &  12 &  4.29 & \  0.30772 \  &  \ 3.1444 \	\\
\hline
 \ t3.009  \ & 4 &  3.8 &  \  0.524159 \  &  \ 1.71804 \  & \ t3.009dual \ &  12 & 4.22 & \    0.312199 \  &  \ 3.09617 \	\\
\hline
 \ t3.010  \ & 5 &  4.33 & \ 0.385082 \  &  \ 2.46298 \  & \ t3.010dual \ &  4 & 3.71  & \  0.443958 \  &  \ 2.09578 \	\\
\hline
 \ t3.011  \ & 4 & 4  & \ 0.422901 \  &  \ 2.21622 \  & \ t3.011dual \ & 6 &  4 & \  0.405579 \  &  \ 2.32391 \	\\
\hline
 \ t3.012  \ & 4 & 4  & \ 0.422918 \  &  \ 2.21611 \  & \ t3.012dual \ &  6 &  4 & \  0.405562 \  &  \ 2.32402 \	\\
\hline
 \ t3.013  \ & 5 &  \ 4.67 \ &  \  0.358238 \  &  \ 2.66764 \  & \ t3.013dual \ &  4 &  3.5 & \    0.472496 \  &  \ 1.94826 \	\\
\hline
 \ t3.014  \ & 6 & 4.8  & \ 0.34806 \  &  \ 2.75304 \  & \ t3.014dual \ &  4 & 3.43  & \  0.483614 \  &  \ 1.89505 \	\\
\hline
 \ t3.015  \ & 6 &  5 & \ $1/3$ \  &  \ 2.88539 \  & \ t3.015dual \ &  4 & 3.33  & \  $1/2$ \  &  \ 1.82048 \	\\
\hline
 \ t3.016  \ & 5 & 4.4 & \ 0.379542 \  &  \ 2.50297 \  & \ t3.016dual \ &  4 & 3.67  & \  0.449756 \  &  \ 2.06443 \	\\
\hline
 \ t3.017  \ & 4 &  4 &  \  0.425345 \  &  \ 2.20167 \  & \ t3.017dual \ &  6 &  4 & \    0.403169 \  &  \ 2.33958 \	\\
\hline
 \ t3.018  \ & 4 &  4 & \ 0.425473 \  &  \ 2.20092 \  & \ t3.018dual \ &  6 &  4 & \  0.403043 \  &  \ 2.3404 \	\\
\hline
 \ t3.019  \ & 5 &  4.8 & \ 0.34806 \  &  \ 2.75304 \  & \ t3.019dual \ &  4 &  3.43 & \  0.483614 \  &  \ 1.89505 \	\\
\hline
 \ t3.020  \ & 6 &  5 & \ $1/3$ \  &  \ 2.88539 \  & \ t3.020dual \ &  4 & 3.33  & \  $1/2$ \  &  \ 1.82048 \	\\
\hline
 \ t3.021  \ & 6 & 5.2  &  \  0.31921 \  &  \ 3.02327 \  & \ t3.021dual \ &  4 & 3.25  & \    0.516058 \  &  \ 1.75146 \	\\
\hline
 \ t3.022  \ & 4 & 4  & \  0.427755 \  &  \ 2.18749 \  & \ t3.022dual \ &  6 &  4 & \  0.4008 \  &  \ 2.35515 \	\\
\hline
 \ t3.023  \ & 5 &  4.89 & \ 0.338834 \  &  \ 2.83467 \  & \ t3.023dual \ &  4 & 3.38  & \  0.493837 \  &  \ 1.84801 \	\\
\hline
 \ t3.024  \ & 6 & 4.42  & \ 0.387232 \  &  \ 2.44774 \  & \ t3.024dual \ &  6 & 3.65  & \  0.44172 \  &  \ 2.10808 \	\\
\hline
 \ t3.025  \ & 5 & 4.25  &  \  0.402408 \  &  \ 2.34456 \  & \ t3.025dual \ &  6 &  3.78  & \    0.426119 \  &  \ 2.1971 \	\\
\hline
 \ t3.026  \ & 4 & 4  & \ 0.431809 \  &  \ 2.16395 \  & \ t3.026dual \ &  6 & 4  & \  0.396834 \  &  \ 2.3816 \	\\
\hline
 \ t3.027  \ & 5 & 4.2  & \ 0.419082 \  &  \ 2.23925 \  & \ t3.027dual \ &  6 & 3.82  & \  0.409362 \  &  \ 2.29967 \	\\
\hline
 \ t3.028  \ & 4 &  4 & \ 0.429929 \  &  \ 2.17481 \  & \ t3.028dual \ &  6 & 4  & \  0.398671 \  &  \ 2.36929 \	\\
\hline
 \ t3.029  \ & 4 &  4 &  \ 0.435421 \  &  \ 2.14332 \  & \ t3.029dual \ &  6 & 4 & \    0.39332 \  &  \ 2.40546 \	\\
\hline
 \ t3.030  \ & 4 &  4 & \ 0.431704 \  &  \ 2.16455 \  & \ t3.030dual \ &  6 & 4  & \  0.396937 \  &  \ 2.38091 \	\\
\hline
\end{tabular}
    \caption{Critical temperatures for the 3-uniform lattices and their duals (``d"). The labeling follows \cite{GalebachWebpage,Portillo} and the critical temperatures are taken from \cite{Portillo}. The maximal coordination numbers are $q_{\rm max}$ and the average coordination numbers are $\bar{q}$. The dual values $t_{\rm c}^{(d)}$ and $\bar{q}^{(\rm d)}$ are determined from Eqs. (\ref{data1}) and (\ref{data2}).}
\label{tcall2}
\end{table*}
\renewcommand{\arraystretch}{1}

\renewcommand{\arraystretch}{1.5}
\begin{table*}[t!]
\begin{tabular}{|c|c|c|c|c||c|c|c|c|c|}
\hline
 \multicolumn{10}{|c|}{3-uniform lattices II/II}\\
 \hline\hline
 \ Lattice \ & \ $q_{\rm max}$ \ & \ $\bar{q}$ \ & \ $t_{\rm c}$ \ & \ $T_{\rm c}/J$ \ & \ Dual Lattice \ & \ $q_{\rm max}^{(\rm d)}$ \ & \ $\bar{q}^{(\rm d)}$ \ & \ $t_{\rm c}^{(\rm d)}$ \ & \ $T_{\rm c}^{(\rm d)}/J$ \ \\
\hline\hline
 \ t3.031  \ & 4 & 4  & \ 0.436159 \  &  \ 2.13914 \  & \ t3.031dual \ & 6 &  4 & \  0.392604 \  &  \ 2.41037 \	\\
\hline
 \ t3.032  \ & 4 & 4  & \ 0.432379 \  &  \ 2.16067 \  & \ t3.032dual \ &  6 &  4 & \  0.396279 \  &  \ 2.38534 \	\\
\hline
 \ t3.033  \ & 5 & 4.5  &  \  0.3837 \  &  \ 2.47285 \  & \ t3.033dual \ &  6 &  3.6 & \    0.4454 \  &  \ 2.08791 \	\\
\hline
 \ t3.034  \ & 6 & 4.42  & \ 0.388976 \  &  \ 2.4355 \  & \ t3.034dual \ &  6 &  3.65 & \  0.43991 \  &  \ 2.11811 \	\\
\hline
 \ t3.035  \ & 5 & 4.2  & \ 0.435064 \  &  \ 2.14534 \  & \ t3.035dual \ &  12 & 3.82 & \  0.393666 \  &  \ 2.40309 \	\\
\hline
 \ t3.036  \ & 6 &  4.8 & \ 0.35466 \  &  \ 2.69714 \  & \ t3.036dual \ &  6 & 3.43  & \  0.476386 \  &  \ 1.92939 \	\\
\hline
 \ t3.037  \ & 6 & 4.8  &  \  0.349193 \  &  \ 2.7433 \  & \ t3.037dual \ &  6 & 3.43  & \    0.482367 \  &  \ 1.90091 \	\\
\hline
 \ t3.038  \ & 6 &  4.8 & \ 0.367133 \  &  \ 2.59669 \  & \ t3.038dual \ &  6 & 3.43  & \  0.462916 \  &  \ 1.99595 \	\\
\hline
 \ t3.039  \ & 4 &  4 & \ 0.440519 \  &  \ 2.11473 \  & \ t3.039dual \ &  6 & 4  & \  0.388389 \  &  \ 2.43961 \	\\
\hline
 \ t3.040  \ & 6 & 4.8 & \  0.3682  \  &  \ 2.5884 \  & \ t3.040dual \ &  6 & 3.43  & \  0.461774 \  &  \ 2.00175 \	\\
\hline
 \ t3.041  \ & 6 &  5 &  \  0.350248 \  &  \ 2.73428 \  & \ t3.041dual \ &  6 &  3.33 & \    0.481209 \  &  \ 1.90637 \	\\
\hline
 \ t3.042  \ & 6 &  \ 4.62 \ & \ 0.381931 \  &  \ 2.4856 \  & \ t3.042dual \ &  6 & 3.53  & \  0.447251 \  &  \ 2.07789 \	\\
\hline
 \ t3.043  \ & 6 &  4.8 & \ 0.363964 \  &  \ 2.62159 \  & \ t3.043dual \ &  6 &  3.43 & \  0.466314 \  &  \ 1.97883 \	\\
\hline
 \ t3.044  \ & 6 &  5.14 & \ $1/3$ \  &  \ 2.88539 \  & \ t3.044dual \ &  6 & 3.27  & \  $1/2$ \  &  \ 1.82048 \	\\
\hline
 \ t3.045  \ & 5 & 4.8  &  \  0.362352 \  &  \ 2.63441 \  & \ t3.045dual \ &  6 & 3.43  & \    0.468049 \  &  \ 1.97017 \	\\
\hline
 \ t3.046  \ & 5 &  4.8 & \ 0.362352 \  &  \ 2.63441 \  & \ t3.046dual \ &  6 &  3.43 & \  0.468049 \  &   \ 1.97017 \	\\
\hline
 \ t3.047  \ & 6 & 4.75  & \ 0.388372 \  &  \ 2.43973 \  & \ t3.047dual \ &  12 &  3.45 & \  0.440536 \  &  \ 2.11463 \	\\
\hline
 \ t3.048  \ & 6 & 4.57  & \  0.40012 \  &  \ 2.35965 \  & \ t3.048dual \ &  12 &  3.56 & \  0.428449 \  &  \ 2.18343 \	\\
\hline
 \ t3.049  \ & 6 & 5.6  &  \  0.292602 \  &  \ 3.31774 \  & \ t3.049dual \ &  4 &  3.11 & \    0.547266 \  &  \ 1.62742 \	\\
\hline
 \ t3.050  \ & 6 &  5.67 & \ 0.288364 \  &  \ 3.36949 \  & \ t3.050dual \ &  4 & 3.09  & \  0.552356 \  &  \ 1.60832 \	\\
\hline
 \ t3.051  \ & 6 &  5.2 & \ 0.31921 \  &  \ 3.02327 \  & \ t3.051dual \ &  4 &  3.25 & \  0.516058 \  &  \ 1.75146 \	\\
\hline
 \ t3.052  \ & 6 &  5.33 & \ 0.310108 \  &  \ 3.11852 \  & \ t3.052dual \ &  4 &  3.2 & \  0.526592 \  &  \ 1.70818 \	\\
\hline
 \ t3.053  \ & 5 & 5  &  \  0.331531 \  &  \ 2.90236 \  & \ t3.053dual \ &  4 &  3.33 & \    0.50203 \  &  \ 1.81154 \	\\
\hline
 \ t3.054  \ & 6 &  5.08 & \ 0.325034 \  &  \ 2.96503 \  & \ t3.054dual \ &  4 &  3.3 & \  0.509396 \  &  \ 1.77963 \	\\
\hline
 \ t3.055  \ & 5 &  5 & \ 0.330162 \  &  \ 2.91537 \  & \ t3.055dual \ &  4 &  3.33 & \  0.503576 \  &  \ 1.80477 \	\\
\hline
 \ t3.056  \ & 5 &  5 & \ 0.335263 \  &  \ 2.86742 \  & \ t3.056dual \ &  6 & 3.33  & \  0.497832 \  &  \ 1.83009 \	\\
\hline
 \ t3.057  \ & 6 &  5.08 &  \  0.323764 \  &  \ 2.97755 \  & \ t3.057dual \ &  4 &  3.3 & \    0.510843 \  &  \ 1.77346 \	\\
\hline
 \ t3.058  \ & 6 & 5.45  & \ 0.306809 \  &  \ 3.15438 \  & \ t3.058dual \ &  6 & 3.16 & \  0.530445 \  & \ 1.69272 \	\\
\hline
 \ t3.059  \ & 6 &  5.6 & \ 0.295728 \  &  \ 3.2805 \  & \ t3.059dual \ &  6 & 3.11  & \  0.543534 \  &  \ 1.64161 \	\\
\hline
 \ t3.060  \ & 6 & 5.67  & \ 0.290831 \  &  \ 3.33919 \  & \ t3.060dual \ &  6 & 3.09  & \  0.549389 \  &  \ 1.61942 \	\\
\hline
 \ t3.061  \ & 6 & 5.14  &  \  0.332211 \  &  \ 2.89594 \  & \ t3.061dual \ &  6 &  3.27 & \    0.501264 \  &  \ 1.8149 \	\\
\hline
\end{tabular}
    \caption{Continuation of Tab. \ref{tcall2}.}
\label{tcall3}
\end{table*}
\renewcommand{\arraystretch}{1}

\end{widetext}

%



\end{document}